\documentclass[12pt]{article}

\usepackage{amsmath}
\usepackage{graphicx}

\usepackage{graphics}
\usepackage{epsfig}

\input{babarsym}
\def\ifmath#1{\relax\ifmmode #1\else $#1$\fi}%

\renewcommand {\rm} {\mathrm}

\newcommand  {\dsk}   {\ifmath{{\mathrm{B_s \rightarrow D_s^\pm K^\mp}}}}
\newcommand  {\dskstar}   {\ifmath{{\mathrm{B_s \rightarrow D_s^{\ast \pm} 
K^\mp}}}}
\newcommand  {\dspi}   {\ifmath{{\mathrm{B_s \rightarrow D_s^\pm \pi^\mp}}}}
\newcommand  {\dspkm}   {\ifmath{{\mathrm{B_s \rightarrow D_s^+ K^-}}}}
\newcommand  {\dspkmbar}   {\ifmath{{\mathrm{\overline{B_s} \rightarrow D_s^+ K^-}}}}

\newcommand  {\dpi}   {\ifmath{{\mathrm{B_d \rightarrow D^\pm \pi^\mp}}}}

\newcommand  {\dpistar}   {\ifmath{{\mathrm{B_d \rightarrow D^{\ast \pm} \pi^\mp}}}}

\catcode`@=11 
\def \gsim{\mathrel{\mathpalette\@versim>}}
\def \lsim{\mathrel{\mathpalette\@versim<}}
\def \@versim#1#2{\lower0.4ex\vbox{\baselineskip\z@skip\lineskip\z@skip
     \lineskiplimit\z@\ialign{$\m@th#1\hfil##\hfil$%
     \crcr#2\crcr\sim\crcr}}}
\catcode`@=12 
\newcommand{\bit}{\begin{itemize}}
\newcommand{\eit}{\end{itemize}}
\newcommand{\beq}{\begin{equation}}
\newcommand{\eeq}{\end{equation}}

\def\dt{\Delta t}

\def\mmiss{m_{\rm miss}}

\def\btodstpipm{\Bz \to \Dstarmp\pi^\pm}

\def\sss{\scriptscriptstyle}
\def\barpd{{\raise.35ex\hbox{${\sss (}$}}--{\raise.35ex\hbox{${\sss )}$}}}
\def\dbarp{\hbox{$D^{0}$\kern-1.25em\raise1.5ex\hbox{\barpd}}}

\def\Dbar{\overline D}
\def\DbarZ{\overline{D}^0}

\def\ra     {\to}

\def\pipi     {\ensuremath{\pi\pi}}

\def\KS {K_S}
\def\Kmp {K^{\mp}}
\def\Km  {K^{-}}
\def\Kp  {K^{+}}
\def\pim {\pi^{-}}
\def\pip {\pi^{+}}
\def\piz {\pi^{0}}

\def\Dz {D^{0}}
\def\Dzb {\bar{D}^{0}}
\def\Bm  {B^-}
\def\Bpm  {B^{\pm}}

\def\Dbar    {\kern 0.2em\overline{\kern -0.2em D}{}\xspace}
\def\dotokspp {\ensuremath {\Dz\to\KS\pim\pip}\xspace}
\def\btodk   {\ensuremath {\Bm\to\Dz\Km}\xspace}
\def \rb {\ensuremath {r_B}\xspace}
\def \rbs {\ensuremath {r^\ast_B}\xspace}
\def \rbbs {\ensuremath {r^{(\ast)}_B}\xspace}
\def \deltab {\ensuremath {\delta_B}\xspace}
\def \deltabs {\ensuremath {\delta^\ast_B}\xspace}
\def \deltabbs {\ensuremath {\delta^{(\ast)}_B}\xspace}

\def \xbbspm {\ensuremath {x_\pm^{(\ast)}}\xspace}
\def \ybbspm {\ensuremath {y_\pm^{(\ast)}}\xspace}

\def \xbbs {\ensuremath {x^{(\ast)}}\xspace}
\def \ybbs {\ensuremath {y^{(\ast)}}\xspace}

\def\bea{\begin{eqnarray}}
\def\eea{\end{eqnarray}}

\newcommand{\resline}[8]{#1\pm#2~^{+#3}_{-#4}~^{+#5}_{-#6}~[#7,#8]}

\newcommand{\reslinep}[7]{#1\pm#2\pm#3~^{+#4}_{-#5}~[#6,#7]}

\newcommand{\reslinepol}[8]{\left(#1\pm#2~^{+#3}_{-#4}~^{+#5}_{-#6}\right)^\circ~[#7^\circ, #8^\circ]}

\newcommand{\reslinepolval}[6]{\left(#1\pm#2~^{+#3}_{-#4}~^{+#5}_{-#6}\right)^\circ}
\newcommand{\reslinepolvalpp}[5]{\left(#1\pm#2~^{+#3}_{-#4}\pm#5\right)^\circ}

\def\mes        {\mbox{$m_{\rm ES}$}\xspace}

\newcommand{\dz}{\ensuremath{\Delta{}z}}
\newcommand{\dm}{\ensuremath{\Delta{}m}}
\def\dsp{\ensuremath{{D^{(\ast)}\pi}}}
\newcommand{\Rd}{\ensuremath{R_{\dsp}}}
\newcommand{\taub}{\ensuremath{\tau_{B^{0}}}}
\newcommand{\Mbc}{\ensuremath{M_{\mathrm{bc}}}}
\newcommand{\DE}{\ensuremath{\Delta{}E}}

\newcommand{\BZ}{\mbox{${B}^0$}}

\def\DstpDstm{\ensuremath{D^{*+} D^{*-} }}
\def\DpDm{\ensuremath{D^{+} D^{-} }}

\def\DstpDm{\ensuremath{D^{*+}D^{-}}}
\def\DstmDp{\ensuremath{D^{*-}D^{+}}}

\def\BDstpDm{\ensuremath{\BZ \to D^{*+}D^{-}}}
\def\BDstmDp{\ensuremath{\BZ \to D^{*-}D^{+}}}

\def\BDbothstpmDbothmp{\ensuremath{\BZ \to D^{(*)\pm}D^{(*)\mp}}}

\def\sss{\scriptscriptstyle}
\def\barpd{{\raise.35ex\hbox{${\sss (}$}}--{\raise.35ex\hbox{${\sss )}$}}}
\def\BorBbar{\hbox{$B$\kern-0.85em\raise1.5ex\hbox{\barpd}\hspace{-0.4mm}$^0$}}

\usepackage{txfonts}            

\def\barD{\overline D{}^0}

\def\Bbar{\overline{B}}
\def\DDbar{D^0-\overline D{}^0}

\def\D0bar{\overline D{}^0}
\def\K0bar{\overline K{}^0}

\def\3bar{\overline{3}}

\def\15bar{\overline{15}}
\def\24bar{\overline{24}}
\def\42bar{\overline{42}}
\def\60bar{\overline{60}}

\def\cal{{\it}}
\def\BR{{\cal B}}
\def\beq{\begin{equation}}
\def\eeq{\end{equation}}
\def\beqa{\begin{eqnarray}}
\def\eeqa{\end{eqnarray}}
\def\bea{\begin{eqnarray}}
\def\eea{\end{eqnarray}}

\def\jpsi{J/\psi~}

\vspace*{-.5in}

\def\uglu{\hskip 0pt plus 1fil
minus 1fil} \def\uglux{\hskip 0pt plus .75fil minus .75fil}

\def\slashed#1{\setbox200=\hbox{$ #1 $}
    \hbox{\box200 \hskip -\wd200 \hbox to \wd200 {\uglu $/$ \uglux}}}

\begin{document}

\begin{center}

  {\Large
    \bf
    \boldmath
    Angles from $B$ Decays with Charm
  }

  \vspace{5ex}

  G.~Cavoto,${}^{1}$
  R.~Fleischer,${}^{2}$
  T.~Gershon,${}^{3,4,5}$
  A.~Soni,${}^{6}$ [conveners] \\

  \vspace{2mm}
  
  K.~Abe,${}^{3}$
  J.~Albert,${}^{7}$
  D.~Asner,${}^{8}$
  D.~Atwood,${}^{9}$
  M.~Bruinsma,${}^{10}$
  S.~Ganzhur,${}^{11}$
  B.~Iyutin,${}^{12}$
  Y.Y.~Keum,${}^{13}$
  T.~Mannel,${}^{14}$
  K.~Miyabayashi,${}^{15}$
  N.~Neri,${}^{16}$
  A.A.~Petrov,${}^{17}$
  M.~Pierini,${}^{18}$
  F.~Polci,${}^{1}$
  M.~Rama,${}^{19}$
  F.~Ronga,${}^{3}$
  L.~Silvestrini,${}^{1}$
  A.~Stocchi,${}^{20}$
  M.H.~Schune,${}^{20}$
  V.~Sordini,${}^{1,20}$
  M.~Verderi,${}^{21}$
  C.~Voena,${}^{1}$
  G.~Wilkinson,${}^{22}$
  J.~Zupan,${}^{23,24}$

  \vspace{5ex}

  ${}^{1}~$\textsl{Universit\`a di Roma La Sapienza, Dipartmento di Fisica and INFN, I-00185 Roma, Italy} \\
  ${}^{2}~$\textsl{CERN, Department of Physics, Theory Division, CH-1211 Geneva 23, Switzerland} \\
  ${}^{3}~$\textsl{High Energy Accelerator Research Organization (KEK), Tsukuba, Japan} \\
  ${}^{4}~$\textsl{Department of Physics, University of Tokyo, Tokyo, Japan} \\
  ${}^{5}~$\textsl{Department of Physics, University of Warwick, Coventry, CV4 7AL, UK} \\
  ${}^{6}~$\textsl{Physics Department, Brookhaven National Laboratory, Upton, New York 11973, USA} \\
  ${}^{7}~$\textsl{California Institute of Technology, Pasadena, California, 91125, USA} \\
  ${}^{8}~$\textsl{Carleton University, Ottawa, Ontario, K1S 5B6, Canada} \\
  ${}^{9}~$\textsl{Dept. of Physics and Astronomy, Iowa State University, Ames, Iowa 50011, USA} \\
  ${}^{10}~$\textsl{University of California at Irvine, Irvine, California 92697, USA} \\
  ${}^{11}~$\textsl{DSM/Dapnia, CEA/Saclay, F-91191, Gif-sur-Yvette, France} \\
  ${}^{12}~$\textsl{Massachusetts Institute of Technology, Cambridge, MA 02139, USA} \\
  ${}^{13}~$\textsl{Department of Physics, National Taiwan University, Taipei, 106 Taiwan, ROC} \\
  ${}^{14}~$\textsl{Theoretische Physik 1, Fachbereich Physik, Universit\"at Siegen, D-57068, Siegen, Germany} \\
  ${}^{15}~$\textsl{Nara Women's University, Nara, Japan} \\
  ${}^{16}~$\textsl{Universit\`a di Pisa, Dipartmento di Fisica, Scuola Normale Superiore and INFN, I-56127 Pisa, Italy} \\
  ${}^{17}~$\textsl{Dept. of Physics and Astronomy, Wayne State University, Detroit, Michigan 48201, USA} \\
  ${}^{18}~$\textsl{University of Wisconsin, Madison, Wisconsin 53706, USA} \\
  ${}^{19}~$\textsl{Laboratori Nazionali di Frascati dell'INFN, I-00044, Frascati, Italy} \\
  ${}^{20}~$\textsl{Laboratoire de l'Acc\'el\'erateur Lin\'eaire, F-91898 Orsay, France} \\
  ${}^{21}~$\textsl{Ecole Polytechnique, LLR, F-91128 Palaiseau, France} \\
  ${}^{22}~$\textsl{University of Oxford, 1, Keble Road, Oxford OX1 3NP, UK} \\
  ${}^{23}~$\textsl{J.~Stefan Institute, Jamova 39, P.O.Box 3000, Ljubljana, Slovenia} \\
  ${}^{24}~$\textsl{Department of Physics, Carnegie Mellon University, Pittsburgh, PA 15213, USA} \\

  \vspace{5ex}

  \textbf{Abstract} \\
\end{center}

\noindent
Proceedings of the CKM 2005 Workshop (WG5), UC San Diego, 15-18 March 2005.

{\renewcommand{\thefootnote}{\fnsymbol{footnote}}}
\setcounter{footnote}{0}

\begin{center}
  {\Large 
    \bf \boldmath
    WG5: Angles from $B$ Decays with Charm
  }
\end{center}

\vspace{5mm}

\noindent
\underline{Charge:} \\
The main goal is the measurement of the angles $\gamma$ and $\beta$
from $B$ decays involving $D$ or charmonium mesons.
The limitations and ways to overcome them will be discussed,
as well as new approaches and high statistics projections.

\vspace{5mm}

\begin{center}
  {\large
    \underline{Introduction}
  }
\end{center}

This report is a summary of the activities of Working Group 5 of the 
CKM2005 Workshop on the Unitarity Triangle (UT), 
which took place in San Diego, California, from 15\textendash18 March 2005. 
The main goal of this Working Group was the determination of the 
UT angles $\beta/\phi_1$ and $\gamma/\phi_3$ from $B$ meson decays into 
final states with charm. 
These determinations play a key role for the experimental testing 
of the Kobayashi\textendash Maskawa (KM) mechanism of CP violation, 
which has already reached an impressive level 
thanks to the great efforts of the BaBar and Belle collaborations 
at the SLAC and KEK $B$ factories, respectively. 

The outline of this report is as follows: 
after this introduction,
a table is provided, listing the talks at the San Diego meeting.
An executive summary provides a brief overview of the activities 
of the Working Group.
This summary describes the current status, both experimental 
and theoretical, and includes an outlook to the future reach of the 
measurements.
More detailed information and references
can be found in the the collated contributions at the end of the document.

\vspace{5mm}

\noindent
\underline{Dictionary:}
\begin{center}
  \begin{minipage}{0.42\textwidth}
    \begin{center}
      \includegraphics[width=0.95\textwidth]{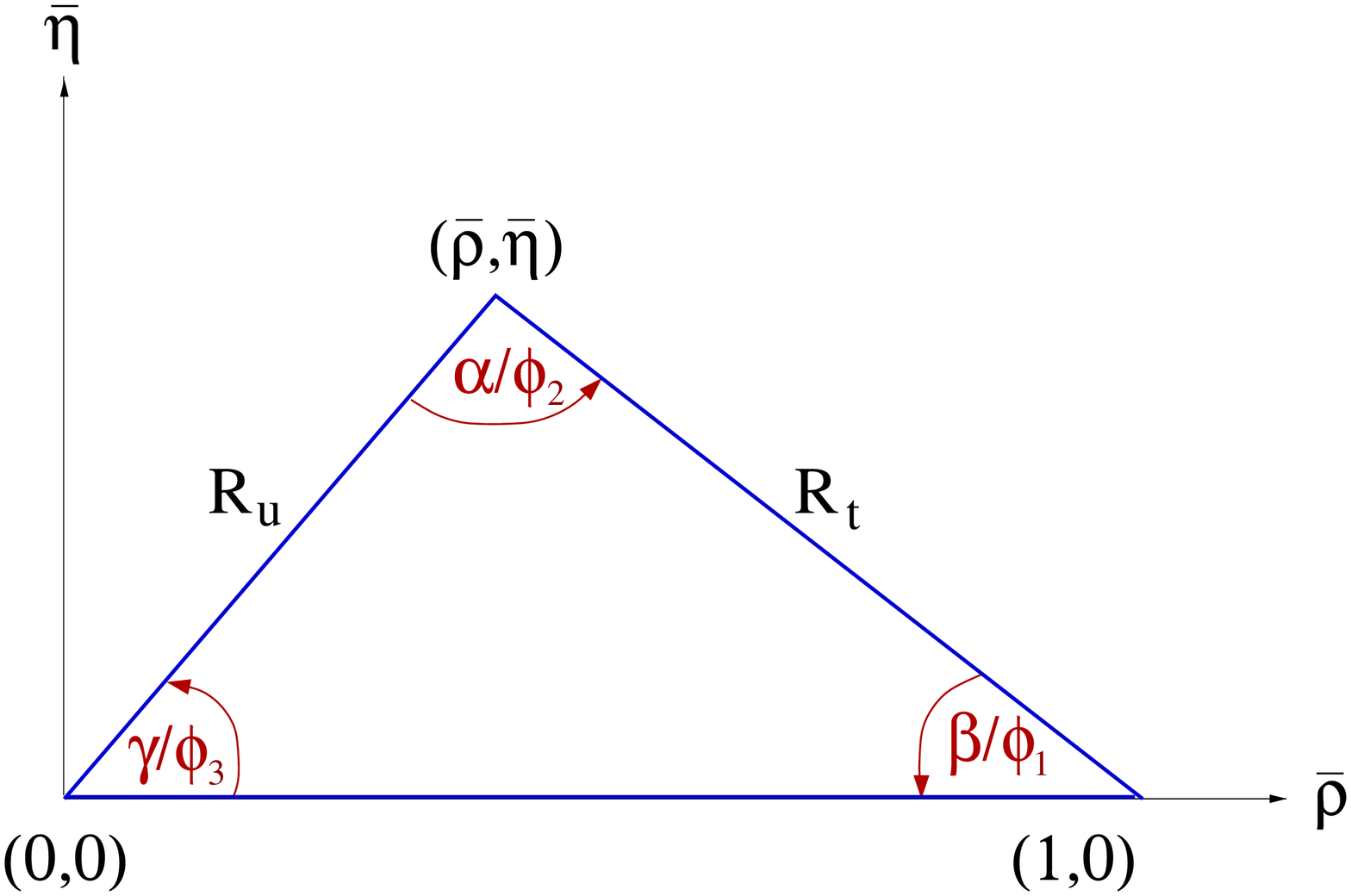}      
    \end{center}
  \end{minipage}
  \begin{minipage}{0.42\textwidth}
    \begin{center}
      $\alpha  \equiv  \phi_2  =
      \arg\left[ - \frac{V_{td}V_{tb}^*}{V_{ud}V_{ub}^*} \right]$

      \vspace{3mm}

      $\beta   \equiv   \phi_1 =
      \arg\left[ - \frac{V_{cd}V_{cb}^*}{V_{td}V_{tb}^*} \right]$

      \vspace{3mm}

      $\gamma  \equiv   \phi_3  =
      \arg\left[ - \frac{V_{ud}V_{ub}^*}{V_{cd}V_{cb}^*} \right]$
    \end{center}
  \end{minipage}
\end{center}

\newpage

\begin{center}
  {\large
    \underline{Talks given in Working Group 5 at the San Diego meeting.}
  }

  \vspace{3mm}

  \begin{tabular}{l@{\hspace{15mm}}l}
    \hline \hline
    \multicolumn{2}{c}{$\sin(2\beta)$ in Charmonium Modes and Discussion of Experimental Analysis Techniques} \\
    T.~Mannel      & Theory of the golden mode \\
    T.~Browder     & Review of charmonium kaon analysis in Belle \\
    D.~Lange       & Review of charmonium kaon analysis in BaBar \\
    W.T.~Ford      & Experimental approaches in BaBar $CP$ measurements \\
    A.~Kusaka      & Vertexing technique in Belle \\
    K.~Sumisawa    & Background suppression in Belle \\
    \hline \hline
    \multicolumn{2}{c}{Measurements related to $\beta$} \\
    M.~Verderi     & $\cos(2\beta)$ with $B_d \to J/ \psi K^*$ Experimental Review \\
    K.~Miyabayashi & $CP$ in $b\to c\bar{c}d$ review \\
    R.~Jesik       & $B_s \to J/ \psi \phi$ status at Tevatron \\
    T.~Gershon     & Feasibility of $\beta$ measurement with $B_d \to D\pi^0$ \\
    M.~Bruinsma    & $B \to$ double charm status in BaBar \\
    B.~Iyutin      & $B_s \to$ double charm status at CDF \\
    J.~Albert      & $\gamma$ from $B \to D^{(*)}D^{(*)}$ \\
    \hline \hline
    \multicolumn{2}{c}{Measurements of $\sin(2\beta+\gamma)$} \\
    S.~Ganzhur     & BaBar status and prospects for $CP$ asymmetry measurements \\
    F.~Ronga       & Belle status and prospects for $CP$ asymmetry measurements \\
    C.~Voena       & Extraction of $2\beta+\gamma$ from $B \to D^{(*)}\pi$ \\
    M.~Baak        & $B \to D^*\rho$ feasibility study \\
    F.~Polci       & $2\beta+\gamma$ from $B \to D^+K_S\pi^-$ \\
    V.~Sordini     & $2\beta+\gamma$ from $B \to D^0K^{(0/+)}$ \\
    Y.Y.~Keum      & $\gamma$ with PQCD \\
    G.~Wilkinson   & $B_s \to D_sK$ {\it etc.} at LHCb \\
    C.~Ferretti    & $\Lambda_b \to J/ \psi \Lambda^0$ feasibility study at ATLAS \\
    \hline \hline
    \multicolumn{2}{c}{$\gamma$ from $B \to DK$ decay modes} \\
    D.~Atwood      & Combined strategies for $\gamma$ \\
    K.~Trabelsi    & Belle status and prospects \\
    M.~Rama        & BaBar status and prospects \\
    Y.~Grossman    & Methods to extract $\gamma$ with $B \to DK$ modes \\
    A.~Petrov      & Charm input for $\beta$ and $\gamma$ measurement, theory \\
    D.~Asner       & Charm input for $\beta$ and $\gamma$ measurement, experiment \\
    R.~Fleischer   & $\gamma$ from $B_c$ decays \\
    \hline \hline
    \multicolumn{2}{c}{$\gamma$ from $B \to DK$ with $D$ Dalitz analysis} \\
    J.~Zupan       & Theory introduction \\
    K.~Abe         & Belle status and prospects \\
    M.H.~Schune    & BaBar status and prospects \\
    N.~Neri        & Statistical treatment in BaBar $DK$ Dalitz analysis \\
    D.~Asner       & Experimental Input from CLEO-c \\
    Q.~Zeng        & $D \to \pi^+\pi^-\pi^0$ feasibility study \\
    \hline \hline
    A.~Soni        & WG5 summary talk \\
    \hline \hline   
  \end{tabular}
    
  \vspace{3mm}
  Links to these talks are available at  
  http://ckm2005.ucsd.edu/WG/WG5/allWG5\_sessions.php

\end{center}

\newpage
\begin{center}
  {\large
    \underline{Executive Summary}
  }
\end{center}

The leading decay of hadrons containing the $b$ quark
is via the $b \to c$ transition.
Therefore, $B$ decays to final states containing either charm or charmonia
are abundant, and provide fertile ground for investigations of 
the phenomenology of the $B$ system, 
including properties of the CKM matrix.
For example, the flavour-specific semileptonic decays $b \to cl\nu$
are used to measure $B$--$\bar{B}$ mixing properties
(in both $B_d$ and $B_s$ systems)
such as the mass difference $\Delta m$ and the parameter
of $CP$ violation in mixing, $\left| q/p \right|$.
Furthermore, $b \to c$ transitions to final states consisting of 
two vector mesons can be used to measure polarization,
search for triple-product correlations,
and, most notably for the decay $B_s \to J/ \psi \phi$,
to measure the lifetime difference $\Delta \Gamma_s / \Gamma_s$.
In addition, the $b \to c$ transition is the principal tool to 
investigate the properties of heavy baryons and hyperons.
While the activity of this working group has touched on these areas,
the charge restricts the scope to the -- nonetheless wide -- subject area of
measurements of angles of the Unitarity Triangle.

\vspace{1ex}

Table~\ref{btoc} lists some of the modes of interest.
These modes include measurements of weak phases
using both mixing-induced and direct $CP$ violation.
At the top of the list is the so-called ``golden mode,'' $B_d \to J/ \psi K_S$.
This mode allows a theoretically clean measurement of $\sin(2\beta)$.
Although there is a penguin contribution,
this amplitude has, dominantly, 
the same weak phase as the leading tree diagram,
and therefore does not cause significant pollution of the result.
There has been some recent activity to try to quantify 
the level of theoretical uncertainty,
due to subleading terms in both mixing and decay amplitudes,
with the outcome that it is below the level
of the current experimental systematic errors \textendash{} about $0.01$ on $\sin(2\beta)$ \textendash{}
and likely to be much smaller.

\begin{table}[htb]
  \caption{
    \label{btoc}
    Quark level transitions of interest.
    For each, an alternative transition which gives the same final state
    quarks is given, and whether this alternative is penguin (P), 
    or tree (T) is noted.
    The CKM suppression, including the phase, of the alternative transition
    is noted, where for penguin amplitudes short-distance dominance
    is assumed (only the top quark in the loop is accounted for).
    Where the alternative is a tree diagram,
    interference can only occur if the ultimately final state can be produced 
    by both intermediate sets of quarks,
    which contain $c\bar{u}$ and $\bar{c}u$, respectively.
  }

  \vspace{1ex}

  \begin{tabular}{
      r@{\extracolsep{0mm}}l
      @{\extracolsep{3mm}}
      r@{\extracolsep{0mm}}l
      @{\extracolsep{3mm}}
      c
      @{\extracolsep{3mm}}
      l
    }

    \hline

    \multicolumn{2}{c}{Transition} & 
    \multicolumn{2}{c}{Alternative} & CKM Suppression & Typical decays \\

    \hline

    $b \to \, $ & $\, c\bar{c}s$ & 
    $b \to \, $ & $\, sc\bar{c}$ (P) & 
    $1$ & 
    $B_d \to J/ \psi K_S$, $B_s \to J/ \psi \phi$, \ldots\\

    & $\, c\bar{c}d$ & 
    & $\, dc\bar{c}$ (P) & 
    $R_t e^{i\beta}$ &
    $B_d \to J/ \psi \pi^0$, $B_d \to D^{(*)+}D^{(*)-}$, \ldots\\

    & $\, c\bar{u}d$ & 
    & $\, u\bar{c}d$ (T) & 
    $\lambda^2 R_u e^{i\gamma}$ & 
    $B_d \to D \pi^0$, $B_d \to D^{(*)\pm} \pi^\mp$, $B_s \to D K^{(*)}$, \ldots \\

    & $\, c\bar{u}s$ &
    & $\, u\bar{c}s$ (T) & 
    $R_u e^{i\gamma}$ &
    $B_{u,d} \to D K^{(*)}$, $B_s \to D \phi$, $B_s \to D_s^{(*)\pm} K^\mp$ \ldots \\

    \hline

  \end{tabular}
\end{table}

\vspace{1ex}

On the experimental side, this mode is equally golden,
with large product branching fractions
and a very clean signal when reconstructed via
$J/ \psi \to l^+l^-$, $K_S \to \pi^+\pi^-$ ($l = e, \mu$).
The precision of the current results from BaBar and Belle,
consequences of the excellent performances of the PEP-II and KEKB accelerators,
is such that the systematic errors are no longer very small compared 
to the statistical errors.
Detailed understanding of the uncertainties due to 
vertexing, resolution, flavour tagging, and so on, 
will thus continue become more and more important.
To reduce these errors, there is particular benefit from 
the continued operation of both $B$ factories,
since the two collaborations use somewhat different techniques,
and much is being learned from discussions and comparisons.
The precision of the \textendash{} already rather mature \textendash{} $\sin(2\beta)$ measurements 
should continue to improve in line with the accumulated luminosity 
over the next few years.
In addition, comparisons of $\sin(2\beta)$ measured with 
different final states of the $b \to c\bar{c}s$ transition 
({\it eg.} $J/ \psi K_S$, $\eta_c K_S$, $\chi_{c0} K_S$)
can test some new physics models.

\vspace{1ex}

Measurements of $\sin(2\beta)$ provide constraints corresponding to
two bands in the $\left(\bar{\rho},\bar{\eta}\right)$ plane.
To remove this ambiguity, $\cos(2\beta)$ should be measured.
In general, where there is interference between $CP$-even and $CP$-odd
final states there will be sensitivity to the cosine of the weak phase.
Several methods based on this concept exist in the literature.
For example, in the decay $B_d \to J/ \psi K^{*}$, where $K^* \to K_S \pi^0$,
different partial waves contributing to the vector-vector final state
have different $CP$ properties, 
and their interference allows measurement of $\cos(2\beta)$.
In order to perform this measurement,
the relative magnitudes and phases of the contributing partial waves
need to be determined, which can be achieved with angular analysis.
Furthermore, to determine the sign of $\cos(2\beta)$,
an ambiguity in the solutions for the strong phase has to be resolved.
This can either be taken from theory, under certain assumptions,
or (as performed by BaBar)
can be extracted from data by exploiting the interference between
$K^*(892) \to K\pi$ and the contribution from the broad $K$\textendash$\pi$ S-wave
in the same invariant mass region of $B \to J/ \psi K\pi$.
Results to date prefer the Standard Model solution $\cos(2\beta)>0$,
albeit with rather large uncertainities.
Recently, a new method has been proposed, 
which utilizes the interference pattern in
$D \to K_S\pi^+\pi^-$ decay following $B_d \to D\pi^0$ (and similar decays).
[This method was first discussed in WG5 at CKM2005.]
In this case the decay model is fixed from studies of flavour tagged
$D$ mesons (from $D^{*\pm} \to D\pi^\pm$) decay,
and the ambiguity is removed by using Breit-Wigner line-shapes.
This method also allows to test the Standard Model prediction
that the weak phase measured in $b \to c\bar{u}d$ transitions 
should be the same as that in $b \to c\bar{c}s$ decays ({\it ie.} $2\beta$).
Results from this method are consistent with the Standard Model,
and $\cos(2\beta)<0$ is now ruled out with greater than $95\%$ confidence.

\vspace{1ex}

In the limit of tree dominance, 
and within the Standard Model,
$B_d$ decays via the $b \to c\bar{c}d$ transition also probe $\sin(2\beta)$.
However, the situation is encumbered by the penguin amplitude,
which is not dominated by a single weak phase
(the contributions with $t$, $c$ and $u$ quarks in the loop
appear at the same level of CKM suppression, 
so the assumption of short-distance dominance is not well justified). 
Studies of decays dominated by the $b \to d$ penguin amplitude,
may provide more information about this contribution.
If there is found to be a non-negligible effect 
due to the weak phase of the penguin,
then direct $CP$ violation may arise,
allowing an additional probe of the weak phases involved.
To date, measurements of time-dependent $CP$ violation in $b \to c\bar{c}d$
transitions have been performed using 
$J/ \psi \pi^0$, $D^+D^-$, $D^{*\pm}D^\mp$ and $D^{*+}D^{*-}$,
all of which are consistent with tree-dominance and with the Standard Model.
Future updates will reduce the statistically dominated errors
on these results, and additional channels may also be added.

\vspace{1ex}

The decay $B_d \to D^{(*)\pm}\pi^\mp$ can proceed either via 
the doubly-Cabibbo-suppressed $b \to u\bar{c}d$ transition,
or by $B_d$\textendash$\bar{B}_d$ mixing followed by the Cabibbo-favoured 
$\bar{b} \to \bar{c} u \bar{d}$ transition.
The interference of these two amplitudes results in sensitivity to $\sin(2\beta+\gamma)$,
and the size of the interference effect depends on the ratio
of the magnitudes of the two amplitudes (usually denoted $R_{D^{(*)}\pi}$).
Although the $D^{(*)\pm}\pi^\mp$ final states are abundant,
the smallness of $R_{D^{(*)}\pi}$ makes the $CP$ violation effect 
hard to measure and, since it must be extracted from a large number of events,
sensitive to systematic errors.
Furthermore, while there are two observables for each final state,
there are also two hadronic parameters
($R_{D^{(*)}\pi}$ and $\delta_{D^{(*)}\pi}$, the strong phase difference 
between the decay amplitudes),
and therefore it is difficult to cleanly extract the weak phase information,
although approaches based on, {\it eg.}, SU(3) symmetry exist.
A further complication arises due to multiple ambiguous solutions.
Nonetheless, these modes are being actively pursued experimentally.
Both BaBar and Belle have results with $D^\pm\pi^\mp$ and $D^{*\pm}\pi^\mp$;
BaBar have also investigated $D^\pm\rho^\mp$.
Both experiments have used partial reconstruction techniques
(in addition to the conventional full reconstruction)
to increase the signal yields in the $D^{*\pm}\pi^\mp$ channel.
Although rather different techniques have been employed,
most notably to deal with the troublesome possibility of $CP$ violation
effects on the flavour tagging side of the $\Upsilon(4S) \to B\bar{B}$ event,
the results are consistent at the current level of precision,
which is starting to probe the region where $CP$ violation effects 
are expected to be found.
Future updates are therefore of great interest,
although it will be difficult to continue reducing the systematic uncertainty.

\vspace{1ex}

Some related modes may also prove useful to measure $2\beta+\gamma$.
$B_d \to D^{*\pm}\rho^\mp$ has a vector-vector final state,
and the interfering amplitudes result in an increased number of observables,
so that all parameters can, in principle, be extracted from the data.
However, this mode is experimentally challenging, 
{\it eg.} the polarization measurement is sensitive to systematic effects,
and recent results suggest that the $CP$ violation effect may 
be even smaller than previously expected.
Larger effects are expected to be found in modes mediated by the 
$b \to c\bar{u}s$ \& $b \to u\bar{c}s$ transitions,
such as $B_d \to DK^{(*)}$ and $B_d \to D^{(*)\pm}K_S\pi^\mp$
(note that for time-dependent $CP$ violation effects to arise,
the kaon must be reconstructed in a strangeness nonspecific final state).
These techniques are currently limited by statistics and, in the latter case,
due to lack of knowledge of the resonant structure of the three-body decay.
Another interesting mode is $B_s \to D_s^{(*)\pm}K^\mp$.
In this case the $B_s$ mixing phase $\phi_s$ replaces $2\beta$, 
so the time-dependence probes $\phi_s+\gamma$.
Since $R_{D_s^{(*)}K}$ is expected to be reasonably large,
there will be sufficient observables to extract all parameters from the data.
The problem of multiple solutions remains, however, this can be addressed
firstly using additional observables which arise 
if $\Delta \Gamma_s/ \Gamma_s$ is not small, and secondly
using U-spin symmetry, which translates $s \leftrightarrow d$,
and thus relates $B_s \to D_s^{(*)\pm}K^\mp$ to $B_d \to D^{(*)\pm}\pi^\mp$.
This symmetry can be applied in various different ways,
and mechanisms exist to quantify the uncertainty due to its breaking.
This approach looks very promising for the LHCb experiment.

\vspace{1ex}

The decays $B_{u,d} \to D^{(*)}K^{(*)}$ 
provide the cleanest method to determine $\gamma$.
The method employs the interference between $b \to c\bar{u}s$ and
$b \to u\bar{c}s$ when the final state is accessible 
to both $D^0$ and $\bar{D}^0$ mesons.
The theoretical uncertainty is completely negligible,
and effects due to mixing and $CP$ violation in the neutral $D$ sector
can be taken into account if they are discovered.
As before, there are important hadronic parameters:
the ratio of the magnitudes of the two amplitudes 
and the strong phase difference between them
(these are usually denoted $r_B$ and $\delta_B$),
that can be extracted from the data.
It is important to emphasise again that this method applies to 
any final state which is accessible to both $D^0$ and $\bar{D}^0$,
and one important way to progress is to add more decay modes.
The theoretical framework exists to do so,
both for exclusive and inclusive final states.
To this end, it is crucial to develop experimental techniques 
to optimally combine the results from the different modes 
and obtain the best sensitivity on $\gamma$.

\vspace{1ex}

The $B$ decay modes which have been exploited to date are 
$B_u^\pm \to DK^\pm$, $D^*K^\pm$ and $DK^{*\pm}$,
and in each case the $D$ decay modes to $CP$ eigenstates 
(principally $K^+K^-$ for $CP$-even and $K_S\pi^0$ for $CP$-odd),
doubly-Cabibbo-suppressed final states ($K^\mp\pi^\pm$),
and multibody final states ($K_S\pi^+\pi^-$) have been used.
Although fairly large samples of $CP$ eigenstate decays have now 
been accumulated,
the statistics are still not sufficient to probe the small $CP$ violation
effect expected ($r_B \sim {\cal O}(0.1)$).
Larger effects are expected using the doubly-Cabibbo-suppressed final state;
however no significant signals in $B_u^\pm \to \left[ K^\mp \pi^\pm \right]_D K^\pm$,
and related modes, have yet been observed.

\vspace{1ex}

One of the major developments of the $B$ factories over the past few years
has been the use of the multibody decay $D \to K_S \pi^+\pi^-$.
The rich interference pattern across the Dalitz plot 
results in regions which are highly sensitive to $\gamma$;
in addition this mode is reasonably clean experimentally,
due to its large product branching fraction and clean signal.
However, in order to perform an unbinned fit
(necessary to extract the maximum possible information from the data)
it is necessary to make 
an assumption about the strong phase variation across the Dalitz plot
(achieved using a sum of resonant amplitudes, 
parametrized using the Breit-Wigner formalism, and a nonresonant term),
which results in model uncertainty, currently estimated to be $\sim10^\circ$ on $\gamma$.
To reduce this, additional studies of the Dalitz plot structure 
are necessary, and the results using $D$ mesons 
coherently produced in $\psi(3770) \to D\bar{D}$ at charm-tau factories
play a particularly crucial r\^ole.
When one $D$ meson is tagged as a $CP$ eigenstate and the 
other reconstructed as $K_S \pi^+\pi^-$,
the Dalitz plot density differs from that in the flavour tagged samples,
and the differences are sensitive to the strong phase at each point.
Thus, the $D$ decay model may be verified by fitting the $CP$ tagged sample,
and if this proves impossible, a model-independent technique
(dividing the Dalitz plot into bins) can be employed.
Additional methods to understand the model uncertainty
({\it eg.} using different multibody $D$ decays) also exist.
The understanding of these issues has been advanced through informal 
discussions between members of various collaborations, and also theorists;
it is expected that this will continue to be a fruitful source of progress.

\vspace{1ex}

By summer 2006, it is to be expected that the available samples 
at the $B$ factories will be approximately double those from CKM2005,
and hopefully the combined data sample will exceed $1 \ \rm{ab}^{-1}$.
This will allow significant progress in many of the channels discussed above.
The precision of the $\sin(2\beta)$ result will continue to improve,
and it should be possible to definitively rule out $\cos(2\beta)<0$.
Significant $CP$ violation effects may be seen in some $b \to c\bar{c}d$
channels, and also in $B_d \to D^{(*)\pm}\pi^\mp$.
Further progress on the extraction of $\gamma$ from 
$B_u^\pm \to D^{(*)}K^{(*)\pm}$ is expected.
Since the precision of the $\gamma$ measurement depends on the $r_B$ values,
it is not possible to say with certainty how the error will evolve,
but there is some optimism that the $CP$ violation effects will cross
some threshold of significance.
Another exciting prospect is that first results from $B_d \to D^{(*)}K^{(*)}$ 
may become available.
In additional to other possible new results from BaBar and Belle,
the output of CLEO-c will be watched closely by those interested in $\gamma$.

\vspace{1ex}

Shortly after, attention will increasingly turn to the start up of the LHC.
It is clear that despite (and, perhaps, partially because of)
the best efforts of the $B$ factories and the Tevatron experiments, 
there will be potential for LHCb to make a big impact, 
in particular in the measurement of $B_s$ mixing parameters and of $\gamma$
(in addition to channels outside the scope of this working group).
As methods exist to measure both $\beta$ and $\gamma$ to within very small
theoretical uncertainty,
the next generation of $B$ physics experiments has a promising future.
Furthermore, these theoretically clean modes act as Standard Model
reference points for other channels ({\it eg.} charmless $B$ decays),
with which LHCb and a Super $B$ Factory may search for new physics.

\vspace{20mm}








\section{\boldmath
  Time-Dependent $CP$ Violation Measurements in $B^0 \to (c\bar{c})K^0$}

The measurement of the mixing-induced $CP$ asymmetry in the so-called
gold-plated mode $B^0 \to J/\Psi\,K_S$ is becoming a precision
measurement. This served as a motivation to revisit the analysis of this 
mode within the standard model \cite{Boos:2004xp}. 

It is well known \cite{BigiSanda} that the time dependent $CP$ asymmetry 
\begin{equation}
  \label{cpasymm0}
 a^{J/\Psi\,K_S}_{CP}  (t) =  
  C_{J/\Psi\,K_S}  \cos(\Delta m \, t)
  - S_{J/\Psi\,K_S} \sin(\Delta m \, t) 
\end{equation}
is very cleanly related to the CKM angle $\beta$, since 
\begin{equation} \label{CSD}
 S_{J/\Psi\,K_S} \approx  \sin (2 \beta) , \quad  C_{J/\Psi\,K_S} \approx 0  \, , 
\end{equation}
where deviations from these relations are expected at the level of a few percent. 

\begin{figure}[b]
\centering
\begin{minipage}{11cm}
  \includegraphics[scale=0.5]{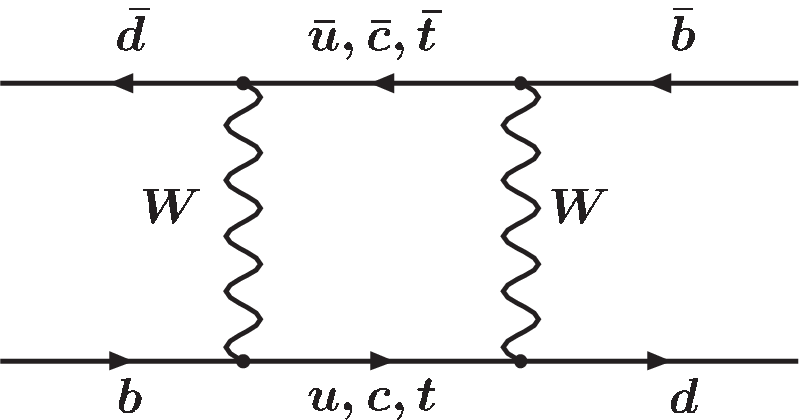} 
  \hfill
  \includegraphics[scale=0.5]{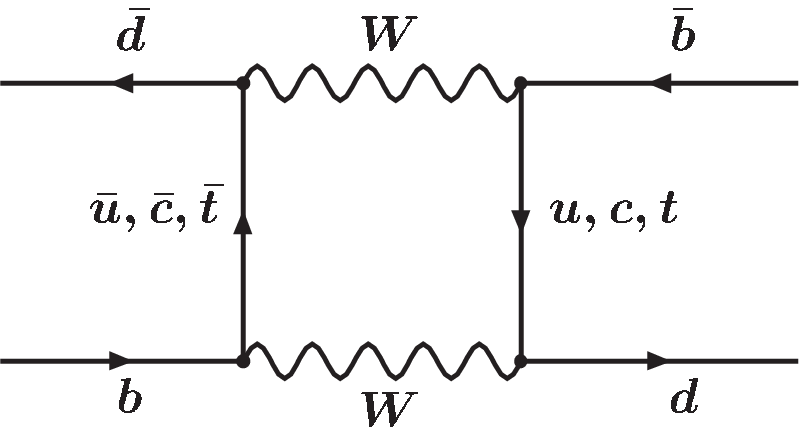} 
\end{minipage}
\caption{Box diagrams mediating  $\Delta B = \pm 2$ transitions.}
\label{fig:boxes}
\end{figure}

The reason for this is summarized as follows. The box diagram shown in fig.~\ref{fig:boxes} 
leads to a $\Delta B = 2$ interaction  transforming a $B^0$ into a $\overline{B}^0$. Thus the 
Eigenstates of the Hamiltonian are (to leading order) the combinations 
\begin{equation}
|B_{\rm{L/H}} \rangle = \frac{p |B^0  \rangle \pm q |\overline{B}^0  \rangle }
{\sqrt{|p|^2+|q|^2}}
\quad \mbox{with} \quad 
\frac{q}{p} \approx - \frac{M_{12}^*}{|M_{12}|}
\end{equation}
where $M_{12}$ is the matrix element of the box diagram contribution between a 
$B^0$ and a  $\overline{B}^0$ state. 
Due to the CKM factors and the large  top-quark mass $M_{12}$ is completely 
dominated by the top quark. 
Thus $M_{12}$  is directly proportional to the  CKM phase
\begin{equation}
M_{12}^* \propto (V_{td} V_{tb}^*)^2 \propto e^{- i 2  \beta}
\quad \mbox{and thus} \quad 
\frac{q}{p} = e^{-i 2 \beta}
\end{equation}

A $CP$ asymmetry is induced by contributions to a decay which have both different 
weak and strong phases. Furthermore, the time evolution also induces a phase 
difference between the two eigenstates $B_{\rm L}$ and $B_{\rm H}$ which is 
$\exp (i \Delta m \, t)$ and which has the same effect as a strong phase 
difference and thus also results in a phase difference. 

The $CP$ asymmetry parameters given in   (\ref{cpasymm0}) are given in terms of the 
mixing parameters, the decay amplitude  $A(B^0 \to J/\Psi K_S)$ and its  
CP conjugate as 
\begin{equation}  \label{CSdef} 
  C_{J/\Psi K_S} = \frac{1 - |\lambda|^2}{1 + |\lambda|^2}
\quad \mbox{and} \quad 
  S_{J/\Psi K_S} = \frac{2 \, \rm{Im}  [\lambda]}{1 + |\lambda|^2} 
\end{equation}
with 
\begin{equation} \label{LamDef}
\lambda = \left( \frac{q}{p} \right)   \frac{A(\overline{B}^0 \to J/\Psi K_S)}{A(B^0 \to J/\Psi
  K_S)} \, . 
\end{equation}
Analyzing the possible contributions to the decay $B^0 \to J/\psi K_s$ one finds that the 
decay amplitude contains a tree contribution proportional to $V_{cb}^* V_{cs}$ which is
real in the standard parametrization. Furthermore, the leading penguin contribution 
is also porportional to the same CKM factor as 
the tree, while the remaining penguin contribution (due to the up quark)  is suppressed 
by two factors of the Wolfenstein parameter. Thus we have 
\begin{equation}
\lambda = - \frac{V_{tb}^* V_{td}}{V_{tb}  V_{td}^*} 
\frac{V_{cb} V_{cs}^*}{V_{cb}^* V_{cs}} = - \exp(-2i \beta)  
\end{equation}
leading to (\ref{CSD}). 

\subsection{\boldmath
  Estimating Corrections to the ``golden relations''  } 

\vspace{+2mm}
\begin{flushright}
 {\it Contribution from T.~Mannel}
\end{flushright}

The corrections arise from two sources: (i) corrections to the mixing and (ii) 
corrections to the 
decay.
The corrections to the mixing amplitude due to contributions which do not carry the phase
$\exp(2i\beta)$ have to be suppressed relative to the leading contribution by a 
factor $m_c^2 / m_t^2 \sim 10^{-4}$ due to the GIM mechanism. More precise 
statements can be made by computing the box diagrams of fig.~\ref{fig:boxes}
\cite{oldstuff,Buras_und_co,NiersteHerrlich}. Since vastly different mass scales are involved, one may
make use of effective field theory methods in which case one may compute $M_{12}$ 
in terms of the mixing of the time-ordered product of two $\Delta B = 1$ interactions
into local $\Delta B = 2$ operators  of dimension 8. The dim-8 operator relevant for the contributions 
with a phase different from $2 \beta$ is given by the mixing 
\begin{equation}
T^{\Delta B=2} = -\frac{i}{2} \int d^4 x \,
T \left[H^{\Delta B=1}  (x) H^{\Delta B=1} 
  (0) \right] \to 
  Q_3 = m_c^2 (\bar{b}_L \gamma_\mu d_L ) (\bar{b}_L \gamma^\mu d_L ) .
\end{equation}
which yields contributions with the CKM factor $ (V_{cb}^* V_{cd})^2 $ and 
$ (V_{cb}^* V_{cd}) (V_{tb}^* V_{td}) $. 

The contributions proportional to $m_c^2$ shift the imaginary part of the ratio 
$M_{12}/|M_{12}|$,
which will contribute to a deviation from the simple relation (\ref{CSD}). One obtains 
\begin{equation}
 \Delta \rm{Im} \left[\frac{M_{12}}{|M_{12}|} \right] =
  -(2.08 \pm 1.23) \cdot 10^{-4}  
\end{equation}
where the uncertainties are due to the input parameters, which are the CKM parameters 
and the masses. 

The second contribution to the deviation from (\ref{CSD}) is much harder to estimate.
It originates from the up quark penguin contribution, which induces a different weak 
phase into the decay. The ratio of matrix elements   
\begin{equation}
 r := \frac{\langle J/\Psi K_S | H (b \to u\bar{u}s)
  |\overline{B}^0 \rangle }{\langle J/\Psi
  K_S | H (b \to c\bar{c}s)|\overline{B}^0 \rangle}
  \frac{|V_{ub}| |V_{us}|}{|V_{cb}|
  |V_{cs}|} \, , 
\end{equation}
where $H(b \to q\bar{q} s)$ denotes the contributions to the effective Hamiltonian due to 
the $b \to q\bar{q} s$ transition, is small due to the ratio of CKM factors. The ratio of 
decay amplitudes appearing in the parameter $\lambda$ of (\ref{LamDef}) can be written in
terms of $r$ as
\begin{equation}
\frac{A(\overline{B}^0 \to J/\Psi K_S)}{A(B^0 \to J/\Psi
  K_S)} = - \frac{1 + r e^{- i \gamma}}{1 + r
  e^{+i \gamma}} \approx - (1 - 2 \, i \, r \sin\gamma)  \, . 
\end{equation}

The ratio $r$ is estimated by using an approach based on naive factorization. The up-quark 
loop  is evaluated perturbatively and yields \cite{BSS} 
\begin{eqnarray} \label{BSS}
&& \langle J/\Psi K_S | H (b \to u\bar{u}s) |B_q \rangle = \\ \nonumber 
 &&  - \frac{G_F}{\sqrt{2}}  \langle J/\Psi K_S |      \Biggl\{
  \frac{\alpha}{3\pi} \left( 
  \overline{s} b \right)_{V-A} \left( \overline{c} c \right)_V + 
\frac{\alpha_s}{3\pi} \left( \overline{s} T^a b \right)_{V-A} \left(
  \overline{c} T^a c \right)_V \Biggr\}  |B_q \rangle  
 \left( \frac{5}{3} - \ln \left(
  \frac{k^2}{\mu^2} \right) + i \pi \right) 
  \end{eqnarray}
where $k^2 \sim m_{J/\Psi}^2$ is an average momentum flowing through the up quark loop and 
$\mu \sim m_b$ is the typical scale of the process. 

The two matrix elements appearing in (\ref{BSS}) are estimated by noting that 
the tree level effective Hamiltonian contains both the color octet and the color 
singlet matrix elements. It is well known that in naive factorization the color 
octet contribution vanishes and that the prediction of naive factorization turns out
to be too small. Hence we estimate the color octet matrix element by ascribing the 
difference of the predicted and the the observed rate to this matrix element. 
In this way we get 
\begin {eqnarray}
  |\langle J/\Psi
   K_S|(\overline{s}b)(\overline{c}c)|\overline{B}^0 \rangle_{\rm{fact}} | 
   &=& (3.96 \pm 0.36) \cdot 10^9 \; \rm{MeV}^3  \\
 |\langle J/\Psi 
   K_S |(\overline{s}T^a b)(\overline{c}T^a c)|\overline{B}^0 \rangle | &=& (1.97
   \, \pm \, 0.64) \cdot 10^8 \, \rm{MeV}^3 
\end{eqnarray}
from which we obtain 
\begin{equation}
  \rm{Re} \left[ r \right] = \left( - 3.62 \, \pm 1.55 \right)
  \cdot 10^{-4} \, , \qquad 
  \rm{Im} \left[ r \right] = \left( - 4.48 \, \pm 1.92 \, \right)
  \cdot 10^{-4} 
\end{equation}
where the imaginary part is assumed to be entirely due to the imaginary part of the loop 
calculated in (\ref{BSS}). 

The effect on the $CP$ asymmetry is given as 
\begin{equation}
a_{CP} (t) = 
  - \left( \sin(2\beta) + \Delta S_{J/\Psi\,K_S} \right)  \sin(\Delta m \, t)  -
  \frac{\delta}{2} \cos(\Delta m \, t)
\end{equation}
and we obtain for the parameter $\delta$ 
\begin{equation}
  \delta = - \left( 1.02 \, \pm \, 0.75 \right) \cdot 10^{-3} 
\end{equation}
The change $\Delta S_{J/\Psi\,K_S}$ in the mixing induced $CP$ asymmetry is 
given as 
\begin{equation}
 \Delta S_{J/\Psi\,K_S} 
  = 2\, \sin\gamma \, \rm{Re} \left[ r \right]
  \cos(2 \beta) - \Delta \rm{Im} \left[
  \frac{M_{12}}{|M_{12}|} \right]
\end{equation}
which yields numerically (for $\sin \gamma =  0.86 \pm 0.12$) 
\begin{equation}
   \Delta S_{J/\Psi\,K_S} = (-2.16 \pm
   2.23) \cdot 10^{-4}
\end{equation} 

In summary it turns out that this naive estimate yields a very small correction
to the relations (\ref{CSD}) from the standard model. However, this statement 
depends crucially on the way the up-quark penguin contribution is estimated, 
while the calculation of the $\Delta B = 2$ mixing is quite reliable. 

On the other hand, even if the up-quark penguin contribution is grossly 
underestimated, the standard model corrections to the simple relations 
(\ref{CSD}) are too small to become visible in the current $B$-factory 
experiments.


\subsection{\boldmath  
  A Method to Measure $\phi_1$ Using 
  $\bar{B}^0 \to D^{(*)} h^0$ With Multibody $D$ Decay}
\vspace{+2mm}
\begin{flushright}
 {\it Contribution from T.~Gershon}
\end{flushright}

The value of $\sin(2\phi_1)$, 
where $\phi_1$ is one of the angles of the Unitarity Triangle~\cite{pdg_review} 
is now measured with high precision: 
$\sin(2\phi_1) = 0.731 \pm 0.056$~\cite{sin2phi1}. 
However, this measurement contains an intrinsic ambiguity: $2\phi_1 \longleftrightarrow \pi-2\phi_1$.
Various methods to resolve this ambiguity 
have been introduced~\cite{phi_ambig}, 
but they require very large amounts of data 
(some impressive first results notwithstanding~\cite{babar_psikstar}).

A new technique based on the analysis of $\bar{B}^0 \to D h^0$.
followed by the multibody decay of the neutral $D$ meson,
has recently been suggested~\cite{bgk}.
Here we use $h^0$ to denote a light neutral meson, such as $\pi^0, \eta, \rho^0, \omega$.
The modes $\bar{B}^0 \to D_{CP} h^0$,
utilizing the same $B$ decay but requiring the $D$ meson to 
be reconstructed via $CP$ eigenstates,
have previously been proposed as ``gold-plated'' modes to search for 
new physics effects~\cite{grossman_worah}.
Such effects may result in deviations from the Standard Model
prediction that $CP$ violation effects in $b \to c\bar{u}d$ transitions
should be very similar to those observed in $b \to c\bar{c}s$ transitions,
such as $\bar{B}^0 \to J/ \psi \KS$.
Detailed considerations have shown that the contributions
from $b \to u\bar{c}d$ amplitudes,
which are suppressed by a factor of approximately $0.02$~\cite{rd},
can be taken into account.
Consequently, within the Standard Model,
studies of $\bar{B}^0 \to D_{CP} h^0$ can give a measurement of $\sin(2\phi_1)$ 
that is more theoretically clean 
than that from $\bar{B}^0 \to J/ \psi \KS$~\cite{fleischer1,fleischer2}.
However, these measurements still suffer from the ambiguity mentioned above.

In the case that the neutral $D$ meson produced in $\bar{B}^0 \to D h^0$
is reconstructed in a multibody decay mode,
with known decay model,
the interference between the contributing amplitudes 
allows direct sensitivity to the phases.
Thus $2\phi_1$, rather than $\sin(2\phi_1)$ is extracted,
and the ambiguity $2\phi_1 \longleftrightarrow \pi-2\phi_1$ can be resolved.
This method is similar to that used to extract $\phi_3$,
using $B^\pm \to D K^\pm$ followed by multibody $D$ decay~\cite{anton,ggsz}.

There are a large number of different final states 
to which this method can be applied.
In addition to the possibilities for $h^0$,
and the various different multibody $D$ decays which can be used,
the method can also be applied to $\bar{B}^0 \to D^* h^0$.
In this case, the usual care must be taken to distinguish 
between the decays $D^* \to D\pi^0$ and $D^* \to D\gamma$~\cite{bg}.
Also, if $h^0$ is not a spinless particle, angular analysis~\cite{dqstl} 
will be required to resolve the contributing amplitudes
to $\bar{B}^0 \to D^* h^0$.

A detailed description of the method and the results of a feasibility
study are given in~\cite{bgk}.
In summer 2005, the Belle Collaboration has presented the first results
on $\phi_1$ using this technique~\cite{belle_d0h0}.

\subsection{\boldmath
  Measurement of $\cos2\beta/2\phi_1$ with $B^0_d\to J/\psi K^*$}

\vspace{+2mm}
\begin{flushright}
 {\it Contribution from M.~Verderi}
\end{flushright}

The measurement of the sign of $\cos2\beta$ provides a direct test of the Standard Model in which $\cos2\beta>0$. The interference of $CP$-odd $(A_\perp)$ and $CP$-even $(A_0,A_\perp)$
amplitudes in the decay $B^0\to J/\psi K^{*0}; K^{*0}\to K_S^0 \pi^0$ generates a $\cos2\beta$ contribution in the time- and angle-dependent distribution of the decay.
This contribution appears as a product of $\cos2\beta$ and cosine of strong phases differences. The strong phases themselves are measured up to a two-fold ambiguity
leading to a sign ambiguity on the cosine of the strong phases, and hence a sign ambiguity on $\cos2\beta$.

BABAR resolves the strong phases ambiguity pointing out the $K\pi$ $S$-wave, beyond the dominant $K^*(892)$ $P$-wave, in the $B\to J/\psi K\pi$ decay and 
exploiting the known behaviour of the $K\pi$ $S$--$P$ relative phase with $K\pi$ mass in the $K^*(892)$ region~\cite{LASS}: among the two strong phases solutions, only one leads to the
physical behaviour of the $K\pi$ $S$--$P$ relative phase with $K\pi$ mass. 

The $\cos2\beta$ parameter is extracted by a time- and angle-dependent analysis of the $B^0\to J/\psi K^{*0}; K^{*0}\to K_S^0 \pi^0$ decay. The BELLE and BABAR measurements
are shown in Table~\ref{tab:cosCPresults}. From MonteCarlo studies, BABAR estimates that the non-standard $\cos2\beta = -\sqrt{1-0.731^2} = -0.67$ solution is excluded at $86\%$ CL.

According to BABAR, 
the prospect for higher luminosities is that $\sigma(\cos2\beta)$ 
empirically scales as $\sigma(\cos2\beta) = 7\cdot({\cal L}/\rm{fb}^{-1})^{-1/2}$ 
for luminosities high enough $(\gtrsim 100 \rm{fb}^{-1})$. 
The $\cos2\beta$ measurement will be  statistically limited for the coming years.

\begin{table}[htb]
\begin{center}
\caption{\label{tab:cosCPresults}BELLE and BABAR measurements of $\sin2\beta$ and $\cos2\beta$ with $B^0\to J/\psi K^{*0}; K^{*0}\to K_S^0 \pi^0$, with both parameters free in the fit
and with $\sin2\beta$ fixed to world average.}
\begin{tabular}{|l|c|c|c|}\hline\hline
         & \# events     & $\sin2\beta$ & $\cos2\beta$ \\ \hline
BELLE~\cite{cosBELLE}    & 354          & $0.30\pm0.32\pm0.02$ & $0.31\pm0.91\pm0.10$ \\
         &               & 0.731 (fixed)& $0.56\pm0.86\pm0.11$ \\ \hline
BABAR~\cite{cosBABAR}    & 104          & $0.10\pm0.57\pm0.14$ & $3.32^{+0.76}_{-0.96}\pm0.27$ \\
         &               & 0.731 (fixed)& $2.72^{+0.50}_{-0.79}\pm0.27$ \\ \hline
\end{tabular}
\end{center}
\end{table}

 \section{\boldmath
  $CP$ Violation in $b \to c\bar{c}d$ Transitions}

Time-dependent $CP$ violation in 
$B$ decays governed by $b \to c \bar{c} d$ transitions 
have been measured in
$B^0 \to J/\psi \pi^0$, $D^{*+} D^{*-}$, $D^{+} D^{*-}$ and $D^{-} D^{*+}$ 
by both Belle and BaBar collaborations.
Since the tree diagrams of $b \to c \bar{c} d$ transitions
contain $V_{cb}$ and $V_{cd}$ which have no KM complex phase, 
$CP$ violation arises from the complex phase in 
$B^0 \overline{B^0}$ mixing ({\it i.e.} $\phi_1$)
if neither penguin nor other contributions are substantial.
These decays are therefore sensitive to the unitarity angle $\beta$ and complement 
the precise measurement of $\sin(2\beta)$ through 
$\BZ \to (c\bar{c}) K^{0(*)}$ decays.
The presence of a gluonic penguin contribution with a different weak phase 
is expected to change the magnitude of the $CP$ asymmetry by not more 
than a few percent~\cite{dstdexpect}. 
However, additional contributions from non-SM processes may lead to shifts 
as large as  $\Delta \beta \approx 0.6$ in some models~\cite{grossman_worah}.
Interference between SM penguin and tree amplitudes can additionally 
provide some sensitivity to the angle 
$\gamma = \arg \left[\, -V_{ud}^{}V_{ub}^* / V_{cd}^{}V_{cb}^*\, \right]$~\cite{DL,ADL}.  

The potential $b \to d$ penguin contribution can be
probed by other processes.
The charge asymmetry between $B^+ \to J/\psi \pi^+$ and 
$B^- \to J/\psi \pi^-$ is one possibility,
although such an asymmetry also depends on the strong phases.
Replacing $c\bar{c}$ by $s\bar{s}$, the $B \to \phi \pi$ decay
generated by $b \to s\bar{s} d$ transition solely takes place 
via penguin transition, so search for such kind of rare $B$ decays 
will give us some information about magnitude of the penguin diagram.

\subsection{\boldmath
  $CP$ Violation in $b \to c\bar{c}d$ Transitions at Belle}
  
\vspace{+2mm}
\begin{flushright}
 {\it Contribution from K.~Miyabayashi}
\end{flushright}

\begin{figure}[!htb]
  \begin{center}
    \begin{tabular}{c@{\hspace{5mm}}c}
      \includegraphics[width=0.32\textwidth,height=0.24\textwidth,bb=280 0 556 302, clip=true]{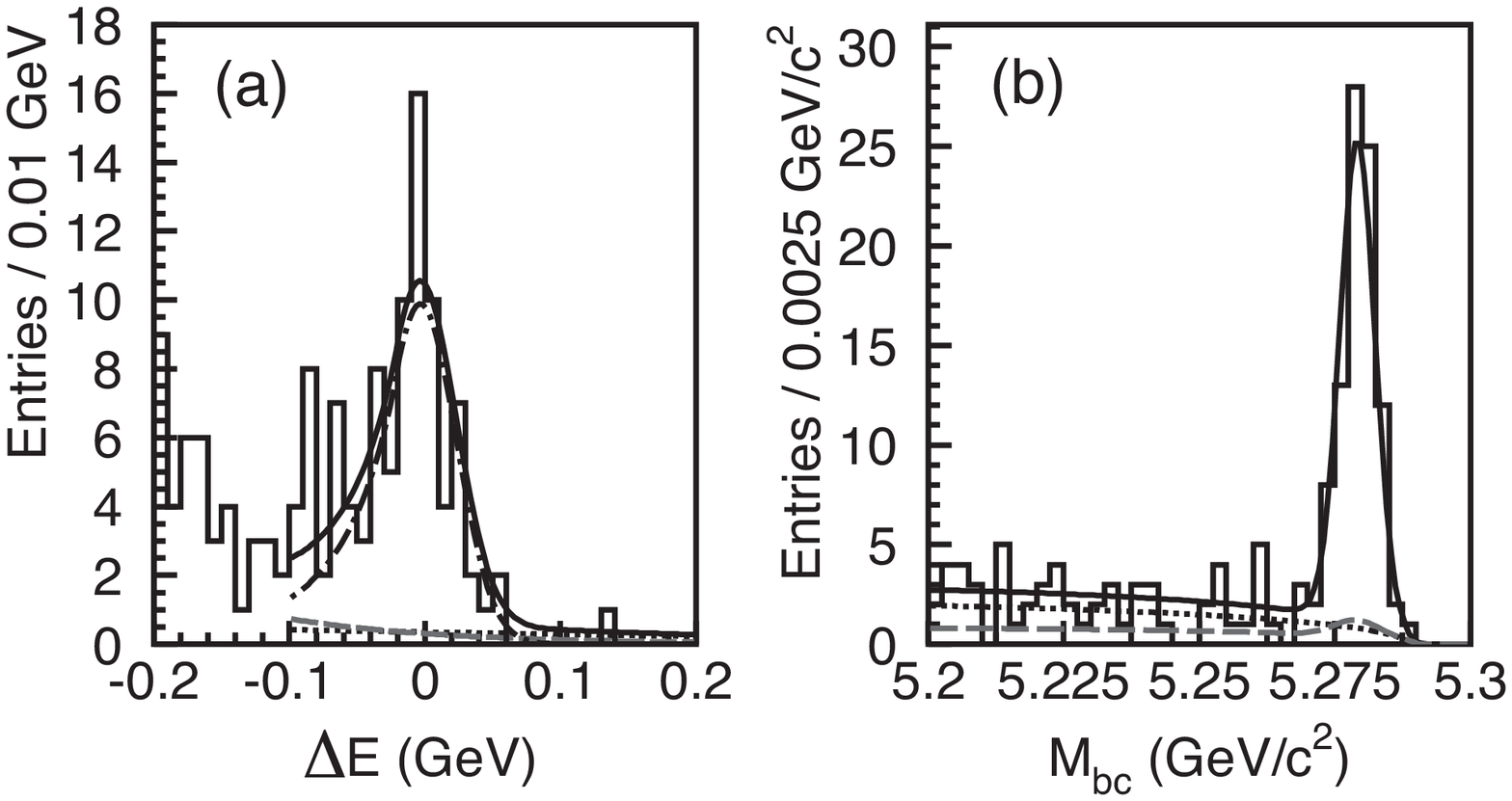}
      &
      \includegraphics[width=0.32\textwidth,height=0.24\textwidth]{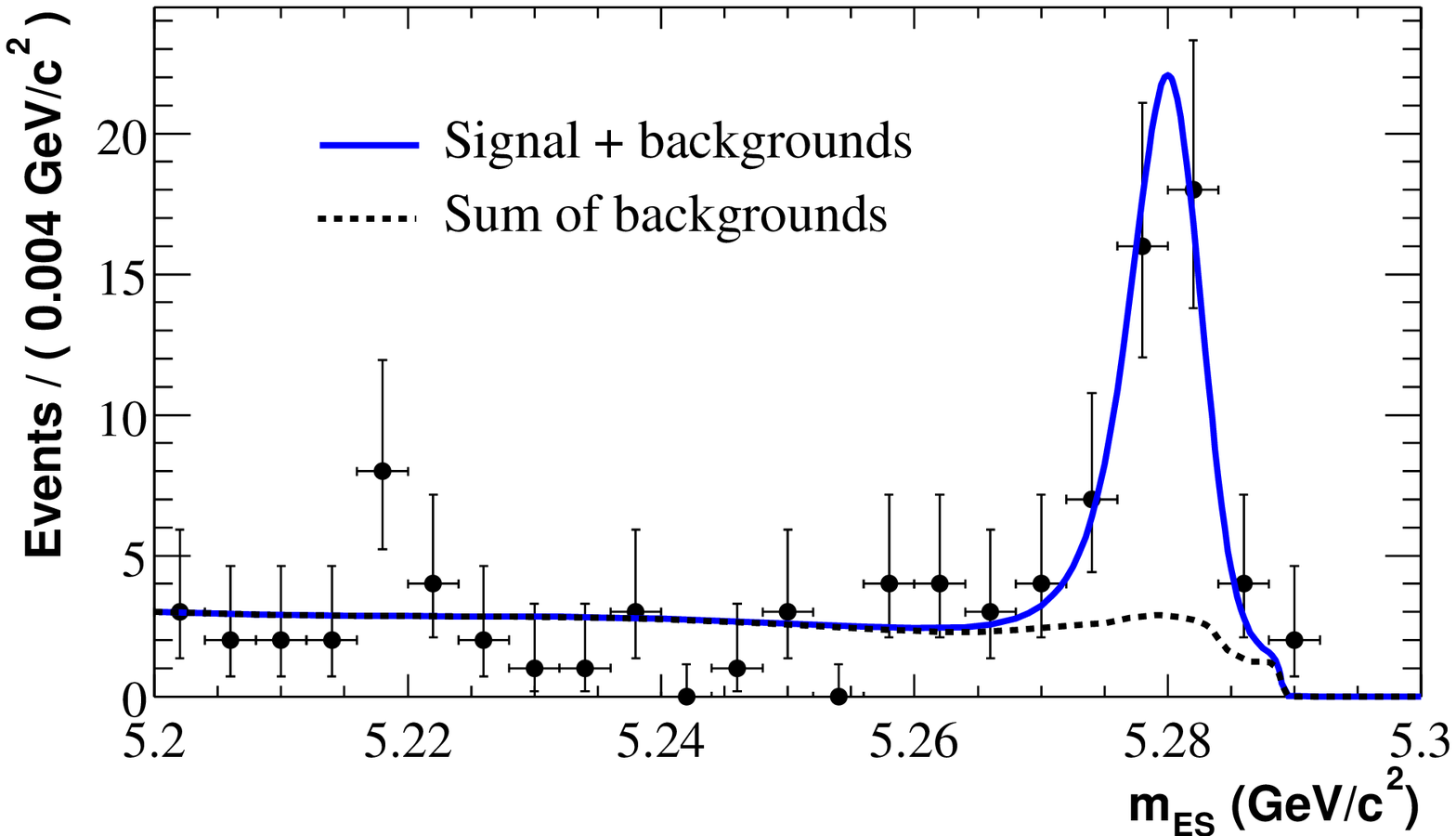} \\
      \includegraphics[width=0.32\textwidth,height=0.32\textwidth]{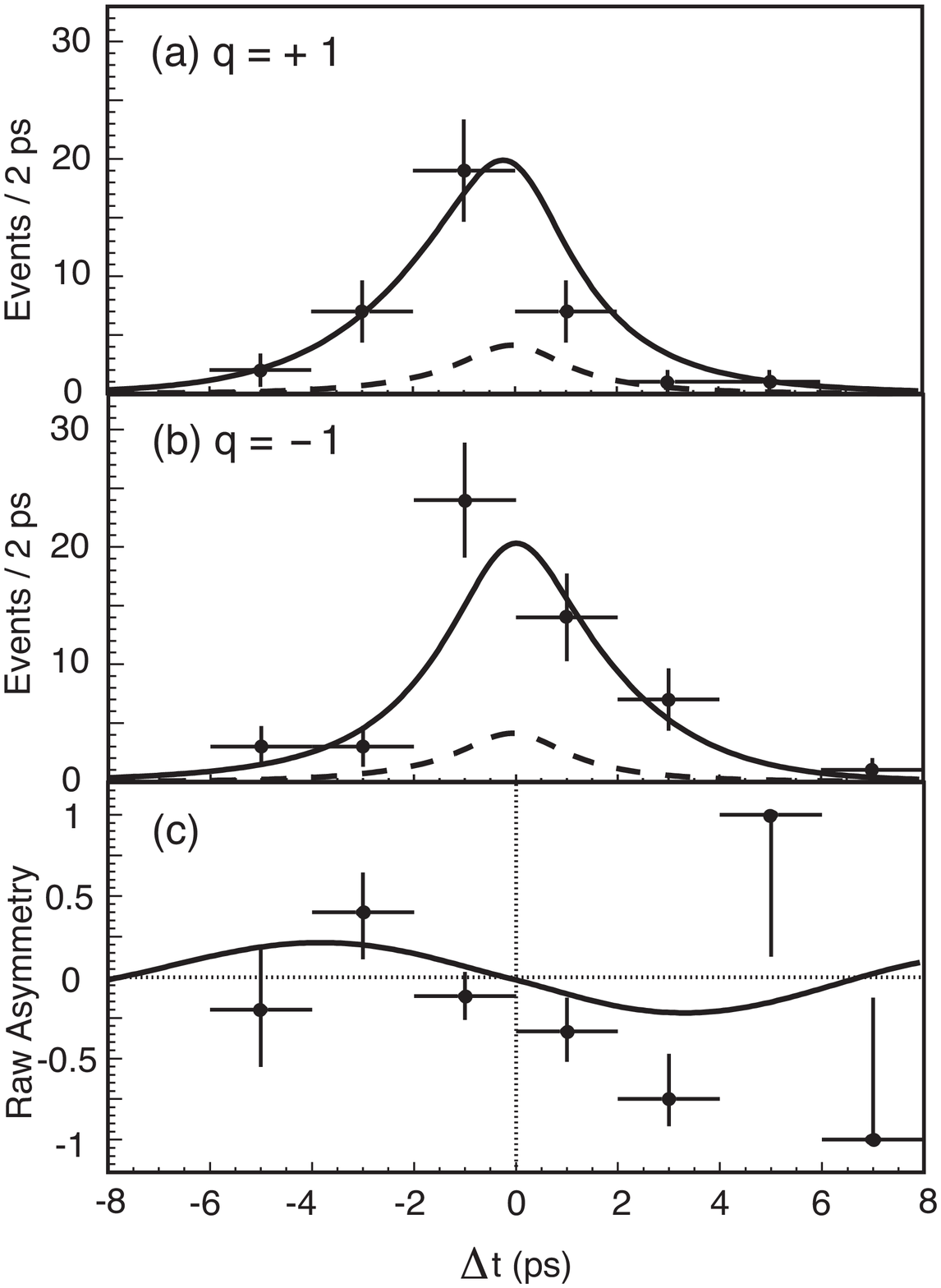}
      &
      \includegraphics[width=0.32\textwidth,height=0.32\textwidth]{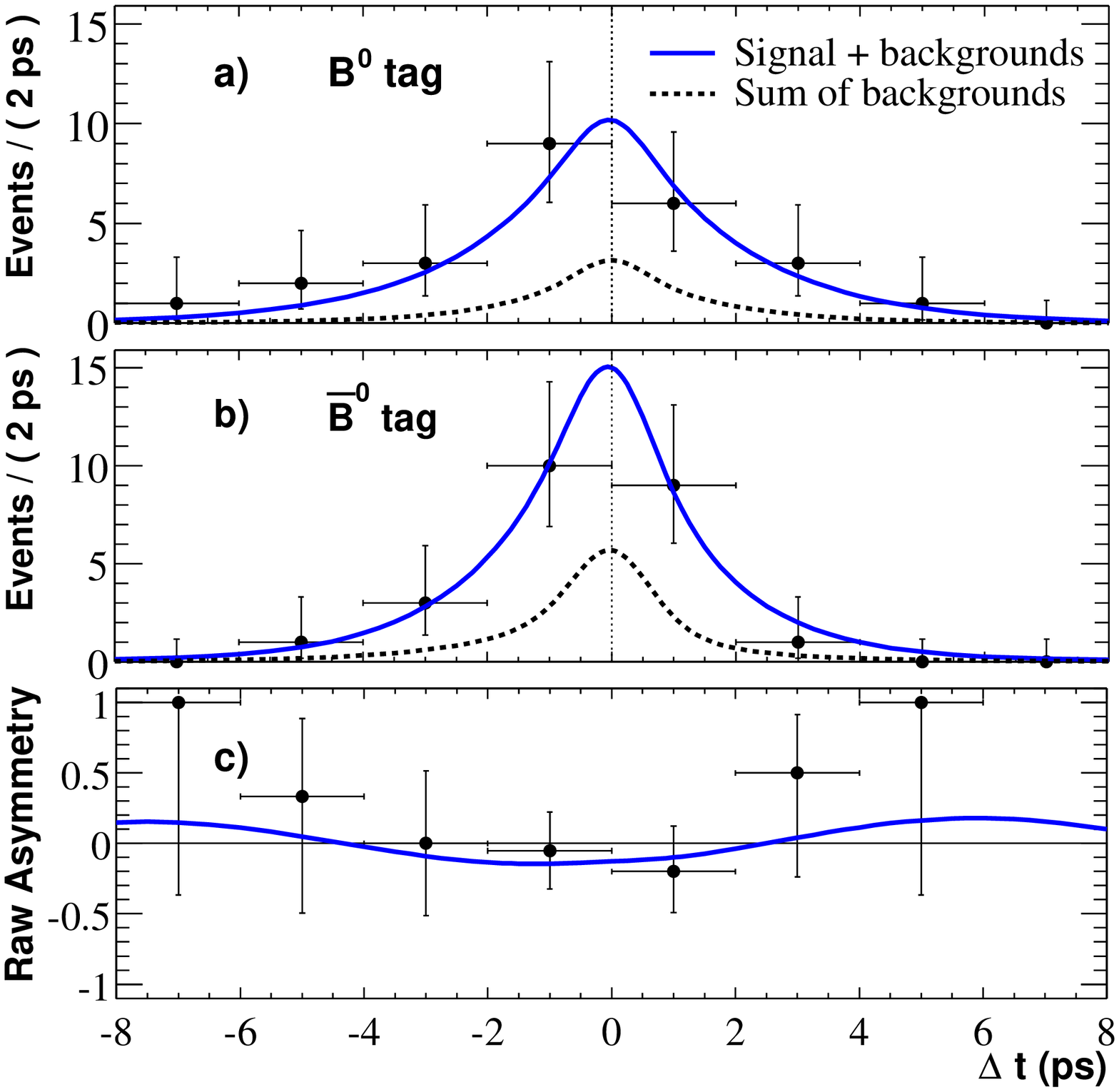}
    \end{tabular}
  \end{center}
  \caption{
    \label{psipi0_fig}
    (Top) $M_{\rm{bc}} / M_{es}$ distributions from 
    (left) Belle and (right) Babar.
    (Bottom) $\Delta t$ distributions and asymmetries.
    Belle plots are from~\cite{belle_psipi0_pub};
    BaBar plots are from~\cite{babar_psipi0_pub}.
  }
\end{figure}

$J/\psi \pi^0$ is a $CP$-even final state, 
and therefore the Standard Model (SM) predicts
(in the limit of tree-dominance)
${\cal{S}}_{J/\psi \pi^0} = - \sin 2 \phi_1$ and 
${\cal{A}}_{J/\psi \pi^0} = 0$.
Results from Belle and BaBar are shown in Fig.~\ref{psipi0_fig}.
In the Belle measurement~\cite{belle_psipi0_pub}, 
91 candidates are found from a data sample containing 
152 M $B\bar{B}$ events.
Backgrounds are estimated separately for $B \to J/\psi X$ and combinatorial. 
The purity is estimated to be $86\pm10\%$.
The BaBar measurement uses 88 M $B\bar{B}$~\cite{babar_psipi0_pub};
49 events are reconstructed in the signal region of which 
37 are estimated to be signal.

The obtained results are (see also Fig.~\ref{cpvinb2ccd})
\begin{equation} 
  \nonumber
  \begin{array}{c@{\hspace{3mm}}c@{\hspace{3mm}}c}
    & {\cal{S}}_{J/\psi \pi^0} & {\cal{A}}_{J/\psi \pi^0} \\
    {\rm{Belle}} & -0.72 \pm 0.42 \stat \pm 0.09 \syst & -0.01 \pm 0.29 \stat \pm 0.03 \syst \\
    {\rm{BaBar}} & +0.05 \pm 0.49 \stat \pm 0.16 \syst & -0.38 \pm 0.41 \stat \pm 0.09 \syst
  \end{array}
  \nonumber
\end{equation}

\begin{figure}[!htb]
  \begin{center}
    \includegraphics[width=0.30\textwidth]{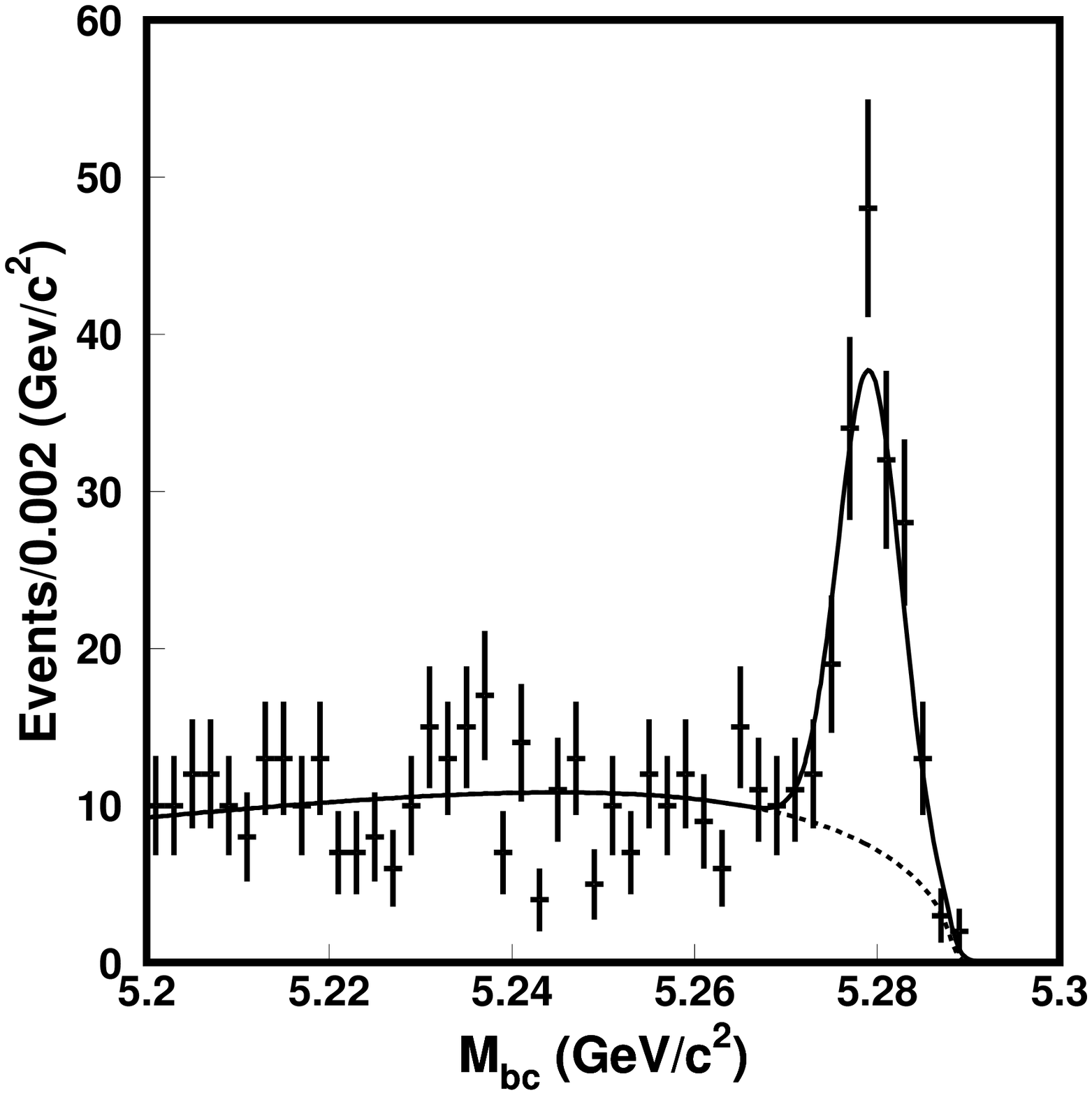}
    \includegraphics[width=0.30\textwidth]{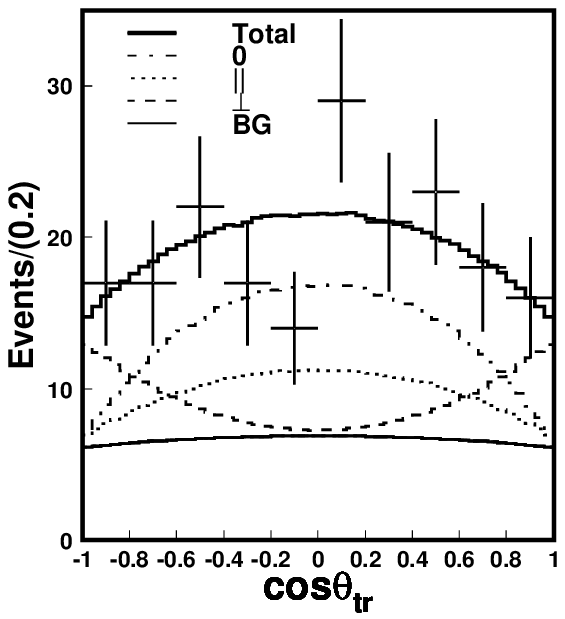}
    \includegraphics[width=0.30\textwidth]{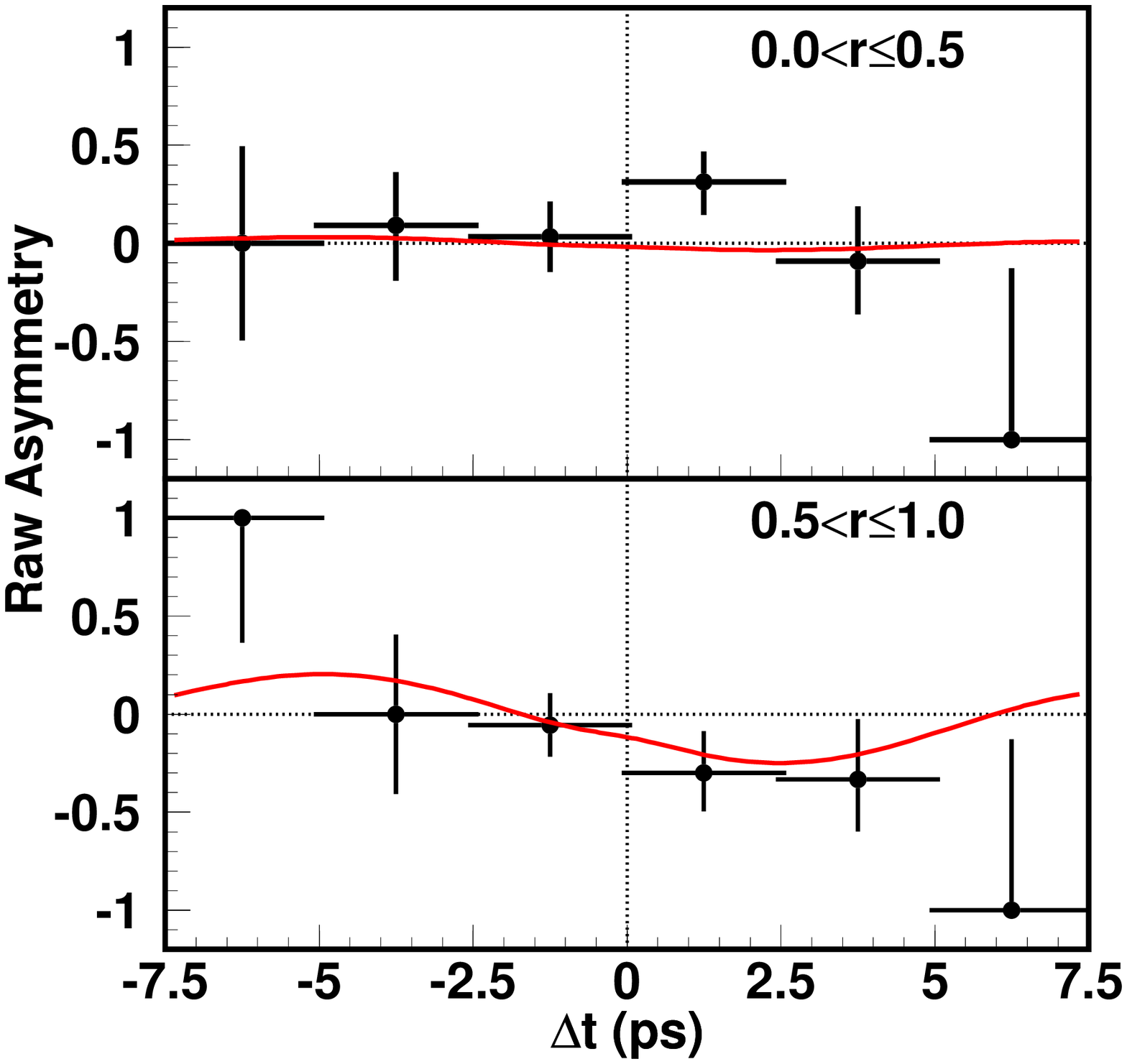}
  \end{center}
  \caption{
    \label{belle_dspdsm_fig}
    Results of the Belle analysis of 
    $B^0 \to D^{*+}D^{*-}$~\cite{belle_dspdsm_pub}.
    (Left to right)
    Distributions of $M_{\rm{bc}}$, $\cos \theta_{\rm{tr}}$
    and raw $\Delta t$ asymmetry separately for 
    (top) poor and (bottom) good quality tags.
  }
\end{figure}

$B^0 \to D^{*+} D^{*-}$ has a final state containing two vector mesons, 
which in general is a combination of $CP$-even and $CP$-odd.
The $CP$-odd fraction can be determined by angular analysis 
of the decay products.
Results from BaBar on this analysis are described elsewhere in this document.
Based on a dataset corresponding to 152 M $B\bar{B}$ pairs, 
Belle reconstructed 194 candidate events 
with $67\pm7\%$ purity in this decay mode~\cite{belle_dspdsm_pub}
The $CP$-odd fraction is found to be $0.19 \pm 0.08 \stat \pm 0.01 \syst$, 
and it is taken into account in the subsequent fit to
obtain the $CP$ violation parameters:
\begin{eqnarray}
  {\cal{S}}_{D^{*+} D^{*-}} &=& -0.75 \pm 0.56 \stat \pm 0.12 \syst \\
  {\cal{A}}_{D^{*+} D^{*-}} &=& -0.26 \pm 0.26 \stat \pm 0.06 \syst.
\end{eqnarray}
These results are shown in Fig.~\ref{belle_dspdsm_fig}.

The final states $D^{*\pm} D^\mp$ are not $CP$ eigenstates,
and hence the phenomenology is more complicated,
though the time-dependent observables are still sensitive to $2\phi_1$.
Results from BaBar on this analysis are described elsewhere in this document.
Belle has reconstructed this decay mode with two different methods,
full and partial reconstruction~\cite{belle_dsd_pub},
using a data set of 152 M $B\bar{B}$ pairs.
The results are summarized in Fig.~\ref{cpvinb2ccd}.

All measurements mentioned above are statistically limited, 
so finding new $b \to c\bar{c}d$ decay modes is also important.
Both Belle~\cite{belle_psipipi_pre} 
and BaBar~\cite{babar_psipipi_pub} collaborations have observed 
$B^0 \to J/\psi \rho^0$ decays;
this final state consists of two vector mesons
and therefore angular analysis is necessary to extract the $CP$ composition.
Furthermore, the contribution of 
non-resonant $B^0 \to J/\psi \pi^+\pi^-$ decays must be well understood.
Recently, the doubly charmed mode 
$B^0 \to D^+D^-$ has been observed by Belle~\cite{belle_dpdm_pub}.
This is a $CP$ eigenstate, and looks quite promising for 
time dependent $CP$ violation measurements in the near future.

\subsection{\boldmath
  Time-Dependent $CP$ Asymmetries in \BDbothstpmDbothmp\ at Babar}
 
\vspace{+2mm}
\begin{flushright}
 {\it Contribution from M.~Bruinsma}
\end{flushright}

Using $(232 \pm 3) \times 10^6$ $\Upsilon(4S)\to B\bar{B}$ decays recorded by
the BaBar detector~\cite{Aubert:2001tu} at the PEP-II  $e^+e^-$ collider, Babar
has measured $CP$ asymmetries in \BZ\ decays to the $CP$ eigenstates \DstpDstm\ and \DpDm and $CP$
'pseudo'-$CP$-eigenstates \DstpDm\ and \DstmDp.  The results are summarized in Table \ref{tab:results} and are consistent with $S=-\sin(2\beta)$ and $C=0$, expected in the SM for a tree-level-dominated 
transition with equal rates for \BDstpDm\ and \BDstmDp. 

\begin{table}[htb]
  \small
  \begin{center}
    \begin{tabular}{lrccc}
      Sample 		& $N_{\rm{sig}}$ & Purity  & $S$ & $C$ \\ \hline \hline
      \DstpDstm	& 396(23)  	  & 0.75(2) &
      &
      $ 0.04\pm0.14 \rm{(stat.)}\pm 0.02 \rm{(syst.)}$ \\
      & & & 
      \hspace{-5mm}
      $-0.65\pm0.26 \rm{(stat.)}^{+0.09}_{-0.07}(R_{T})\pm 0.04 \rm{(syst.)}$ 
      \hspace{-5mm}
      \\
      \DstpDm	& $145(16)$	& $0.49(3)$ &  
      $-0.54\pm0.35 \rm{(stat.)}\pm 0.07 \rm{(syst.)}$ & 
      $0.09 \pm 0.25 \rm{(stat.)} \pm 0.06 \rm{(syst.)}$ \\
      \DstmDp	& $126(16)$	& $0.49(3)$ & 
      $-0.29\pm0.33 \rm{(stat.)}\pm 0.07 \rm{(syst.)}$ & 
      $0.17 \pm 0.24 \rm{(stat.)} \pm 0.04 \rm{(syst.)}$ \\
      \DpDm	& $\;54(11)$	& $0.37(6)$ & 
      $-0.29\pm0.63 \rm{(stat.)}\pm 0.06 \rm{(syst.)}$ &
      $0.11 \pm 0.35 \rm{(stat.)} \pm 0.06 \rm{(syst.)}$ \\ \hline
    \end{tabular}
  \end{center}
  \caption{
    Signal yield and purity for each of the \BDbothstpmDbothmp\ samples,
    and results for the parameters $S$ and $C$ 
    describing the time-dependent $CP$-asymmetries. 
    The purity is defined as the fraction of signal events 
    in the region $m_{ES} > 5.27$~GeV.
  }
  \label{tab:results}
\end{table}

The vector-vector final state \DstpDstm\ is predominantly $CP$-even, but contains also 
$CP$-odd contributions. The $CP$-odd fraction for this decay is measured in a time-integrated 
angular analysis to be $R_t=0.124\pm 0.044 (\rm{stat.}) \pm 0.007 (\rm{syst.})$.
The $S$ coefficient for \DstpDstm\ in table \ref{tab:results} is corrected for this $CP$-odd dilution.
The results for \DstpDstm\ are preliminary while those for \DstmDp, \DstpDm and \DpDm are final\cite{newresults}.

\subsection{\boldmath
  $B_s \to$ double charm status at CDF}
 
\vspace{+2mm}
\begin{flushright}
 {\it Contribution from B.~Iyutin}
\end{flushright}

With  $243 \ \rm{pb}^{-1}$ of CDF displaced track trigger \cite{YB}  data 
the ratio of branching fractions for fully reconstructed hadronic modes is measured     
$Br(B^0 \to D_s^+ D^-) / Br(B^0\to D^-3\pi)$ where 
$D^+ \to K\pi\pi$ and $D_s \to \phi\pi (K^*K, \pi\pi\pi)$. 

The optimization of selection cuts was done maximizing the significance 
of the signal Monte Carlo over combinatorial background from data. 
The signal region was blinded.

\begin{figure}
\begin{center}
\epsfig{ file = 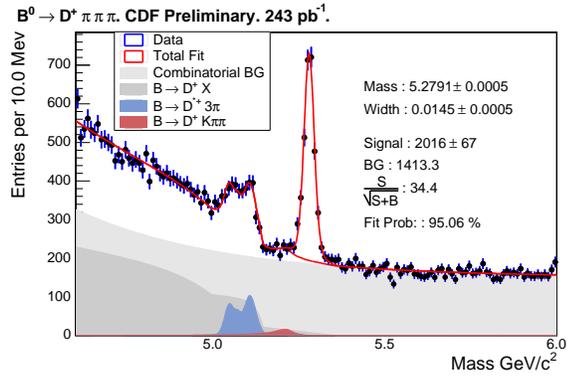, width = 0.5 \textwidth}\nolinebreak
\epsfig{ file = 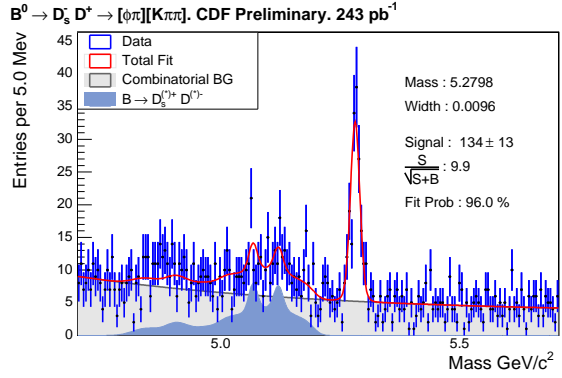, width = 0.5 \textwidth}
\epsfig{ file = 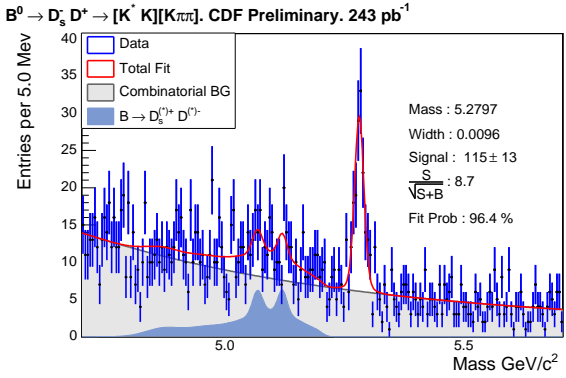, width = 0.5 \textwidth}\nolinebreak
\epsfig{ file = 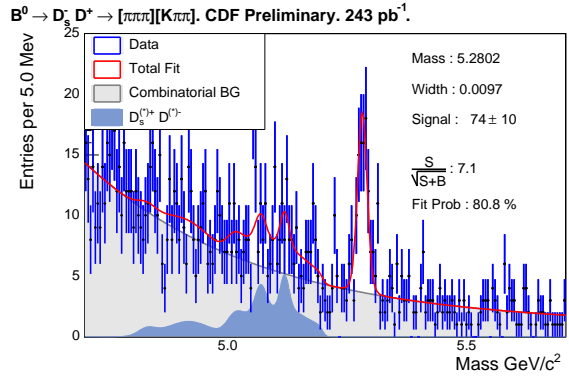, width = 0.5 \textwidth}
\end{center}
\caption{
Data fits of $B^0\to D^-3\pi$ and $B^0\to D_s^+ D^-, D_s^+ \to \phi\pi$ (top),  $B^0\to D_s^+ D^-, D_s^+ \to K^*K$ and $B^0\to D_s^+ D^-, D_s^+ \to \pi\pi\pi$ (bottom). 
}
\label{fig:fitData}
\end{figure}

 Monte Carlo simulation is used to determine shapes describing signal and partially 
reconstructed decays. The shapes were fixed while doing the mass fit 
(figure \ref{fig:fitData}). Only relative normalizations were allowed to float. 
The mass of the signal was also allowed to vary as well as parameters 
describing combinatorial background. Signal yields were obtained integrating 
 over a $\pm 2.5 \sigma$ window, where $\sigma$ means the fitted signal width.   

The analysis efficiency was extracted from signal Monte Carlo.
To reduce systematic   Monte Carlo mass distribution was fitted 
and integrated it in a $2.5 \sigma$ window the way it was done for data. 

Among the systematic uncertainties  an important one is the $B^0\to D^-3\pi $ sample 
composition uncertainty. It arises due to the poor knowledge of the $3\pi$ 
sub-resonant structure. Cut efficiency uncertainty is assigned due to the 
difference between data and Monte Carlo distributions for variables we 
cut on. A fit uncertainty is due to the fit instability with respect 
to fixed parameters. For $D_s^- D^+$, where $D_s \to 3\pi$, there is also 
a sample composition systematic due to the poor knowledge of $D_s$ 
sub-resonant structure. All of these combined are well below
the statistical uncertainty. 

 The ratios of branching fractions corresponding to the three different 
$D_s$ modes:

\vspace{0.2in}
{\large  $\frac{Br(B^0 \to~D_s^+ D^-,{D_s\to\phi\pi})}
                            { Br(B^0\to~D^-3\pi)}$}
  = $ {1.95}\pm0.20(stat) \pm 0.12 (syst) {\pm 0.49 (BR_1)}$

\vspace{0.2in}
  {\large {$\frac{Br(B^0 \to D_s D^+, {D_s\to K^*K})}
                             { Br(B^0 \to D^- 3\pi)}$}}
  =  ${1.83}\pm 0.22 (stat) \pm 0.11 (syst){\pm 0.46 (BR_1)}\pm~0.17 (BR_2)$

\vspace{0.2in}
  {\large   {$\frac{Br(B^0 \to D_s D^+, {D_s\to\pi\pi\pi})}
                             { Br(B^0 \to D^- 3\pi)}$}}
  = $ {2.46} \pm0.34  (stat) \pm 0.17 (syst) {\pm~0.62~(BR_1)}\pm~0.34~(BR_3)$

\vspace{0.2in}
Branching fraction uncertainties are quoted separately:

\begin{itemize}
\item {{$BR_1$} is due to {$Br(D_s \to \phi\pi)$}}

\item {$BR_2$} is due to {\large {$\frac{Br(D_s \to K^*K)}{Br(D_s \to \phi\pi)}$ }}

\item {$BR_3$} is due to {\large {$\frac{Br(D_s \to \pi\pi\pi)}{Br(D_s \to \phi\pi)}$ }}
\end{itemize}
where branching fractions from PDG (2004) were used.
In combining results several assumptions were made. The systematic 
uncertainty for all three modes was considered fully correlated, and  
``$BR_2$'' and ``$BR_3$'' uncertainties were assumed to be non-correlated.  
 The final value is:

\begin{eqnarray}
  \frac{Br(B^0 \to D_s^+ D^-)}{Br(B^0\to D^-3\pi)}
  =  2.00\pm 0.16 (NC) \pm0.12(syst)\pm 0.50(BR_1)  \nonumber
\end{eqnarray}
where only one number  is quoted
for the non-correlated error (marked {\em ``NC''}). 
The current world average  from the PDG (2004) is:  
\begin{eqnarray}
  \frac{Br(B^0 \to D_s^+ D^-)}{Br(B^0\to D^-3\pi)} =  1.00\pm 0.39  \nonumber
\end{eqnarray}
which is consistent with CDF result within $2\sigma$. The error 
on $D_s \to \phi\pi$ branching fraction $(3.6\% \pm 0.9\%)$, which  should be 
greatly reduced in coming years, dominates the measurement for now.   
This analysis prepares the way  to measure 
$Br(B_s\to D_s D_s)/Br(B^0 \to D_s^+ D^-)$.

\vspace{2ex}

\begin{figure}[!htb]
  \begin{center}
    \includegraphics[width=0.40\textwidth]{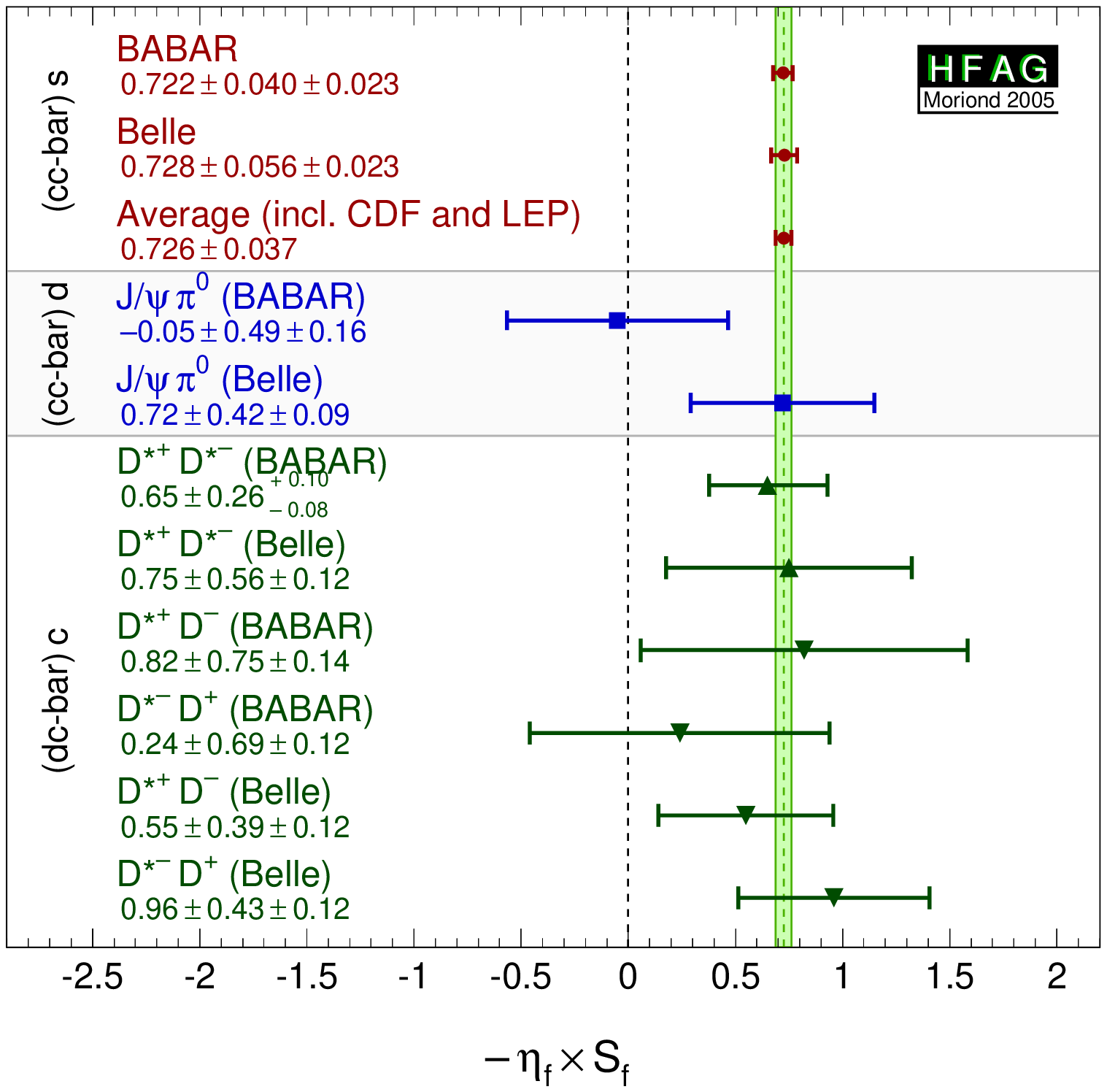}
    \includegraphics[width=0.40\textwidth]{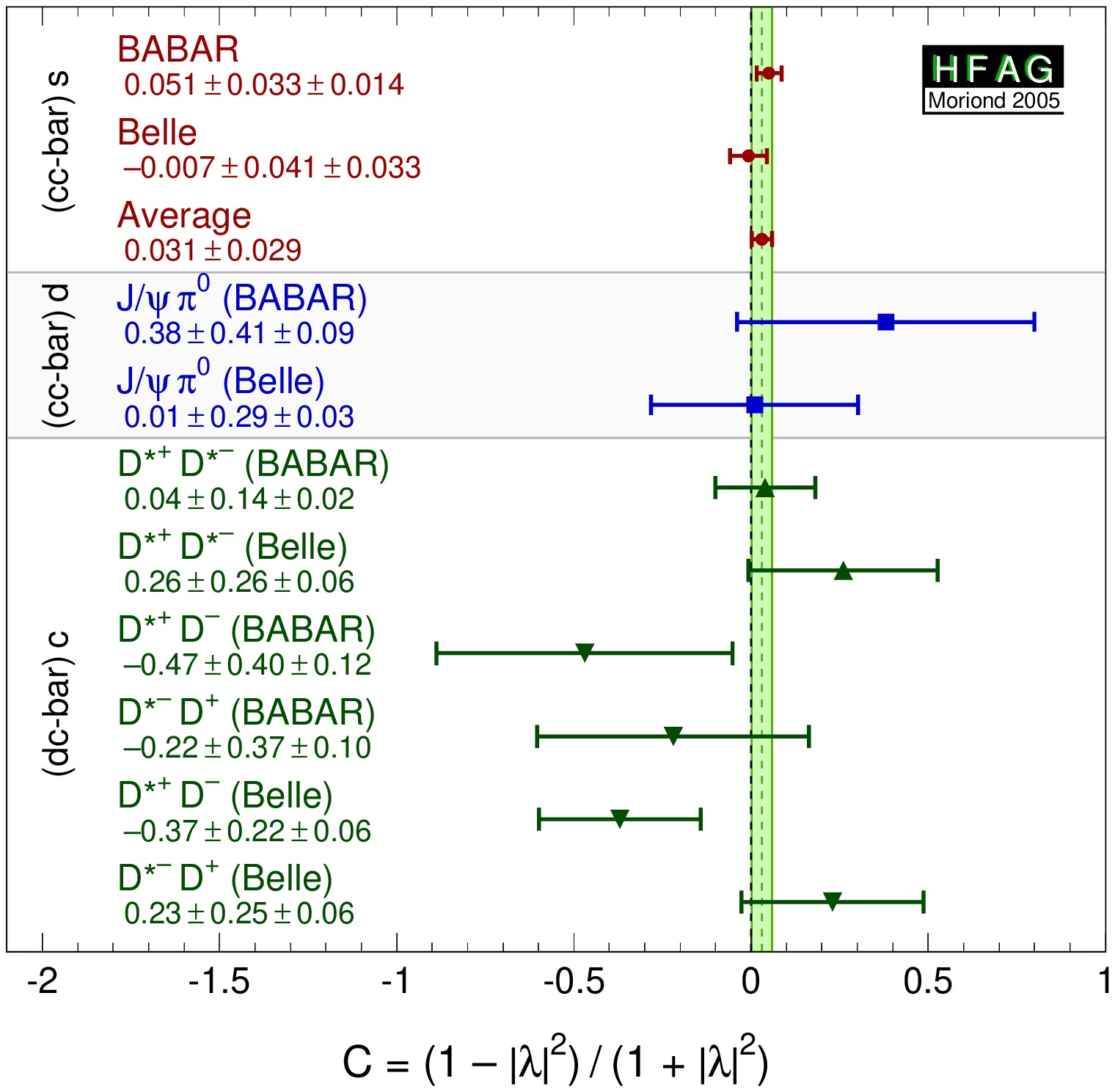}
  \end{center}
  \caption{
    \label{cpvinb2ccd}
    Summary of time-dependent $CP$ violation measurements in 
    $b \to c\bar{c}d$ transitions.
    (Left) $-\eta_f {\cal S}_{f_{CP}}$ 
    where $\eta_f$ is the $CP$ eigenvalue of the final state. 
    (Right) ${\cal C}_{f_{CP}} = - {\cal A}_{f_{CP}}$.
  }
\end{figure}

To summarize this section,
time dependent $CP$ violation in $b \to c\bar{c}d$
transitions have been measured using $B^0 \to J/\psi \pi^0$ and 
doubly charmed modes.
The measurements performed so far are summarized in 
Fig.~\ref{cpvinb2ccd}.
All these measurements are statistically limited.
Note that no direct $CP$ violation has been observed so far.
With $5\times10^8$ or $10^9$ $B\bar{B}$, 
we would expect to be able to observe non-zero $CP$ violation,
and in order to check the deviation from $b \to c\bar{c}s$ modes, 
much more data is needed.
Therefore, it is interesting to continue to hunt for new decay modes.
In order for better understanding of the $b \to d$ penguin contribution,
it is also important to search for rare decay modes
and $CP$ asymmetries in relevant charged $B$ decay modes.

\subsection{\boldmath
  $\gamma$ from $B \to D^{(*)} \Dbar^{(*)}$ and $B \to D^{(*)}_S \Dbar^{(*)}$}

\vspace{+2mm}
\begin{flushright}
  {\it Contribution from J.~Albert}
\end{flushright}

One can combine information from 
$B \to D^{(*)} \Dbar^{(*)}$ and $B \to D^{(*)}_S \Dbar^{(*)}$ branching fractions,
along with \CP asymmetry measurements in $B \to D^{(*)} \Dbar^{(*)}$, 
to obtain information on $\gamma$~\cite{DL,ADL} (see also~\cite{RF_dsds}).
An extraction of constraints on $\gamma$
may be obtained by using an SU(3) relation between the $B \to D^{(*)} \Dbar^{(*)}$ and $B \to D^{(*)}_S \Dbar^{(*)}$ decays.  
In this technique, the breaking of SU(3) is parametrized via the ratios of decay
constants $f_{D_s^{(*)}}/f_{D^{(*)}}$, which are quantities measured in lattice QCD~\cite{lattice}.
The quality of the constraint on $\gamma$ using this technique is highly dependent on the difference in the
values of the \CP asymmetries in $B \to D^{(*)} \Dbar^{(*)}$ from the value of $\sin 2\beta$ as measured in charmonium decays,
as this difference is what provides information on the additional weak phase.

\section{\boldmath
  Measurements of  $\sin(2\beta+\gamma)$/$\sin(2\phi_1+\phi_3)$}

The decay modes $\Bz \to D^{(*)\mp} \pi^{\pm}$ have been
proposed  to measure
$\sin(2\beta+\gamma)$~\cite{ref:sin2bg_th}.
In the Standard Model the decays
$\Bz \to D^{(*)+} \pi^-$ and $\Bzb \to D^{(*)+} \pi^-$
proceed through the $\overline{b} \to \overline{u}  c  d $ and
$\b\to c$ amplitudes $A_u$ and $A_c$, respectively.
The relative weak phase between these two amplitudes
is $\gamma$. When combined with $\Bz \Bzb$ mixing, this yields a weak phase
difference of $2\beta+\gamma$ between the interfering amplitudes.

The decay rate distribution for $B \to {D^{(*)}}^\pm\pi^\mp$ is described by 
an equation similar to \ref{cpasymm0}, where
the parameters $C$ and $S$ are given by
\[
C \equiv {1 - \frac{r^2}{1 + r^2}}\, , \ \ \ \
S^\pm \equiv {\frac{2 r}{1 + r^2}}\, \sin(2 \beta + \gamma \pm \delta).
\]
Here $\delta$ is the strong phase difference
between $A_u$  and  $A_c$ and $r \equiv |A_u / A_c|$.
Since $A_u$ is doubly CKM-suppressed with respect
to $A_c$, one expects $r$ to be small of order 2\%.
Due to the small value of $r$, large data samples
are required for a statistically significant measurement of $S$.

\subsection{\boldmath  
  Belle Measurements of $\sin(2\phi_1+\phi_3)$}
\vspace{+2mm}
\begin{flushright}
{\it Contribution from F.~Ronga}
\end{flushright}

Belle has probed the $CP$-violating parameter $\sin(2\phi_{1}+\phi_{3})$ 
using full reconstruction of $B^{0}\to D^{(*)}\pi$ decays~\cite{belle_dsp_full} 
and partial reconstruction of $B^{0}\to D^{*}\pi$ decays~\cite{belle_dsp_partial}.
Both analyses are based on a sample of 152~million $B\bar B$ pairs.

Belle expresses the time-dependent decay probabilities
for these modes as
\begin{eqnarray}
  P(B^0\to D^{(*)\pm}\pi^{\mp} &\approx& 
  \frac{1}{8\taub}e^{-|\dt|/\taub}[1\mp\cos(\dm\dt)-S^{\pm}\sin(\dm\dt)]\nonumber\\
  P(\overline{B}{}^0\to{}D^{(*)\pm}\pi^{\mp} &\approx& 
  \frac{1}{8\taub}e^{-|\dt|/\taub}[1\pm\cos(\dm\dt)-S^{\pm}\sin(\dm\dt)]
\end{eqnarray}
where $S^{\pm} = (-1)^{L}2\Rd\sin(2\phi_{1}+\phi_{3}\pm\delta_\dsp)$.
$L$ is the angular momentum of the final state 
(1 for $D^{*}\pi$)~\cite{fleischer2},
\Rd\ is the ratio of magnitudes of the suppressed and favoured amplitudes,
and $\delta_{\dsp}$ is their strong phase difference. It is
assumed that \Rd\ is small~\cite{rd}.

In the full reconstruction analysis, 
the following decay chains are reconstructed
(charge conjugate modes are included):
$B\to D^{*+}\pi^{-}, D^{*+}\to D^{0}\pi^{+}_{s}, 
D^{0}\to K^{-}\pi^{+}, K^{-}\pi^{+}\pi^{0}, D^{0}\to K^{-}\pi^{+}\pi^{+}\pi^{-}$
and 
$B\to D^{+}\pi^{-}, D^{+}\to K^{-}\pi^{+}\pi^{+}$.
The event yields are summarized in Table~\ref{full-yield-tab}.

\begin{table}[htb]
\begin{center}
  \begin{tabular}{lccc}
    \hline
    Decay mode   & Candidates & Selected & Purity \\
    \hline
    $B \to D\pi$     & 9711	 & 9351		& 91\%\\
    $B \to D^{*}\pi$ & 8140	 & 7763		& 96\% \\
    \hline
  \end{tabular}
  \caption{
    \label{full-yield-tab}
    Number of reconstructed candidates, 
    selected candidates (after tagging and vertexing) and purity, 
    extracted from the fit to $(\DE,\Mbc)$.
  }
\end{center}

\end{table}
The standard Belle tagging algorithm is used to identify the flavour
of the accompanying $B$ meson. 
It returns the flavour and a tagging quality $r$ 
used to classify events in six bins. 
The standard Belle vertexing algorithm is then used to
obtain the proper-time difference $\dt$. 
A fit to the $\dt$ distribution of events in the $(\DE,\Mbc)$ side-band
region is performed to get the parameters of the background PDFs.
A lifetime fit is the performed on all components together to get the
resolution parameters; the wrong-tag fractions for the six $r$ bins
are also determined from the data.

\begin{figure}[!htb]
  \begin{minipage}{0.24\textwidth}
    \includegraphics[width=\textwidth]{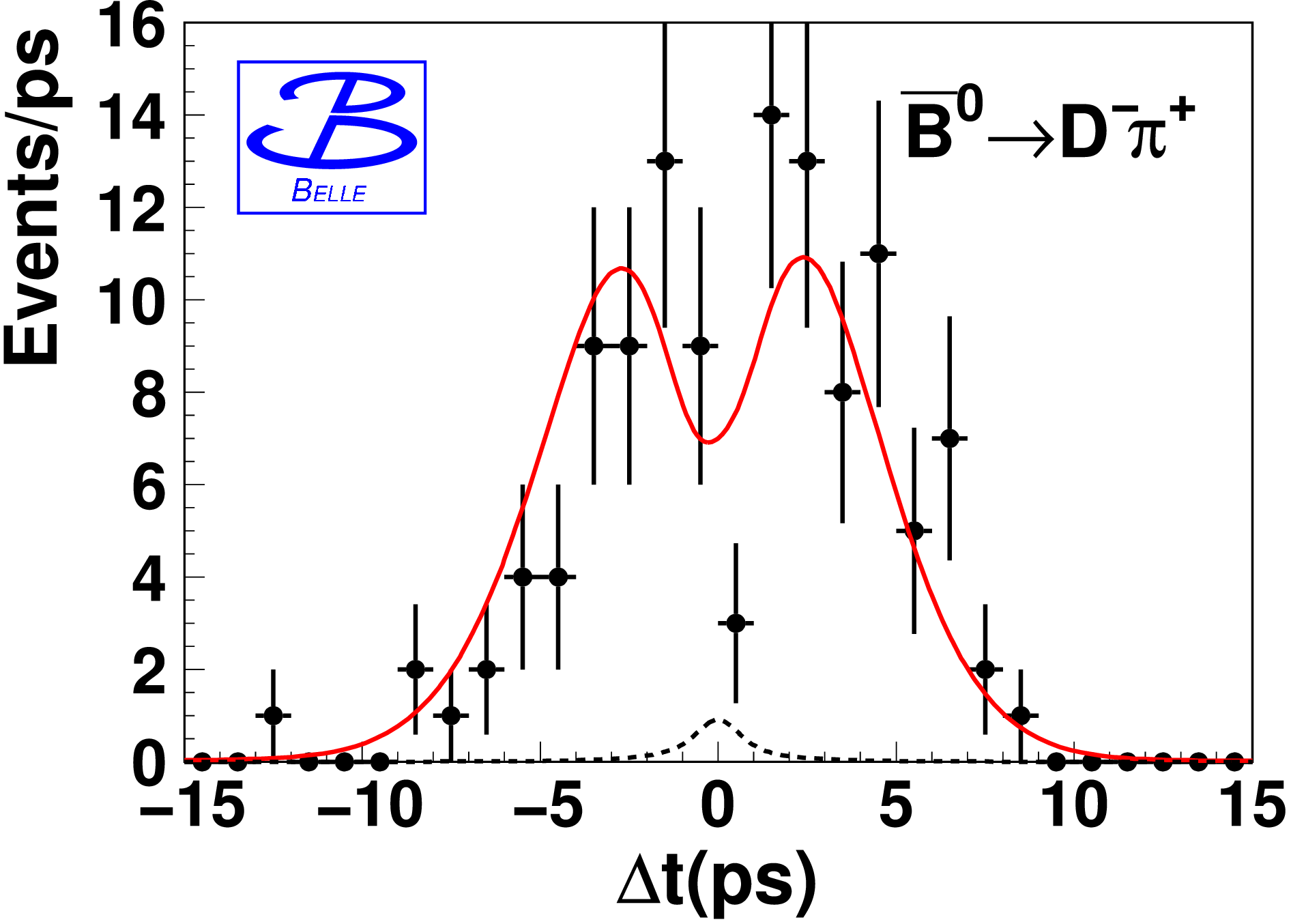}
  \end{minipage}
  \begin{minipage}{0.24\textwidth}
    \includegraphics[width=\textwidth]{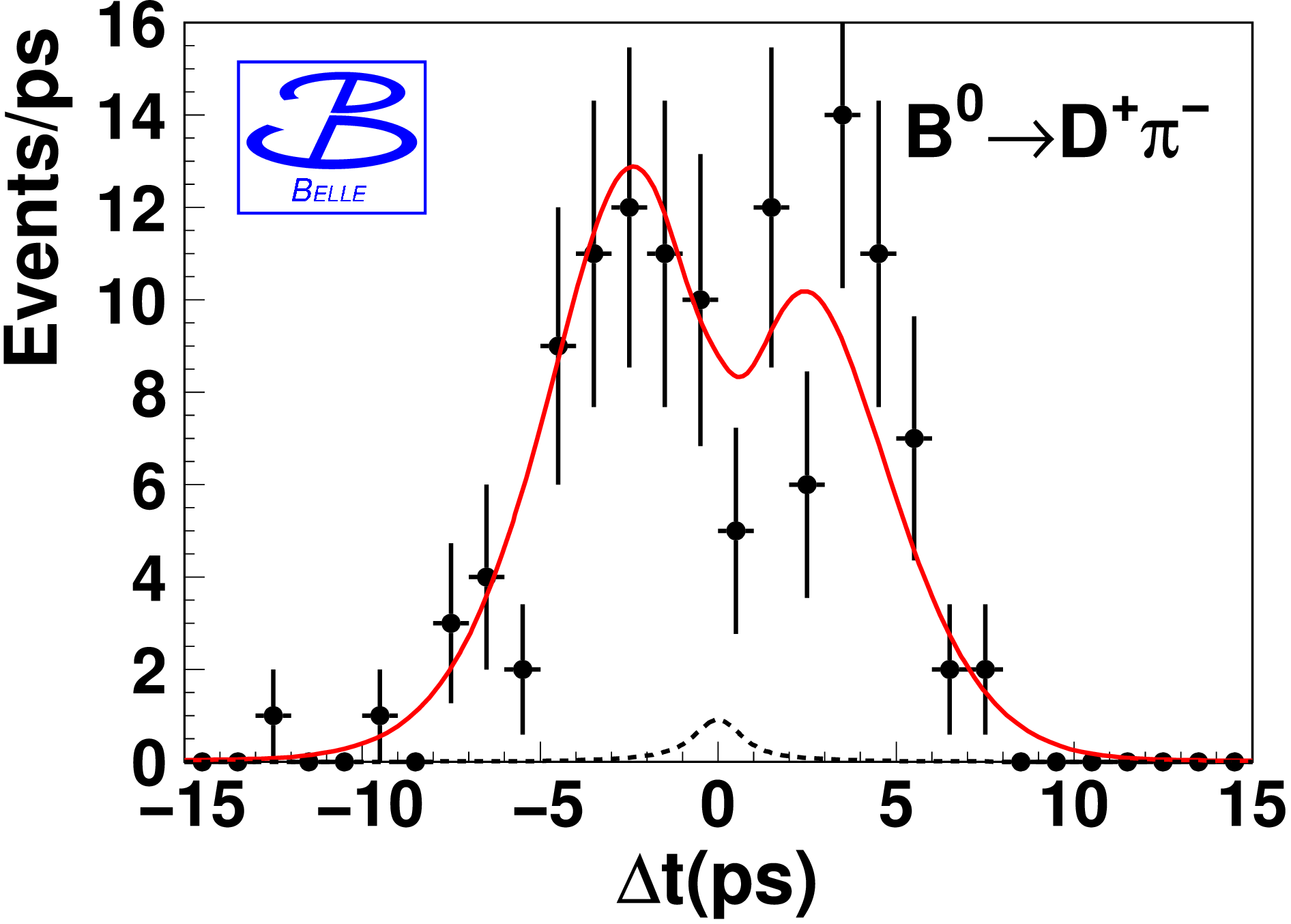}
  \end{minipage}
  \begin{minipage}{0.24\textwidth}
    \includegraphics[width=\textwidth]{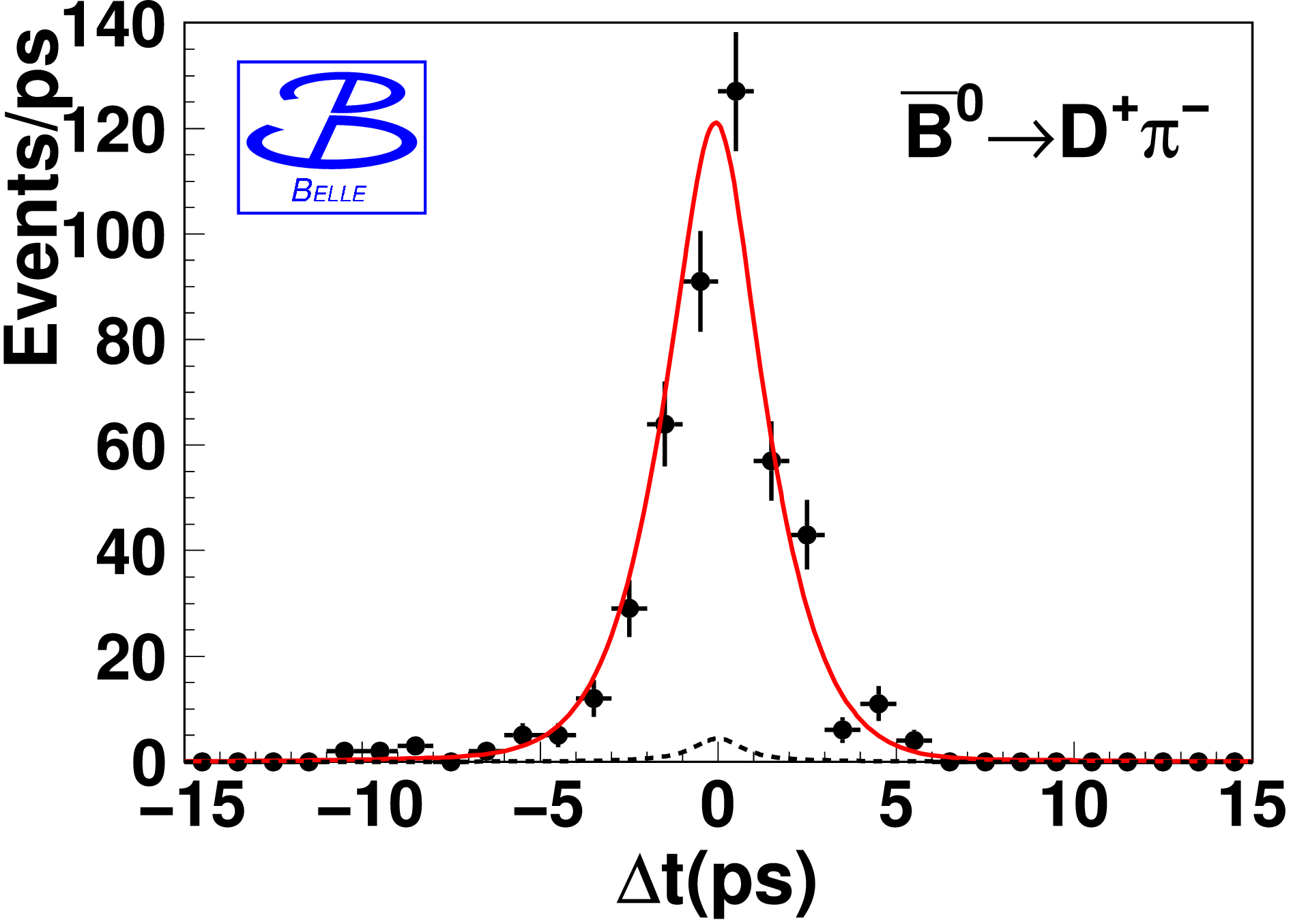}
  \end{minipage}
  \begin{minipage}{0.24\textwidth}
    \includegraphics[width=\textwidth]{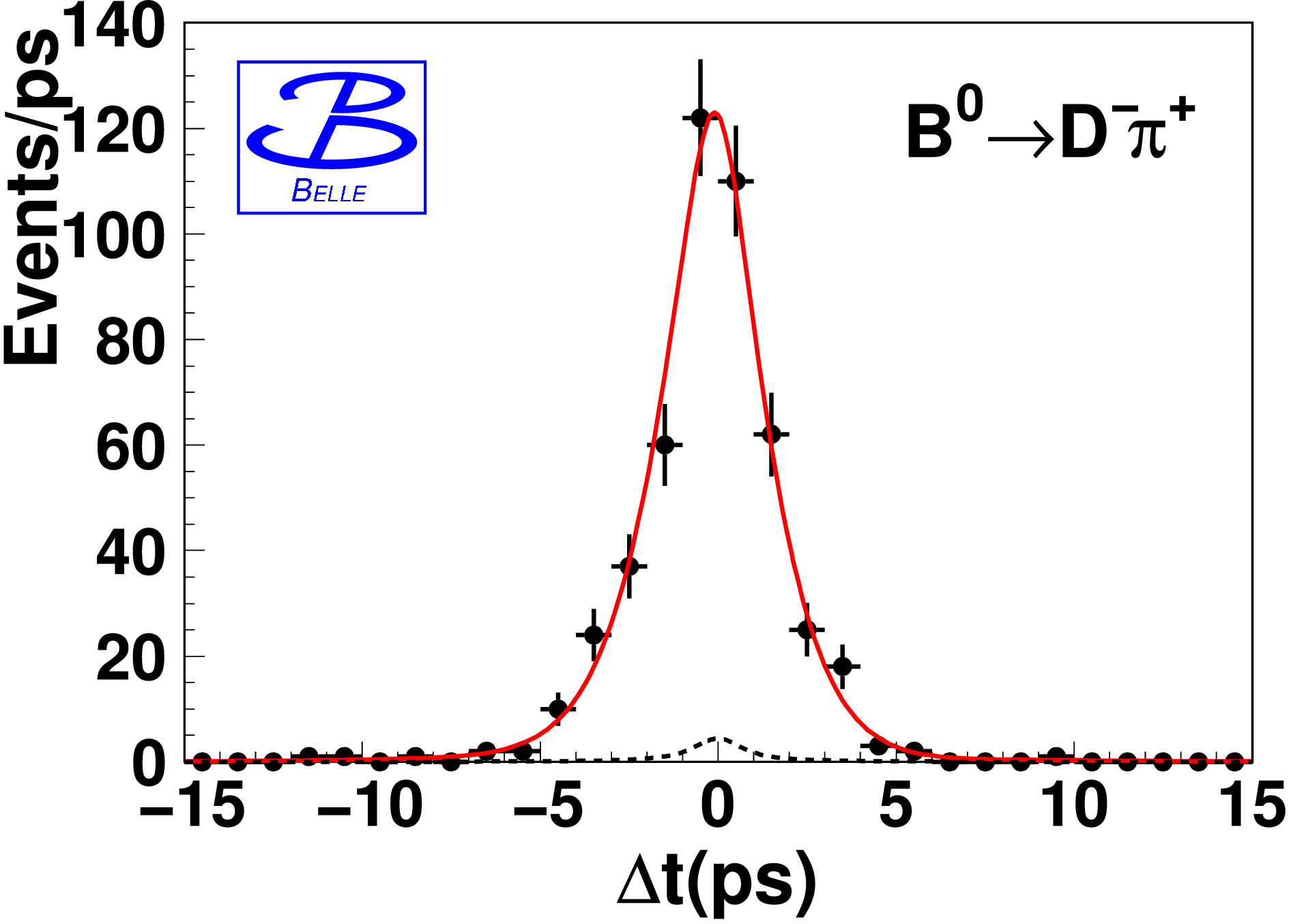}
  \end{minipage}
  
  \begin{minipage}{0.24\textwidth}
    \includegraphics[width=\textwidth]{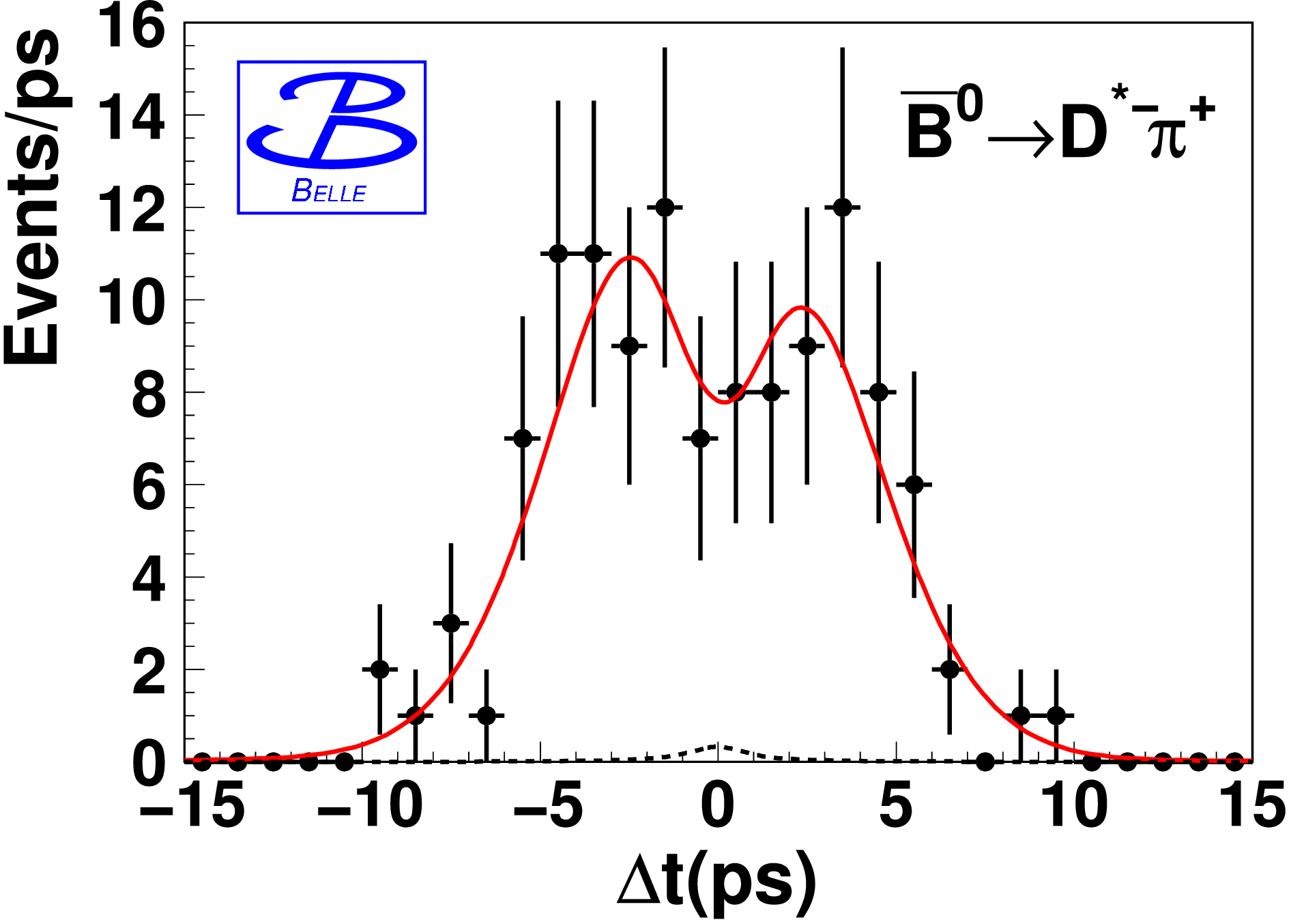}
  \end{minipage}
  \begin{minipage}{0.24\textwidth}
    \includegraphics[width=\textwidth]{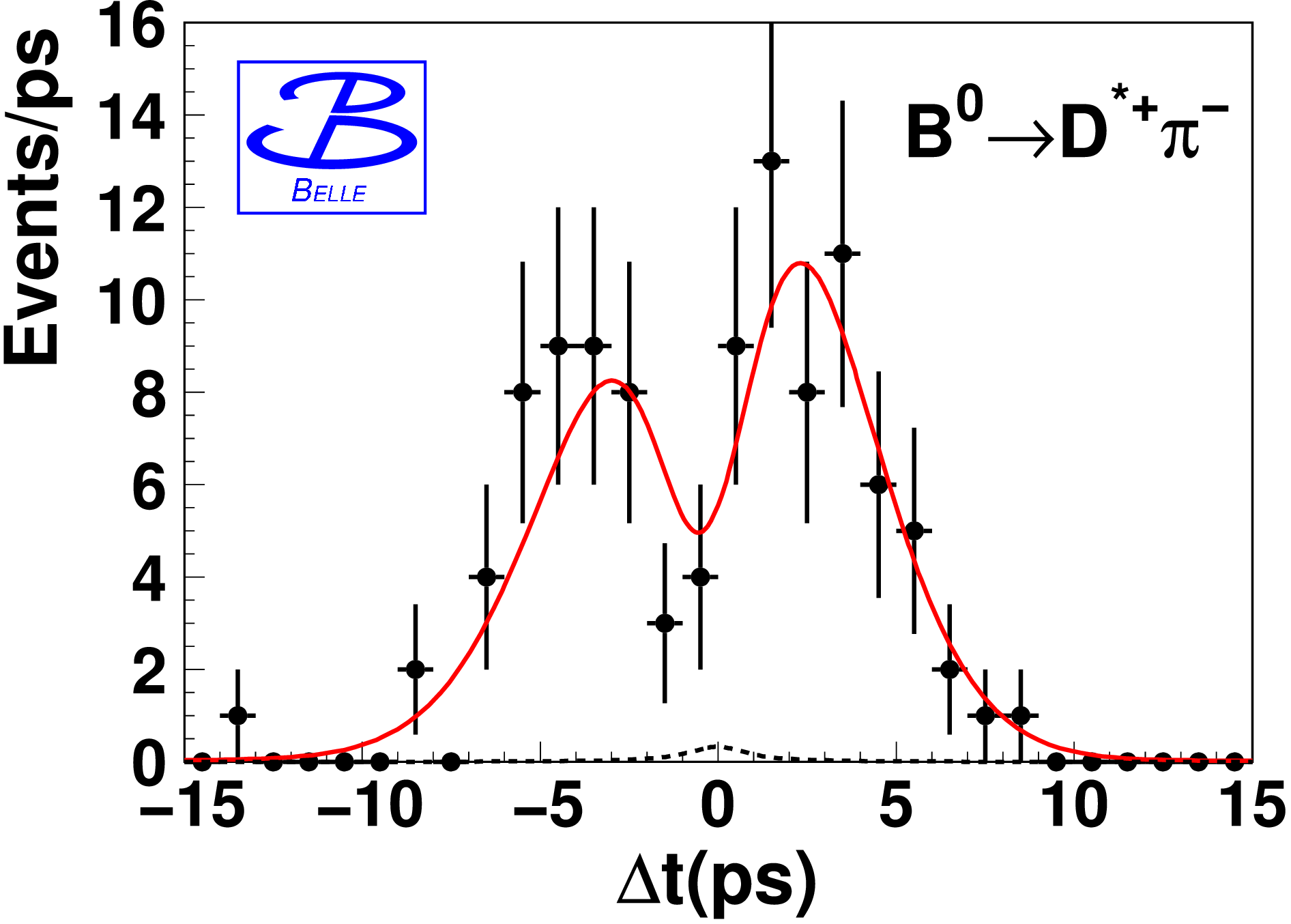}
  \end{minipage}
  \begin{minipage}{0.24\textwidth}
    \includegraphics[width=\textwidth]{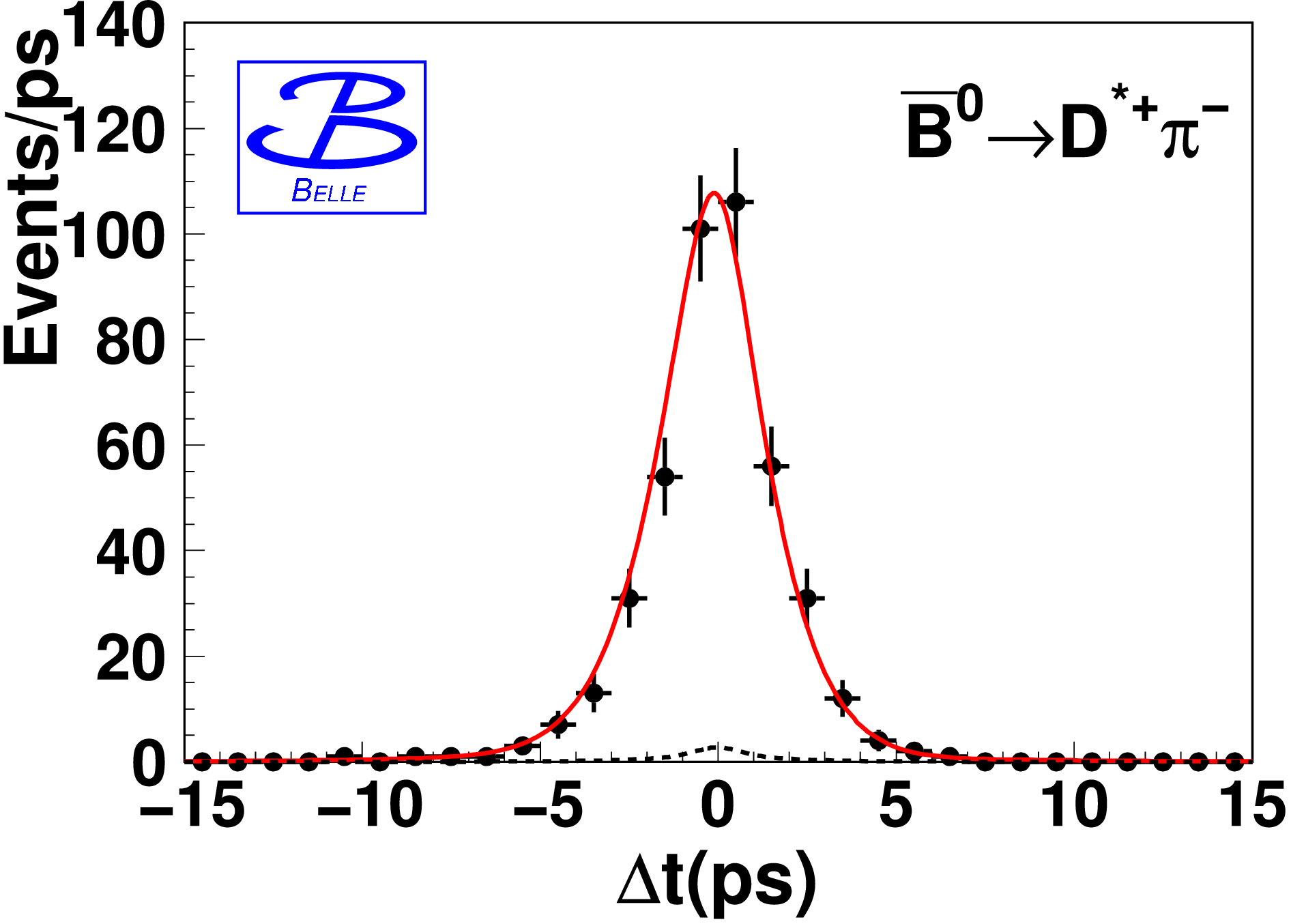}
  \end{minipage}
  \begin{minipage}{0.24\textwidth}
    \includegraphics[width=\textwidth]{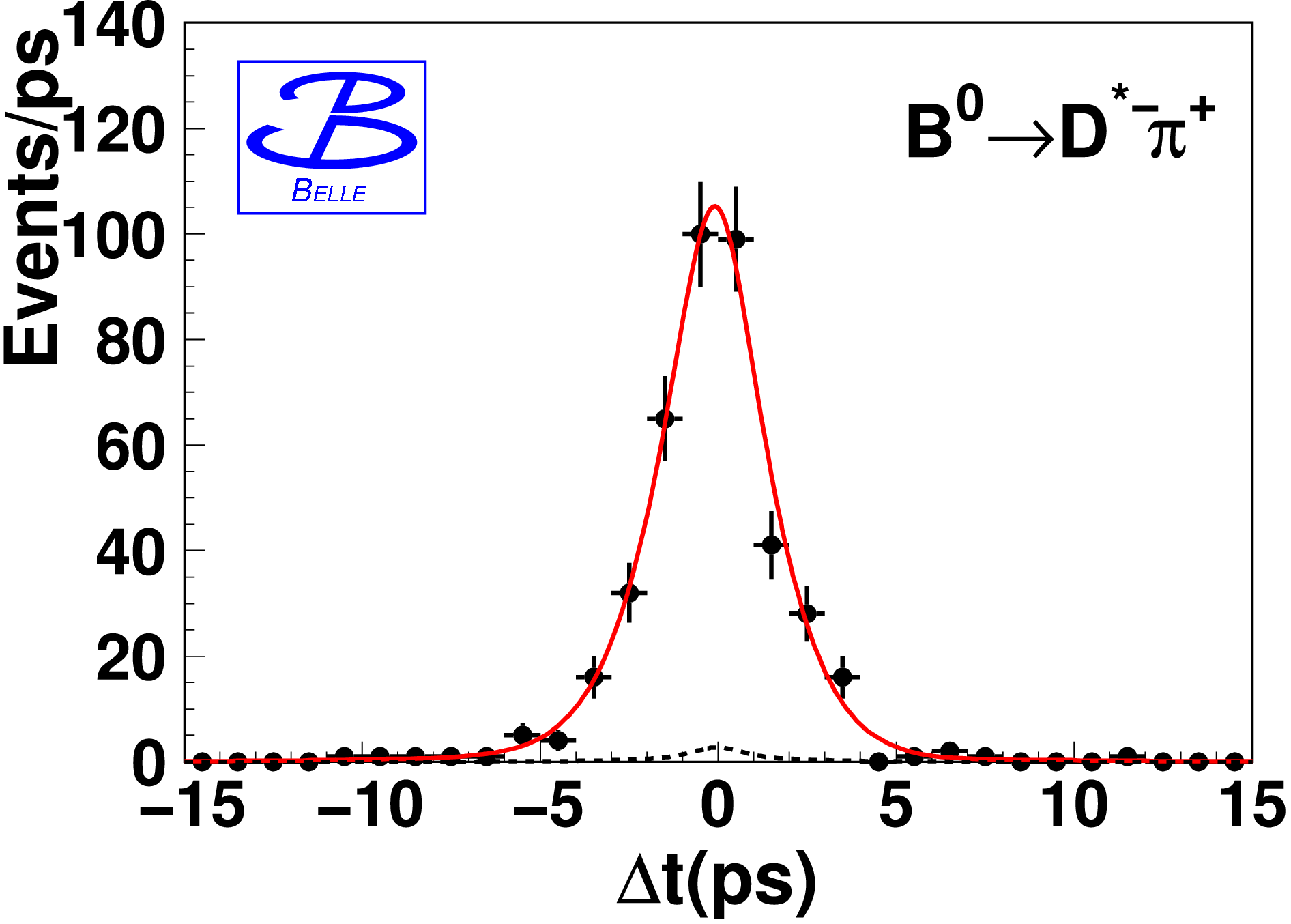}
  \end{minipage}
  \caption{
    \label{full-fit-fig}
    $\dt$\ distributions for the $D\pi$ events (top) 
    and $D^{*}\pi$ (bottom) events with
    the best quality flavour tagging.
  }
\end{figure}

The $S^{\pm}$ have to be corrected to take into account possible 
tag-side interference due to tagging on $B^{0}\to DX$ decays. 
Effective corrections $\{S^{\pm}_{\mathrm{tag}}\}^{\mathrm{eff}}$ 
are determined for each $r$~bin by a fit to 
fully reconstructed $D^{*}\ell\nu$ events, 
where the reconstructed side asymmetry is known to be zero. 
Finally, a fit is performed to determine $S^{\pm}$, 
with \dm\ and \taub\ fixed to the world average, 
and the wrong-tag fractions and $\{S^{\pm}_{\mathrm{tag}}\}^{\mathrm{eff}}$
for each $r$ bin fixed to the values determined previously. 
We obtain:
\begin{eqnarray}
 2R_{D\pi}\sin(2\phi_{1}+\phi_{3}+\delta_{D\pi}) = 0.087 \pm 0.054 \stat \pm 0.018 \syst 
 \nonumber \\
 2R_{D\pi}\sin(2\phi_{1}+\phi_{3}-\delta_{D\pi}) = 0.037 \pm 0.052 \stat \pm 0.018 \syst 
 \nonumber \\
 2R_{D^{*}\pi}\sin(2\phi_{1}+\phi_{3}+\delta_{D^{*}\pi}) = 0.109 \pm 0.057 \stat \pm 0.019 \syst
 \nonumber \\
 2R_{D^{*}\pi}\sin(2\phi_{1}+\phi_{3}-\delta_{D^{*}\pi}) = 0.011 \pm 0.057 \stat \pm 0.019 \syst
\end{eqnarray}
The systematic errors come from the uncertainties of parameters that
are constrained in the fit and uncertainties on the tagging side asymmetry.
The result of the fit for the sub-samples having the best 
quality flavour tagging is shown on Fig.~\ref{full-fit-fig}.

The partial reconstruction of $B \to D^{*+}\pi_f$, $D^{*+}\to D^{0}\pi_{s}$
is performed by requiring a fast pion $\pi_{f}$ and a slow pion $\pi_{s}$,
without any requirement on the $D^{0}$. 
The candidate selection exploits the 2-body kinematics of the decay 
using 3 variables: the fast pion CM momentum $p^{*}_{f}$; 
the cosine of the angle between the fast pion direction
and the opposite of the slow pion direction in the CM ($\cos\delta_{fs}$);
the angle between the slow pion direction and the opposite of the $B$
direction in the $D^{*}$ rest frame ($\cos\theta_\mathrm{hel}$). 
Yields are extracted from a 3D fit
to these variables (see Table~\ref{part-rec-tab}). 
The flavour of the
accompanying $B$ meson is identified by a fast lepton, $\ell_{\mathrm{tag}}$.
The proper time $\dt$ is obtained from the $z$ coordinate
of $\pi_{f}$ and $\ell_\mathrm{tag}$ constrained to the
$B$-lifetime smeared beam profile.

The resolution function is modeled by a convolution of three Gaussians
whose parameters are determined by a fit to a $J/\psi\to\mu^{+}\mu^{-}$
sample selected the same way as the signal sample. 
In order to correct for possible bias due to 
tiny misalignment in the tracking devices that would mimic $CP$ violation, 
the mean of the Gaussian resolution is allowed to be slightly offset. 
The vertex position is also corrected to account for possible misalignments. 
A fit to events where the two pions have same sign is performed 
to determine the shape of uncorrelated background $\dt$ distribution, 
while a fit to the  $(p^{*}_{f},\cos\theta_\mathrm{hel})$ side-bands 
provides that of the correlated background. 
The $D^{*}\rho$ component is modeled the same
way as the signal, with $S^{\pm}$ fixed to zero.

A fit for \dm\ and \taub\ is performed to check the fit procedure.
A fit to a $D^*\ell\nu$ sample selected similarly to the signal sample
is performed to check the bias correction. 
Several MC samples with $S^{\pm} \neq 0$ are fitted to 
check possible fit bias in the extraction of $S^{\pm}$.

\begin{figure} [!htb]
\begin{center}
  \begin{minipage}{0.33\textwidth}
    \includegraphics[width=\textwidth]{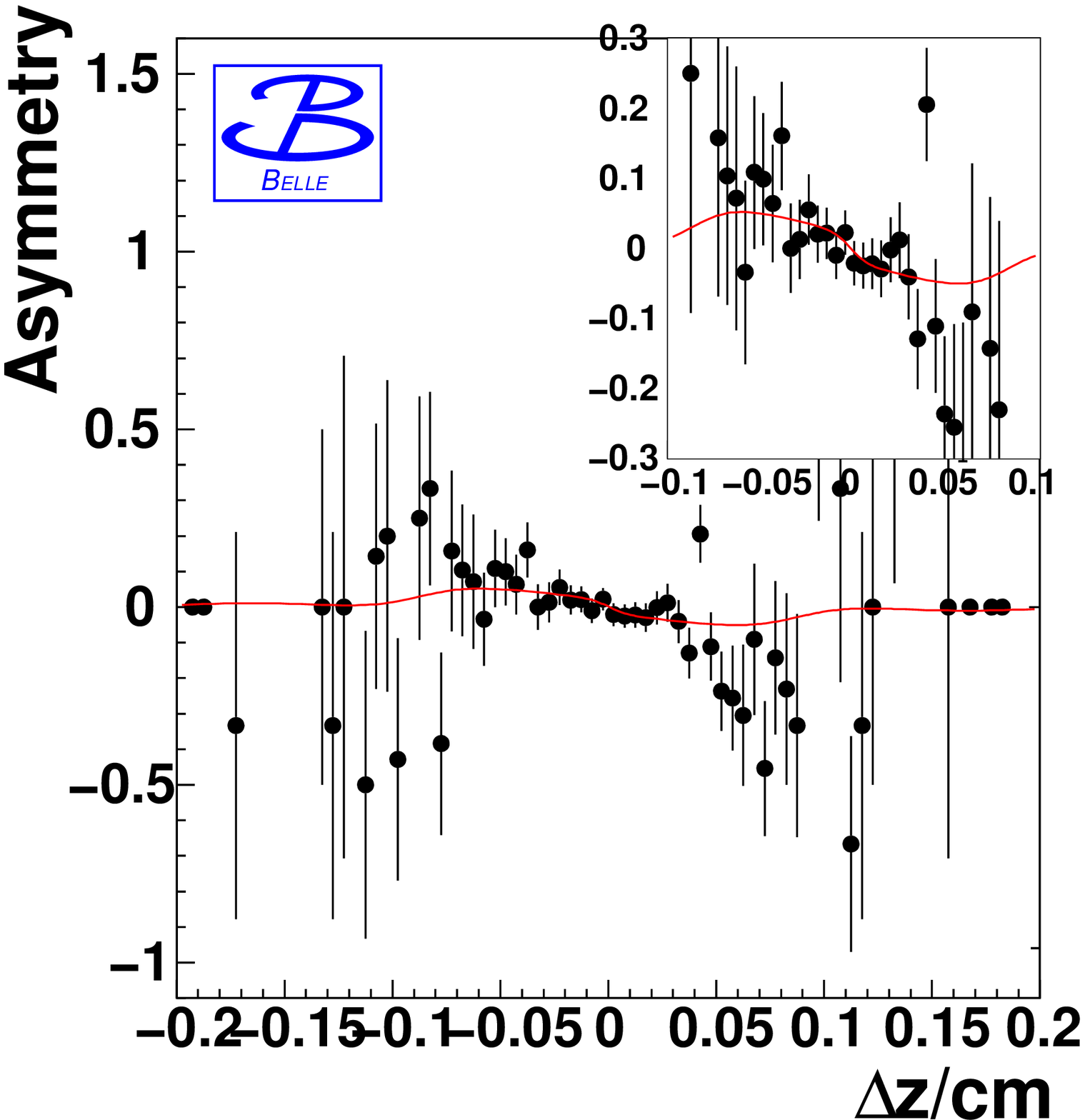}
  \end{minipage}\
  \begin{minipage}{0.33\textwidth}
    \includegraphics[width=\textwidth]{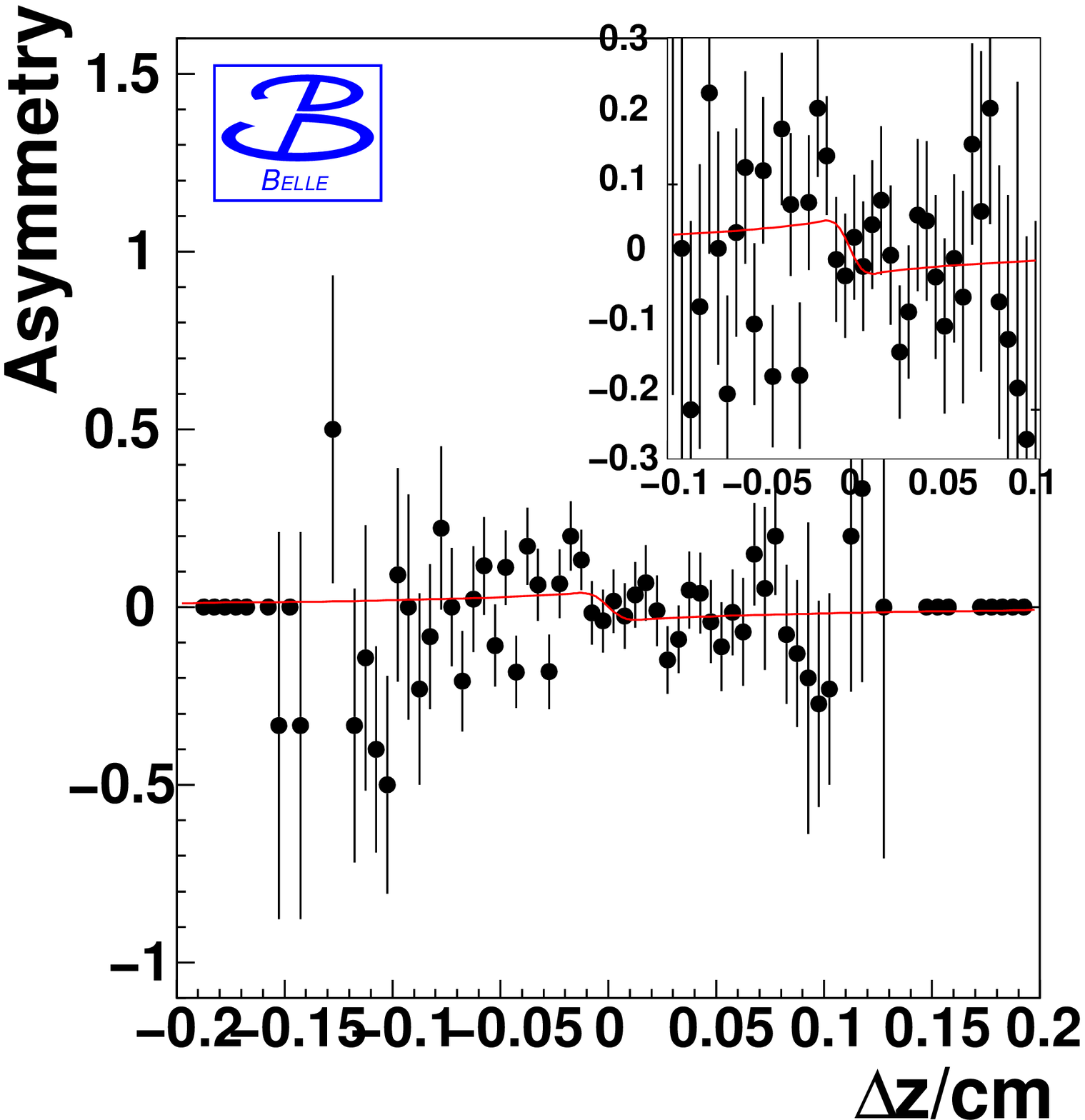}
  \end{minipage}
 \caption{
   \label{part-fit-fig}
   Results of the partial reconstruction fit shown as
   asymmetries for (left) SF events and (right) OF events. 
   The inset plots magnify the region around $\dz=0$
 }
\end{center}

\end{figure}

An unbinned maximum likelihood fit with \dm\ and \taub\ fixed to the
world average, and $S^{\pm}$, $\dt$\ offsets and wrong-tag fractions
floated yields:
\begin{eqnarray}
  2R_{D^{*}\pi}\sin(2\phi_{1}+\phi_{3}+\delta_{D^{*}\pi})= -0.035 \pm 0.041 \stat \pm 0.018 \syst 
  \nonumber \\
  2R_{D^{*}\pi}\sin(2\phi_{1}+\phi_{3}-\delta_{D^{*}\pi})= -0.025 \pm 0.041 \stat \pm 0.018 \syst
\end{eqnarray}
The main systematic errors come from the background fractions, the
background shapes, the resolution function and the offsets. The result
of the fit is shown as asymmetries on Fig.~\ref{part-fit-fig}.
Asymmetries are defined as:
\begin{eqnarray}
  {\cal A}^{\mathrm{SF}} &=& 
  (N_{\pi^{-}\ell^{-}}-N_{\pi^{+}\ell^{+}})/(N_{\pi^{-}\ell^{-}}+N_{\pi^{+}\ell^{+}}) \nonumber \\
  {\cal A}^{\mathrm{OF}} &=& 
  (N_{\pi^{+}\ell^{-}}-N_{\pi^{-}\ell^{+}})/(N_{\pi^{+}\ell^{-}}+N_{\pi^{-}\ell^{+}}) \nonumber
\end{eqnarray}

\begin{table}
\begin{center}
  \begin{tabular}{lrrrrr}
    \hline
    Mode & Data & Signal & $D^{*}\rho$	& Corr. bkg	& Uncorr. bkg \\
    \hline
    \hline
    SF	  & 2823 & 1908	  & 311		& ---		& 637         \\
    OF	  & 10078& 6414	  & 777		& 928		& 1836        \\
    \hline
  \end{tabular}
  \caption{
    \label{part-rec-tab}
    Fit yield for the signal and the various types of
    background in same-flavour (SF) and opposite-flavour (OF)
    events.}
 \end{center}
\end{table}

Increase of the available data and addition of more modes
in the full reconstruction, as well as tuning of the selection
and vertexing on more Monte Carlo and data, will help reduce
both statistical and systematic errors. 
A reduction by a factor 0.3 for the former and 0.5 for the latter 
is expected with 1~ab${}^{-1}$.

\subsection{\boldmath
  Measurements of $\sin(2\beta+\gamma)$ at BaBar}

\subsubsection{\boldmath
  Status and Prospects for \CP Asymmetry Measurements}
 
\vspace{+2mm}
\begin{flushright}
 {\it Contribution from S.~Ganzhur}
\end{flushright}

Two different analysis techniques, full reconstruction~\cite{ref:run1-2-breco} were used for 
and partial reconstruction~\cite{ref:run1-2-ihbd}  the $\sin(2\beta+\gamma)$ measurement with 
$\Bz \to D^{(*)\mp} \pi^{\pm}$.

The full reconstruction technique is used to measure the \CP asymmetry in 
$\Bz \to D^{(*)\mp} \pi^{\pm}$ and  $\Bz \to D^{*\mp} \rho^{\pm}$ decays. 
The result with 110 million \BB pairs is 
\begin{eqnarray}
2 r^{D\pi} \sin(2\beta+\gamma) \cos \delta^{D\pi} &=& -0.032\pm0.031\pm 0.020 \, \nonumber \\
2 r^{D\pi} \cos(2\beta+\gamma) \sin \delta^{D\pi} &=& -0.059\pm0.055\pm 0.033 \, \nonumber \\
2 r^{D^*\pi} \sin(2\beta+\gamma) \cos \delta^{D^*\pi} &=& -0.049\pm0.031\pm 0.020 \, \nonumber \\
2 r^{D^*\pi} \cos(2\beta+\gamma) \sin \delta^{D^*\pi} &=& +0.044\pm0.054\pm 0.033 \, \nonumber \\
2 r^{D\rho} \sin(2\beta+\gamma) \cos \delta^{D\rho} &=& -0.005\pm0.044\pm 0.021 \, \nonumber \\
2 r^{D\rho} \cos(2\beta+\gamma) \sin \delta^{D\rho} &=& -0.147\pm0.074\pm 0.035,
\label{math:dstpi_fullreco}
\end{eqnarray}
where the first error is statistical and the second is systematic. 
The systematic error for  $\Bz \to D^{*\mp} \rho^{\pm}$ includes the maximum bias of
asymmetry parameters due to possible dependence of $r$ on the $\pi\pi^0$ invariant mass. 
For the measurement of $2r\cos(2\beta+\gamma)\sin\delta$ parameter only the lepton-tagged events are used due to 
a presence of tag-side \CP violation effect~\cite{ref:abc}.

\begin{figure}
\begin{minipage}{0.48\textwidth}
         \includegraphics[width=0.95\textwidth]{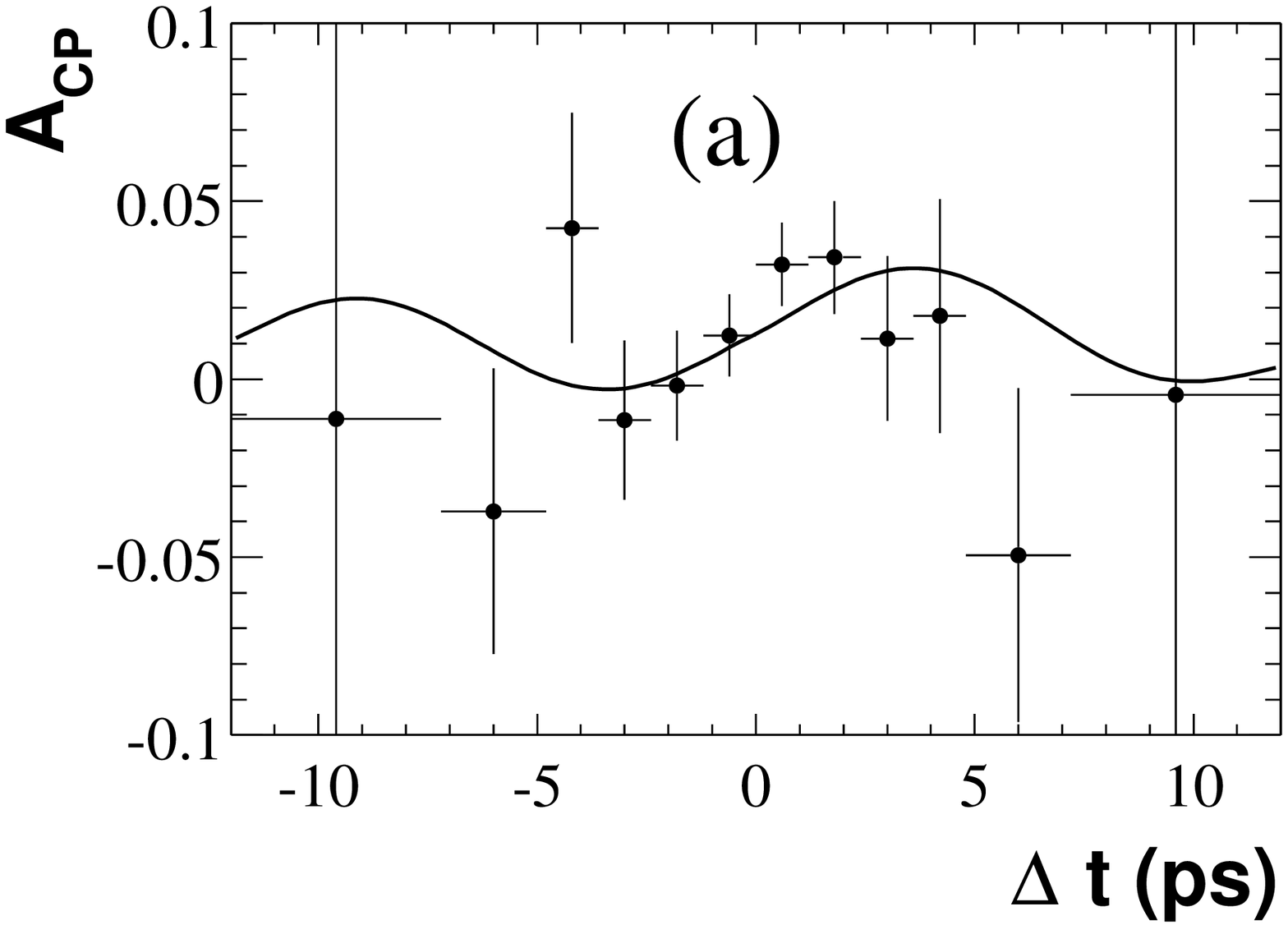}
\end{minipage}
\hfill
\begin{minipage}{0.48\textwidth}
    \includegraphics[width=0.95\textwidth]{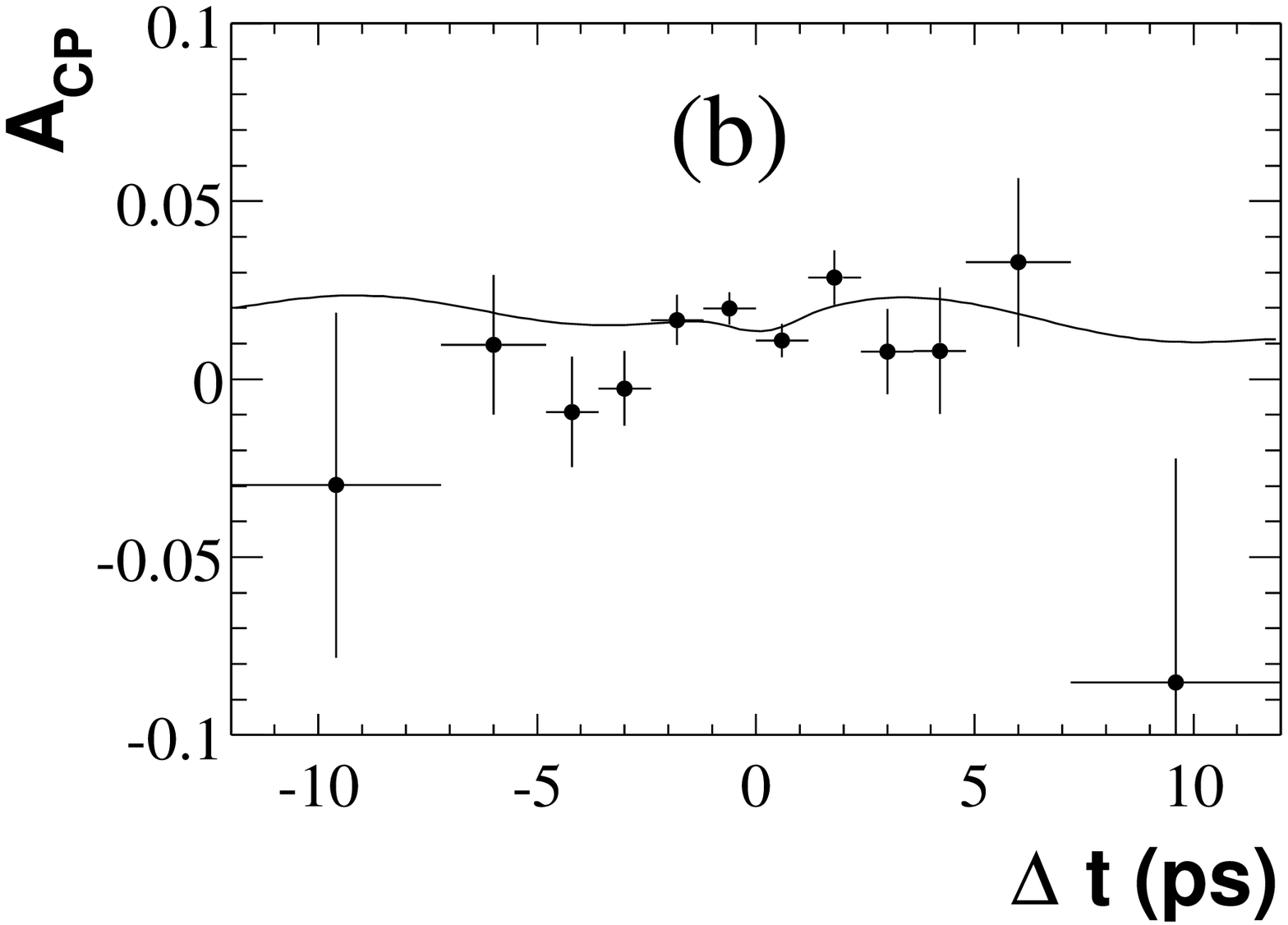}
\end{minipage}
\caption{Raw asymmetry for (a) lepton-tagged and (b) kaon-tagged
  events of $\btodstpipm$ decay mode using the method of the partial reconstruction. 
  The curves represent  the projections of the PDF for the raw
  asymmetry.}
\label{fig:sin2bg_asym}
\end{figure}

In the partial reconstruction of a $\btodstpipm$ candidate,
only the hard (high-momentum) pion track $\pi_h$ from the $B$ decay and the
soft (low-momentum) pion track $\pi_s$ from the decay
$D^{*-}\to \Dzb \pi_s^-$ are used.
Applying kinematic constraints consistent with the signal decay mode,
the four-momentum of the non-reconstructed, ``missing''
$D$ is calculated. Signal events are peaked
in the $\mmiss$ distribution at the nominal $\Dz$ mass.
This method eliminates the efficiency loss associated with the
neutral $D$ meson reconstruction. The \CP asymmetry independent 
on the assumption on $r^*$ measured with this technique by \babar\ using 232 million produced \BB pairs is
\begin{eqnarray}
2 r^{D^*\pi} \sin(2\beta+\gamma) \cos \delta^{D^*\pi} = -0.034\pm0.014\pm 0.009 \, \nonumber \\
2 r^{D^*\pi} \cos(2\beta+\gamma) \sin \delta^{D^*\pi} = -0.019\pm0.022\pm 0.013,
\label{math:dstpi_partial}
\end{eqnarray}                                                                                                                                        
where the first error is statistical and the second is systematic. This measurement deviates from zero by 2.0 standard deviations.
Figure~\ref{fig:sin2bg_asym} shows the raw, time-dependent \CP asymmetry
\[
A(\dt) = \frac{N_{\Bz}(\dt) - N_{\Bzb}(\dt)}{N_{\Bz}(\dt) + N_{\Bzb}(\dt)}
\]
In the absence of background and with high statistics, perfect tagging, and
perfect $\dt$ measurement, $A(\dt)$ would be a sinusoidal
oscillation with amplitude $2r\sin(2\beta+\gamma)\cos\delta$.

Recently it was proposed to consider the $\Bz \to D^{(*)\mp} a_{0(2)}^{\pm}$ decays for measurement of $\sin(2\beta+\gamma)$~\cite{ref:th-dsta0_a2}.
The decay amplitudes of \B mesons to light scalar or tensor mesons such as $a_0^+$ or $a_2^+$, emitted from a weak current, are significantly suppressed 
due to the small decay constants $f_{a_{0(2)}}$. Thus, the absolute value of the CKM-suppressed and favored amplitudes become comparable and the \CP asymmetry in such decays is expected to be large. However, the theoretical predictions of the branching fractions for  $\Bz \to D^{(*)\mp} a_{0(2)}^{\pm}$
is expected of the order of $(1\div4)\cdot 10^{-6}$~\cite{ref:br-dsta0_a2}. 
 One way to verify the expectations and test a validity of the factorization approach is to measure the branching  fractions for the more abundant decay modes $\Bz\to D_s^{(*)+}a_{0(2)}$. Using a sample of about 230 millon $\FourS\to\BB$ no evidence for these decays were observed. 
This allowed one to set  upper limits at 90\% C.L. on the branching fractions $\BR(\Bz\to D_s^+a^-_0)<4.0\times10^{-5}$ and  $\BR(\Bz\to D_s^+a^-_2)<2.5\times10^{-4}$.

The decay modes $\Bzb\to D^{(*)0}\bar{K}^0$ have been proposed for determination of $\sin(2\beta+\gamma)$ from measurement 
of time-dependent \CP asymmetries~\cite{GronauLondon,ref:th-dst0k0,fleischer1,fleischer2}. 
In the Standard Model the decays of \Bz and \Bzb mesons into final state 
$D^{(*)0}\KS$ proceed through the $\b \to \c$ and $\overline{b}\to \overline{u}$ amplitudes, respectively. Due to relatively large \CP asymmetry 
($r\equiv|A(\Bzb\to \bar{D}^{(*)0}\bar{K}^0)|/|\Bzb\to D^{(*)0}\bar{K}^0)|\simeq 0.4$) these decay channels look very  attractive for such a measurement. 
Since the parameter $r$ can be measured by fitting the $C$ coefficient  in time distributions, the measured asymmetry 
can be interpreted in terms of $\sin(2\beta+\gamma)$ without additional assumptions. However, the branching fractions of such decays are relatively small 
$\sim5\cdot10^{-5}$ and a large data sample is therefore still required. Moreover, the \B decay dynamics can lower the expectation for the ratio $r$. The magnitude 
of this ratio can be probed by measuring the rate for the decays  $\Bzb\to D^{(*)0}\bar{K}^{*0}$ and  $\Bzb\to \bar{D}^{(*)0}\bar{K}^{*0}$. From the measured 
branching fractions~\cite{ref:dst0kst0}, one obtains $r<0.8$ at the 90\% C.L. from a central value of $r=0.4\pm0.2(stat.)\pm0.2(syst.)$

\subsubsection{\boldmath
  $2\beta + \gamma$ from $B^0 \to D^+ K^0 \pi^-$ decays.}

\vspace{+2mm}
\begin{flushright}
 {\it Contribution from F.~Polci, M.-H.~Schune \& A.~Stocchi}
\end{flushright}

The use of the $B^0 \to D^+ K^0 \pi^-$ decays for the extraction of $2\beta + \gamma$
has been proposed  in \cite{Aleksan:2002mh}. 
In these decays the sensitivity to the weak phase $2\beta + \gamma$ comes from
the interference between the Cabibbo allowed and the Cabibbo suppressed  
amplitudes leading to the same final state through the $B^0$\textendash$\bar{B}^0$ mixing. 
Here the main advantage comes from the possibility of performing a 
Dalitz analysis, which allows to reduce the eight fold ambiguity in 
the determination of $2\beta + \gamma$  
to only a two fold ambiguity~\cite{Aleksan:2003fm}.

To explore the potentiality of this approach, 
a study of the sensitivity to $2\beta + \gamma$ has been performed. 
The distribution of the invariant masses 
$M_{K^0\pi^-}$ and $M_{D^+ K^0}$ has been parametrized 
with a model where the decay amplitude in each point $i$ of the 
Dalitz plot is a linear combination with complex parameters of 
resonances described by Breit-Wigner functions.
This model realistically reproduces the distribution obtained by 
the BaBar collaboration~\cite{Aubert:2004at}.

The time evolution can be written as:
\begin{equation}
  P(B^{0} \to D^+ K^0 \pi^-)=
  \frac{A_{c_i}^2+A_{u_i}^2}{2} e^{-\Gamma t} 
  \{ 1 + 
  (\frac{|r_i|^2-1}{|r_i|^2+1}) \cos(\Delta m t) + 
  (\frac{2\,{\rm Im}(r_i)}{|r_i|^2+1}) \sin(\Delta m t) \}
\end{equation}
where $A_{c_i}$ ($A_{u_i}$) is the magnitude of the 
Cabibbo allowed (suppressed) amplitude 
and  $r_i$  the ratio between the Cabibbo suppressed and the Cabibbo allowed 
amplitudes which  varies across the $M_{K^0\pi^-}$-$M_{D^+ K^0}$ plane. 
Analogous expressions can be written for $\bar{B^{0}} \to D^+ K^0 \pi^-$, 
$B^{0} \to D^- K^0 \pi^+$ and $\bar{B^{0}} \to D^- K^0 \pi^+$ decays.

.
 The regions showing the highest sensitivity to $2\beta + \gamma$ are the ones with interference between $\bar{B^0} \to D^{**0} K^0$ and the Cabibbo suppressed $\bar{B^0} \to \bar{D}^{**0} K^0$ and between $\bar{B^0} \to D^{+} K^{*-}$ and the Cabibbo suppressed $\bar{B^0} \to \bar{D}^{**0} K^0$ and it depends on the actual $D^{**0}$ component.

The conclusion of a feasibility study is  that with $500 \ \rm{fb}^{-1}$ the relative error will lie between  25\% and 50\%.

\subsubsection{\boldmath
  $\sin(2\beta+\gamma)$ constraint from $CP$ asymmetries in  $B^0 \to D^{(*)}\pi/\rho$ decays}
 
\vspace{+2mm}
\begin{flushright}
 {\it Contribution from C.~Voena}
\end{flushright}

The $CP$ parameters extracted from the time-dependent evolutions of 
$B^0 \to D^{(*)}\pi/\rho$ decays that have been studied at the 
$B$-factory experiments Babar and Belle can be written as:
\begin{eqnarray}\nonumber
&a_j&=\ 2r_j\sin(2 \beta+\gamma)\cos\delta_j\,, \\ \nonumber
&b_i&=\ 2r^\prime_i\sin(2 \beta+\gamma)\cos\delta^\prime_i\,, \\
&c_{i,j}&=\ 2\cos(2 \beta+\gamma) (r_j\sin\delta_j-r^\prime_i\sin\delta^\prime_i)\,.
\label{acdep1}
\end{eqnarray}
where $i$ is the tagging category and $j$ is the reconstructed $B$ decay
($j$=$D\pi$, $D^*\pi$, $D\rho$). The parameter $r_j$ is the ratio of the suppressed over the allowed amplitude contributing to the corresponding decay and $\delta_j$ is a strong phase. The primed parameters are
the corresponding quantities related to $CP$ violation on the tagging side.
We expect a very small asymmetry in these decays, $r_j$ is expected to be $\sim0.02$ (with $r'_i \leq r_j$).
In the extraction of $\sin(2\beta+\gamma)$ we make use of the $CP$ parameters
free of the unknown tag-side interference:
\begin{eqnarray}\nonumber
&a_j&=\ 2r_j\sin(2 \beta+\gamma)\cos\delta_j\,, \\ \nonumber
&c_{lep,j}&=\ 2\cos(2 \beta+\gamma) r_j\sin\delta_j\,.
\label{acdep2}
\end{eqnarray}
For each mode we have two observables and three unknowns,  
external inputs are therefore needed  to extract $\sin(2\beta+\gamma)$. 
$SU(3)$ symmetry is currently used to  estimate the $r_j$ parameters. 
The relation (for  $B^0 \to D^+\pi^-$)  is:
\begin{equation}
  r_{D\pi} = \tan \theta_{C}
  \sqrt{
    \frac{BR(B^0 \to D_s^{+}\pi^-)}{BR(B^0 \to D^{-}\pi^+)}}\frac{f_{D}}{f_{{D_s}}
  }
~\cite{ref:sin2bg_th}
\label{lambdadef}
\end{equation}
Similar relations have been used for the other two decay modes.
Equation~\ref{lambdadef} has been obtained with two approximations.
In the first approximation, the exchange diagram amplitude contributing to the decay $B^0\to D^{+} \pi^-$ has been neglected and only the
tree-diagram amplitude has been considered.
No reliable estimate of the exchange term for these decays exists although there are experimental hints that it suppressed with respect to the tree term.
The second approximation involves the use of the ratio
of decay constants
$f_{D} / f_{D_s}$ to take into account
SU(3) breaking effects and assumes factorization.
We attribute
a $30\%$ relative error to the theoretical assumptions
involved in obtaining the value of $r_j$.

Using the current experimental inputs we obtain:
\begin{eqnarray}\nonumber
  &r_{D\pi}   & =\ 0.020 \pm 0.003 \pm 0.006 (th.),, \\ \nonumber
  &r_{D^*\pi} & =\ 0.015^{+0.004}_{-0.006} \pm 0.005 (th.), \\ \nonumber
  &r_{D\rho}   & =\ 0.006 \pm 0.003 \pm 0.002 (th.)\,.
\label{lambdameas2}
\end{eqnarray}

It has been suggested by theorists~\cite{dspi0} 
to use the decay $B^+ \to D^+\pi^0$ to determine $r_{D\pi}$ 
(and similar for the other two modes) 
making use only of isospin symmetry (and not of SU(3) symmetry).
The measurement of these branching ratios seems however challenging 
since we expect few events even at high luminosity 
($\sim6$ in $500 \ \rm{fb}^{-1}$).
Using the current world averages from HFAG for the $a_j$ and $c_{lep,j}$ 
parameter and the $r_j$ parameters obtained  above
we determine constraints on $\sin(2\beta+\gamma)$~\cite{RF} 
using two different statistical approaches: 
a bayesian~\cite{utfit} approach which gives:
\begin{eqnarray}
  2\beta+\gamma = (88^{+40}_{-39})^o 
  \label{sin1}
\end{eqnarray}
and a frequentistic~\cite{freq} approach which gives:
\begin{center}
  $|\sin(2\beta+\gamma)|> 0.49~(0.27)$ at 68\% (90\%) CL.
\end{center}

Figure shows the corresponding bounds obtained in the 
unitarity $(\bar{\rho},\bar{\eta})$  plane.

\begin{figure}
\begin{minipage}{0.48\textwidth}
         \includegraphics[width=0.95\textwidth]{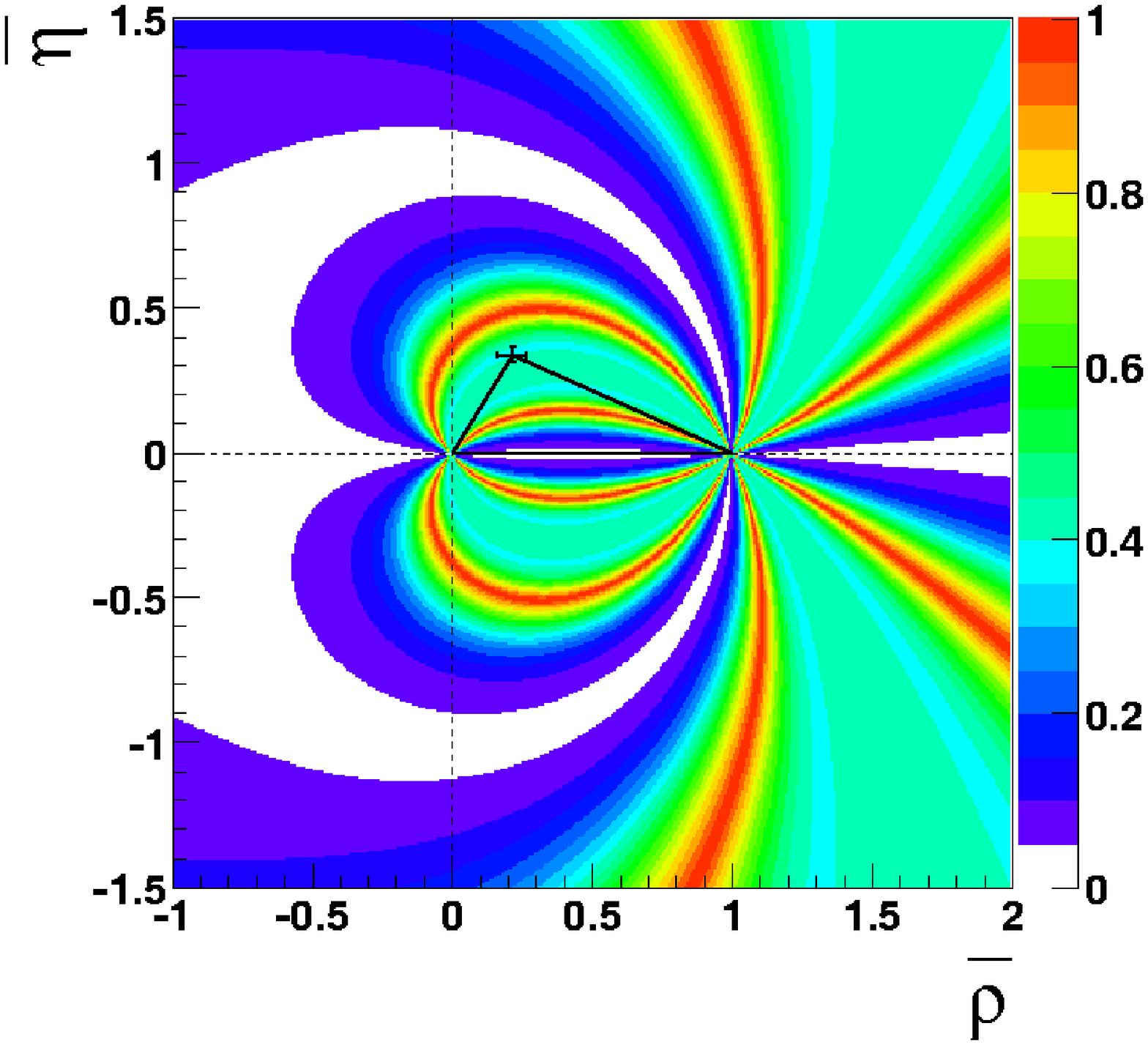}
\end{minipage}
\hfill
\begin{minipage}{0.48\textwidth}
    \includegraphics[width=0.95\textwidth]{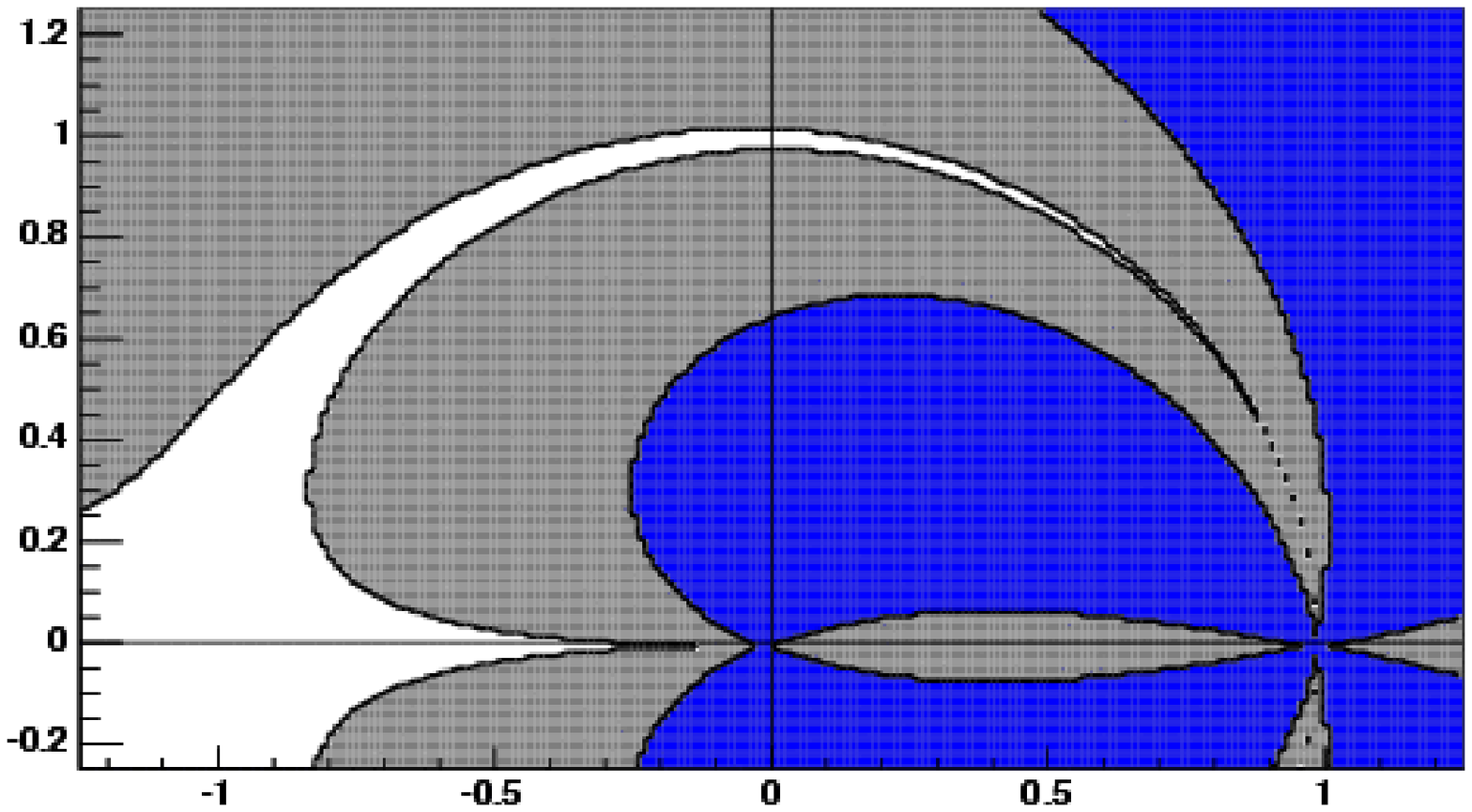}
\end{minipage}
\caption{Bound in the unitarity $(\bar{\rho},\bar{\eta})$ plane from $\sin(2\beta+\gamma)$ in the frequentistic (left) and in the bayesian approach (right). For the bayesian approach, the blue region corresponds to the 68$\%$ confidence level and the gray region to the  90$\%$ confidence level.}
\label{fig:figs}
\end{figure}

Projecting the error on $2 \ \rm{ab}^{-1}$, assuming the same measured central values for the $a_j$ and $c_{lep,j}$ parameters and no improvement on the knowledge of $r_{j}$ gives an error on $2\beta+\gamma$ of about 30$^o$.

\subsubsection{\boldmath
  $\sin(2\beta+\gamma)$ from   $B^0 \to D^{0} K^{(0/+)}$ decays}

\vspace{+2mm}
\begin{flushright}
 {\it Contribution from V.~Sordini, M.~Pierini, L.~Silvistrini \& A.~Stocchi}
\end{flushright}

In terms of the Operator Product Expansion, 
we can express the amplitudes for the decays of the $B$ to neutral $D$-charged $K$:
\begin{eqnarray*}
A \, (B^{+} \to \overline{D^{0}} K ^{+} )\, &=& \,T + C \\
A \, (B^{+} \to D^{0} K ^{+} )\, &=&\, \overline{C} + A
\end{eqnarray*}
where T stays for ``tree contribution'' \begin{math}(\, T =\,V_{cs} \,V_{ub}^{*} \, E_{1}\, (s, l, c, K, D) \, ) \end{math} , A for ``annihilation contribution'' \begin{math}(  A =\,V_{us} \,V_{cb}^{*} \, A_{1}\, (s, l, c, K, D) \,)\end{math} , C and $\overline{C}$ are ``colour-suppressed contributions'' \begin{math}( \,C = \,V_{cs} \,V_{ub}^{*} \,E_{2}\, (c, l, s, D, K) \,, \,\overline{C} = \,V_{us} \,V_{cb}^{*} \, E_{2}\, (l, c, s, D, K) \,) \end{math}
The parameters $E_{1}\,$ ,$E_{2}\,$ and $A_{1}$ are renormalization scheme and scale independent.
We define $r_{D^{0}K^{+}}$ as the amplitudes ratio: 
\begin{displaymath}
r_{D^{0}K^{+}} =  \frac{|A \, (B^{+} \to \overline{D^{0}} K ^{+} )| }{|A \, (B^{+} \to D^{0} K ^{+} )|} 
\end{displaymath}
Similarly for the decays of the $B$ to neutral $D$-neutral $K$ we can write the amplitudes in terms of the same parameters :
\begin{eqnarray*}
A \, (B^{+} \to \overline{D^{0}} K ^{0} )\, &=& \,C\\
A \, (B^{+} \to D^{0} K ^{0} )\, &=&\,\overline{C}
\end{eqnarray*}
and define the amplitudes ratio:
\begin{displaymath}
r_{D^{0}K^{0}} =  \frac{|A \, (B^{0} \to \overline{D^{0}} K ^{0} )| }{|A \, (B^{0} \to D^{0} K ^{0} )|} 
\end{displaymath}
C is known since the \begin{math}BR(B^{+} \to \overline{D^{0}} K^{0} )\end{math} is measured and writing the two ratios in terms of the OPE parameters
\begin{eqnarray*}
r_{D^{0}K^{+}} &=& \frac{|\overline{C} + A|}{|T + C|} \\
r_{D^{0}K^{0}} &=& \frac{|\overline{C}|}{|C|}
\end{eqnarray*}
we get for $\overline{C}$ the equation:
\begin{eqnarray*}
|\overline{C}| = -A \cos\delta \pm \sqrt{ A^{2} (\cos^{2}\delta - 1) + r_{D^{0}K^{+}}^{2} BR(B^{+} \to D^{0} K^{+} ) }\\ \mbox{where the phase space term cancels and } \delta \mbox{ is the strong phase .}
\end{eqnarray*}
There are two solutions for $|C|$, these are both acceptable only if:
\begin{eqnarray*}
\cos(\delta) < 0
\mbox{    and    }
A^{2} &>& r_{D^{0}K^{+}}^{2} BR(B^{+} \to D^{0} K^{+} )
\end{eqnarray*}
If we use the actual experimental values:
\begin{eqnarray*}
 BR(B^{+} \to \overline{D^{0}} K ^{+}) &=& (3.7 \pm 0.6)\cdot 10^{-4}  \\
 BR(B^{+} \to \overline{D^{0}} K ^{0}) &=& (5.0 \pm 1.4)\cdot 10^{-5}  \\
 BR(B^{+} \to D^{+} K ^{0}) &<& 5.0 \cdot 10^{-6}  \mbox{     @ 90\% probability}
\end{eqnarray*}
( the $ BR(B^{+} \to D^{+} K ^{0})  $is useful to determine the annihilation parameter A since:\begin{math}\, A\,(\,B^{+} \to D^{+} K ^{0} ) = A  \,\end{math} )
and the average for $r_{D^{0}K^{+}}$ from $UT_{\emph{fit}}$ :
\begin{eqnarray*}
r_{D^{0}K^{+}} = 0.10 \pm 0.04   \mbox{  @ 68\% \mbox{probability} }
\end{eqnarray*}
and we decide to accept, in case of ambiguity, both the solutions (the one with the sign + and the one with the sign -), we get :
\begin{displaymath}
r_{D^{0}K^{0}} = 0.27 \pm 0.18 \, \,(\, @ \, 68\% \, \mbox{probability})
\end{displaymath}
Where, of the error on $r_{D^{0}K^{0}}$, a contribution of 0.08 is due to the error on $r_{D^{0}K^{+}}$ and 0.16 to the other uncertainties. 
For the ratio $\frac{r_{D^{0}K^{0}}}{r_{D^{0}K^{0}}}$ :
\begin{displaymath}
  \frac{r_{D^{0}K^{+}}}{r_{D^{0}K^{0}}} = 0.43 \pm 0.19 \,( @ 68\% \mbox{probability})
\end{displaymath}
If we extrapolate the errors on the measurements to a statistics of $500 \ \rm{fb}^{-1}$ we would have ambiguity in 7.2\% of cases.
In this situation we would get:
\begin{displaymath}
  r_{D^{0}K^{0}} = 0.24 \pm 0.15 \, \,(\, @ \,68\% \, \mbox{probability})
\end{displaymath}
Where, of the error on $r_{D^{0}K^{0}}$, a contribution of 0.07 is due to the error on $r_{D^{0}K^{+}}$ and 0.13 to the other uncertainties.
For the ratio $\frac{r_{D^{0}K^{0}}}{r_{D^{0}K^{0}}}$ :
\begin{displaymath}
  \frac{r_{D^{0}K^{+}}}{r_{D^{0}K^{0}}} = 0.44 \pm 0.18 \, \,(\, @\, 68\% \mbox{probability})
\end{displaymath} 
In principle, if one could determine the ratio $r_{D^{0}K^{0}}$, 
this would be a useful input in the analysis of $\sin(2\beta + \gamma)$.
We made a study to see the sensitivity to $\sin(2\beta + \gamma)$.
This study was made assuming the information we get on $r_{D^{0}K^{0}}$ 
in the case of {\bf no ambiguity} ($\sim93\%$ of cases for $500 \ \rm{fb}^{-1}$), 
for which we get an error on $r_{D^{0}K^{0}}$ of 0.12.
Assuming:
\begin{displaymath}
  r_{D^{0}K^{0}} = 0.30 \pm 0.15 \, \,(\, @ \,68\% \, \mbox{probability})
\end{displaymath}
and errors on the observables S and C \cite{rahatlou_bad}  
\begin{displaymath}
  \delta S = 0.6 \; \; \delta C = 0.5
\end{displaymath}
we get an error  of 60\% on the determination of $(2\beta + \gamma)$ through this method.
The method doesn't seem much sensitive to $(2\beta + \gamma)$: 
the same exercise made assuming a statistic of $ 1 \ \rm{ab}^{-1}$ 
and a value of $r_{D^{0}K^{0}} = 0.40$ returns an error on $(2\beta + \gamma)$ of 42\%.

\section{\boldmath Measurements of $\gamma$/$\phi_3$}

\subsection{\boldmath
  Combined Strategies for extraction of a clean $\gamma$}  

\vspace{+2mm}
\begin{flushright}
 {\it Contribution from D.~Atwood}
\end{flushright}

A promising feature of clean and precise extraction of $\gamma$
from $B \to K D $ type of modes is that a multitude of strategies
exist which are very effective when used in combination\cite{path}.
In here we focus on the use of direct $CP$ from $B^- \to "K^-" D^0/\bar
D^0$ though it is also possible to use time dependent $CP$ violation
via $B^0 \to "K^0" D^0$.

\begin{figure}
\epsfxsize 5 in
\mbox{\epsfbox{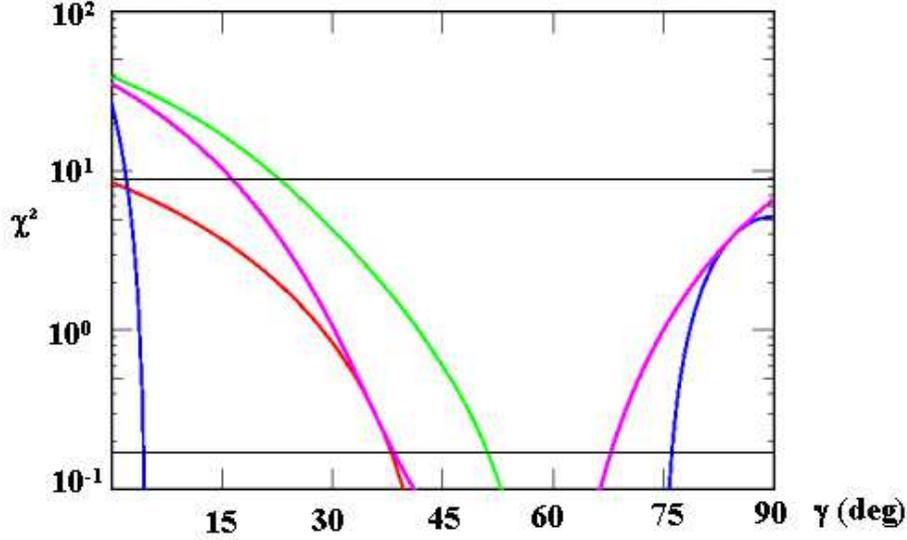}}
\caption{\it
 $\gamma$ determination with incomplete input ({\it i.e.}
 cases when the number of observables is less than the number
 of unknown parameters). The upper horizontal line corresponds
 to low-luninosity {\it i.e.} around current B-factories
 whereas the lower horizontal curve is relevant for
 a SBF. Blue uses all CPES modes of $D^0$, 
red is with only $K^+ \pi^-$ and purple uses combination
 of the two. Green curve again uses on
 $D^0$, $\bar D^0 \to K^+ \pi^-$ but now includes $K^{*-}$ and 
 $D^{*0}$; see text for details.     
\label{un_det} }
\end{figure}

Fig~\ref{un_det} illustrates several of the important features. Here $\chi^2_{\rm{min}}$
versus $\gamma$ is compared for various methods and input data sets.
Let us focus first specifically to $B^- \to K^- D^0$, $\bar D^0$ with
$D^0$, $\bar D^0$ decays to common modes 
such as (CPNES) $K^+ \pi^-$ (ADS\cite{ads2})
or (CPES) $K_S^0 \pi^0$ (GLW\cite{GronauWyler}). Then for each such common mode
(say $K^+ \pi^-$)  
there are basically three unknowns: $\delta_{st}$, $r_B$ and $\gamma$
where $\delta_{st}$ is the strong phase and $r_B = Br(B^- \to K^- \bar
D^0)/Br(B^- \to K^- D^0)$ and two observables: the rate for
$B^-$ and for $B^+$. Therefore as such we cannot hope to
solve for $\gamma$.  
However, as we add another 
common mode of $D^0$, $\bar D^0$, say $K^+ \rho^-$,
then one is adding one new unknown (a strong phase) but 2 more
observables. So now there are 4 observables and 4 unknowns and the
system, is in principle, soluble though in practice
presence of discrete ambiguities complicates the solution.
In general for N such common modes, there will
be N + 2 unknowns and 2N observables and as more modes are added
$\gamma$ can be solved for very effectively.

\begin{figure}
\epsfxsize 5 in
\mbox{\epsfbox{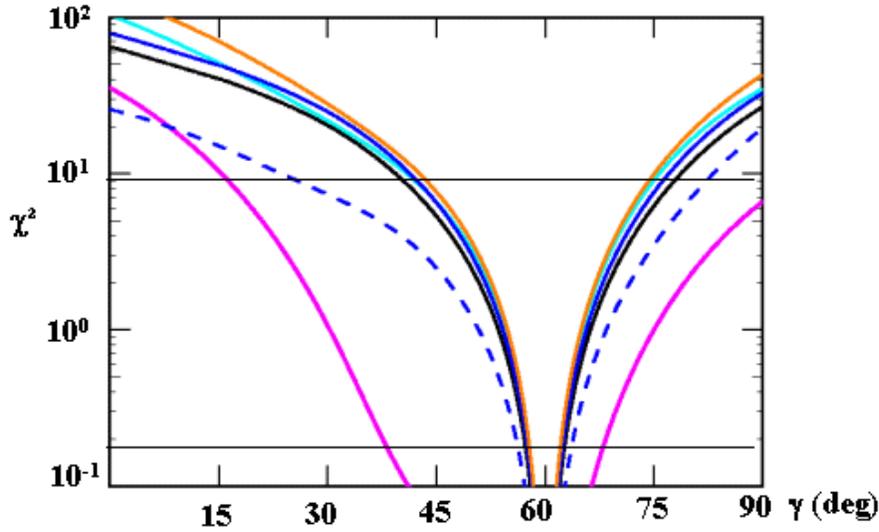}}
\caption{\it
  $\gamma$ extraction with over-determined cases. Purple curve shows
 the effect
  of combining GLW (all CPES modes) with one ADS ($K^+ \pi^-$) mode; black
  curve differs from purple only in that it also
  includes $D^0$ from $D^{*0}$;
  blue curves show the effect of properly including the correlated
  strong phase between $D^{0*} \to D^0 + \pi$ and
  $D^{0*} \to D^0 + \gamma$. Orange curve includes lot
  more input including Dalitz and multibody modes.
  see text for details (See also Fig~\ref{un_det}).  
\label{over} }
\end{figure}

Fig.~\ref{un_det} and Fig.~\ref{over}
show situation with regard to under-determined and
over-determined cases respectively. The upper horizontal
line corresponds roughly to the low luminosity i.e. comparable
to the current $B$-factories\cite{sbf12} whereas the lower horizontal
curve is relevant for a super $B$-factory. In Fig.~\ref{un_det} in blue
is shown the case when only the input from (GLW) CPES modes
of $D^0$ is used; note all the CPES modes are included here.
You see that the resolution on $\gamma$
then is very poor. In particular, this method is rather 
ineffective in giving a lower bound; its upper bound 
is better.

In contrast, a single ADS mode ($K^+ \pi^-$) is very effective
in so far as lower bound is concerned, 
but it does not yield an effective upper bound (red).  

Note that in these two cases one has only two observables and 3
unknowns.
In purple is shown the situation when these two methods are combined
Then at least at high luminosity there is significant improvement
in attaining a tight upper bound; lower bound obtained by ADS alone
seems largely unaffected.

Shown in green is another under determined case consisting of
the use of a single ADS mode, though it includes $K^{*-}$ as well
$D^{*0}$; this again dramatically improves the lower bound.
From an examination of these curves it is easy to see that
combining information from different methods and modes
improves the determination significantly\cite{path}.

Next we briefly discuss some over determined cases (Fig.~\ref{over}).
In purple all the CPES modes of $D^0$ are combined with just
one doubly cabibbo-suppressed (CPNES) mode. Here there are 4 observables
for the 4 unknowns and one gets a reasonable solution at least
especially for the high luminosity case.

The black curve is different from the purple one in only
one respect; the black one also includes the $D^{0*}$ from 
$B^- \to K^- D^{0*}$ where subsequently the $D^{0*}$ gives
rise to a $D^0$. Comparison of the black one with the purple
shows the remarkable improvement by including the $D^{0*}$.
In this case the number of observables (8) exceeds
the number of unknowns (6).

Actually, the $D^{0*}$ can decay to $D^0$ via two modes:
$D^{0*} \to D^0 + \pi$ or $D^0 + \gamma$. Bondar and Gershon \cite{bg}have
made a very nice observation that the strong phase for the 
$\gamma$ emission is opposite that of the $\pi$
emission. Inclusion (blue curves)
of both types of emission increases the number
of observables to 12 with no increase in number of unknowns.
So this improves the resolving power for $\gamma$ even more. 

The orange curves show the outcome when a lot more input
is included; not only $K^-$, $K^{-*}$, $D^0$, $D^{0*}$  
but also Dalitz and multibody decays of $D^0$ are included.
But the gains now are very modest; those once the number of observables
exceeds the number of unknowns by a few (say O(3)) further 
increase in input only has a minimal impact.

\subsection{\boldmath
  Measurements of $\gamma$ at BaBar} 

The measurement of the angle $\gamma$ of the unitarity triangle
through $B^-\to D^{(*)}K^{(*)-}$ requires the combination of as many 
$B$ and $D$ modes as possible to reduce the statistical uncertainty.
Though the most statistically powerful approach analyzed so
far is the one exploiting the Dalitz plot analysis of $D^0\to
K^0_S\pi^-\pi^+$ from $B^-\to D^{(*)0} K^-$, the GLW and ADS
methods also provide useful information.

\subsubsection{\boldmath
  BaBar GWL and ADS}

\vspace{+2mm}
\begin{flushright}
 {\it Contribution from M.~Rama}
\end{flushright}

\begin{figure}
  \begin{center}
    \psfig{figure=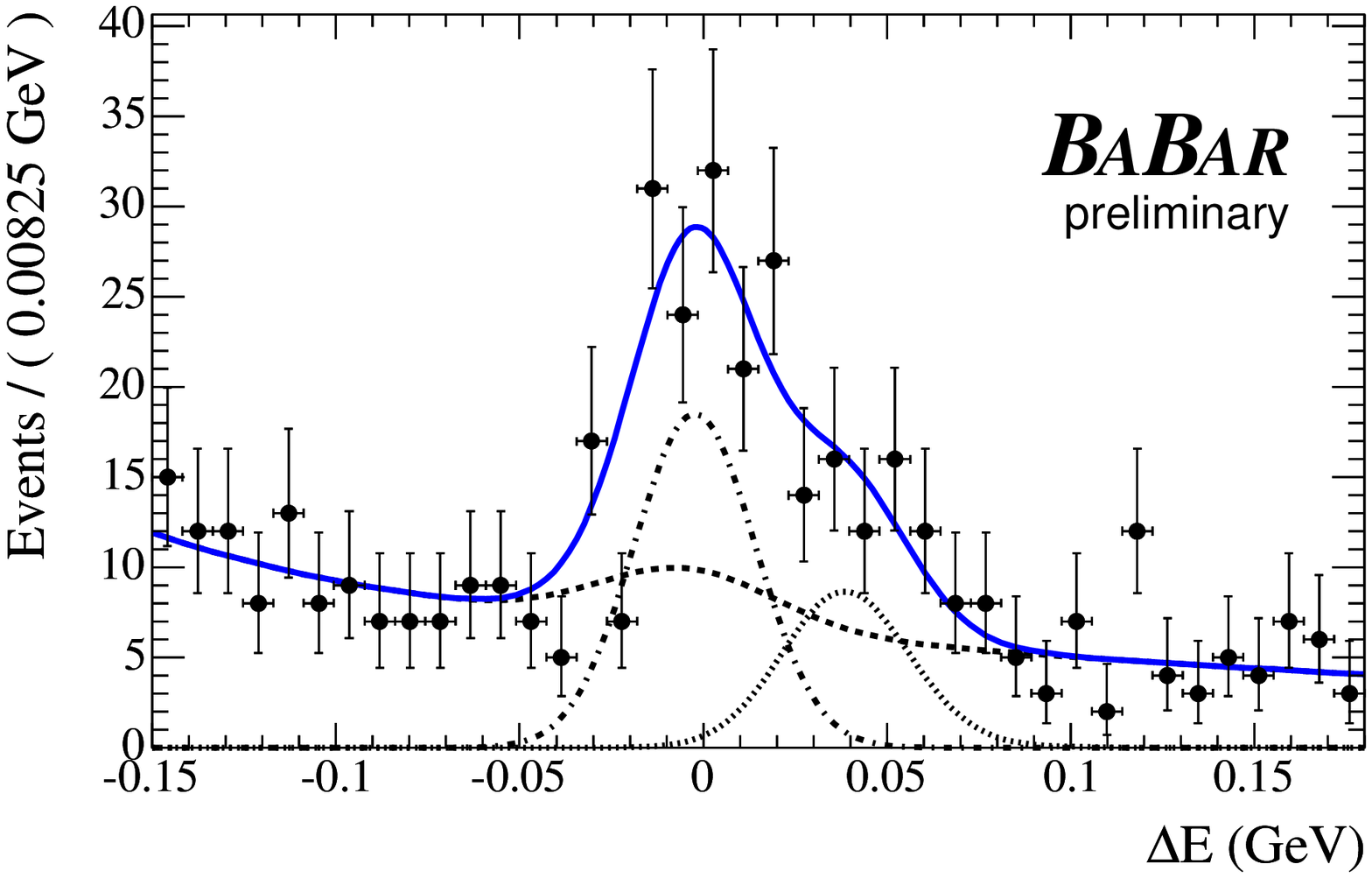,height=4cm}
    \psfig{figure=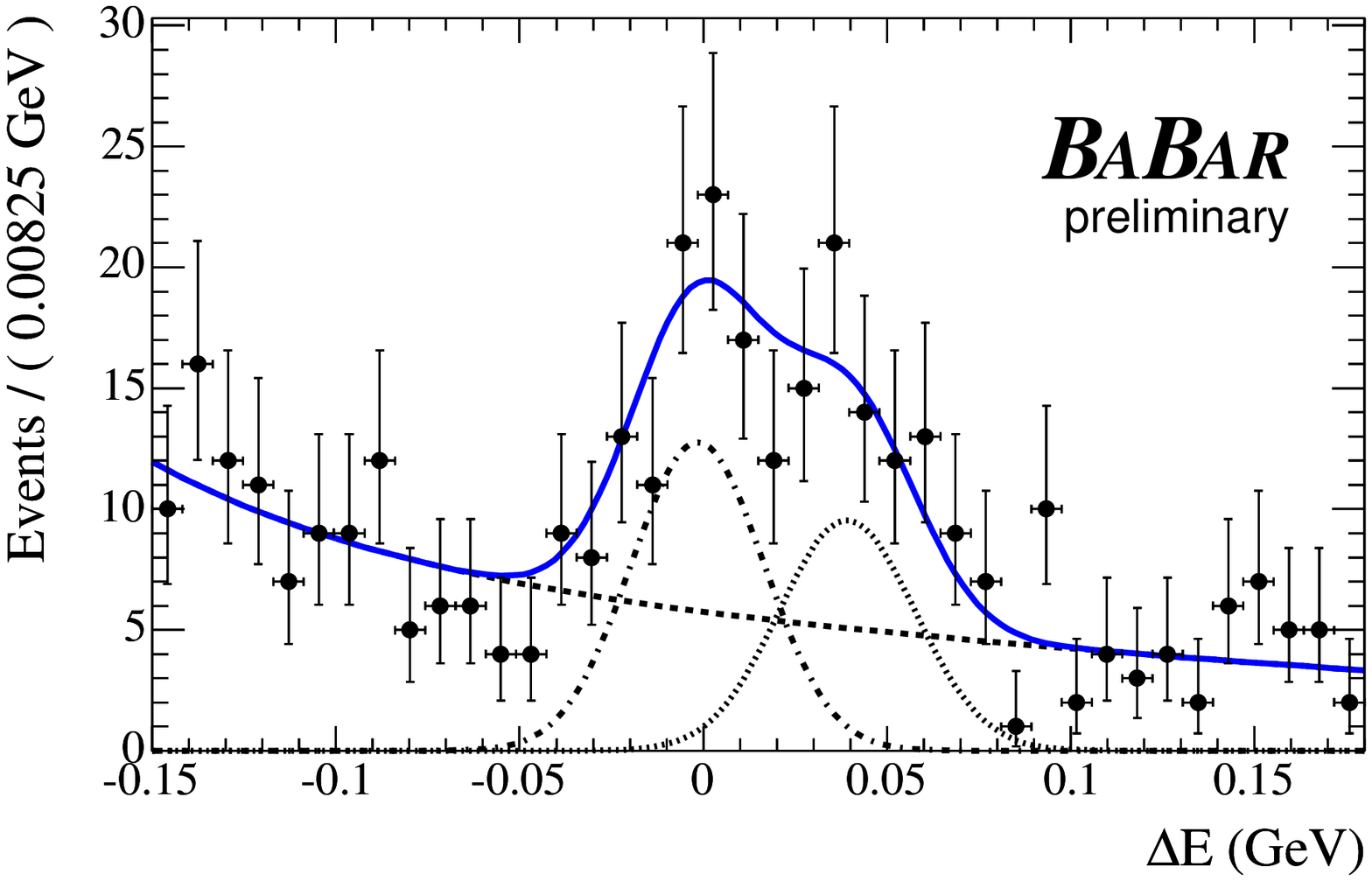,height=4cm}
    \caption{
      Distributions of $\Delta E$ for events enhanced in $B\ra D^0K$ signal. 
      Left: $CP+$; right: $CP-$. Solid curves represent projections of 
      the maximum likelihood fit; dashed-dotted, dotted and dashed curves 
      represent the $B\to D^0K$, $B\to D^0\pi$ and background contributions.
    }\label{fig:babar_glw}
  \end{center}
\end{figure}

BaBar has studied the decays $B^\mp \to D^{(*)0}K^{(*)\mp}$
using $D^{*0}\to D^0\pi^0$ and $K^{*-}\to K^0_S\pi^-$.
The $D^0$ meson is reconstructed
in the CKM-favored modes $K^-\pi^+$, $K^-\pi^+\pi^+\pi^-$ and $K^-\pi^+\pi^0$;
the $CP+$ eigenstates $K^-K^+$ and $\pi^-\pi^+$;
and the $CP-$ eigenstates $K^0_S\pi^0$, $K^0_S\phi$, $K^0_S\omega$.
Figure~\ref{fig:babar_glw} shows the distributions of $\Delta E$ of
the $B\to D^0K$ events ($CP+$ and $CP-$ modes). 
The signal yields are extracted through a maximum likelihood fit that uses as
input $\Delta E$ and the Cherenkov angle of the bachelor track $K$. 
Similar techniques are used to select $B\to D^{*0}K$ and
$B\to D^0K^{*}$~\cite{babar_glw_btdk1,babar_glw_btdk2,babar_glw_btdstark,babar_glw_btdkstar}.
The measurements of $A_{CP\pm}$ and $R_\pm$ allow to constrain the
unknowns $r_b$, $\delta_b$ and the CKM angle $\gamma$. The results of
the measurements are reported in Table~\ref{tab:babar_glw_results},
where $R_{flav}\equiv BF(B^-\to D^{(*)0}K^-)/BF(B^-\to D^{(*)0}\pi^-)$. 
Particular care is needed when evaluating the systematic
uncertainties associated to peculiar sources of background. The
$B^-\to D^0_{CP+}K^-$ decays are affected by the charmless 3-body background
$B^-\to K^-h^-h^+$ ($h=\pi,K$), characterized by the same
\mes and $\Delta E$ distribution as the signal. 
The $B^-\to K^-\pi^-\pi^+$ and $B^-\to K^-K^-K^+$
backgrounds ( $4\pm 4$ and $29\pm 7$ events, respectively) are
subtracted from the $B^-\to D^0[\to \pi^-\pi^+]K^-$
and $B^-\to D^0[\to K^-K^+]K^-$ signals ($18\pm 7$ and $75\pm 13$
events, respectively). Similarly, the
$B^-\to D^0[\to K^0_S a_0]K^{*-}$ is evaluated and
subtracted from the $B^-\to D^0[\to K^0_S \phi]K^{*-}$
signal. In this case, due to the different spin-parity of $a_0$ with
respect to $\phi$, the background has an opposite $CP$ asymmetry with
respect to the signal.

On a datasample of 227 million $\Upsilon(4S)\to B\bar{B}$
decays we have searched for $B^- \to \dbarp~ K^-$ followed by 
$\dbarp~\to K^+\pi^-$, as well as the charge conjugate sequences. In
these processes, the favored $B$ decay followed by the doubly
CKM-suppressed $D$ decay interferes with the suppressed $B$ decay
followed by the CKM-favored $D$ decay. The yields of the signal mode
and the normalization mode ($B^-\to D^0 K^-$ with 
$D^0\to K^-\pi^+$) are extracted through a fit on $\mes$,
after requiring that the events have a $\Delta E$ value consistent
with zero within the resolution. We find a total of
$4.7^{+4.0}_{-3.2}$ signal events and $356\pm26$ normalization events, 
that are used to evaluate the charge-integrated ratio 
$R_{K\pi}=[\BR(B^-\to[K^+\pi^-]_DK^-)+\BR(B^+\to[K^-\pi^+]_DK^+)]/(2\BR(B^\mp\to[K^\mp\pi^\pm]_DK^\mp))=r_B^2+r_D^2+2r_B\,r_D\,\cos\delta_B\cos\gamma$.
The resulting limit is $\mathcal{R}_{K\pi}<0.030$ at 90\% CL. Using 
$r_D=0.060\pm0.003$\cite{ref:pdg2004}, and allowing the variation of
$\gamma$ and $\delta_B$ on the full range $0^\circ -180^\circ$, we set
the limit $r_B<0.23$ at 90\% CL~\cite{babar_ads_btdk}.

\begin{table}[htb]
  \caption{
    Measured $CP$ asymmetries $A_{CP\pm}$ and ratios $R_\pm$. 
    The first error is statistical, the second is systematic.
  }\label{tab:babar_glw_results}
  \vspace{0.4cm}
  \begin{center}
    \begin{tabular}{|lccc|}
      \hline
      &$B^-\ra D^0 K^-$ & $B^-\ra D^0 K^{*-}$ & $B^-\ra D^{*0} K^-$\\
      \hline
      $N_{B\bar{B}}$~$(\times 10^6)$ & 214 & 227 &123\\
      $A_{CP+}$ & $0.40\pm0.15\pm0.08$ &  $-0.09\pm0.20\pm0.06$  & $-0.02\pm0.24\pm0.05$ \\
      $A_{CP-}$ & $0.21\pm0.17\pm0.07$ & $-0.33\pm0.34\pm0.10$~\footnotemark[1] & $1.09\pm0.26^{+0.10}_{-0.08}$\\
      $R_+$ & $0.87\pm0.14\pm0.06$ & $1.77\pm0.37\pm0.12$ & ***\\
      $R_-$ & $0.80\pm0.14\pm0.08$ & $0.76\pm0.29\pm0.06$~\footnotemark[1] & ***\\ 
      \hline
    \end{tabular}
  \end{center}
\end{table}
\footnotetext[1]{
  Additional biases $\delta A_{CP-}=0.15\pm0.10 \cdot (A_{CP-}-A_{CP+})$ and 
  $\delta R_-=^{-0.04}_{-0.14}$ are quoted by the authors, 
  reflecting possible interference effects between the $\phi$ and $\omega$ 
  resonances and the background.
}

\section{\boldmath 
  Measurements of  $\gamma$/$\phi_3$ with  $D^0$ Dalitz analysis }

The most precise determination of the standard model CKM phase 
in the long run is provided by the methods
based on the interference between 
$b\to c \bar{u} s$ and $b\to u \bar{c} s$ \cite{GronauWyler,GronauLondon}.  
In the case of charged $B$ decays this means that the interference is between 
$B^-\to D K^-$ followed by $D\to f$ decay and 
$B^-\to \bar{D} K^-$ followed by $\bar{D}\to f$, where 
 $f$ is any common final state of $D$ and $\bar{D}$ \cite{GronauWyler,GronauLondon,bg}. 
Here we will restrict ourselves to the case, 
where $f$ is a multibody final state. 
For concreteness we focus on the following cascade decay \cite{ggsz}
\begin{equation}
  B^- \to D K^- \to (K_S \pi^- \pi^+)_D K^-.
\end{equation}
For the $B$ decay amplitudes we define
\begin{eqnarray}
  A(B^- \to D^0 K^-)&\equiv& A_B \label{AB},\\ 
  A(B^- \to {\DbarZ} K^-) &\equiv&
  A_B r_B e^{i(\delta_B-\gamma)}.
  \label{weakphase}
\end{eqnarray}
Here $\delta_B$ is the difference of strong phases and $A_B$ is taken to be positive.
The same definitions apply to the amplitudes for the $CP$ conjugate
cascade $B^+ \to D K^+ \to (K_S \,\pi^+\pi^-)_D K^+$, 
except that the weak phase flips the sign $\gamma\to -\gamma$ in (\ref{weakphase}).  
For the  $D$ meson decay  we further define
\begin{eqnarray}
  \label{CP-for-D}
  A_D(s_{12},s_{13}) \equiv A_{12,13}\,e^{i\delta_{12,13}}&\equiv&  A(D^0 \to K_S(p_1) \pi^-(p_2) \pi^+(p_3))\\
  & =&A(\DbarZ \to K_S(p_1) \pi^+(p_2) \pi^-(p_3)),\nonumber
\end{eqnarray}
where $s_{ij}=(p_i+p_j)^2$, 
and $p_1,p_2,p_3$ are the momenta of the $K_S, \pi^-,\pi^+$ respectively. 
Note that in the last equation $CP$ symmetry was used. 

Obviously, if $A_D(s_{12}, s_{13})$ were known, 
one could extract $\gamma$ from $B^\pm\to D K^\pm$ decay widths. 
One can measure $A_D(s_{12}, s_{13})$ from tagged $D\to K_S\pi^+\pi$ by modeling it 
with a sum of Breit-Wigner forms and fitting the parameters 
from the corresponding Dalitz plot. 
This is the approach used at present by both BaBar~\cite{Aubert:2004kv} 
and Belle~\cite{Abe:2004gu,Abe:2005ct}, 
where the modeling error on extracted value of $\gamma$ is estimated 
to be around $10^\circ$. 
In the future this modeling error can be avoided by performing a 
model independent analysis~\cite{ggsz,Atwood:2003mj}.

Once the function ${\cal A}_D(m^2_-,m^2_+)$ is fixed 
using a model for the $D^0 \to K_s \pi^+ \pi^-$ decay,
the Dalitz distributions can be fitted simultaneously 
using the expressions for the two amplitudes and 
$r_B$, $\delta$ and $\gamma$ can be obtained.

\begin{figure}[!htb]
\begin{center}
\includegraphics[height=8cm]{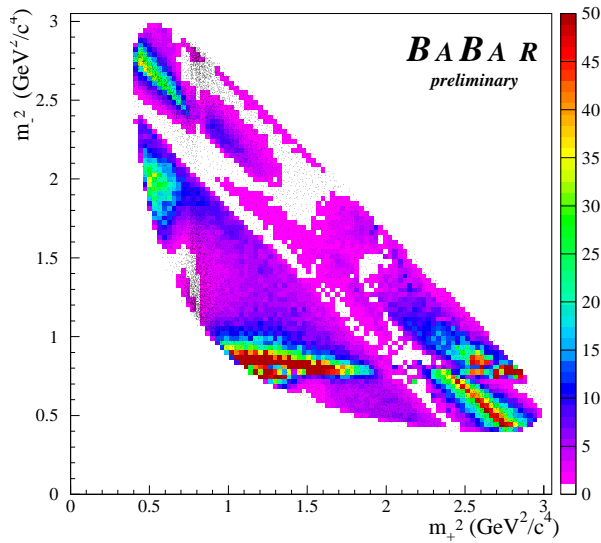}
\caption{$\overline{D}^0 \ra K_S \pi^+ \pi^-$ Dalitz distribution of simulated $ (B^+ \ra \overline{D}^0  K^+)$  events.  Each event is given a weight $\frac{d^2\ln {\cal L}}{d^2\gamma}$. The weight scale is indicated on the right of the plot. The black points represent the same events with weight equal to unity.}
\label{fig:gammasens}
\end{center}
\end{figure}

To illustrate the region of the Dalitz plot most sensitive to $\gamma$ measurement,  
we show in Fig.~\ref{fig:gammasens}   
the distribution of simulated $ B^-\ra D^{0}  K^-$ 
events based on our Dalitz model, where  each event is given a weight of
$\frac{d^2\ln {\cal L}}{d^2\gamma}$.
Here ${\cal L}$ is the likelihood function described in the following section.
The regions of interference between doubly Cabibbo suppressed 
and Cabibbo allowed decays and $CP$ eigenstate decays 
are clearly the most sensitive ones: they exhibit the highest weights. 

\subsection{\boldmath  
  Belle Measurement of $\phi_3$ From $B^\pm \to D^{(*)}K^{(*)-}$, 
  $D \to K_S \pi^+\pi^-$ Dalitz Analysis}
  
\vspace{+2mm}
\begin{flushright}
  {\it Contribution from K.~Abe}
\end{flushright}

\begin{figure} 
  \begin{center}
    \begin{tabular}{c@{\vspace{-0.1mm}}c@{\vspace{-0.1mm}}c}
      $D K^\pm$ & $D^* K^\pm$ & $D K^{*\pm}$ \\
      \includegraphics[width=0.33\textwidth]{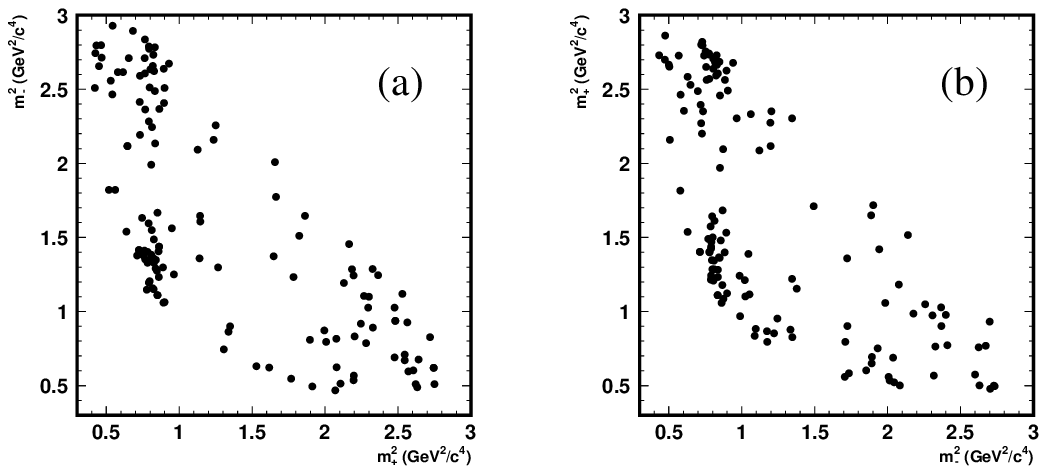} &
      \includegraphics[width=0.33\textwidth]{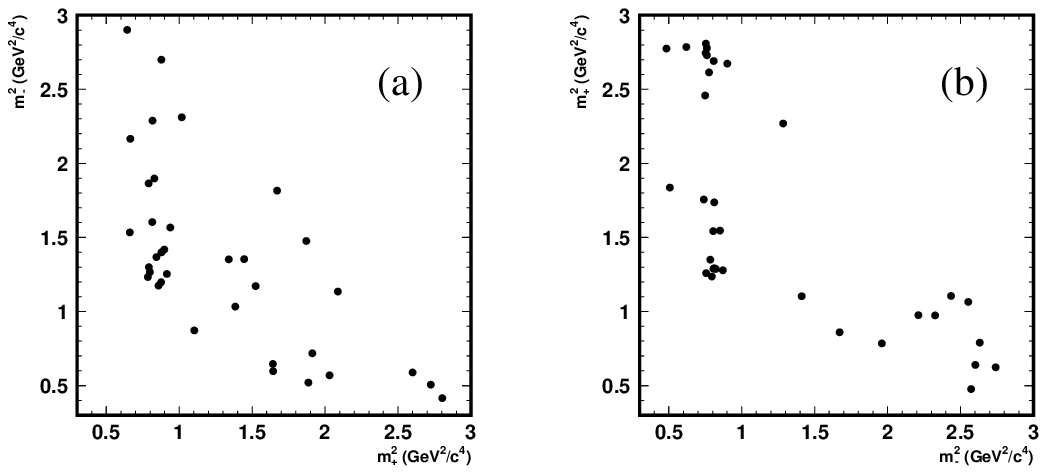} &
      \includegraphics[width=0.154\textwidth]{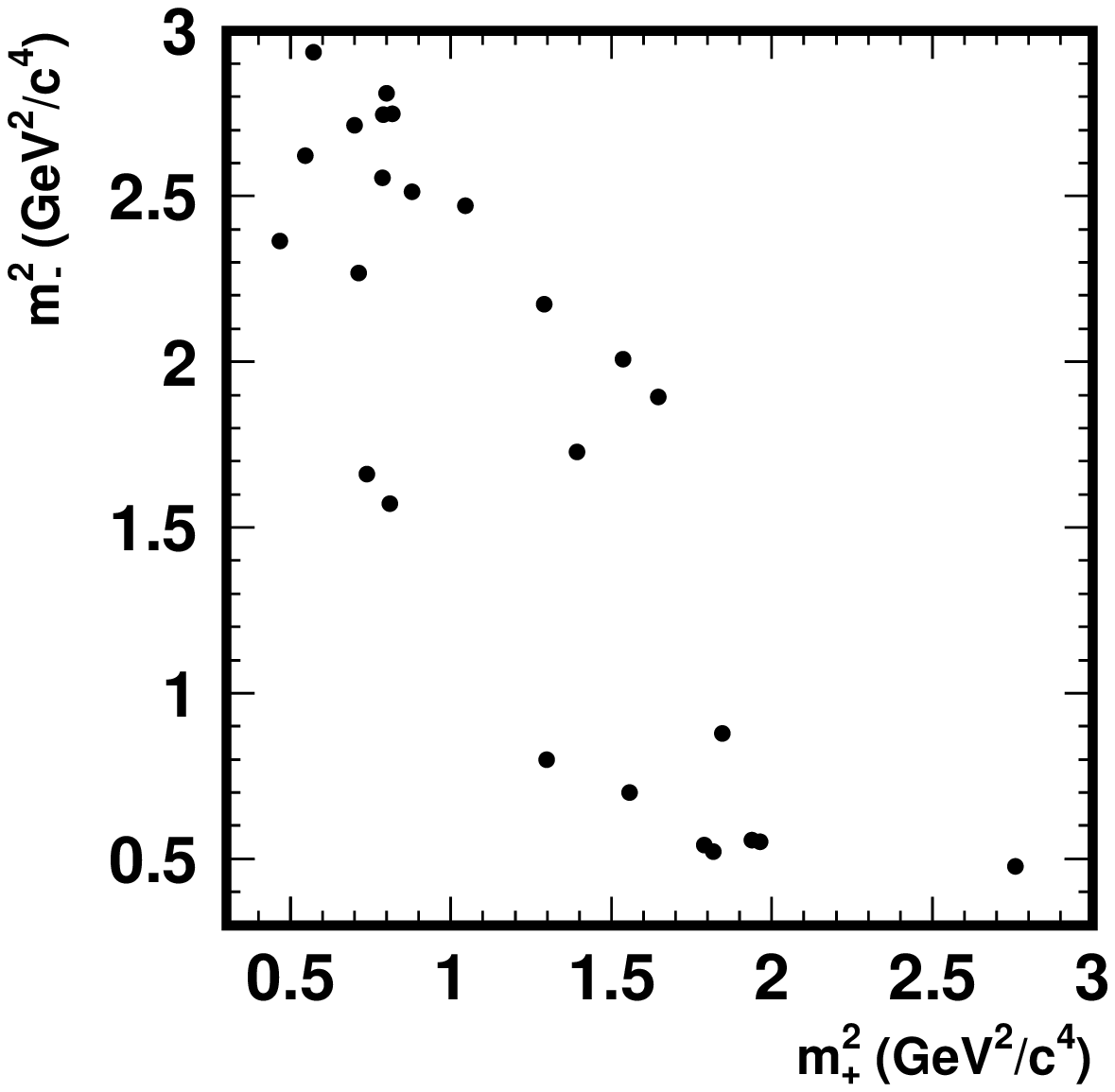}      
      \includegraphics[width=0.154\textwidth]{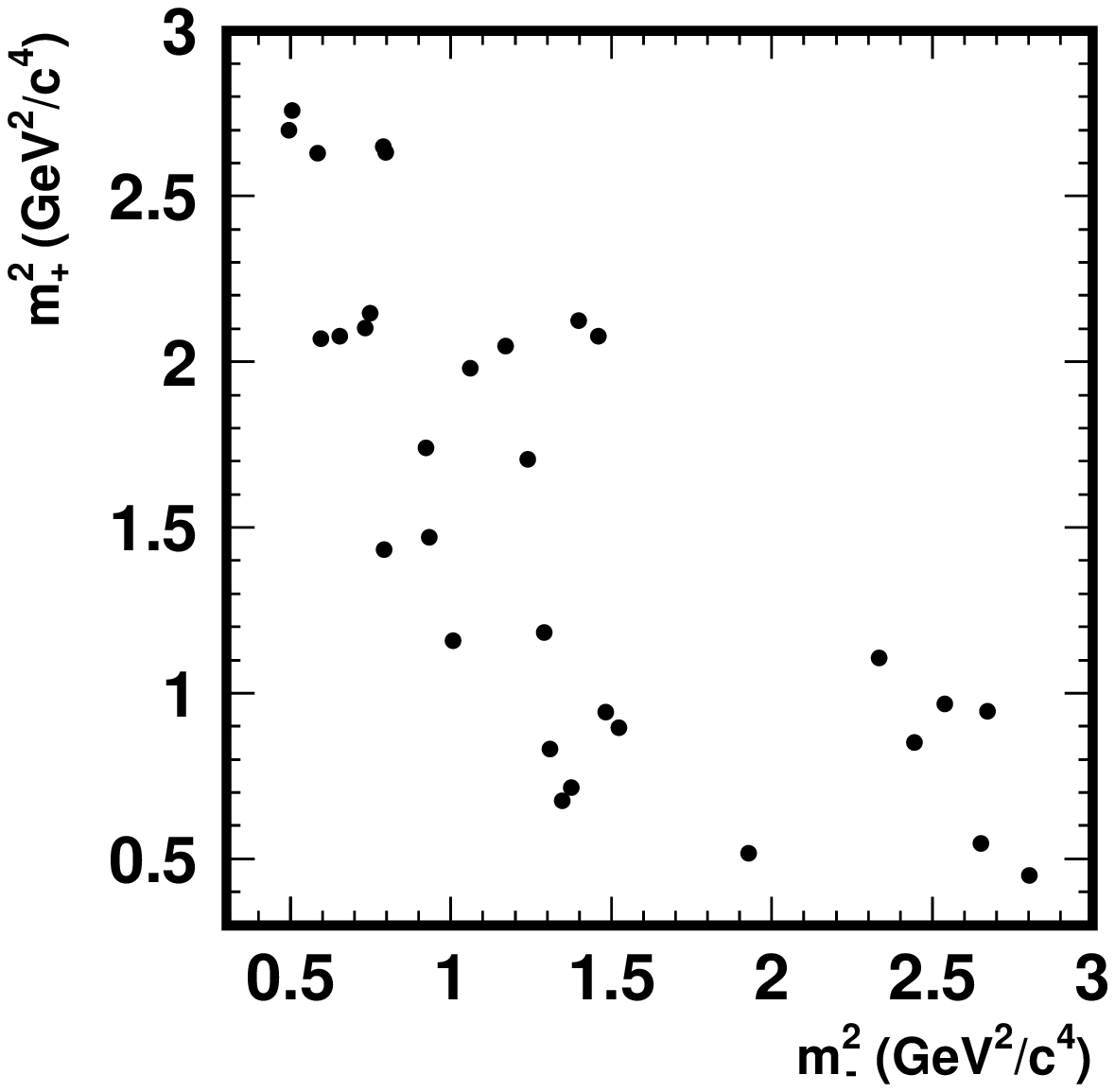}
    \end{tabular}
    \caption{
      \label{belle_dalitz}
      Dalitz distributions from Belle, 
      separately for $B^+$ and $B^-$ decays, for 
      (left) $B^\pm \to DK^\pm$, (middle) $B^\pm \to D^*K^\pm$, (right) $B^\pm \to DK^{*\pm}$.
      In each pair, the left plot contains the $B^+$ candidates,
      while the right shows those from $B^-$.
    }
  \end{center}
\end{figure}

Belle have measured $\phi_3$ using the Dalitz analysis method 
in $B^\pm \to D^{(*)} K^{(*)\pm}$ using a data sample of 
275 million $B \bar{B}$ pairs. 
The subdecays $D^* \to D\pi^0$, $K^{*\pm} \to K_S\pi^\pm$, $D \to K_S\pi^+\pi^-$ are used.
The numbers of candidate events are 
$209 \pm 16$ (with a background fraction of $25 \pm 2\%$) for $B^{\pm} \to D K^{\pm}$,  
$58 \pm 8$ ($13 \pm 2\%$) for $B^{\pm} \to D^* K^{\pm}$, 
and $36 \pm 7$ ($27 \pm 5\%$) for $B^{\pm} \to D K^{*\pm}$. 
The Dalitz distributions of candidate events 
are shown in Fig.~\ref{belle_dalitz}.

We use continuum $D^{*\pm} \to [K_S \pi^+ \pi^-]_D \pi_s^\pm$ decays
to obtain a flavour tagged $\bar{D^0} \to K_S \pi^+ \pi^-$ sample and  
express the $\bar{D^0} \to K_S \pi^+ \pi^-$ amplitude as a sum of 18 resonant 
and one non-resonant amplitudes. 
\begin{equation}
  f(m_{+}^2, m_{-}^2) = 
  \sum_{j=1}^{N} a_j e^{i\alpha_j} A_j(m_+^2, m_-^2) + b e^{i\beta}.
\end{equation}
The expected $\bar{D^0} \to K_S \pi^+ \pi^-$ Dalitz distribution, 
with which we compare the data, is then given by 
\begin{equation}
  p(m_+^2,m_-^2) = 
  \epsilon (m_+^2, m_-^2) 
  \int_{-\infty}^{+\infty} 
  |f(m_+^2 +\mu^2, m_-^2 +\mu^2)|^2 {\rm exp}(-\frac{\mu^2}{2 \sigma_m^2 (m_{\pi \pi}^2)}) 
  d\mu^2 + B(m_+^2, m_-^2)
\end{equation}
with efficiency $\epsilon$, $\pi\pi$ mass resolution $\sigma_m$, and background $B$. 
We obtain the parameters of the Dalitz model by minimizing
\begin{equation}
  -2 \mathrm{log} L =
  -2\left[ 
    \sum_{i=1}^{n} 
    \mathrm{log}~ p(m_{+,i}^2, m_{-,i}^2) -
    \mathrm{log} \int_{D} p(m_+^2,m_-^2) dm_+^2 dm_-^2 
  \right]
\end{equation}
with free parameters $a_j$ and $\alpha_j$ for each resonance
(except for $K_S \rho$ for which $a_j = 1$, $\alpha_j = 0$ are reference values), 
$b$, $\beta$, and the masses and widths of $\sigma_1$ and $\sigma_2$.
The fit results are listed in Table~\ref{kspipi_model}. 

\begin{table}
  \begin{center}
    \begin{tabular}{|l|c|c|c|} 
      \hline
      Intermediate state & Amplitude        & Phase ($^{\circ}$) & Fit fraction \\ 
      \hline      
      $K_S \sigma_1$          & $1.57\pm 0.10$     & $214\pm 4$     & 9.8\% \\
      $K_S\rho^0$           & $1.0$ (fixed)    & 0 (fixed)    & 21.6\% \\     
      $K_S\omega$             & $0.0310\pm 0.0010$ & $113.4\pm 1.9$ & 0.4\% \\      
      $K_S f_0(980)$     & $0.394\pm 0.006$   & $207\pm 3$     & 4.9\% \\      
      $K_S \sigma_2$          & $0.23\pm 0.03$     & $210\pm 13$    & 0.6\% \\      
      $K_S f_2(1270)$    & $1.32\pm 0.04$     & $348\pm 2$     & 1.5\% \\      
      $K_S f_0(1370)$    & $1.25\pm 0.10$     & $69\pm 8$      & 1.1\% \\      
      $K_S \rho^0(1450)$    & $0.89\pm 0.07$     & $1\pm 6$       & 0.4\% \\      
      $K^*(892)^+\pi^-$    & $1.621\pm 0.010$   & $131.7\pm 0.5$ & 61.2\% \\       
      $K^*(892)^-\pi^+$    & $0.154\pm 0.005$   & $317.7\pm 1.6$ & 0.55\% \\
      $K^*(1410)^+\pi^-$   & $0.22\pm 0.04$     & $120\pm 14$    & 0.05\% \\
      $K^*(1410)^-\pi^+$   & $0.35\pm 0.04$     & $253\pm 6$     & 0.14\% \\
      $K_0^*(1430)^+\pi^-$ & $2.15\pm 0.04$     & $348.7\pm 1.1$ & 7.4\% \\
      $K_0^*(1430)^-\pi^+$ & $0.52\pm 0.04$     & $89\pm 4$      & 0.43\% \\
      $K_2^*(1430)^+\pi^-$ & $1.11\pm 0.03$     & $320.5\pm 1.8$ & 2.2\% \\
      $K_2^*(1430)^-\pi^+$ & $0.23\pm 0.02$     & $263\pm 7$     & 0.09\% \\
      $K^*(1680)^+\pi^-$   & $2.34\pm 0.26$     & $110\pm 5$     & 0.36\% \\
      $K^*(1680)^-\pi^+$   & $1.3\pm 0.2$       & $87\pm 11$     & 0.11\% \\
      non-resonant       & $3.8\pm 0.3$       & $157\pm 4$     & 9.7\%  \\ 
      \hline
    \end{tabular}
    \caption{
      \label{kspipi_model}
      Results of the fit to obtain the parameters of the 
      $\bar{D}^0 \to K_S \pi^+\pi^-$ decay model, from Belle.
    }
  \end{center}
\end{table}

The fitting procedure for $B^\pm \to D^{(*)} K^{*\pm}$ 
is similar to that for the $\bar{D^0} \to K_S \pi^+ \pi^-$ fit except 
$f(m_+^2, m_-^2)$ is replaced with  
$f(m_+^2, m_-^2) + r e^{i(+\phi_3 + \delta)} f(m_-^2,m_+^2)$. 
The efficiency, resolution, and background distributions 
are also replaced with those relevant for the $B$ decay.  

We use $B^{\pm} \to D^{(*)} \pi^{\pm}$
as control samples to test the analysis procedures. 
Separate fits to the $B^+$ and $B^-$ data give 
$r_+ = 0.039 \pm 0.021$, $\theta_+ = 240^\circ \pm 28^\circ$, and 
$r_- = 0.047 \pm 0.018$, $\theta_- = 193^\circ \pm 24^\circ$ for 
the $D\pi^{\pm}$ samples, and 
$r_+ = 0.015 \pm 0.042$, $\theta_+ = 169^\circ \pm 186$, and  
$r_- = 0.086 \pm 0.049$, $\theta_- = 280^\circ \pm 30^\circ$ for 
the $D^* \pi^{\pm}$ samples. 
These results are not inconsistent with 
the expectation of $r \sim |V_{ub}V^*_{cd}/V_{cb}V^*_{ud}| \sim 0.01 - 0.02$ 
although $\sim 2\sigma$ deviation is possible, which we include in 
the systematic  error. 

\begin{figure}
  \begin{center}
    \includegraphics[width=0.67\textwidth]{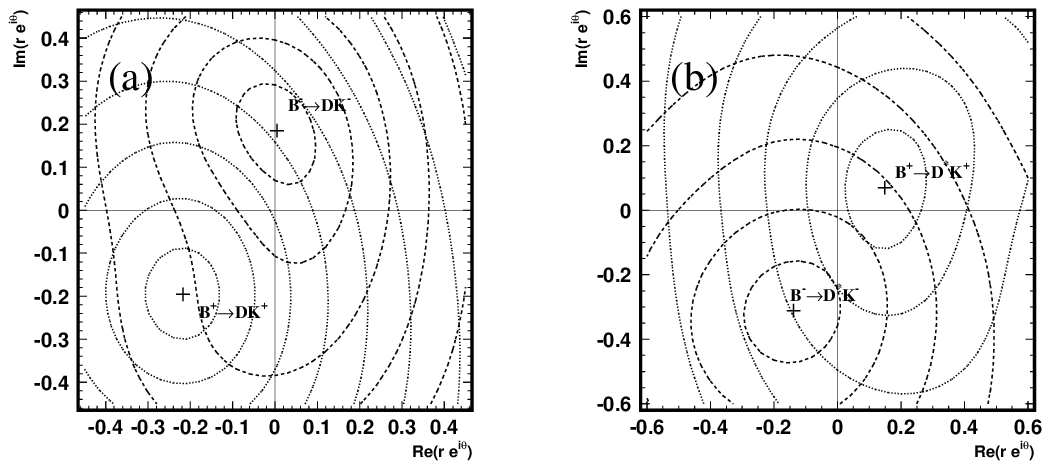}
    \includegraphics[width=0.31\textwidth]{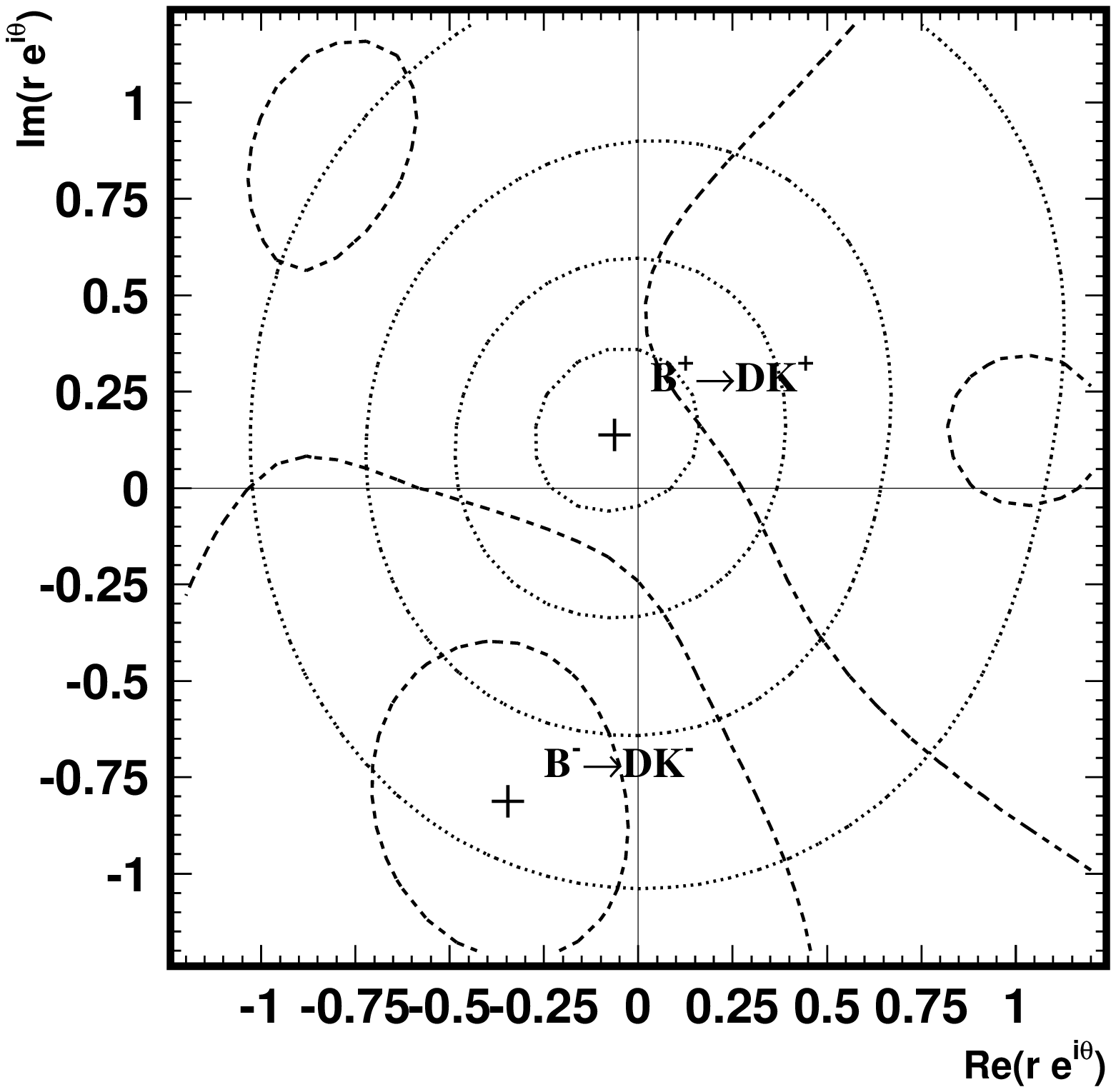}
    \caption{
      \label{belle_dalitz_cartesian}
      Results from Belle of separate $B^+$ and $B^-$ fits in terms of 
      $Re(r e^{i\theta})$-$Im(re^{i\theta})$ for
      (left) $B^\pm \to DK^\pm$, (middle) $B^\pm \to D^*K^\pm$, (right) $B^\pm \to DK^{*\pm}$.
      $CP$ violation is expected to appear as a relative rotation
      around the origin of the $B^+$ and $B^-$ contours.
    }
  \end{center}
\end{figure}

Fig.~\ref{belle_dalitz_cartesian} shows the results of the fits
in terms of $Re(r e^{i\theta})$-$Im (re^{i\theta})$ 
separate $B^+$ and $B^-$ fits for $DK^\pm$, $D^* K^\pm$, and $D K^{*\pm}$ 
samples, respectively. 

We use the frequentist technique for determining the confidence 
region. Here we determine PDFs using a toy Monte Carlo and calculate 
the confidence level and estimate the true $(r, \phi_3,\delta)$ and 
their errors. This procedure removes possible bias which arises 
from an assumption of $r$ being positive-definite. 
For the $\delta$ and $\phi_3$, the difference from unbinned fits are small. 
However, $r$ is changed by a small amount as expected.
The results are shown in Table.~\ref{belle_dalitz_results}.

Systematic errors arise from, 
i) background fractions and their shapes in the Dalitz plane, 
ii) efficiency shapes, 
iii) momentum resolution shapes, and 
iv) possible bias in the analysis method. 
In addition, another error is assigned due to the model-dependence
of the variation of the complex phase in the 
$\bar{D^0} \to K_S \pi^+ \pi^-$ amplitude.
Furthermore, for the $D K^{*\pm}$ mode, 
we include an additional systematic error 
due to the possible presence of non-resonant $D K_S \pi$ component which 
can give interference with different values of $r$ and $\delta$.  

\begin{table}
  \begin{center}
    \begin{tabular}{|l|c|c|c|}
      \hline \hline
      & $r$    &   $\phi_3~(^\circ)$  &  $\delta~(^\circ)$ \\
      \hline
      $D K^\pm$ & 
      $0.21 \pm 0.08 \pm 0.03 \pm 0.04$ &
      $64 \pm 19 \pm 13 \pm 11$ &
      $157 \pm 19 \pm 11 \pm 21$ \\
      $D^{*} K^\pm$ & 
      $0.12 ^{+0.16}_{-0.11} \pm 0.02 \pm 0.04$ &
      $75 \pm 57 \pm 11 \pm 11$ &
      $321 \pm 57 \pm 11 \pm 21$ \\
      $D K^{*\pm}$  & 
      $0.25^{+0.17}_{-0.18} \pm 0.09 \pm 0.04 \pm 0.08$ &
      $112 \pm 35 \pm 9 \pm 11 \pm 8$ &
      $353 \pm 35 \pm 8 \pm 21 \pm 49$ \\
      \hline
    \end{tabular}
    \caption{
      \label{belle_dalitz_results}
      Results of the Belle $B^\pm \to D^{(*)}K^{(*)\pm}$,
      $D \to K_S\pi^+\pi^-$ measurements.
    }
  \end{center}
\end{table}

When the $D K^\pm$ and $D^* K^\pm$ modes are combined, we obtain
$\phi_3 = {68^\circ} ^{+14^\circ}_{-15^\circ} \pm 13^\circ \pm 11^\circ$. 
In the near future, 
the statistical error can be improved by combining all modes, 
$D K^\pm$, $D^* (D \pi) K^\pm$, $D K^{*\pm}$, $D^* (D \gamma) K^\pm$, 
and by exploring other $D$ decay modes such as 
$D^0 \to \pi^+ \pi^- \pi^0$ and $D^0 \to K^+ K^- K_S$. 
As the data size increases, the experimental systematic error, 
which is presently dominated by the $D^{(*)} \pi$ control sample size, will 
also improve. 
We hope to improving the $D^0$ decay model
by studying the $CP$-tagged $D \to K_S \pi^+ \pi^-$ data from 
$\psi (3770) \to D^0 \bar{D^0}$ at CLEO-c or BES.
These data will provide $|f(m_+^2, m_-^2) \pm f(m_-^2, m_+^2)|^2$ 
and therefore a direct measurement of the phases, 
thus dramatically reducing the model uncertainty.

\subsection{\boldmath
  Extracting $\gamma$ from $D$ Dalitz analysis at BaBar }

\vspace{+2mm}
\begin{flushright}
 {\it Contribution from M.H.~Schune}
\end{flushright}

\subsubsection{\boldmath
  The Dalitz Model}

\begin{figure}[!t]
\begin{center}
\begin{tabular} {rr}  
\includegraphics[height=3.8cm]{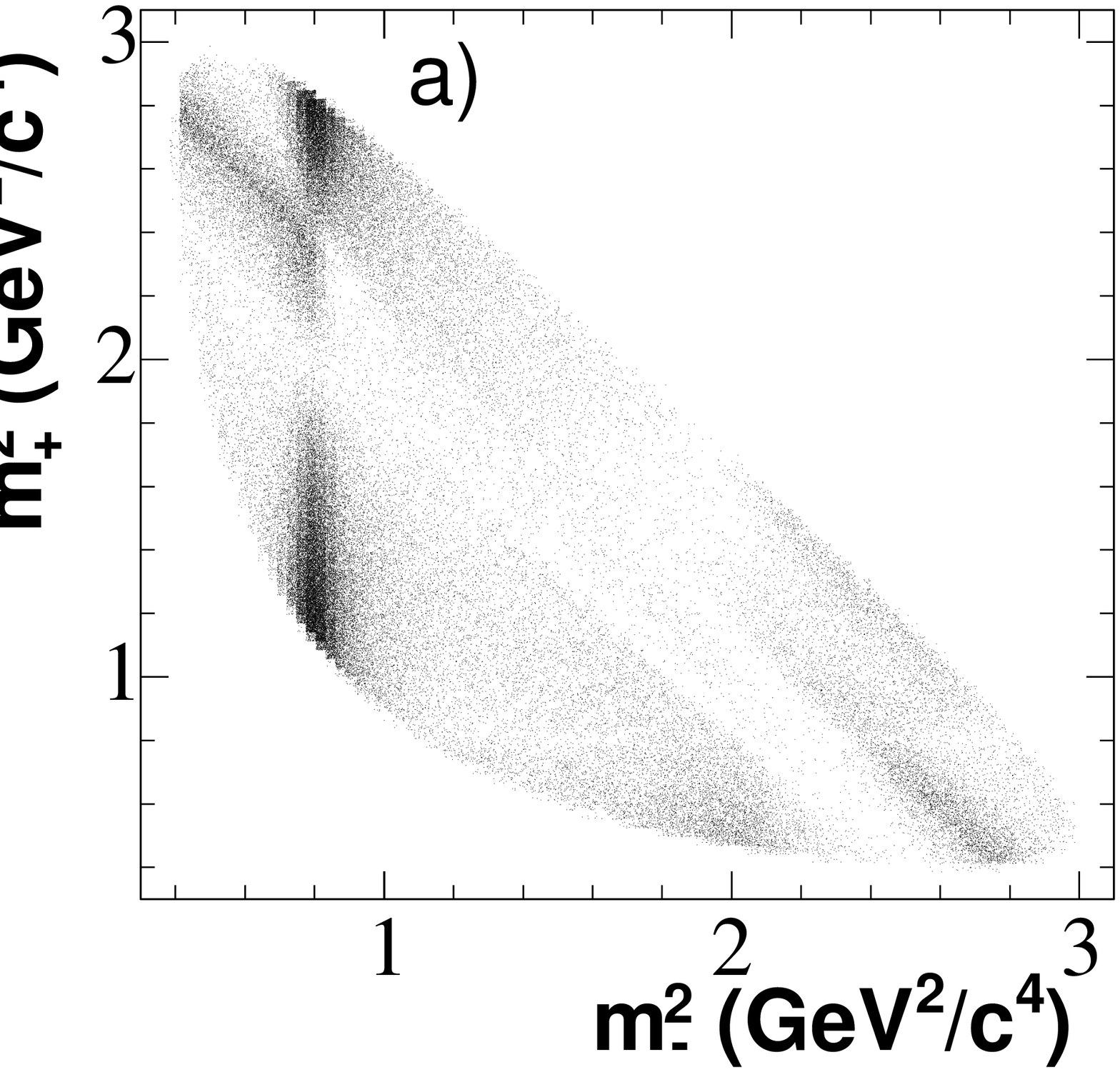} &
\includegraphics[height=3.8cm]{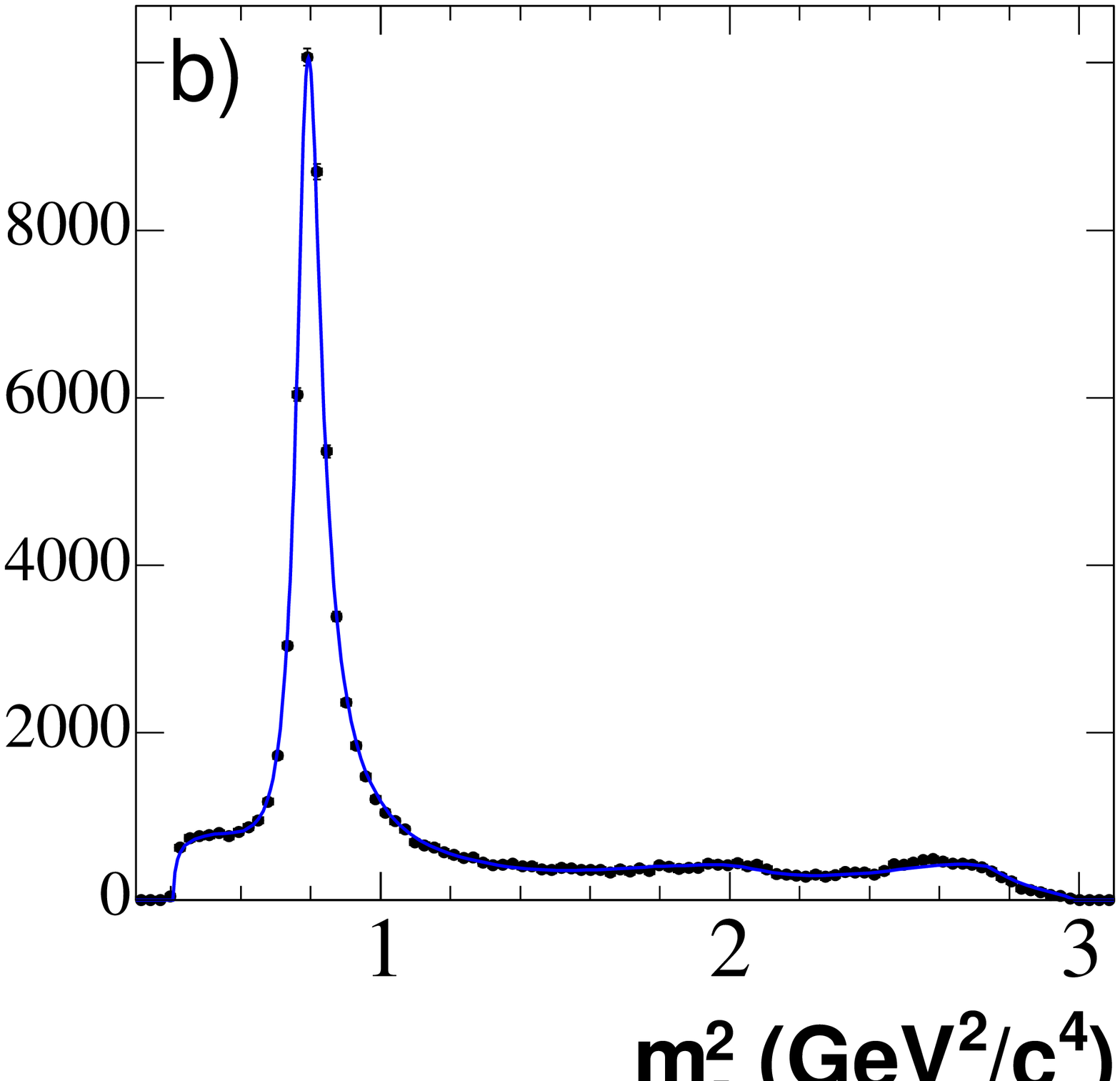} \\
\includegraphics[height=3.8cm]{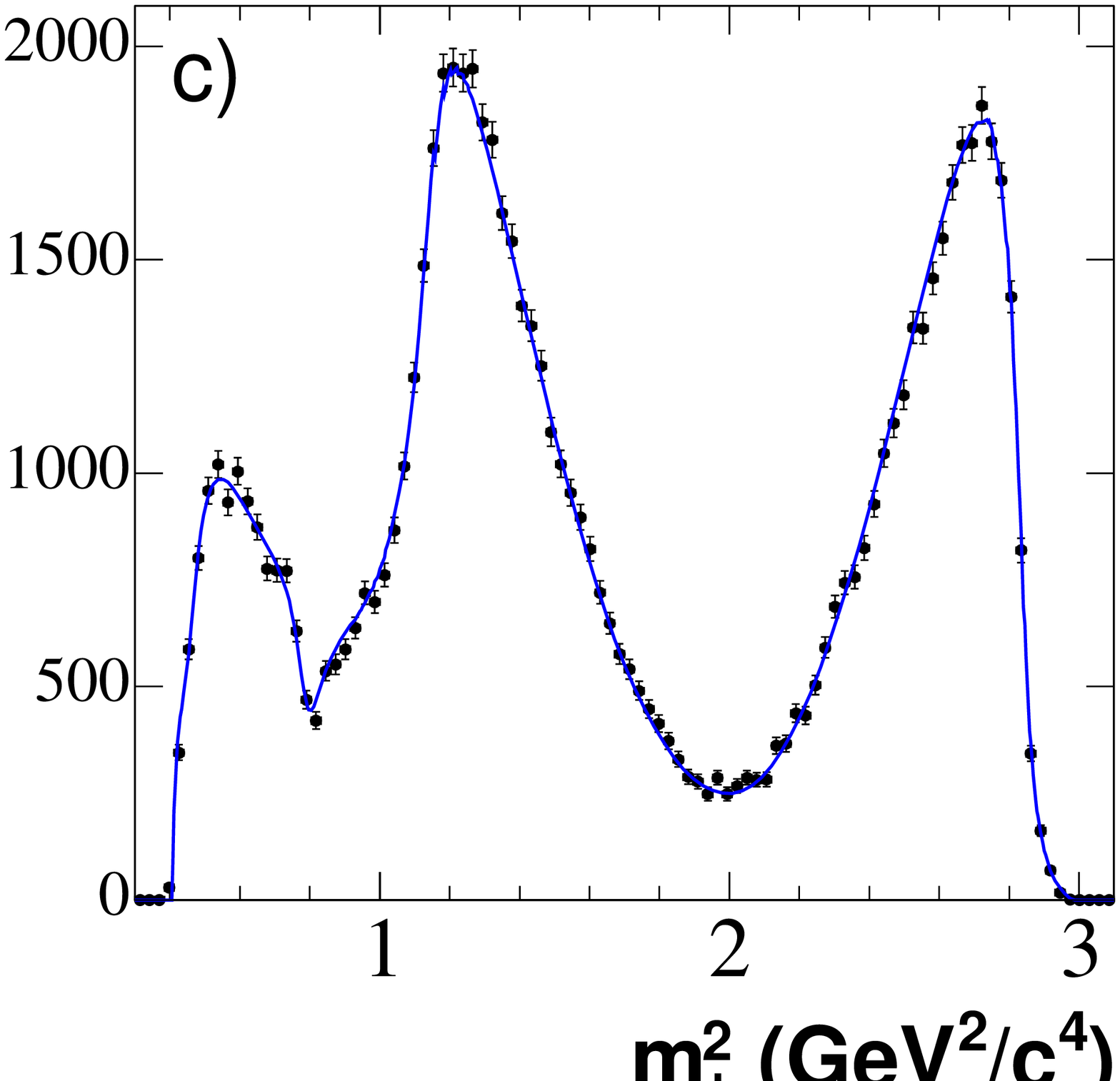} &
\includegraphics[height=3.8cm]{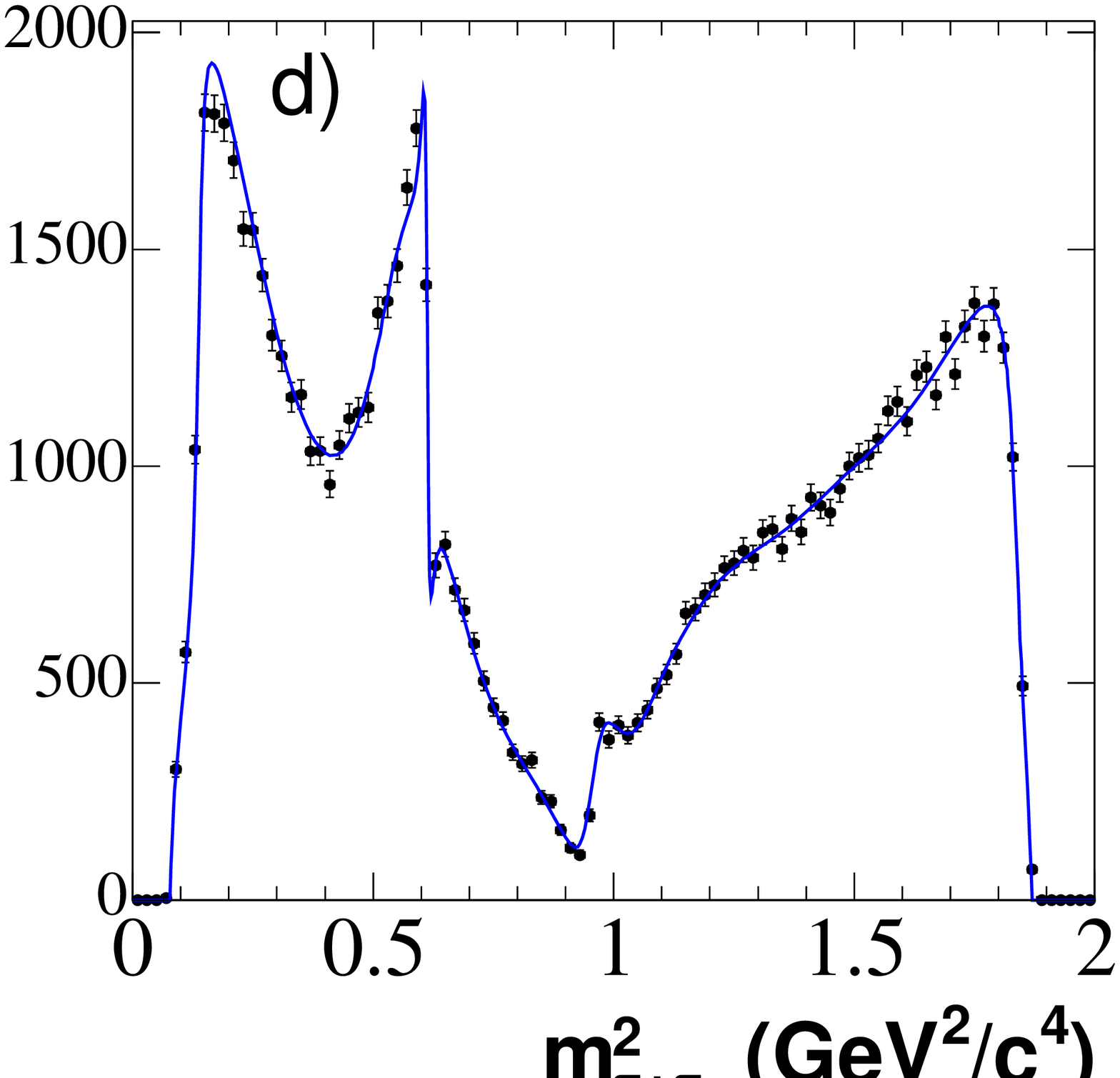} \\
\end{tabular}   
\caption{
  (a) $\Dz \to \KS \pim \pip$ Dalitz distribution from 
  $\Dstarp \to \Dz \pip$ events, and projections
  on (b) $m^2_-$, (c) $m^2_+$, and (d) $m^2_{\pip\pim}$; from \babar. 
  The curves are the fit projections.}
\label{fig:dalmkspidcs}
\end{center}
\end{figure}

The function ${\cal A}_D$  has been obtained using the flavour tagged $D$ 
meson sample from the continuum decays $D^{*\pm} \to D\pi_s^{\pm}$.
The Dalitz $(m^2_\pm,m^2_\mp)$ distribution 
(Fig.~\ref{fig:dalmkspidcs}) is fitted in the context 
of the isobar formalism described in \cite{ref:cleomodel}. 
In this formalism the amplitude $f$ can be written as a sum 
of two-body decay matrix elements and a non-resonant term according 
to the  expression:  
${\cal A}_D = 
a_{nr} e^{i\phi_{nr}} + \Sigma_r a_r e^{i\phi_r} {\cal A}_{s}(K_S \pi^- \pi^+ | r)$.
Each term of the sum is parameterized  with an amplitude and a phase. 
 We fit the Dalitz distribution with a model consisting of 13 resonances  
leading to 16 two-body decay amplitudes and phases 
(Table~\ref{tab:fitreso-likelihood}). 
Of the 13 resonances eight involve a $K_S$ plus a $\pi\pi$ resonance 
and the remaining five are made of a ($K_S \pi^-$) resonance plus a $\pi^+$. 
We also include the corresponding doubly Cabibbo-suppressed amplitudes  
for most of the ($K_S \pi^-$) $\pi^+$ decays.  
All the  resonances considered in this model are well established except 
for the two scalar $\pi\pi$ resonances, $\sigma_1$ and $\sigma_2$, 
whose masses and widths are obtained  from our sample. 

An unbinned  maximum likelihood fit is performed to measure the amplitudes $a_{nr}, a_{r}$ and the phases $\phi_{nr}, \phi_r$. 
components. The results of the fit are shown in Fig.~\ref{fig:dalmkspidcs}. Amplitudes, phases and fit fractions  
as obtained by the  likelihood fit are  reported in Table \ref{tab:fitreso-likelihood}.

\begin{table}[htbp]
\begin{center}
\begin{tabular}{lccc}
    Resonance  &  Amplitude & ~~Phase (deg)  & Fit fraction \\ 
\hline
$K^{*}(892)^-$        &    $  1.781      \pm  0.018   $   &  $ \phantom{-0}131.0   \pm  0.8$           &       $0.586$    \\ 
$K^{*}_0(1430)^-$     &    $  2.447      \pm  0.076  $    &  $  \phantom{00}-8.3       \pm  2.5\phantom{..}$   &$0.083$    \\ 
$K^{*}_2(1430)^-$     &    $  1.054      \pm  0.056   $   &  $  \phantom{.}-54.3 \pm  2.6\phantom{.}$   &       $0.027$    \\ 
$K^{*}(1410)^-$       &    $  0.515      \pm  0.087  $    &  $  \phantom{-0.}154 \pm  20\phantom{.}$  &         $0.004$    \\ 
$K^{*}(1680)^-$       &    $  0.89       \pm  0.30  $     &  $  \phantom{.}-139 \pm  14\phantom{.}$   &         $0.003$    \\ 
\hline
$K^{*}(892)^+$     &    $  0.1796     \pm  0.0079  $   &  $ \phantom{.}-44.1   \pm  2.5\phantom{.}$  &       $0.006$     \\ 
$K^{*}_0(1430)^+$  &    $  0.368      \pm  0.071   $   &  $  \phantom{.0}-342 \pm  8.5\phantom{..}$   &      $0.002$    \\ 
$K^{*}_2(1430)^+$  &    $  0.075      \pm  0.038   $   &  $  \phantom{.}-104 \pm  23\phantom{.}$    &        $0.000$    \\ 
\hline 
$\rho(770)$         &    1 (fixed)                      &    \phantom{00}0 (fixed)                                &        $0.224$    \\ 
$\omega(782)$       &    $  0.0391     \pm  0.0016   $  &  $  \phantom{-}115.3  \pm  2.5$            &        $0.006$     \\ 
$f_0(980) $         &    $  0.4817     \pm  0.012\phantom{0}$   &  $-141.8     \pm  2.2$   &        $0.061$     \\ 
$f_0(1370) $        &    $  2.25       \pm  0.30   $    &  $  \phantom{-}113.2     \pm  3.7$          &       $0.032$    \\ 
$f_2(1270) $        &    $  0.922      \pm  0.041  $    &  $  \phantom{0}-21.3  \pm  3.1\phantom{..}$   &     $0.030$     \\ 
$\rho(1450)$        &    $  0.516      \pm  0.092   $   &  $  \phantom{-00.}38 \pm  13\phantom{.}$   &        $0.002$    \\ 
$\sigma$            &    $  1.358      \pm  0.050   $   &  $  -177.9     \pm  2.7  $                &         $0.093$    \\ 
$\sigma'$           &    $  0.340      \pm  0.026   $   &  $ \phantom{-}153.0      \pm  3.8   $               &         $0.013$     \\ 
\hline 
Non Resonant        &    $  3.53       \pm  0.44  $     &  $ \phantom{-}127.6      \pm  6.4   $               &         $0.073$  \\ 
\end{tabular}
\caption{
  Amplitudes, phases and fit fractions as obtained by 
  the likelihood fit on the tagged $D$ sample, from \babar.}
\label{tab:fitreso-likelihood}
\end{center}
\end{table}

We estimate the goodness of fit through a $\chi^2$ test 
and obtain $\chi^2= 3824$ for $3054-32$ degrees of freedom.

\subsubsection{\boldmath
  Results of the $CP$ fit}

BaBar  reconstruct the decays  $B^-\to D K^-$ and 
$B^-\to D^{\ast} K^-$ with $D^{\ast} \to D\pi^0 \, , \,  D\gamma$.
$\Bm$ candidates are formed by combining a mass-constrained 
$D^{(*)}$ candidate with a track identified as a kaon~\cite{Aubert:2001tu}.

We simultaneously fit the $\Bm\to D^{(*)} \Km$ samples using
an unbinned extended maximum-likelihood fit to extract the 
$CP$-violating parameters along with the signal and background yields.

In the sample of 211 million $B\bar B$ events,
the following  signal yields are obtained 
\begin{eqnarray}
   N(B^- \ra D^0 K^-)                 & = & 261 \pm 19 ,\\ \nonumber
   N(B^- \ra D^{*0}(D^0 \pi^0) K^-)   & = & 83 \pm  11 ,\\ \nonumber
   N(B^- \ra D^{*0}(D^0 \gamma) K^-)  & = & 40  \pm 8 .
\end{eqnarray}

The results for the $CP$-violating parameters $\xbbspm$ and $\ybbspm$, are defined as the real and 
imaginary parts of the complex amplitude ratios $\rbbs e^{i(\deltabbs \pm \g)}$, 
respectively.

\begin{figure}[!t]
\begin{center}   
\begin{tabular} {cc} 
 \includegraphics[height=3.9cm]{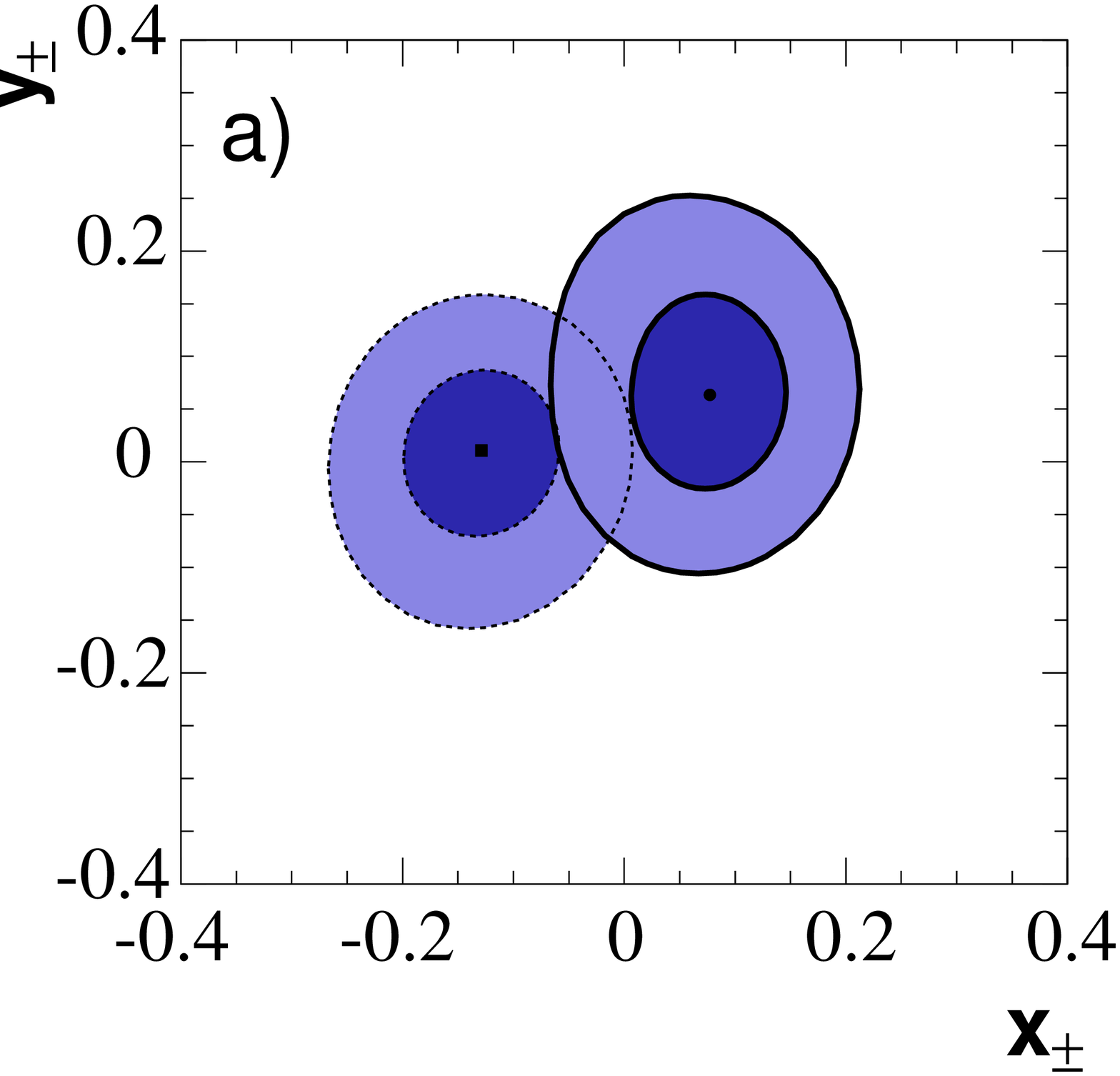} &
 \includegraphics[height=3.9cm]{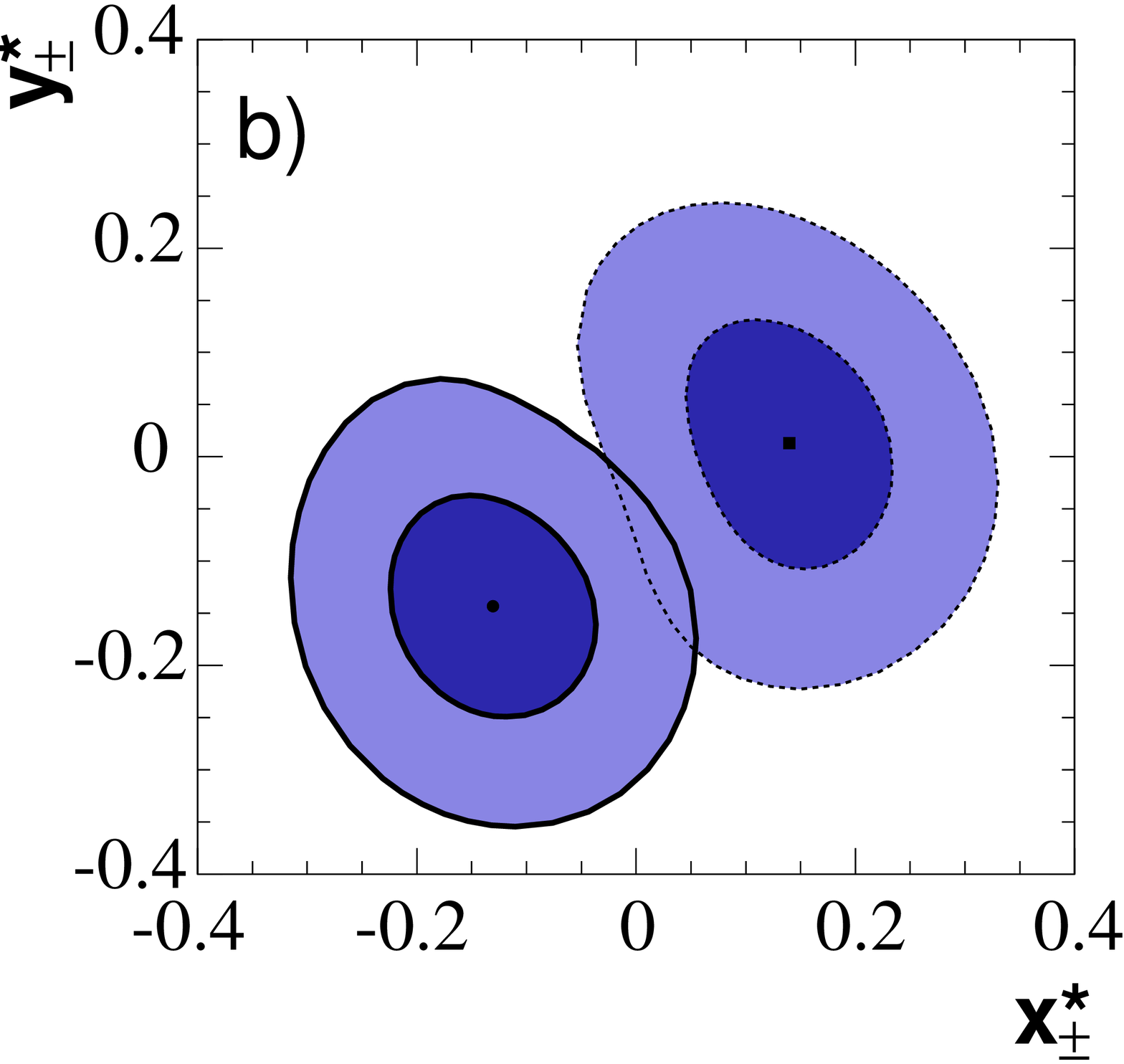} \\
 \includegraphics[height=3.9cm]{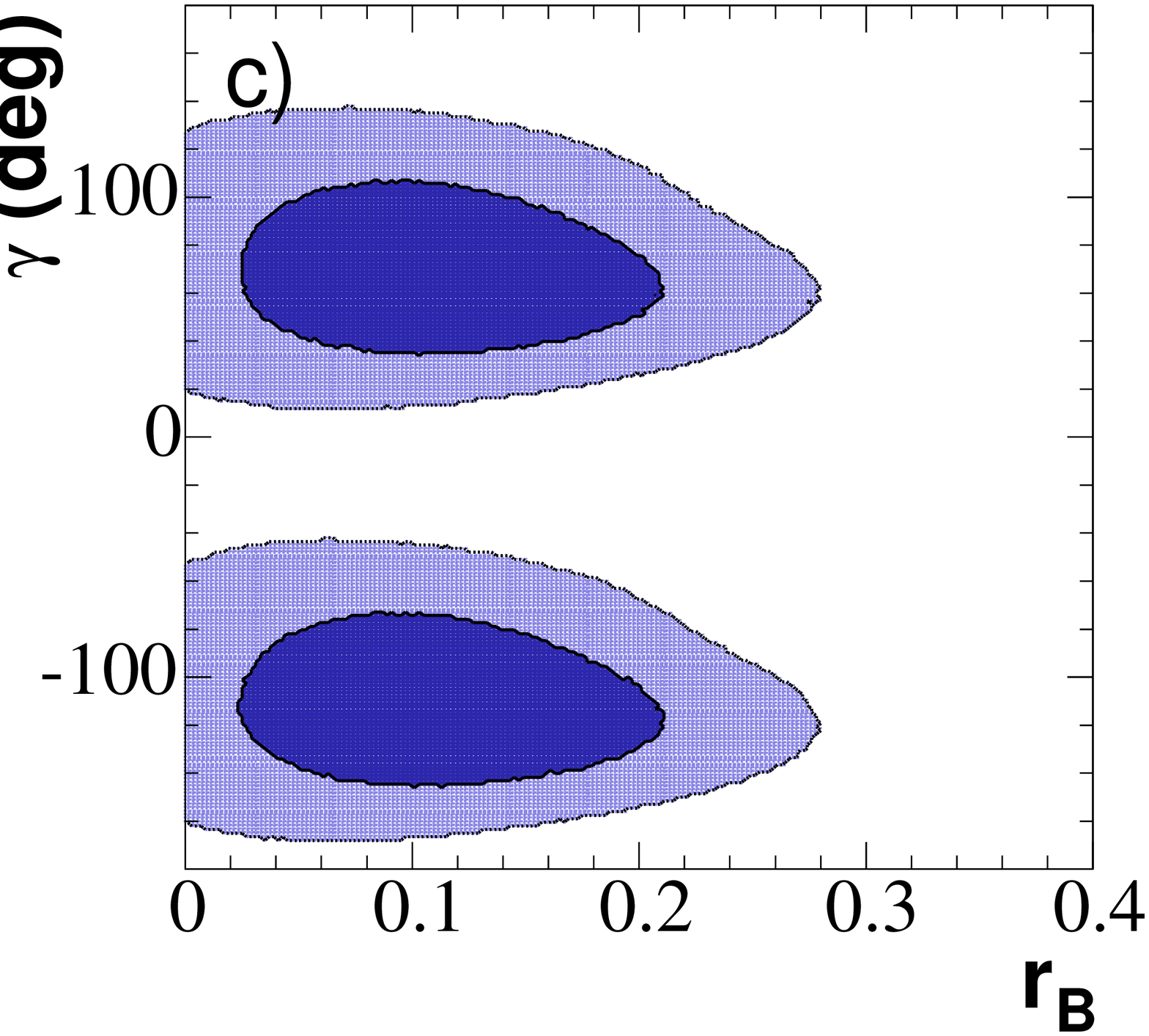} &
 \includegraphics[height=3.9cm]{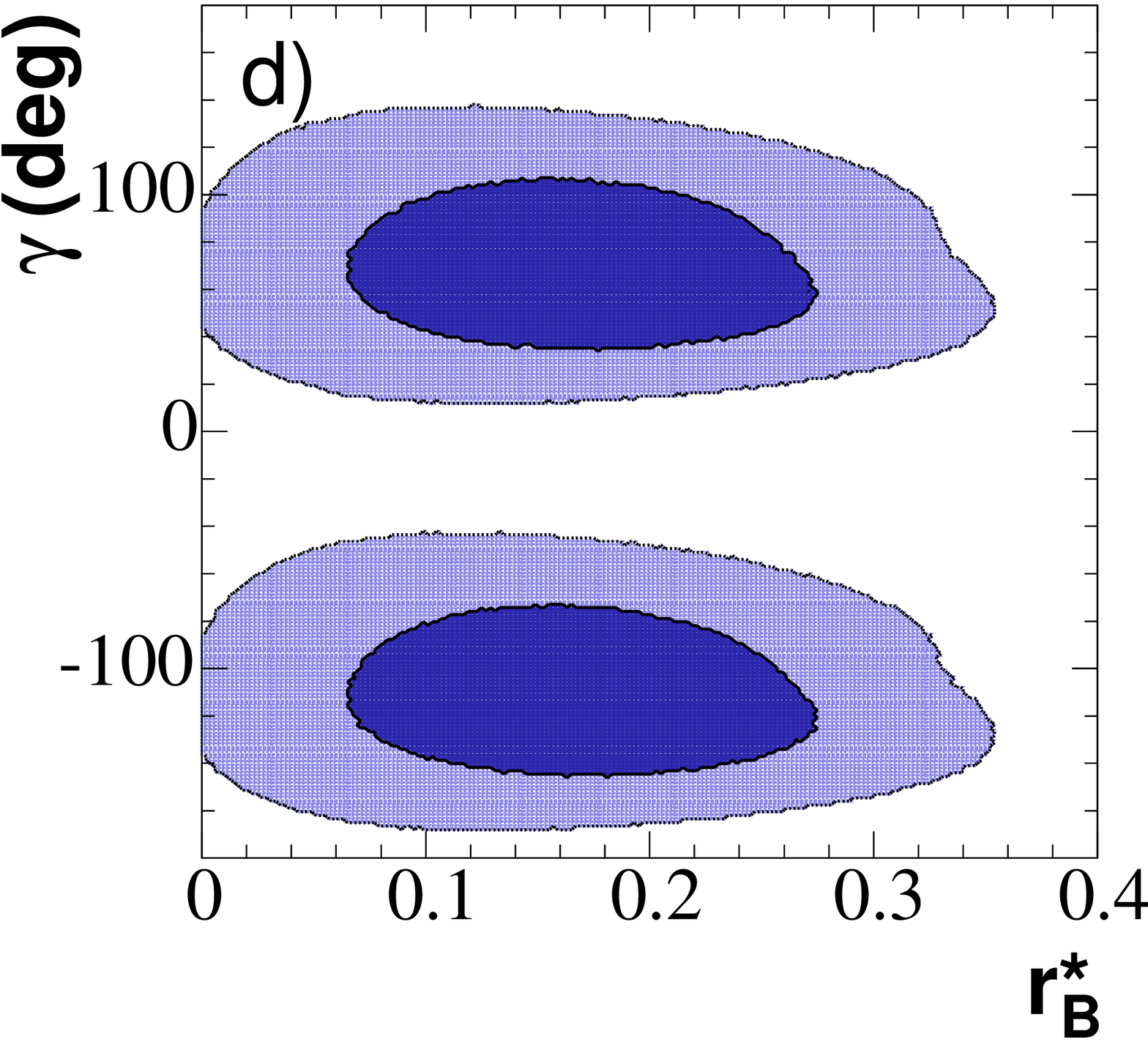}
\end{tabular}   
\caption{Contours at 68.3\% (dark) and 95\% (light) confidence level (statistical only) in the
$(\xbbs,\ybbs)$ planes for (a) $D\Km$ and (b) $D^*\Km$, separately for $\Bm$ (thick and solid) 
and $\Bp$ (thin and dotted). 
Two-dimensional projections in the $\rbbs-\g$ planes of the five-dimensional one- (dark) and two- (light)
standard deviation regions, for (c) $D\Km$ and (d) $D^*\Km$.  
}
\label{fig:contours}
\end{center} 
\end{figure}

The $\xbbspm$ and $\ybbspm$ variables are more suitable fit parameters than 
\rbbs, \deltabbs and \g because they are better behaved near the origin, especially in low-statistics samples.
Figures~\ref{fig:contours}(a,b) show the 68.3\% and 95\% confidence-level contours (statistical only)
in the $(\xbbs,\ybbs)$ planes for $D\Km$ and $D^*\Km$, and separately for $\Bm$ and $\Bp$. 
The separation between the $\Bm$ and $\Bp$ regions in these planes is an indication of direct $CP$ violation.

\indent
The largest contribution to the systematic uncertainties in the $CP$ parameters
comes from the choice of the model used to describe 
the $\Dz \to \KS \pim \pip$ decay amplitudes. 
To evaluate this uncertainty we use the nominal Dalitz model 
(Table~\ref{tab:fitreso-likelihood}) to generate
large samples of experiments and we compare experiment by experiment 
the values of \xbbspm and \ybbspm obtained from
fits using the nominal model and a set of alternative models.
Models where one or both of the $\sigma$ resonances are removed lead to a
significant increase in the $\chi^2$ of the fit. 
We use the average variations of \xbbspm and \ybbspm 
corresponding to this second set of alternative 
models as the systematic uncertainty due to 
imperfect knowledge of ${\cal A}_D$.

A frequentist construction of the confidence regions of ${\bf p} \equiv (\rb,\rbs,\deltab,\deltabs,\g)$
based on the constraints on \xbbspm and \ybbspm has been adopted
(see below and Ref.~\cite{ref:pdg2004}) 
and allows to extract $\gamma$, $\rb$, $\rbs$, $\deltab$ and $\deltabs$: \\
$\gamma = \reslinepol{70}{31}{12}{10}{14}{11}{12}{137}$, 
$\rb = \reslinep{0.118}{0.079}{0.034}{0.036}{0.034}{0}{0.279}$, 
$\rbs = \resline{0.169}{0.096}{0.030}{0.028}{0.029}{0.026}{0}{0.354}$,
$\deltab=\reslinepolval{104}{45}{17}{21}{16}{24}$, 
$\deltabs=\reslinepolvalpp{296}{41}{14}{12}{15}$.
The first error is statistical, the second is the experimental
systematic uncertainty and the third reflects the Dalitz model uncertainty. 
The values inside square brackets indicate the two-standard-deviation intervals.

\subsection{\boldmath
  $\gamma$ extraction in $B^\pm \to D^{(*)} K^{\pm}$ Dalitz: a frequentist statistical treatment}

\vspace{+2mm}
\begin{flushright}
 {\it Contribution from N.~Neri}
\end{flushright}

\subsubsection{Cartesian coordinates as fit parameters}

The likelihood function used for the measurement of the angle $\gamma$, 
reconstructing $\Dz$ three-body decays (such as $\Dz \to \KS \pi^+ \pi^-$) 
from $B^{-} \to D^{(*)} K^-$ decays through a Dalitz analysis~\cite{ggsz}, 
is affected by biased values and non-Gaussian errors 
if we fit directly for the parameters 
$r_{B}$, $\gamma$ and $\delta_{B}$. 
We find that {\it cartesian coordinates}:
\begin{eqnarray}
  x_\pm = r_{B} \cos(\gamma\pm\delta_{B}) \nonumber \\ 
  y_\pm = r_{B} \sin(\gamma\pm\delta_{B}) \nonumber
\end{eqnarray}
are more suitable set of parameters because they are better behaved
near the origin expecially in low statistics sample. 
In addition the cartesian coordinates are largely uncorrelated. 
If we represent in the $(x_\pm,y_\pm)$ plane the results for $B^\pm$ decays, 
the distance of the two points $d$ is
\begin{eqnarray}
  d  =  [(x_- - x_+)^2+(y_- - y_+)^2]^{1/2} = 2 r_B |\sin\gamma| \nonumber~. 
\end{eqnarray}
A non null distance means evidence of direct $CP$ violation. 

\subsubsection{Frequentist procedure}

In the case of a single signal sample\footnote{
  A similar procedure has been used for combined measurements 
  of two and three signal $B^\pm \to D^{(*)}K^{(*)\pm}$ samples.
} $B^\pm \to DK^\pm$, 
we define a 4-dimensional PDF, 
of the fit parameter as a function of the true parameter, as:

\begin{eqnarray}
  \frac{d^4 P}{d^2{\bf z}_+ d^2{\bf z}_-}({\bf z}_+,{\bf z}_- | \bf{p}^t) 
  & = & 
  G_2 \left( 
    {\bf z}_+; r^t_\B \cos(\delta^t+\gamma^t), r^t_\B \sin(\delta^t+\gamma^t), \sigma_{x_+}, \sigma_{y_+}, \rho_+ 
  \right) ~\times \nonumber \\
  & & 
  G_2 \left( 
    {\bf z}_-; r^t_\B \cos(\delta^t-\gamma^t), r^t_\B \sin(\delta^t-\gamma^t), \sigma_{x_-}, \sigma_{y_-}, \rho_- 
  \right) \nonumber
  \label{eq:d4pdf}
\end{eqnarray}
where
\begin{eqnarray}
  G_2 \left( {\bf z}; \mu_x, \mu_y, \sigma_x, \sigma_y, \rho \right) 
  & = & 
  \frac{1}{2\pi\sigma_x \sigma_y \sqrt{1-\rho^2}} 
  e^{-\frac{1}{2(1-\rho^2)} 
    \left[ 
      \frac{(x-\mu_x)^2}{\sigma_x^2} + \frac{(y-\mu_y)^2}{\sigma_y^2} 
      -\frac{2\rho(x-\mu_x)(y-\mu_y)}{\sigma_x \sigma_y}  
    \right] }  \nonumber
\end{eqnarray}
and
${\bf z}_\pm = (x_\pm,y_\pm)$ and ${\bf p} = (r_\B,\gamma,\delta)$. 
The vectors ${\bf z}_\pm^t$ and ${\bf p}^t$, 
defined equivalently to ${\bf z}_\pm$ and ${\bf p}$ respectively, 
are the corresponding parameters in the truth parameter space. 
The Gaussian widths $(\sigma_{x_\pm}$, $\sigma_{y_\pm})$ and the 
correlation distributions ($\rho_\pm$) can obtained either from Toy MC experiments 
or from the fit to the data sample itself.

The confidence level $1-\alpha$ for each set of true parameters ${\bf p}^t$ 
is calculated as
\begin{eqnarray}
  \alpha({\bf p}^t) & = & 
  \int_D \frac{d^4 P}{d^2 {\bf z}_+ d^2 {\bf z}_-} ({\bf z}_+,{\bf z}_- | {\bf p}^t ) d^2 {\bf z}_+ d^2 {\bf z}_-~, \nonumber
  \label{eq:CLdef}
\end{eqnarray}
where the integration domain $D$ (the confidence region) 
is given by the condition
\begin{eqnarray}
  \frac{d^4 P}{d^2 {\bf z}_+ d^2 {\bf z}_-} ({\bf z}_+,{\bf z}_- | {\bf p}^t ) \geq \frac{d^4 P}{d^2 {\bf z}_+ d^2 {\bf z}_-} ({\bf z}_+^{\rm data},{\bf z}_-^{\rm data} | {\bf p}^t )~, \nonumber
  \label{eq:Ddef}
\end{eqnarray}
{\it i.e.} it includes all points in the fit parameter space 
closer to the truth point than the data point ${\bf z}^{\rm data}$.
If we are interested in building a 3-dimensional region of joint probability
$1-\alpha_0$, then we select only those points for which $\alpha({\bf p}^t) \leq \alpha_0$. 
The 2-dimensional and 1-dimensional contours are then built by projecting the 3-dimensional 
joint probability regions. The correct coverage of the method has been verified.

Finally, the significance of $CP$ violation can be determined by 
finding the confidence level $1-\alpha_{CP}$ for the most
probable \CP conserving point ${\bf p}^t_{\CP}$, 
{\it i.e.} the ${\bf p}^t$ point with $r^t_\B=0$ or $\gamma^t=0$.

\subsubsection{\boldmath
  Perspectives}
A preliminary study has been performed for the $D^0 \to K_s K^+ K^-$ decay mode. 
The $\Bm$ candidate is reconstructed using similar selection of the $D^0 \to K_s \pi^- \pi^+$ mode
Using this event yield and a study of the $\Dz \to \KS \Kpm \Kmp$ Dalitz plot, toy studies indicate a marginal gain of few percent on the $\gamma$ uncertainty. However, it is worthwhile noting that the systematic uncertainty due  to the Dalitz model will be totally independent of the one of the \dotokspp analysis. 

A first analysis of the \btodk mode using the $\Dz \to \pi^+ \pi^- \piz$ decay mode
has been performed. 
This channel is affected by a larger background that could dilute 
the sensitivity to $\gamma$. 
Nevertheless, adding this channel is certainly worthwhile in a 
strategy of a combined fit with other channels.

Using the present cartesian central values and a Bayesian technique we have 
computed the expected uncertainty for $\gamma$ for various integrated luminosities. 
The systematic uncertainty has been (conservatively) assumed
to stay constant. It has been shown  that, up to 500 \invfb, 
the result is yet not dominated by the Dalitz model systematic uncertainty. 
However, in a view of a larger statistics accumulated by $B$-factory  
it is probably worthwhile performing  a model independent analysis~\cite{ggsz}.

 \subsection{\boldmath
  Model independent approach } 

\vspace{+2mm}
\begin{flushright}
  {\it Contribution from J.~Zupan}
\end{flushright}

 The first step in the model independent approach is to 
partition the Dalitz plot into $2k$ bins placed symmetrically 
about the $12 \leftrightarrow 13$ line. The
$k$ bins lying below the symmetry axis are denoted by index $i$,
while the remaining bins  are indexed with $\bar{i}$.
The $\bar{i}$-th bin is obtained by  mirroring the $i$-th bin over
the axis of symmetry. For bins below the symmetry axis we define
\begin{eqnarray}
c_i &\equiv& \int_i dp\; 
A_{12,13}\,A_{13,12}\,\cos(\delta_{12,13}-\delta_{13,12}),
\label{ci}\\
s_i &\equiv& \int_i dp\;  
A_{12,13}\,A_{13,12}\sin(\delta_{12,13}-\delta_{13,12}),
\label{si}\\
T_i &\equiv& \int_i dp\;  
A_{12,13}^2, \label{Bi}
\end{eqnarray}
where the integrals are done over the phase space of the $i$-th
bin. The variables $c_i$ and $s_i$ contain differences of strong phases and
are therefore unknowns in the analysis. The variables $T_i$, on
the other hand, can be measured from the flavor tagged $D$ decays, and are assumed to be known inputs into the analysis.
The variables
$c_i, s_i$ of the $i$-th bin are related to the variables of the
$\bar{i}$-th bin by
\begin{equation}
c_{\,\bar{i}}=c_i, \qquad s_{\,\bar{i}}=-s_i, \label{vanish}
\end{equation}
while there is no relation between $T_i$ and $T_{\bar{i}}$. 

Together with the information available from the $B^+$ decay, we 
arrive at a set of $4k$ equations
\begin{eqnarray}
\label{11a}
\hat\Gamma^-_i &\equiv&
\int_i d\hat\Gamma(B^- \to (K_S \pi^-\pi^+)_D K^-)=\nonumber\\
&&T_i  +r_B^2 T_{\bar{i}}\; + 
2 r_B [\cos(\delta_B-\gamma) c_i + \sin(\delta_B-\gamma) s_i],
\end{eqnarray}
where we display only the first $k$ equations, while the other expressions for $\hat\Gamma^-_{\bar i}$, $\hat\Gamma^+_i$, 
$\hat\Gamma^+_{\bar i}$ can be obtained from the above by  $12\leftrightarrow
13$ and/or $\gamma\leftrightarrow -\gamma$ replacements \cite{ggsz}. We have
$2k + 3$ unknowns: $c_i, ~s_i, ~r_B, ~\delta_B, ~\gamma$ for $4k$ observables, so that the set of equations is solvable 
for $k\geq 2$, i.e., if Dalitz plot is divided into at least four bins. The whole approach has been extended also to neutral 
$B$ decays \cite{Gronau:2004gt}. 

An important input can be provided by 
charm factories~\cite{ggsz,Atwood:2003mj}. 
Namely, the parameters $c_i$ and $s_i^2$ describing $D$ decay can be measured 
at charm factories working at $\psi(3770)$. 
This greatly reduces the number of unknowns needed to be fit from $B$ decays 
(which is the limiting statistical factor). 
Another observation is that $c_i, s_i$ are bounded from above and below, 
$|s_i|,\;|c_i| \leq \int_i dp\; A_{12,13} A_{13,12} \leq \sqrt{T_i \,T_{\bar{i}}}$.  
This can prove useful in the actual implementation of the method. 
An  important question for the implementation of the method is also, 
how small need the bins be, not to average out sensitivity to $\gamma$? 
Answer to this question is rather complicated and 
difficult to answer without a Monte Carlo study. 

Another interesting observation is, that to leading order in 
$x=\Delta m_D/\Gamma_D$, $y=\Delta \Gamma_D/ (2 \Gamma_D)$ 
CP conserving $D-\bar D$ mixing does not change the formalism~\cite{work}. 
Also, the uncertainty due to $CP$ violation in $D$ sector is $\lambda^6\sim 10^{-4}$ 
suppressed in SM and therefore completely negligible. 
However, even these effects can be included in the analysis~\cite{work}.

\subsection{\boldmath
  Charm phenomenology and extraction of the CKM angle $\gamma$}
\vspace{+2mm}
\begin{flushright}
  {\it Contribution from A.A.~Petrov}
\end{flushright}

The Standard Model is a very constrained system, which implements a remarkably simple
and economic description of all $CP$-violating processes in the flavor sector by 
a single $CP$-violating parameter, the phase of the CKM matrix. This fact relates 
all $CP$-violating observables in bottom, charm and strange systems and provides an 
excellent opportunity for searches of physics beyond the Standard Model.
One of the major goals of the contemporary experimental $B$ physics program is an accurate 
determination of the CKM parameters. As we shall see below, inputs from charm decays are 
important ingredient in this program, both in the extraction of the angles and sides 
of the CKM unitarity triangle.

The cleanest methods of the determination of the CKM phase 
$\gamma=\arg\left[-V_{ud}V_{ub}^*/V_{cd}V_{cb}^*\right]$ involve
the interference of the $b \to c\bar{u}s$ and $b \to u\bar{c}s$ quark-level
transitions~\cite{Intro}. A way to arrange for an interference 
of those seemingly different 
processes was first pointed out in~\cite{GronauWyler,GronauLondon}. 
It involves interference of two {\it hadronic} decays 
$B^+ \to D^0 K^+ \to f K^+$ and $B^+ \to \barD K^+ \to f K^+$, 
with $f$ being any common final state for $D^0$ and $\barD$ decays. 
Since then, many different methods have been proposed, mainly differing by the 
$f K$ final state (i.e. with $f$ being a $CP$-eigenstate or not a $CP$-eigenstate) and paths of 
reaching it: a combination of the Cabibbo-favored (CF) and doubly-Cabibbo suppressed decays (DCSD)
$D^0(\barD) \to K^+ \pi^-$~\cite{ads2}, singly-Cabibbo suppressed decays (SCSD)
$D^0(\barD) \to K K^*$~\cite{Grossman:2002aq}, Cabibbo-favored decays employing
large $K^0-\overline{K^0}$ mixing 
transitions~\cite{ggsz}. 
For similar methods involving interference of the 
{\it initial} state, see~\cite{Falk:2000ga}.

All these methods produce expressions that depend on $\gamma$ and several
hadronic parameters, such as ratios of hadronic amplitudes and strong phases.
These parameters cannot be reliably computed at the moment, so their values must
be fixed from experimental data. 
In principle, all of the hadronic parameters in these methods can be obtained 
from the measurements of $B$-decays only. However, accuracy of these methods
can be significantly improved if some measurements of charm-related parameters 
are performed separately. A good example is provided by the original
GW method~\cite{GronauWyler,GronauLondon}, 
which does not take into account the possibility of 
relatively large $\DDbar$ mixing~\cite{Silva:1999bd}. 
This method relies on the simple triangle amplitude relation 
\beq
\sqrt{2} A(B^+ \to D_\pm K^+)= A(B^+ \to D^0 K^+) \pm A(B^+ \to \barD K^+),
\eeq
which follows from the relation which neglects $CP$-violation in charm,
\beq \label{CPeig}
\sqrt{2} | D_\pm \rangle = | D^0 \rangle \pm | \barD \rangle.
\eeq
An amplitude $A(B^+ \to D_\pm K^+)$ is measured with $D$ decaying to a 
particular $CP$-eigenstate. Neglecting $\DDbar$ mixing, angle $\gamma$ can then be extracted, 
up to a discrete ambiguity, from the measurements of 
$B^\pm \to f_{CP} K^\pm$ and $B^\pm \to D^0,\barD K^\pm$. In particular,
\beq
\Gamma[B^\pm \to f_{CP} K^\pm] \propto 1 + r_B^2 + 2 r_B c_\pm, 
\eeq
where $c_\pm=\cos (\gamma \pm \delta_B)$ and $r_B$ and $\delta_B$ are defined from
\beq\label{BDratio}
A(B^+ \to D^0 K^+)/A(B^+ \to \barD K^+)=r_B e^{i \left(\gamma+\delta_B\right)},
\eeq
Then,
\beq
\sin^2 \gamma = \frac{1}{2} \left[
1-c_+ c_- \pm \sqrt{\left(1-c_+^2\right) \left(1-c_-^2\right)}
\right].
\eeq
It is easy to see that $\DDbar$ mixing, if not properly accounted for, can
affect the results of this analysis. Indeed, 
taking $\DDbar$ mixing into account results in the modification of 
the definitions of $c_\pm$,
\bea
c_\pm\to\cos (\gamma \pm \delta_B) \mp \frac{x}{2 r_B} \sin 2\theta_D 
- \frac{y}{2 r_D}
\left[
2 \eta_f r_D \cos\left(\gamma+2\theta_D\pm\delta_B\right)+\cos 2\theta_D
\right],
\eea
where 
\begin{eqnarray} \label{definition}
x \equiv \frac{m_2-m_1}{\Gamma}, ~~
y \equiv \frac{\Gamma_2 - \Gamma_1}{2 \Gamma},
\end{eqnarray}
with $m_{1,2}$ and $\Gamma_{1,2}$ being the masses and widths of D-meson mass eigenstates
$D_{1,2}$ (which reduce to $D_\pm$ in the $CP$-invariance limit), $\eta_f$ is a $CP$-parity of 
$f_{CP}$, and $\theta_D$ is a $CP$-violating phase of $\DDbar$ mixing.
It is easy to see that $y \sim 1\%$ can impact the determination of $\gamma$ from these modes.
Thus, separately constraining $\DDbar$ mixing parameters will be helpful. 

The current experimental upper bounds on $x$ and $y$ are on the order of 
a few times $10^{-2}$, and are expected to improve significantly in the coming
years. As was recently shown~\cite{Falk:2001hx}, in the Standard Model, 
$x$ and $y$ are generated only at second order in SU(3)$_F$ breaking, 
\beq
x\,,\, y \sim \sin^2\theta_C \times [SU(3) \mbox{ breaking}]^2\,,
\eeq
where $\theta_C$ is the Cabibbo angle.  Therefore, predicting the
Standard Model values of $x$ and $y$ depends crucially on estimating the 
size of SU(3)$_F$ breaking. Although $y$ is expected to be determined
by the Standard Model processes, its value can affect the sensitivity 
to new physics of experimental analyses of $D$ mixing~\cite{Bergmann:2000id}.

Presently, experimental information about the $\DDbar$ mixing parameters 
$x$ and $y$ comes from the time-dependent analyses that can roughly be divided
into two categories. First, one can look at the time
dependence of $D \to f$ decays, where $f$ is the final state that tags 
the flavor of the decayed meson. The most popular is the
non-leptonic doubly Cabibbo suppressed decay $D^0 \to K^+ \pi^-$.
Time-dependent studies allow one to separate the DCSD from the mixing 
contribution $D^0 \to \D0bar \to K^+ \pi^-$,
\begin{eqnarray}\label{Kpi}
\Gamma[D^0(t) \to K^+ \pi^-]
=e^{-\Gamma t}|A_{K^-\pi^+}|^2 \qquad\qquad\qquad\qquad\qquad
\nonumber \\
\times ~\left[
R+\sqrt{R}R_m(y'\cos\phi-x'\sin\phi)\Gamma t
+\frac{R_m^2}{4}(y^2+x^2)(\Gamma t)^2
\right],
\end{eqnarray}
where $R$ is the ratio of DCS and Cabibbo favored (CF) decay rates. 
Since $x$ and $y$ are small, the best constraint comes from the linear terms 
in $t$ that are also {\it linear} in $x$ and $y$.
A direct extraction of $x$ and $y$ from Eq.~(\ref{Kpi}) is not possible due 
to unknown relative strong phase $\delta_D$ of DCS and CF 
amplitudes~\cite{Falk:1999ts}, 
as $x'=x\cos\delta_D+y\sin\delta_D$, $y'=y\cos\delta_D-x\sin\delta_D$. 
As discussed below, this phase can be measured 
independently~\cite{Silva:1999bd,GGR,GP}. 
Second, $D^0$ mixing can be measured by comparing the lifetimes 
extracted from the analysis of $D$ decays into the $CP$-even and $CP$-odd 
final states. This study is also sensitive to a {\it linear} function of 
$y$ via
\beq
\frac{\tau(D \to K^-\pi^+)}{\tau(D \to K^+K^-)}-1=
y \cos \phi - x \sin \phi \left[\frac{R_m^2-1}{2}\right].
\eeq
Time-integrated studies of the semileptonic transitions are sensitive
to the {\it quadratic} form $x^2+y^2$ and are not 
competitive with the analyses discussed above. 

Charm factories (ChF) such as CLEO-c and BES-III introduce new {\it time-independent} 
methods that are sensitive to a linear function of $y$. There, one can use the 
fact that heavy meson pairs produced in the decays of heavy quarkonium 
resonances have the useful property that the two mesons are in the $CP$-correlated 
states~\cite{AtwoodPetrov},
\begin{eqnarray} \label{TCFinitial}
|D \barD \rangle_L = \frac{1}{\sqrt{2}} 
\left \{
| D^0 (k_1)\barD (k_2) \rangle +
(-1)^L | D^0 (k_2)\barD (k_1) \rangle
\right \},
\end{eqnarray}
where $L$ is the relative angular momentum of two $D$ mesons, 
$CP$ properties of the final states produced in
the decay of $\psi(3770)$ are anti-correlated,
one $D$ state decayed into the final state with
definite $CP$ properties immediately identifies or tags
$CP$ properties of the state ``on the other side.'' 
That is to say, if one state decayed into, say $\pi^0 K_S$ 
with $CP=-1$, the other state is ``$CP$-tagged'' as being in the 
$CP = +1$ state. By tagging one of the mesons as a $CP$ eigenstate, a lifetime difference 
may be determined by measuring the leptonic branching ratio of the other meson.
Its semileptonic {\it width} should be independent of the $CP$ quantum number 
since it is flavor specific, yet its {\it branching ratio} will be inversely 
proportional to the total width of that meson. Since we know whether this $D(k_2)$ state is 
tagged as a ($CP$-eigenstate) $D_\pm$ from the decay of $D(k_1)$ to a 
final state $S_\sigma$ of definite $CP$-parity $\sigma=\pm$, we can 
easily determine $y$ in terms of the semileptonic branching ratios of $D_\pm$. This 
can be expressed simply by introducing the ratio
\beq \label{DefCor}
R^L_\sigma=
\frac{\Gamma[\psi_L \to (H \to S_\sigma)(H \to X l^\pm \nu )]}{
\Gamma[\psi_L \to (H \to S_\sigma)(H \to X)]~Br(H^0 \to X l \nu)},
\eeq
where $X$ in $H \to X$ stands for an inclusive set of all
final states. A deviation from $R^L_\sigma=1$ implies a
lifetime difference. Keeping only the leading (linear) contributions
due to mixing, $y$ can be extracted from this experimentally obtained 
quantity,
\begin{eqnarray}
  y\cos\phi= (-1)^L {\sigma} \frac{R^L_\sigma-1}{R^L_\sigma}
  \label{y-cos-phi}.
\end{eqnarray}
Theoretical predictions of $x$ and $y$ within and beyond
the Standard Model span several orders of magnitude~\cite{Nelson:1999fg}.
Roughly, there are two approaches, neither of which give very reliable
results because $m_c$ is in some sense intermediate between heavy and
light.  The ``inclusive'' approach is based on the operator
product expansion (OPE).  In the $m_c \gg \Lambda$ limit, where
$\Lambda$ is a scale characteristic of the strong interactions, $\Delta
M$ and $\Delta\Gamma$ can be expanded in terms of matrix elements of local
operators~\cite{Inclusive}.  Such calculations yield $x,y < 10^{-3}$.  
The use of the OPE relies on local quark-hadron duality, 
and on $\Lambda/m_c$ being small enough to allow a truncation of the series
after the first few terms.  The charm mass may not be large enough for these 
to be good approximations, especially for nonleptonic $D$ decays.
An observation of $y$ of order $10^{-2}$ could be ascribed to a
breakdown of the OPE or of duality.
The ``exclusive'' approach sums over intermediate hadronic
states, which may be modeled or fit to experimental data~\cite{Exclusive}.
Since there are cancellations between states within a given $SU(3)$
multiplet, one needs to know the contribution of each state with high 
precision. However, the $D$ is not light enough that its decays are dominated
by a few final states.  While most 
studies find $x,y < 10^{-3}$, Refs.~\cite{Exclusive} obtain $x$ and 
$y$ at the $10^{-2}$ level by arguing that SU(3)$_F$ violation is of order
unity, but the source of the large SU(3)$_F$ breaking is not made explicit.
It was shown that phase space effects alone provide enough SU(3)$_F$ 
violation to induce $y\sim10^{-2}$~\cite{Falk:2001hx}. Large effects in $y$ 
appear for decays close to $D$ threshold, where an analytic expansion in 
SU(3)$_F$ violation is no longer possible.
Thus, theoretical calculations of $x$ and $y$ are quite uncertain, and the values
near the current experimental bounds cannot be ruled out. 

Another example of a method that will benefit from the separately performed charm decay 
measurements is the ADS method~\cite{ads2}. In its original form, it seeks to obtain
$\gamma$ from two sets of measurements, $B^\pm \to D^0(\to K^\mp \pi^\pm) K^\pm$, i.e. 
use decays of $D$ into non-$CP$-eigenstate final states,
\begin{eqnarray} 
  \label{ADS}
  R_{ADS}&=&\frac{\Gamma(B^-\to D^0_{\to K^+\pi^-}K^-)+\Gamma(B^+\to D^0_{\to K^-\pi^+}K^+)}
  {\Gamma(B^-\to D^0_{\to K^-\pi^+}K^-)+\Gamma(B^+\to D^0_{\to K^+\pi^-}K^+)} 
  \nonumber \\
  &=&r_D^2+2r_Br_D\cos\gamma \cos\left(\delta_B+\delta_D\right)+r_B^2, \\
  A_{ADS}&=&\frac{\Gamma(B^-\to D^0_{\to K^+\pi^-}K^-)-\Gamma(B^+\to D^0_{\to K^-\pi^+}K^+)}
  {\Gamma(B^-\to D^0_{\to K^-\pi^+})+\Gamma(B^+\to D^0_{\to K^+\pi^-}K^+)} 
  \nonumber \\
  &=&2r_Br_D\cos\gamma \cos\left(\delta_B+\delta_D\right)/R_{ADS},
\end{eqnarray}
where one parameterizes the ratio of a DCS to CF decays, 
$A(D^0 \to K^+ \pi^-)/A(D^0 \to K^- \pi^+)=r_D e^{i \delta_D}$.
The accuracy of this method will be greatly improved if $r_D$ and $\delta_D$ are measured 
separately. While the separate measurement of $r_D$ is already available,
determination of $\delta_D$ will became possible at the charm factories  
at Cornell and Beijing. This allows one to measure $\cos \delta_D$.
In order to see this, let us write a triangle relation,
\begin{equation}
\sqrt{2} A(D_{\pm} \to K^- \pi^+) = A(D^0 \to K^- \pi^+) \pm 
A(\barD \to K^- \pi^+),
\end{equation}
which follows from the fact that, in the absence of $CP$-violation in charm,
mass eigenstates of the neutral $D$ meson coincide with its $CP$-eigenstates, as in
Eq.~(\ref{CPeig}). This implies a relation for the branching ratios,
\begin{equation} \label{cos1}
1 \pm 2 \cos \delta_D \sqrt{r_D} = 
2 \frac{Br(D_{\pm} \to K^- \pi^+)}{Br(D^0 \to K^- \pi^+)}, 
\end{equation}
where we used the fact that $r_D \ll \sqrt{r_D}$ and neglected 
$CP$ violation in mixing, which could undermine the
$CP$-tagging procedure by splitting the $CP$-tagged state
on one side into a linear combination of $CP$-even and
$CP$-odd states. Its effect, however, is completely negligible here.
Notice that the phase $\delta_D$ in Eq.~(\ref{cos1}) is the same
as the one that appears in Eq.~(\ref{Kpi}).
Now, if both decays of $D_+$ and $D_-$ are measured, 
$\cos \delta_D$ can be obtained from the asymmetry
\begin{equation} \label{cos2}
\cos \delta_D = 
\frac{Br(D_{+} \to K^- \pi^+)-Br(D_{-} \to K^- \pi^+)}
{2 \sqrt{r_D} Br(D^0 \to K^- \pi^+)}. 
\end{equation}
Both Eqs.~(\ref{cos1}) and (\ref{cos2}) can be used to extract 
$\delta_D$ at ChF. Similar measurements are possible for other 
$D$ decays~\cite{Rosner:2003yk}. 

One potential problem in measuring the strong phase this way is 
the need for high-statistics measurement which might not be
possible at CLEO-c, provided if phase turns out to be small.
One indication of that is the fact that 
$A(D^0 \to K^+ \pi^-)=\lambda^2 A(\barD \to K^+ \pi^-)=
\lambda^2 A(D \to K^- \pi^+)$ in the flavor SU(3) limit, which implies 
that $\delta_D=0$ in this limit. Therefore, calculation of
the value of $\delta_D$ is equivalent to computation of 
an SU(3)-breaking correction. This, however, does not imply that
it is very small~\cite{Falk:2001hx}, as SU(3) breaking in charm decays
can be significant.

Another possibility would be to use other common modes of $D^0$ and
$\barD$, such as $D \to K^{*+}\pi^-,\rho^-K^+,\rho^+\pi^-,$ etc.
The same arguments leading to Eq.~(\ref{cos2}) can be applied there as 
well and the resulting equation would look identical. One added benefit is that
SU(3) symmetry arguments alone do not force the strong phase to be zero, so its
value could in principle be larger. One can speculate that if the chiral
symmetry in QCD is realized in the ``vector'' mode~\cite{Georgi}, than 
$A(D^0 \to K^{*+} \pi^-)=\lambda^2 A(D^0 \to K^{*-} \pi^+)$ in the limit of
``vector'' SU(3)$_L \times$ SU(3)$_R$ symmetry, so the relevant strong phase is 
still zero. It is however not clear if this realization of chiral symmetry is relevant 
to charm decays (except, may be in the large $N_c$ limit). In any case, 
vector symmetry is badly broken in $D$ decays~\cite{Georgi}, so the resulting strong 
phase can still be large. Clearly, only experimental measurements would settle
these issues.

\subsection{\boldmath  
  CLEO-c Impact on $\gamma/\phi_3$ Measurements}
\vspace{+2mm}
\begin{flushright}
  {\it Contribution from D.~Asner}
\end{flushright}

The CLEO-c research program~\cite{bib:cleoc} consists of 
studies of hadronic, semileptonic and leptonic charm meson decay 
which include measurements of doubly-Cabibbo processes, 
studies of charm Dalitz plot analyses, 
and tests for physics beyond the Standard Model 
including searches for charm mixing.
Precision determination of $\gamma/\phi_3$ depends upon constraints on 
charm mixing amplitudes, measurements of doubly-Cabibbo suppressed amplitudes 
and relative phases, 
and studies of charm Dalitz plots tagged by flavor or $CP$ content.

The CLEO-c physics program~\cite{bib:cleoc} includes a variety of measurements
that will improve the determination of $\gamma/\phi_3$ from the $B$-factory experiments.
The total number of charm mesons accumulated at CLEO-c will be much smaller
than the samples already accumulated by the $B$-factories. 
However,  the quantum correlations in the 
$\psi(3770) \to D\bar{D}$ system provide a unique laboratory in which to study charm.

\subsubsection{\boldmath  
  Components of CLEO-c Physics Program Pertinent to $\gamma/\phi_3$}

Neutral flavor oscillation in the $D$ meson system is highly suppressed
within the Standard Model. The time evolution of a particle produced as a $D^0$
or $\bar{D}^0$, in the limit of $CP$ conservation, is governed by four parameters:
$x=\Delta m/\Gamma$ and $y=\Delta \Gamma/2\Gamma$ which 
characterize the mixing matrix, 
$\delta$ the relative strong phase
between Cabibbo favored (CF) and doubly-Cabibbo suppressed (DCS) amplitudes and 
$R_D$ the DCS decay rate relative to the CF decay rate. The mixing rate $R_M$ is defined as $\frac{1}{2}(x^2+y^2)$~\cite{bib:asner}.
Standard Model based predictions for $x$ and $y$, as well as a variety of non-Standard 
Model expectations, span several orders of magnitude~\cite{bib:Nelson}.
It is reasonable to expect that $x\approx y \approx 10^{-3}$ in the Standard Model.
The mass and width differences $x$ and $y$ can be measured in a variety of ways.
The most precise limits are obtained by exploiting the time-dependence of 
$D$ decays~\cite{bib:asner}. Time-dependent analyses are not feasible at 
CLEO-c; however,
the quantum-coherent $D^0\bar{D}^0$ state provides time-integrated sensitivity 
to $x$, $y$ at ${\cal O}(1\%)$ level and $\cos\delta\sim 0.1$~\cite{bib:cleoc,GGR}
in $1 \ \rm{fb}^{-1}$ of $\psi(3770)\to D\overline D$. 
Although CLEO-c does not have sufficient sensitivity to observe Standard
Model charm mixing the projected sensitivity in $1 \ \rm{fb}^{-1}$ at $\psi(3770)$
compares favorably with current experimental results; 
see Fig.~1 in Ref.~\cite{bib:asner}.

\subsubsection{\boldmath
  Targeted Analyses\label{sec:target}}

Measurements of each of the four parameters that describe $CP$ conserving
mixing can in principle be determined individually by a series of ``targeted''
analyses. The techniques and projected sensitivities in $1 \ \rm{fb}^{-1}$ are described briefly in the subsections that follow. Greater detail can be found in Ref.~\cite{bib:cleoc}.  

\vspace{1ex}
\indent \underline{\boldmath $R_M$}
The measurement of $R_M$ can be determined unambiguously by considering the
decays $\psi(3770)\to (K^-\pi^+)(K^-\pi^+)$ and 
$\psi(3770)\to (\ell^\pm(KX)) (\ell^\pm(KX))$. The hadronic final state cannot
be produced from DCS decay due to a quantum statistics argument -
 the $C$-odd initial state cannot produce the 
symmetric final state required by Bose statistics
if both the $D^0$ and $\bar{D}^0$ decay into the same final state ($K\pi$). 
A fit
utilizing six constraints make this channel effectively background free.
The double semileptonic channel has fewer constraints due to the two neutrinos.
The number of ``right-sign'' $(K^-\pi^+)(K^+\pi^-)$ and 
$(\ell^\pm(KX)) (\ell^\mp(KX))$ events in the two channels combined produced in 
$1 \ \rm{fb}^{-1}$ is expected to be $\sim 20,000$ corresponding 
to $\sqrt{2 R_M}<1.7\%$ @95\% Confidence Level (C.L.).

\vspace{1ex}
\indent \underline{\boldmath $\cos\delta$}
Final states where one $D$ decays to $K\pi$ and the other to a $CP$ eigenstate
can be used to probe $\cos\delta$. The $CP$-even eigenstates accessible to 
CLEO-c include $K^+K^-$, $\pi^+\pi^-$ and $K^0_S\pi^0\pi^0$ and the $CP$-odd
eigenstates include $K^0_S\pi^0$, $K^0_S\eta$, $K^0_S\omega$, $K^0_S\phi$, and
$K^0_S\eta^\prime$. The $CP$ content of the $K^0_S\pi^+\pi^-$ and 
$\pi^+\pi^-\pi^0$ Dalitz plots will also be utilized. Sensitivity to 
$\cos\delta\sim$$1/(2 \sqrt{R_D N})$ where $N$, the total number of $CP$
tagged $K\pi$, will be $\sim$9200 and $R_D$ is already well measured, 
thus leading to an expected precision of $\pm 0.09$ in $\cos\delta$.

\vspace{1ex}
\indent \underline{\boldmath $y$}
Tagging one of the $D$ mesons as a $CP$ eigenstate, $y$
can be determined by measuring the flavor specific branching ratios of the 
other meson.
The flavor tag width is independent of the $CP$ quantum number;
however the branching ratio is inversely proportional to the total width. 
Consequently, charm threshold experiments have 
time-integrated sensitivity to $y$. 
Neglecting factors related to DCS decays 
(described in detail in Ref.~\cite{asnersun}), 
we can construct the ratio
\begin{equation}
\frac{2\Gamma[\psi\!\to\!(D\!\to\! CP)(D \!\to\! {\rm{flavor}})]}
{\Gamma[\psi \!\to\! (D\!\to\! CP)(D \!\to\! X) Br(D \!\to\! {\rm{flavor}})]}\!\sim\!(1\!\pm\!y)
\label{eq1}
\end{equation}
where $X$ represents an inclusive set of all states. A positive $y$ would make
the above ratio $>1$ for $CP+$ tags.

The decay $D^0 \to K^0_S \pi^+\pi^-$ is measured with a Dalitz plot analysis 
to 
proceed through intermediate states that are $CP+$ eigenstates, such as $K^0_S f_0$, $CP-$ such as $K^0_S\rho$ and flavor eigenstates such as $K^{*-}\pi^+$~\cite{bib:asner2}. 
The presence of mixing through $y$ would introduce an intensity modulation across the Dalitz plot as a function of the $CP$ of the contributing amplitudes.
The sensitivity at CLEO-c to $y$ with Dalitz plot analyses has not yet been fully evaluated. 
Preliminary estimates suggests a limit of $y\sim$ few\% @95\% C.L. is attainable with the CLEO-c data.

\vspace{1ex}
\indent \underline{\boldmath $x$}
Separate measurement of $y$ and $R_M$ as outlined above allow the value $x$ to be determined. The upper limit on $R_M$ corresponds to a limit of
$|x|<1.7\%$ @95\% C.L..

\vspace{1ex}
\indent \underline{\boldmath $R_D$}
Although $R_D$ is determined by the ratio of $(K^-\pi^+)(K^-\ell^+\nu)$ to
$(K^-\pi^+)(K^+\ell^-\overline \nu)$ up to corrections that are second order
in $x$ and $y$, a more precise measurement is attainable using a $D^{*+}$ tag
to measure $D^0\to K^+\pi^-$ relative to $D^0 \to K^-\pi^+$~\cite{bib:asner}.

\subsubsection{Comprehensive Analysis\label{sec:comp}}

The comprehensive analysis simultaneously
determines mixing and doubly-Cabibbo suppressed parameters by examining various
single tag and double tag rates.
Due to quantum correlations in the $C=-1$ and $C=+1$ $D^0\bar{D}^0$ pairs
produced in the reactions $e^+e^-\to D^0\bar{D}^0(\pi^0)$ and
$e^+e^-\to D^0\bar{D}^0\gamma$, respectively, the time-integrated
$D^0\bar{D}^0$ decay rates are sensitive to interference between amplitudes for
indistinguishable final states.  This size of this interference is governed by
the relevant amplitude ratios and can include contributions from
$D^0$-$\bar{D}^0$ mixing. 

We consider in the comprehensive analysis the following categories of 
final states:
\begin{itemize}
\item $f$ or $\bar f$: hadronic states that can be reached from either $D^0$ or
$\bar{D}^0$ decay but that are not $CP$ eigenstates.  An example is $K^-\pi^+$,
which results from Cabibbo-favored $D^0$ transitions or doubly
Cabibbo-suppressed $\bar{D}^0$ transitions.  We include in this category
Cabibbo-suppressed transitions as well as
self-conjugate final states of mixed $CP$, such as $K^0_S\pi^+\pi^-$.
\item $\ell^+$ or $\ell^-$: semileptonic or purely leptonic final states,
which, in the absence of mixing,
tag unambiguously the flavor of the parent $D$.
\item $S_+$ or $S_-$: $CP$-even and $CP$-odd eigenstates, respectively.
\end{itemize}
We calculate decay rates for $D^0\bar{D}^0$ pairs to all possible combinations
of the above categories of final states in Ref.~\cite{asnersun}, 
for both $C=-1$ and $C=+1$,
reproducing the work of Refs.~\cite{GGR,AtwoodPetrov}.  
Such $D^0\bar{D}^0$ combinations, where both $D$ final states are specified, 
are referred to as double tags (or DT).
In addition, we calculate rates for single tags (or ST), where either the
$D^0$ or $\bar{D}^0$ is identified and the other neutral $D$ decays generically.
Any DT event also contains two ST decays, and ST rates are obtained by
summing the corresponding DT rates.

Experimental measurements of $D^0$ branching fractions rely on determining
yields of ST decays assuming knowledge of the luminosity and $D^0\bar{D}^0$ 
cross section.
If, in addition, one also measures DT yields, then the luminosity and
$D^0\bar{D}^0$ cross section need not be
known~\cite{markiii-1,markiii-2,cleo-c}.
However, in correlated $D^0\bar{D}^0$ decay, the ST and DT rates no longer
depend solely on the branching fractions of interest; the rates are modified by
$x$, $y$, and the magnitudes and phases of various amplitude ratios.
Therefore, one must correct the measured ST and DT yields in order to extract
the branching fractions.  Conversely, as shown in Ref.~\cite{asnersun}, 
using these yields, it is possible to determine simultaneously the mixing and amplitude ratio parameters as well as the relevant branching fractions. 
The estimated uncertainties on the fit parameters based on approximately
$3\times 10^6$ $D^0\bar{D}^0$ pairs are presented in Table~\ref{tab:sensitivity}
using efficiencies and background levels similar to those found at CLEO-c.

\begin{table}[tb]
  \caption{
    Estimated uncertainties (statistical and systematic, respectively)
    for different $C$ configurations, with $r_{D}$ and branching
    fractions to $CP$ eigenstates constrained to the world averages.
  }
  \label{tab:sensitivity}
  \begin{center}
    \begin{tabular}{c|c|ccc}
      \hline\hline
      Parameter & Value & 
      $\Gamma_{D^0\bar{D}^0}^{C=-1} = 3\times 10^6$ &
      $\Gamma_{D^0\bar{D}^0}^{C=+1} = 3\times 10^6$ &
      $\Gamma_{D^0\bar{D}^0}^{C=-1} = 10\cdot\Gamma_{D^0\bar{D}^0}^{C=+1} = 3\times 10^6$
      \\
      \hline
      $y$
      & $0$
      & $\pm 0.015\pm 0.008$
      & $\pm 0.007\pm 0.003$
      & $\pm 0.012\pm 0.005$ \\
      $x2$
      & $0$
      & $\pm 0.0006\pm 0.0006$
      & $\pm 0.0003\pm 0.0003$
      & $\pm 0.0006\pm 0.0006$ \\
      $\cos\delta$
      & $1$
      & $\pm 0.15\pm 0.04$
      & $\pm 0.13\pm 0.05$
      & $\pm 0.13\pm 0.03$ \\
      $x\sin\delta$
      & $0$
      & ---
      & $\pm 0.010\pm 0.003$
      & $\pm 0.024\pm 0.005$ \\
      \hline\hline
    \end{tabular}
  \end{center}
\end{table}

\subsubsection{Dalitz Plot Analyses}

A Dalitz plot analysis of multibody final states measures amplitudes and phases
rather than the decay rates. A better understanding of final state interactions in
exclusive weak decays is needed in order to elucidate the origin of 
$CP$ violation in the $B$ sector~\cite{ggsz}.
Currently, the most important three-body decay for $\gamma/\phi_3$ determination
is $D^0 \to K^0_S \pi^+\pi^-$. 
Recently Babar~\cite{babargamma} and Belle~\cite{Abe:2005ct}
have reported $\gamma = (70\pm 31\!^{+12}_{-10}\!^{+14}_{-11})^\circ$ and 
$\phi_3 = (77^{+17}_{-19} \pm 13 \pm 11)^\circ  $, respectively, 
where the third error is the
systematic error due to modeling of the Dalitz plot. 
A fit which includes only established resonances, 
similar to that of CLEO~\cite{bib:asner2} 
which used Breit-Wigner (BW) line shapes, 
provides very poor description of the data. 
An improved description requires the inclusion of 
two ad-hoc $\pi\pi$ $S$-wave resonances which account for $\sim$10\% of the fit area.

It may be possible to reduce the systematic error due to model uncertainty
by using the K-matrix formalism to describe the 
$\pi\pi$ $S$-wave contribution to $D^0 \to K^0_S\pi^+\pi^-$. 
The FOCUS collaboration has used a hybrid BW/K-Matrix model 
where the higher spin resonances are described using BW line shapes 
and the $\pi\pi$ $S$-wave is parameterized using the K-matrix model of
Anisovich and Sarantsev~\cite{sarantsev} 
to describe $D^+\to \pi^+\pi^-\pi^+$~\cite{focus3pi}. 
Similarly, the CLEO collaboration has searched for $\pi\pi$ S-wave in 
$D^0\to\pi^+\pi^-\pi^0$~\cite{cleopipipi0} following the 
$K$-matrix formalism of Au, Morgan, and Pennington~\cite{aumope}. 
Currently, the CLEO collaboration is considering K-matrix 
descriptions of both the $\pi\pi$ $S$-wave and the 
$K\pi$ $P$-wave contributions to $D^0 \to K^0_S\pi^+\pi^-$. 
This more sophisticated description of the $K^0_S\pi^+\pi^-$ decay
may lead to a smaller systematic error in the determination of $\gamma/\phi_3$.

Both $D^0$ and $\bar{D}^0$ populate the Dalitz plots $K^0_S\pi^+\pi^-$, 
$\pi^+\pi^-\pi^0$, $K^+K^-\pi^0$ and $K^0_S K^\pm\pi^\mp$ and so can be used in
the determination of $\gamma/\phi_3$ which exploit the interference between
$b \to c \bar u s$ ($B^- \to D^0 K^-$) and 
$b \to u \bar c s$ ($B^- \to \bar{D}^0 K^-$) 
where the former process is real and the later is $\propto e^{-i\gamma}$~\cite{bigi}. 
Studying $CP$ tagged Dalitz plots allows
a model independent determination of the relative $D^0$ and $\bar{D}^0$
phase across the Dalitz plot. 
Consider $D^0 \to K^0_S\pi^+\pi^-$  decay which proceeds through 
intermediate states that are $CP+$ eigenstates, such as $K^0_S f_0$, 
$CP-$ such as $K^0_S\rho$ and 
flavor eigenstates such as $K^{*-}\pi^+$~\cite{bib:asner2}. 
The Dalitz plots for $\psi(3770) \to D^0\bar{D}^0 \to S_+K^0_S\pi^+\pi^-$ and 
$\psi(3770) \to D^0\bar{D}^0 \to S_-K^0_S\pi^+\pi^-$ will be distinct 
and the Dalitz plot for the untagged sample 
$\psi(3770) \to D^0\bar{D}^0 \to X K^0_S\pi^+\pi^-$ will be different 
from that observed with uncorrelated $D$ mesons 
from continuum production at $\sim 10$~GeV~\cite{bib:asner2}.
The CLEO collaboration will perform a simultaneous fit to $CP+$, $CP-$, and 
flavor tag samples with BW/K-matrix hybrid models. 
A good fit will validate Dalitz plot model and 
is expected to reduce the model dependent systematic error 
on the $\gamma/\phi_3$ measurements to a few degrees.  
If a good model cannot be constructed,
a model independent result can be obtained from a binned analysis 
of the $CP$ tag and flavor tag Dalitz plots.
It is noteworthy that, although the statistical sample is smaller, 
the negligible $\pi\pi$ $S$-wave contribution to
$D \to \pi^+\pi^-\pi^0$ \emph{may} lead to a better decay model and smaller
uncertainty on $\gamma/\phi_3$ relative to that obtained with $D \to K^0_S\pi^+\pi^-$.

\subsubsection{Summary}

CLEO-c measurements are important inputs for each of the methods proposed to
determine $\gamma/\phi_3$. 
The Gronau-London-Wyler and Atwood-Dunietz-Soni methods will 
benefit from improved mixing parameters and DCS parameters, respectively.
The Dalitz plot method will benefit from improved models of three-body charm
decay.

The quantum correlated $D^0\bar{D}^0$ system from the decay of $\psi(3770)$ 
provides time-integrated sensitivity to the mixing parameters 
$x$ and $y$ and the doubly-Cabibbo suppressed parameters $R_D$ and $\delta$. 
Targeted analyses will provide the first measurement of $\cos\delta$ 
and sensitivity to $R_M$ and $y$ competitive with $B$-factory measurements. 
A comprehensive analysis of single tag and double tag yields 
allows simultaneous determination of hadronic and semileptonic 
branching fractions, mixing parameters, and doubly-Cabibbo suppressed 
parameters with sensitivity similar to the collection of targeted analyses.

The decays $B^- \to D^{(*)}K^{-(*)}$ followed by a three-body decay of the $D$
such as $K^0_S\pi^+\pi^-$ or $\pi^+\pi^-\pi^0$ currently provide the greatest sensitivity
to the CKM angle $\gamma/\phi_3$. 
The precision of these measurements will eventually be limited by the 
understanding of the $D \to K^0_S\pi^+\pi^-$ Dalitz plot. 
K-matrix descriptions of the $\pi\pi$ S-wave may yield improved models of 
charm Dalitz plots and these models will be tested using the 
$CP$ tagged sample of charm decays at \hbox{CLEO-c}. 
The model uncertainty, which is currently $\pm 10^\circ$, 
may be reduced to a few degrees with sufficient data.

\subsection{\boldmath
  $B_c$ Mesons: Another Probe of $CP$ Violation}

\vspace{+2mm}
\begin{flushright}
 {\it Contribution from R.~Fleischer}
\end{flushright}

Many strategies to explore $CP$ violation through the ``conventional'' 
charged $B_u$ and the neutral $B_{d,s}$ mesons 
were proposed in the literature. There is, however, another
species of $B$ mesons, the $B_c$-meson system, which consists of
$B_c^+\sim c\overline{b}$ and $B^-_c\sim b\overline{c}$. These mesons were 
observed by the CDF collaboration through their decay 
$B_c^+\to J/\psi \ell^+ \nu_\ell$, with the following mass and lifetime 
\cite{CDF-Bc}:
\begin{equation}
M_{B_c}=(6.40\pm0.39\pm0.13)\,\mbox{GeV}, \quad
\tau_{B_c}=(0.46^{+0.18}_{-0.16}\pm 0.03)\,\mbox{ps}.
\end{equation}
Recently,  D0 reported the measurement of
$B_c^+\to J/\psi\,\mu^+ X$ \cite{D0-Bc}, yielding 
\begin{equation}
M_{B_c}=(5.95^{+0.14}_{-0.13}\pm0.34)\,\mbox{GeV}, \quad
\tau_{B_c}=(0.448^{+0.123}_{-0.096}\pm 0.121)\,\mbox{ps}.
\end{equation}
Moreover, there is now also evidence for the decay $B_c^+\to J/\psi \pi^+$ 
from the CDF collaboration \cite{CDF-Bc-new}, allowing the extraction of
\begin{equation}
M_{B_c}=(6.2870 \pm 0.0048  \pm 0.0011)\,\mbox{GeV}.
\end{equation}

\begin{figure}[htbp] 
   \centering
   \includegraphics[width=11.0truecm]{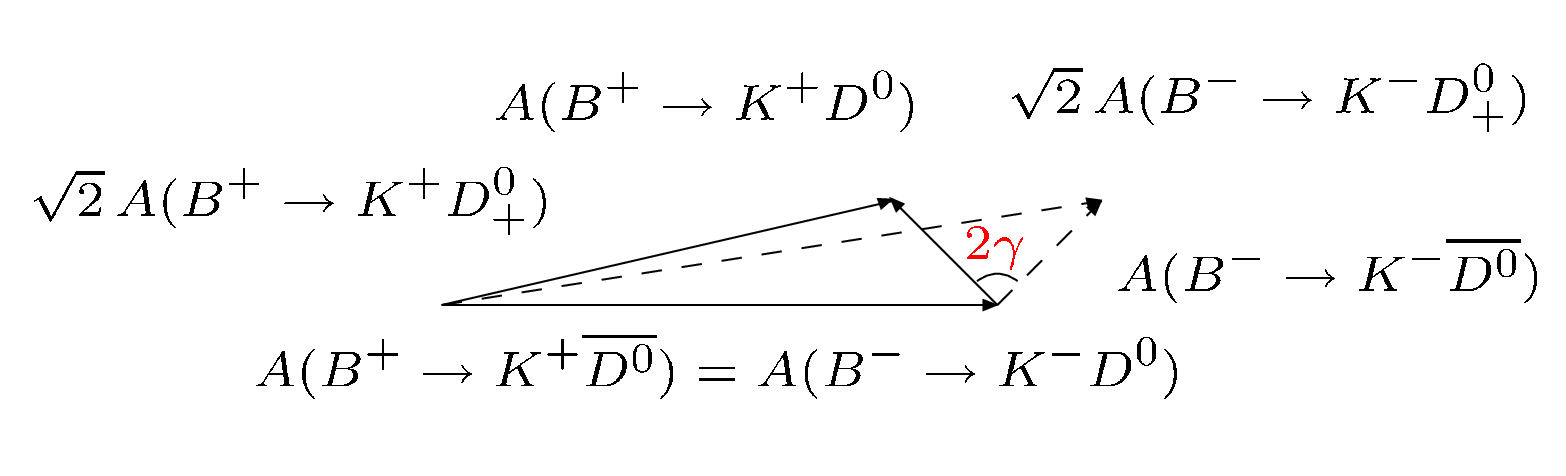} 
   \caption{The extraction of $\gamma$ from 
   $B^\pm\to K^\pm\{D^0,\bar D^0,D^0_+\}$ 
   decays.}
   \label{fig:Bu-triangles}
\end{figure}

As run II of the Tevatron will provide further insights into $B_c$ physics and
a huge number of $B_c$ mesons will be produced at the LHC, the natural
question arises whether insights into $CP$ violation can also be obtained from 
$B_c$-meson decays. In order to address this question, we have to 
consider non-leptonic $B_c$ decays, as in the case of the exploration of
CP violation through decays of the conventional $B_{u,d,s}$ mesons. Since
the $B_c$ mesons are charged particles, i.e.\ do not exhibit mixing 
effects, we have obviously to rely on direct $CP$ violation. It is well-known
that the extraction of $\gamma$ from such effects suffers from hadronic
uncertainties, and that an elegant solution of this problem is offered
in the case of the $B^\pm_u$ mesons through the ``triangle approach''
illustrated in Fig.~\ref{fig:Bu-triangles} \cite{GronauWyler}.

\begin{figure}
   \centering
   \includegraphics[width=11.8truecm]{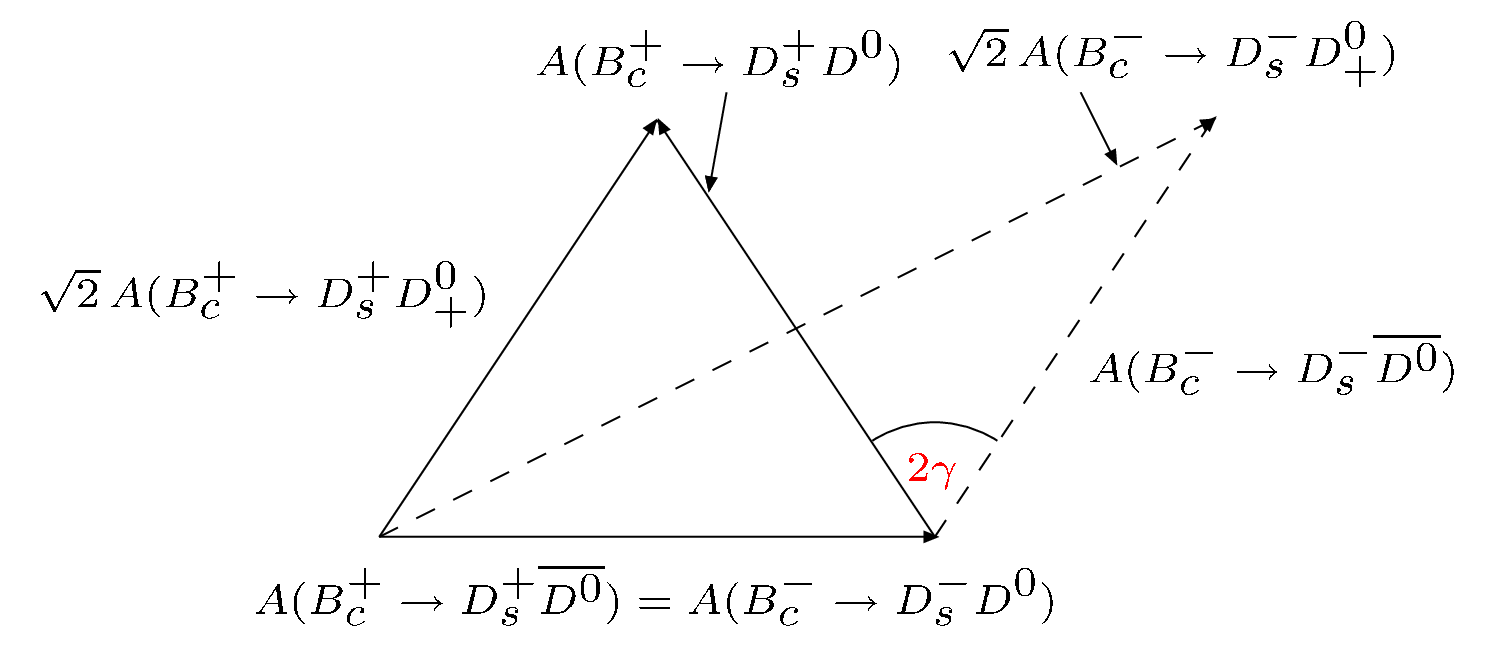} 
   \caption{The extraction of $\gamma$ from 
   $B_c^\pm\to D^\pm_s\{D^0,\bar D^0,D^0_+\}$ decays.}
   \label{fig:Bc-triangles}
\end{figure}

In the $B_c$-meson system, such a strategy for the determination of
$\gamma$ is offered by the decays $B_c^\pm\to D_s^\pm D$, which are 
the $B_c$-meson counterparts of the $B_u^\pm\to K^\pm D$ modes 
used in the conventional triangle method \cite{GronauWyler}, 
and satisfy the following amplitude relations \cite{masetti}:
\begin{eqnarray}
\sqrt{2}A(B_c^+\to D_s^+D^0_+)&=&A(B_c^+\to D_s^+D^0)+
A(B_c^+\to D_s^+\bar D^0)\\
\sqrt{2}A(B_c^-\to D_s^-D^0_+)&=&A(B_c^-\to D_s^-\bar D^0)+
A(B_c^-\to D_s^-D^0),
\end{eqnarray}
with
\begin{eqnarray}
A(B^+_c\to D_s^+\bar D^0)&=&A(B^-_c\to D_s^-D^0)\\
A(B_c^+\to D_s^+D^0)&=&A(B_c^-\to D_s^-\bar D^0)\times e^{2i\gamma}.
\end{eqnarray}
At first sight, everything is completely analogous to the $B_u^\pm\to K^\pm D$
case~\cite{GronauWyler}. 
However, there is an important difference \cite{fw,RF-Phys-Rep}, 
which becomes obvious by having a look at the corresponding Feynman
diagrams: in the $B_c^\pm\to D_s^\pm D$ system, the amplitude with the 
rather small CKM matrix element $V_{ub}$ is {\it not} colour-suppressed, while 
the larger element $V_{cb}$ comes with a colour-suppression factor. Therefore, 
we obtain
\begin{equation}\label{Bc-ratio1}
\left|\frac{A(B^+_c\to D_s^+ D^0)}{A(B^+_c\to D_s^+ 
\bar D^0)}\right|=\left|\frac{A(B^-_c\to D_s^-\bar D^0)}{A(B^-_c\to D_s^- 
D^0)}\right|\approx\frac{1}{\lambda}\frac{|V_{ub}|}{|V_{cb}|}
\times\frac{a_1}{a_2}\approx0.4\times 3 = {\cal O}(1),
\end{equation}
and conclude that the two amplitudes are similar in size. In contrast 
to this favourable situation, in the decays $B_u^{\pm}\to K^{\pm}D$, 
the matrix element $V_{ub}$ comes with the colour-suppression factor, 
resulting in a very stretched triangle. The extraction of $\gamma$ from 
the $B_c^\pm\to D_s^\pm D$ triangles is illustrated in 
Fig.~\ref{fig:Bc-triangles}, which should be compared with the
squashed $B^\pm_u\to K^\pm D$ triangles shown in 
Fig.\ \ref{fig:Bu-triangles}. Another important advantage is that 
the interference effects arising from $D^0,\bar D^0\to\pi^+K^-$, which
were pointed out in the context of the  $B_u^\pm\to K^\pm D$
case in \cite{ads2}, are practically unimportant for the measurement of 
BR$(B^+_c\to D_s^+ D^0)$ and BR$(B^+_c\to D_s^+ \bar D^0)$ since the 
$B_c$-decay amplitudes are of the same order of magnitude. This feature
implies also that the sensitivity to $D^0$--$\bar D^0$ mixing, which provides
a nice avenue for new physics to enter these triangle strategies \cite{D-NP},
is considerably smaller than in the $B_u$ case \cite{RF-Phys-Rep}.

Consequently, the $B_c^\pm\to D_s^\pm D$
decays provide -- from the theoretical point of view -- the ideal
realization of the ``triangle'' approach to determine $\gamma$ \cite{fw}. 
On the other hand, the practical implementation still appears to 
be challenging, although detailed experimental feasibility studies for
LHCb are strongly encouraged. Using a relativistic quark model \cite{IKP}, 
which predicts the branching ratios for $B^+\to \overline{D^0}e^+\nu_e$,
$B^+\to K^+ \overline{D^0}$ and $B^+\to D_s^{+}\overline{D^{0}}$
in good accordance with experiment, the branching ratios for non-leptonic
$B_c$ decays were recently predicted, yielding in fact a pattern in
nice accordance with (\ref{Bc-ratio1}) \cite{IKP}. In another study \cite{IKS},
also semi-leptonic $B_c$ decays were investigated, which give a nice testing 
ground.

In addition to $CP$ violation, there are several other interesting aspects
of $B_c$ physics. Since these mesons are the lowest lying bound states
of two heavy quarks, $\overline{b}$ and $c$, the QCD dynamics of the
$B_c^+$ mesons is similar to quarkonium systems, 
such as $\overline{b}b$ and  $\overline{c}c$, which are approximately
non-relativistic. However, there is an important difference: as the 
$B_c$ mesons contain open flavour, they are stable under strong
interactions. The quarkonium-like $B_c$ mesons provide an 
important laboratory to explore exciting topics such as heavy-quark 
expansions (HQE), non-relativistic QCD (NRQCD) or factorization 
in a setting that is complementary to ``conventional'' weak hadron decays 
\cite{LHCWKSHP}, 
and should provide further valuable insights into the interplay 
of strong and and weak interactions.

\subsection{\boldmath  
  Extraction of $\gamma$ at LHCb with a Combined \dsk\ and \dpi\ Analysis}
  
\vspace{+2mm}
\begin{flushright}
  {\it Contribution from G.~Wilkinson}
\end{flushright}

\subsubsection{Introduction}

The potential of  \dsk, \dpistar\ and \dpi\ decays for extracting the CKM 
angle $\gamma$ is well known~\cite{LHCWKSHP}.  An analysis based on any
of these modes in isolation, however,  suffers from the problem of discrete
ambiguities,   and in the case of $\mathrm{B_d}$ decays,  of difficulties posed by
very small interference effects.     This report explains how a 
combined analysis of,  for example,  \dsk\ and \dpi\  under the assumption
of U-spin symmetry, circumvents these problems.   This strategy was 
proposed in~\cite{RF} and is here explored in the context of the expected
performance in these decays at LHCb.

\subsubsection{Formalism}

From measuring the two flavour tagged decay modes, \dspkm, 
as a function of proper time, $t$,
the $CP$ asymmetry ${\cal A}_{CP} (\rm{D_s^+K^-})(t)$ may be constructed:
\begin{eqnarray}
{\cal A}_{CP} (\rm{D_s^+K^-})(t) & \equiv & \frac {\dspkm(t) \,-\, \dspkmbar(t)}
{\dspkm(t) \,+\, \dspkmbar(t)}. \nonumber 
\end{eqnarray}
\noindent This has the following dependence:
\begin{eqnarray}
{\cal A}_{CP} (\rm{D_s^+K^-})(t) & = & 
\frac{C_s \cos\Delta m_s t \,+\, S_s \sin\Delta m_s t}
{\cosh(\Delta \Gamma_s t /2) \, - \, A_{\Delta \Gamma_s} \sinh(\Delta \Gamma_s t /2) }, \nonumber
\end{eqnarray}
where $\Delta m_s$ and $\Delta \Gamma_s$ are the mass and lifetime
difference between the heavy and light $\mathrm{B_s}$ eigenstates,  which
for the purposes of this discussion are assumed to be known. The three 
observables $C_s,\, S_s$ and $A_{\Delta \Gamma_s}$ can then be fitted
from the data. (In doing this the full flavour untagged statistics may be used to
fix $A_{\Delta \Gamma_s}$.)  From performing an equivalent analysis 
for the $\mathrm{D_s^-K^+}$ final state three additional observables,
$\overline{C_s}, \, \overline S_s$  and  $\overline{A_{\Delta \Gamma_s}}$,
can be obtained.   The observables depend on the underlying physics
parameters in the following manner:
\begin{eqnarray}
C_s , \, (\overline{C_s}) & = & 
- \,(+) \left( \frac{1 - r_s^2}{1 + r_s^2} \right),  \label{eq:cpasymms1} \\
S_s , \, (\overline{S_s}) & = &
\frac { 2r_s \, \sin (\phi_s + \gamma +\, (-)\, \delta_s) }{ 1 \, + \, r_s^2 }, 
\label{eq:cpasymms2} \\
A_{\Delta \Gamma_s} , \, (\overline{A_{\Delta \Gamma_s}} ) & = &
- \frac { 2r_s \, \cos (\phi_s + \gamma +\, (-)\, \delta_s) }{ 1 \, + \, r_s^2 }. 
\label{eq:cpasymms3}
\end{eqnarray}
Here $r_s$ is the ratio of amplitudes between the interfering tree
diagrams,  $\delta_s$ is a possible $CP$ conserving strong phase difference
between the diagrams and $\phi_s$ is the $CP$ violating weak phase associated
with the $\mathrm{B_s}$\textendash$\overline{\mathrm{B_s}}$ oscillations, 
believed to be very small in the Standard Model.   It is assumed that
$\phi_s$ can be constrained from other measurements,  such as those 
in $\mathrm{B_s} \to J/\psi \phi$ decays.   

Measurement of the six observables $C_s, \, \overline{C_s}, \, S_s, \, 
\overline{S_s}, \, A_{\Delta \Gamma_s}$ and $\overline{A_{\Delta \Gamma_s}}$
allows $r_s$, $\delta_s$ and $\gamma$ to be determined.   The same information 
may be extracted by making a simultaneous event-by-event fit to the four
decay distributions.

Exactly parallel relations hold in the $\mathrm{B_d}$ system, 
for analysis of \dpi\ or \dpistar\ decays.   
In this case however,  the negligible width
splitting between the mass eigenstates,  means that there are effectively 
only four observables: $C_d, \, \overline{C_d},  \, S_d$ and $\overline{S_d}$.
These observables depend on $r_d$, $\phi_d$ (measured from 
$\mathrm{B_d \to J/\psi K^0}$ and equal to 
$2\beta$ in the Standard Model), $\delta_d$ and $\gamma$.
Note that the value of these observables will in general be different between
\dpi\ and \dpistar,   
firstly because of the possibility of different values of 
$r_d$ and $\delta_d$ for the two cases,  and secondly because the $l=1$
state of the \dpistar\ decay introduces some sign flips in 
expressions~\ref{eq:cpasymms2} and~\ref{eq:cpasymms3}
(see~\cite{RF} for more details).

\subsubsection{LHCb Event Yields and Performance}
\label{sec:perf}

LHCb has reported simulation studies of \dsk\ and \dpistar\ in~\cite{LHCBLITE}.
The results are summarised here,  together with estimates from a preliminary
study of \dpi.

The isolation of \dsk\ decays is experimentally challenging,  
because of the low branching ratio and background from 
the order-of-magnitude more prolific \dspi\ decay mode.   
The LHCb trigger system gives good performance for fully hadronic
modes and selects \dsk\ events with an efficiency of $30\%$.   
The $\pi-K$ discrimination of the RICH system reduces the \dspi\ contamination 
to $\sim10\%$.
It is estimated that the experiment will accumulate 5.4k
events per year of operation (with a year defined as $2 \, \rm{fb}^{-1}$
of integrated luminosity), with a background to signal level of $<$~0.5.   
The excellent $\sim 40 \, \rm{fs}$
proper time precision provided by the silicon Vertex Locator will ensure 
that - provided $\Delta m_s$ is not far in excess of expectation - 
the  $B_s$ oscillations will be well resolved, 
and hence the $CP$ asymmetries can be measured.   
It is estimated that the statistical precision on $\gamma$ from this channel alone 
will be $14^\circ$ for $2 \, \rm{fb}^{-1}$,  
assuming $\Delta m_s = 20 \, \rm{ps}^{-1}$, 
$\Delta \Gamma_s / \Gamma_s = -0.10$ and taking plausible values
of $\gamma$ and $\delta_s$.  

The channel \dpistar\ has been investigated through the sub-decay 
$\rm{D^{\ast \pm} \to D^0 (\to K \pi) \pi^\pm}$, in which
it is estimated 206k events will be accumulated per year, with a background
to signal level $<$~0.3.   Earlier studies
using inclusive $\rm D^0$ decays~\cite{JONAS} suggest that these statistics 
can be increased still further.

Studies are underway to investigate the feasibility of reconstructing 
\dpi,   with $\rm D^\pm \to \rm{K^\pm \pi^\mp \pi^\pm}$.   The preliminary
indications are that 210k events will be collected each year, with a background
to signal ratio of around 0.3.

\subsubsection{\boldmath  
  Extraction of $\gamma$ from modes in isolation}

In extracting $\gamma$ from the observables of a single decay mode,  two
problems are encountered,  one general, and one specific to the $\mathrm{B_d}$
decays.

\begin{enumerate}
\item
  The extraction of $\gamma$ from
  $S_{s(d)}$ and $\overline{S_{s(d)}}$ yields 8 possible solutions.
  The same is true of calculating $\gamma$ from $A_{\Delta \Gamma_s}$ and
  $\overline{A_{\Delta \Gamma_s}}$.  
  Figure~\ref{fig:exampsols} illustrates this by plotting all possible 
  solutions for $\gamma$ and $\delta_s$ in the case where the true
  values are assumed to be $\gamma=60^\circ$ and $\delta=60^\circ$.
  Therefore the study of \dpi\ or \dpistar\
  in isolation results in an 8-fold ambiguity   for $\gamma$.
  The extra observables available in the $\mathrm{B_s}$ system mean that
  in principle there is only a 2-fold ambiguity,   but in practice measurement
  errors may make it difficult to exclude local minima coming from the
  additional bogus solutions.  This problem will be accentuated if the magnitude
  of $\Delta \Gamma_s/\Gamma_s$ is small, hence making $A_{\Delta \Gamma_s}$ and 
  $\overline{A_{\Delta \Gamma_s}}$ difficult to measure.  
  
  \begin{figure}[htb]
    \begin{center}
      \includegraphics[width=0.40\textwidth]{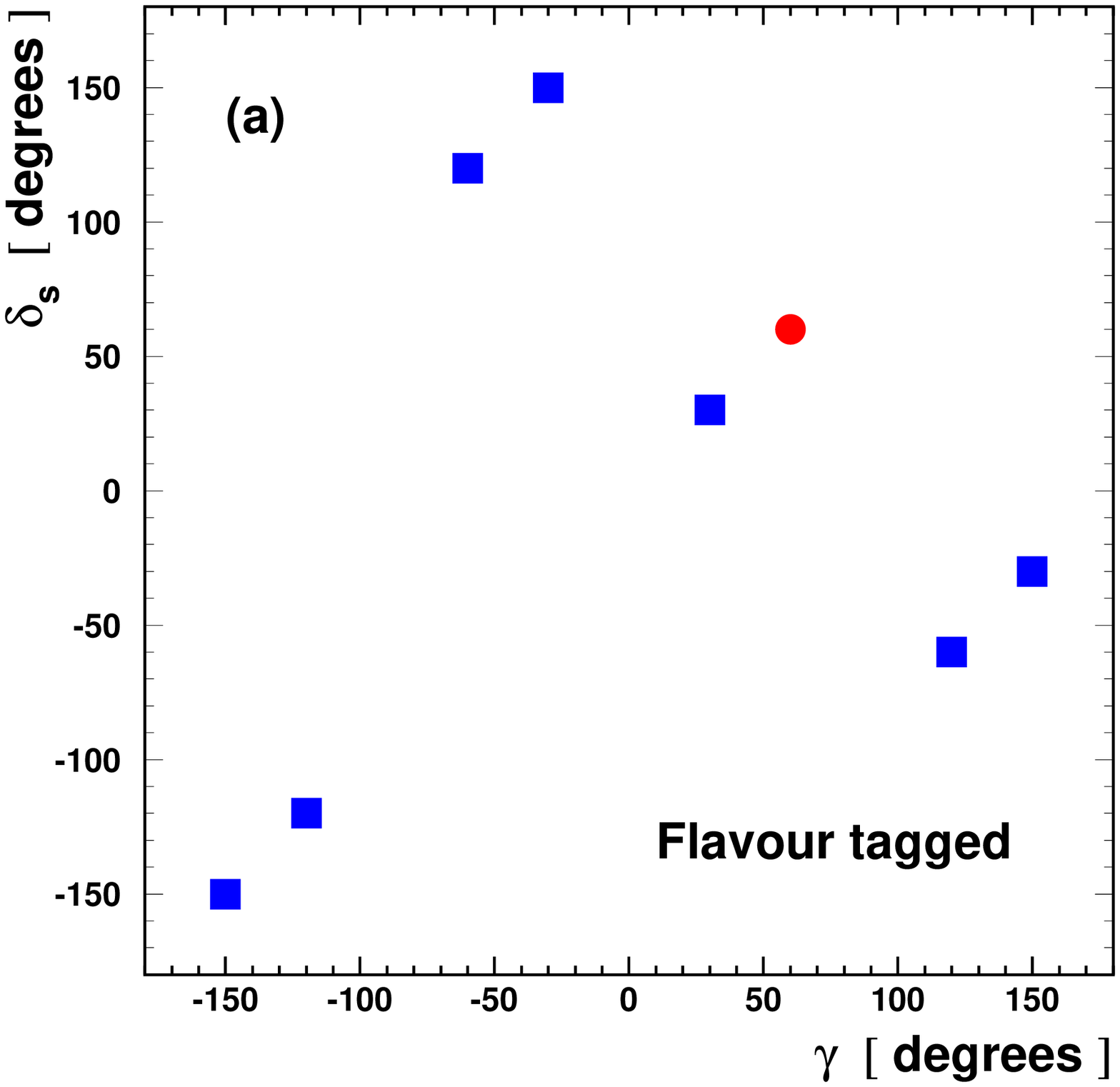}
      \includegraphics[width=0.40\textwidth]{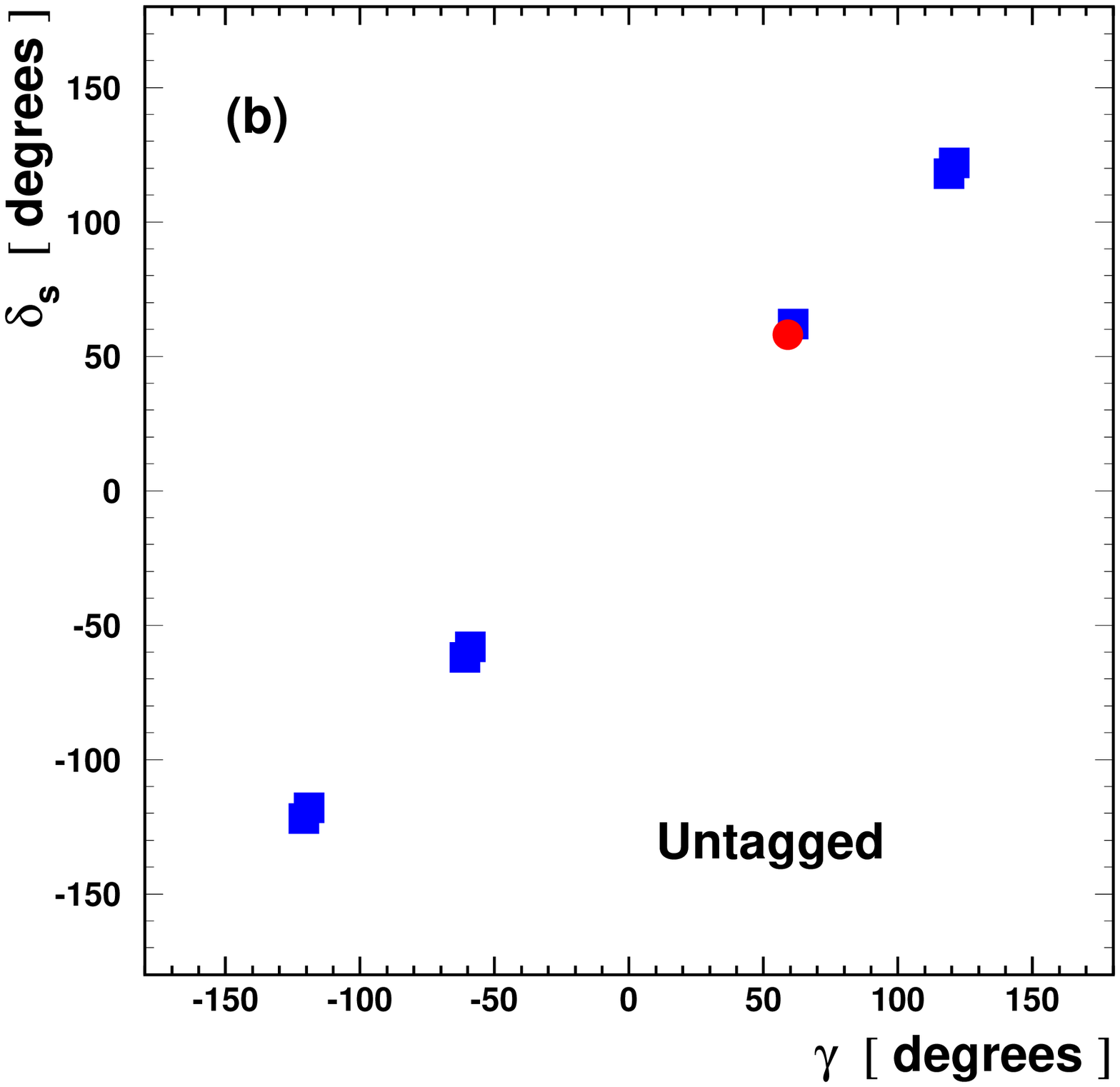}
      \caption{
        \label{fig:exampsols}
        Valid solutions for $\delta_s$ and $\gamma$ for 
        (a) the flavour tagged $S_s, \, \overline{S_s}$ observables and 
        (b) the untagged $A_{\Delta \Gamma_s}, \, \overline{A_{\Delta \Gamma_s}}$ observables.
        In both cases there are eight solutions 
        (although, in the case of (b), these are two-fold degenerate). 
        The  solid circle indicates the true solution $\gamma=60^\circ, \delta_s=60^\circ$.
      }
    \end{center}
  \end{figure}

\item
  As is clear from expressions~\ref{eq:cpasymms2} and~\ref{eq:cpasymms3},
  extracting $\gamma$ from $S_{s(d)}$ and $\overline{S_{s(d)}}$, or
  $A_{\Delta \Gamma_s}$ and
  $\overline{A_{\Delta \Gamma_s}}$ require that $r_{s(d)}$ be known.
  From comparing the CKM elements in the interfering diagrams it is expected
  that $r_s \sim 0.4$ and $r_d \sim 0.02$.  Therefore $|C_s|$ will
  be significantly different from 1, and measurements of $C_s$ and 
  $\overline{C_s}$ will allow $r_s$ to be extracted from the data --
  indeed this is what is done in the present LHCb simulation studies.  This
  will not be possible, however, for $r_d$.   Instead this parameter has
  to be set using external assumptions~\cite{RD}. These assumptions introduce a
  troublesome systematic error to the analysis.  

\end{enumerate}

\noindent 
Both of these problems may be tackled by making a
combined analysis of U-spin related $\mathrm{B_d}$ and $\mathrm{B_s}$ modes.

\subsubsection{\boldmath  
  A Combined U-Spin Analysis of \dsk\ and \dpi}

The decays \dsk\ and \dpi\ are identical under U-spin symmetry,
{\it ie.} the exchange of $d$ and $s$ quarks.   
This symmetry allows the observables
in both decays to be combined in a manner to yield certain relations,
which then give $\gamma$ under the assumption that the a priori unknown
hadronic contributions to the observables are identical in both channels.
These unknowns are the strong phases $\delta_d$ and $\delta_s$,
and $a_d$ and $a_s$, the hadronic contributions to $r_d$ and $r_s$, 
defined by $r_{d,s} = a_{d,s} \, f^{\rm{CKM}}_{d,s}$,  where the 
CKM factors, $f^{\rm{CKM}}_{d,s}$, are easily calculable.

The following analysis follows the strategy proposed in~\cite{RF}.  
The example plots and numbers assume the scenario $\gamma=60^\circ$,
$\delta_{d,s}=60^\circ$,  $a_{d,s}=1$,  $\phi_d=47^\circ$ and $\phi_s=0^\circ$.
The experimental contours assume that in one year LHCb can measure 
$S_d$ and $\overline{S_d}$ with
an uncertainty of $0.014$, and $S_s$ and $\overline{S_s}$ with
an uncertainty of $0.14$ (results consistent with the performance 
figures quoted in section~\ref{sec:perf}).   
It is also useful to assume that in the early
year of operation the analysis can benefit from studies of \dpi\ made at
the $B$-factories.   
Taking existing measurements~\cite{belle_dsp_full},  
and scaling to $2500 \ \rm{fb}^{-1}$ to represent a plausible 
$B$-factory integrated luminosity in 2008 
gives an error on $S_d$ and $\overline{S_d}$ of $0.014$.

Using 
expression~\ref{eq:cpasymms2}, the sine observables for the $\mathrm{B_s}$ and
$\mathrm{B_d}$ channels may be combined to give the following exact relations:
\begin{eqnarray}
\left( \frac{a_s \, \cos \delta_s}{a_d \, \cos \delta_d} \right) & = & 
- \frac{1}{R} \left( \frac{\sin(\phi_d \,+\, \gamma)}{\sin(\phi_s \, + \, \gamma)} \right)
\left(\frac{S_s \, + \, \overline{S_s}}{S_d \, + \, \overline{S_d}} \right), \label{eq:full1}
\end{eqnarray}
\begin{eqnarray}
\left( \frac{a_s \, \sin \delta_s}{a_d \, \sin \delta_d} \right) & = & 
- \frac{1}{R} \left( \frac{\cos(\phi_d \,+\, \gamma)}{\cos(\phi_s \, + \, \gamma)} \right)
\left(\frac{S_s \, - \, \overline{S_s}}{S_d \, - \, \overline{S_d}} \right). \label{eq:full2}
\end{eqnarray}
Here $R =  \left( \frac{1 \, - \, \lambda^2}{\lambda^2} \right) 
\left(\frac{1 \,+\, r_d^2}{1 \, + \, r_s^2}\right)$, where $\lambda$ is
the sine of the Cabibbo angle.  Because $r_d << 1$ 
$R \simeq \frac{1 \, - \, \lambda^2}{ \lambda^2 \,(1 \, + \, r_s^2 )}$ 
and hence these relations may be exploited without the need to measure $r_d$.
In the limit of full U-spin symmetry, the left hand sides of 
equations~\ref{eq:full1} and~\ref{eq:full2} are equal to unity,  and both
relations give a determination of $\gamma$.   

\begin{figure}[htb]
  \begin{center}
    \includegraphics[width=0.40\textwidth]{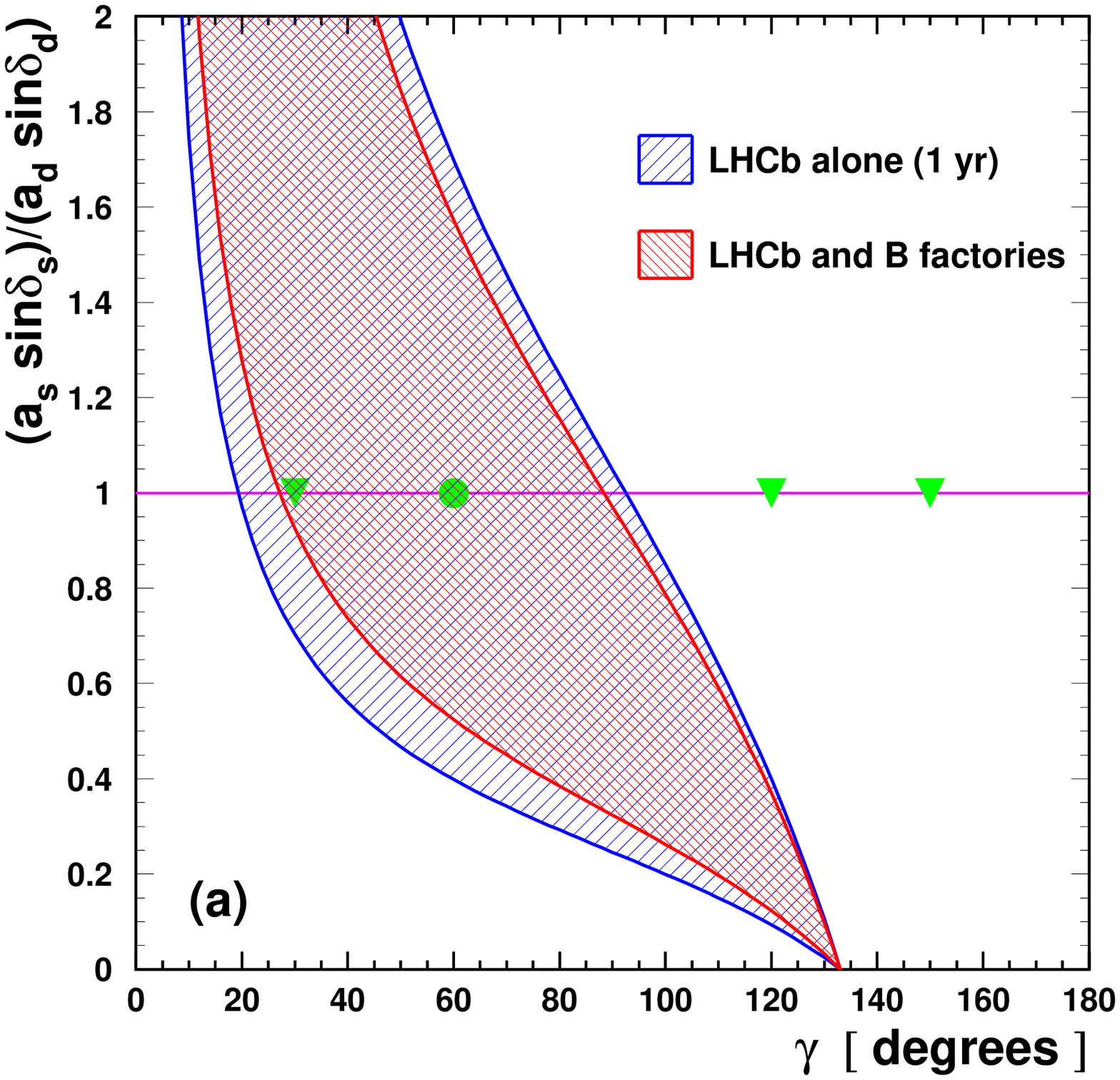}
    \includegraphics[width=0.40\textwidth]{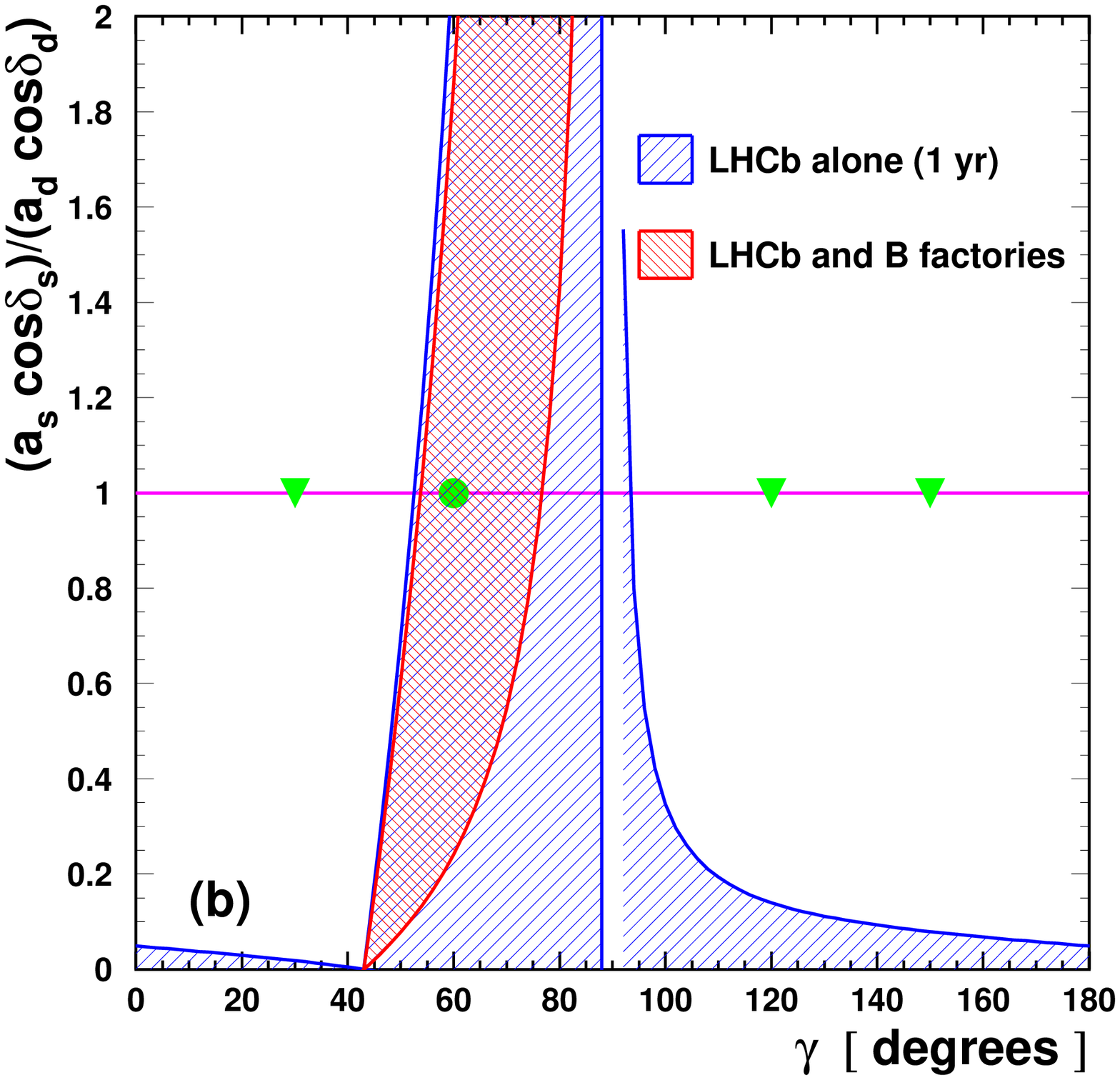}
    \caption{
      \label{fig:fulluspin}     
      Contours formed from 
      (a) expression~\ref{eq:full1} and (b) expression~\ref{eq:full2}
      showing the 1 sigma contours for one year of LHCb data,  
      and one year of LHCb data together with the measurements
      which may be available from the $B$-factories by 2008.
      The solid circle indicates the true solution; 
      the inverted solid triangles indicate fake solutions from an
      analysis of \dsk\ alone.   
      Exact U-spin symmetry corresponds to a value
      of unity for the ordinate in both plots.
    }
  \end{center}
\end{figure}

Fig.~\ref{fig:fulluspin}
show the one sigma contours which will be obtained with one year of LHCb 
operation, together with the improvement possible if $B$-factory data
are also included.   
The ambiguous solutions from the `conventional' analysis are indicated.  
It can be seen that these bogus solutions are disfavoured by 
the combined analysis.
There is, however, a mirror solution at $-180^\circ+\gamma$ 
outside the region of the plots.   
This possibility can be rejected either by
making assumptions about the orientation of the unitarity triangle, 
or by noting that such
a solution is accompanied by a very sizable strong phase difference, 
which is unlikely to be the case.

The precision achievable on $\gamma$ is in general different for the
two expressions.  
In the chosen scenario, relation~\ref{eq:full2} gives the best result,
returning an uncertainty of $\sigma_\gamma = ^{+16}_{-7}$~degrees.
In the one year analysis the contribution of the $B$-factory data is significant.
With five years of data,  the precision improves to $5^\circ$. 

The plots also allow any biases from U-spin breaking effects to be assessed.
It can be seen that relation~\ref{eq:full2} has the steepest contour
and hence exhibits the best robustness.  For example, a 20\% deviation in 
$a_s \cos \delta_s / a_d \cos \delta_d$ from unity leads to a $3^\circ$ bias in $\gamma$.   

In order to further assess the reliability the $\gamma$ extraction,
the U-spin dependence may be weakened.   This can be done by combining
relations~\ref{eq:full1} and~\ref{eq:full2} into a single expression
which involves {\it either} $\delta_d$ and $\delta_s$ {\it or}
$a_s$ and $a_d$.   For example,  the latter exercise yields 
the equation
\begin{eqnarray}
\frac{a_s}{a_d} & = & 
\pm \, \frac{1}{R} \, \frac{\sin 2(\phi_d \,+\, \gamma)}{\sin 2(\phi_s \,+\, \gamma)} 
\sqrt{ 
\frac{(S^+_s)^2 \cos^2(\phi_s \,+\,\gamma) \,+\, (S^-_s)^2 \sin^2(\phi_s \,+\,\gamma) }
{(S^+_d)^2 \cos^2(\phi_d \,+\,\gamma) \,+\, (S^-_d)^2 \sin^2(\phi_d \,+\,\gamma) }}. \label{eq:weakuspin}
\end{eqnarray}
It is now possible to determine $\gamma$ by demanding that $a_s = a_d$, but 
making no assumption  about $\delta_d$ and $\delta_s$.  
With this approach, the statistical precision from five years of data
is about 6 degrees.
Again the dependence is sufficiently steep that deviations in $a_s/a_d$ from
unity coming from U-spin breaking give relatively small biases in
the result.

It should be emphazised that these analyses only make use of the flavour
tagged observables in \dsk\ and \dpi.   If the magnitude of 
$\Delta \Gamma_s$ is sufficiently large, then measurements of 
the untagged observables $A_{\Delta \Gamma_s}$ and 
$\overline{A_{\Delta \Gamma_s}}$ will provide additional information which
will help further in the exclusion of ambiguous solutions, and add to
the ultimate precision on $\gamma$.

\subsubsection{Conclusions}

LHCb will accumulate large samples of \dsk,  \dpistar\ and \dpi\ events.
Independently each of these channels may be used to extract the CKM
angle $\gamma$,  although in the case of the $B_d$ channels this
requires making assumptions about $r_d$, the relative magnitude of the 
interfering
tree diagrams.  When using the flavour-tagged observables alone, 
this $\gamma$ determination carries with it an 8-fold ambiguity, which
compromises the usefulness of the measurement.

A combined  analysis of \dsk\ and \dpi\ under the assumption of 
U-spin symmetry allows the true solution to be isolated with only
a 2-fold ambiguity.  Plausible assumptions about the size of the
strong phase difference or the orientation of the unitarity triangle
allow the remaining bogus solution to be excluded.  This U-spin analysis
has the further benefit of exploiting the $\mathrm{B_d}$ data without the
need to know $r_d$.

The intrinsic precision of the combined analysis is competitive
with other approaches.  For example,  in the example scenario considered,
a $5^\circ$ uncertainty is achievable after five years of operation.
The combined analysis does not make use of the untagged observables available
in  \dsk,  which provide additional information which will improve
the precision still further.    

The systematic error associated with the assumption of U-spin symmetry can
be transparently assessed through studying the contours associated
with the measurements.  In some cases these offer a very robust $\gamma$
extraction.  Furthermore,  a variety of separate $\gamma$ determinations
may be performed,  each with different U-spin symmetry assumptions.
Comparison between the results will help in assigning the systematic error.

Finally,  an analogous exercise can be performed from considering the 
U-spin related channels \dpistar\ and \dskstar.  The reconstruction
of the latter channel at LHCb is under investigation.

\section{\boldmath  
  Summary \& (near) Outlook for $\beta$ \& $\gamma$} 
\vspace{+2mm}
\begin{flushright}
  {\it Contribution from A.~Soni}
\end{flushright}

While the direct measurement of $\beta$ are now in  a ``matured"
stage since the post-$B$-factory era of first  
determinations in ~2001-2, past year has seen significant progress in
extracting $\gamma$. Of course, the errors are still rather large but
we expect reduction of these to take place quite rapidly in the
near future.  
Table~\ref{summ} summarizes the current status and the prospects
for the next few years for $\beta$ and $\gamma$.
\begin{table}[htbp]
\caption{Summary of status and prospects for $\beta$ \& $\gamma$;
luminosity per experiment is given in units of $10^8 B \bar B$ pairs.
Note current central values, $\beta = 0.725$ \& $\gamma = 69^{\circ}$.
IEE is irreducible experimental error and ITE is irreducible 
theory error \label{summ}}  
\begin{center}
\begin{tabular}{|c|c|c|c|c|c|}
\hline
Qty & ~CKM05 & ~SUM06 & ~End08 & IEE05 & ITE \\
\hline
Lumin/(expt) & ~2 & ~5 & ~10 & & \\
$\delta (\sin2 \beta)$ & 0.037 & ~0.028 & ~0.02 & 0.02 & ~0.001\\ 
$\delta \gamma^{\circ}$ & $ \pm13 \pm 9 \pm 13$ & $\pm 9 \pm 6 \pm 7$ & 
~5 to 2 & ? &
.05 \\ 
\hline
\end{tabular}
\end{center}
\end{table}

On $\beta$ the current error of ~0.037\cite{hfag}
should decrease
to about 0.02 in about 3 years but after that the
currently estimated
irreducible experimental error (IEE), which
both BABAR \& BELLE agree 
to be around 0.02 (see talks by Browder and Lange),
will start to become the limiting factor.
Since the irreducible theory error (ITE) on $\beta$ is estimated
to be  $\approx$ 0.01\cite{Boos:2004xp}(and talk by Mannel) it is
important that effort is put into reducing the IEE  
so that it is comparable to ITE. 

Regarding $\gamma$, so far the  $B \to D K$ with Dalitz analysis of the  
multi-body mode $D^0 \to K_s \pi^+ \pi^-$ proposed by~\cite{ggsz} 
and independently in~\cite{anton} has been very successful. 
Using about $3 \times 10^8$ B-pairs
BELLE gets $\gamma = 68 \pm 15 \pm 13 \pm 11^{\circ}$ (see  talk by Abe),
where the first error is statistical, the second is experimental
systematics and the third is modelling error of the Dalitz
study. BABAR's combined use of GLW\cite{GronauWyler,GronauLondon}, 
ADS~\cite{ads2} and this  multi-body mode in about $2 \times 10^8$ $B$-pairs 
leads to $\gamma = 70 \pm 31 \pm 12 \pm 13^{\circ}$ (see talk by Schune).
Combining these we arrive at the current error on
$\gamma$ in the Table \ref{summ} of about $\pm 13 \pm 9 \pm 13$.
We think it is a bit too premature to add the modelling
and the systematic errors
in quadrature to the statistical error; especially the modelling
error ought to be left separate for now.  
Prospects for appreciably reducing the error on
$\gamma$ in the next three  years or so seems to be excellent;
there
are many reasons for this optimism.  
In fact is well known there are many common modes
CPES(GLW), single-Cabibbo suppressed~\cite{Grossman:2002aq} (CPNES \& CPES)
as well as double
Cabibbo suppressed ones~\cite{ads2} (CPNES \& CPES). 
With increasing luminosity from
about $10^8$ to about $10^9$ B-pairs as more and more of these
channels get added to the analysis, it is anticipated that 
the statistical error on $\gamma$ will reduce faster than
$1/\sqrt{N}$. 
This is the basis for guess-estimating the error in
2006 to get around $\pm 9 \pm 6 \pm 7$. 
As we enter the era of $1 \ \rm{ab}^{-1}$/expt around 2008 then we expect
the error on $\gamma$ to be around 5 to 2 degrees. 
Input from charm factories can help a lot in reducing the error
on $\gamma$ (see talks by Asner, Atwood and Petrov). 
The optimistic error of $2^{\circ}$ may well be plausible by 2008 if sufficient
information from charm factory is forthcoming. 

Another important way to
get these angles is, of course,  by studying time-dependent (i.e. MIXCP) CP
violation via $B^0 \to D^{0(*)}$``$K^0$''. 
Once again, all the common decay modes of $D^0$ and $\bar D^0$
can be used just as in the case of direct CP studies involving $B^{\pm}$
decays.  Therefore, needless to say  
input from charm factory also becomes desirable for MIXCP
studies of $B^0 \to D^{0(*)}$``$K^0$''. 
It is important to stress that
this method give not only the combinations of the angles 
(2$\beta +\gamma \equiv \alpha - \beta + \pi$) but also
in addition this is another way to get $\beta$ 
cleanly~\cite{Kayser:1999bu,fleischer1,bgk}. 
In fact whether one uses $B^{\pm}$ with DIRCP or $B^0- \bar B^0$ with
TDCP these methods are very clean with (as indicated above)
the ITE of $\approx 0.01$. However, the TDCP studies
for getting $\gamma$ (with the use of $\beta$ 
as determined from $\psi K_s$ ) is less efficient
than with the use of DIRCP involving $B^{\pm}$. However,
once we go to luminosities $\geq 1 \ \rm{ab}^{-1}$ then 
the two methods for $\gamma$ should become competitive. 
This method for getting $\beta$ is significantly less efficient than from
the $\psi K_s$ studies~\cite{Kayser:1999bu}.


\begin{thebibliography}{99}

\bibitem{Boos:2004xp}
  H.~Boos, T.~Mannel and J.~Reuter,
  Phys.\ Rev.\ D {\bf 70}, 036006 (2004).

\bibitem{BigiSanda}
  I.~I.~Y.~Bigi and A.~I.~Sanda,
  Nucl.\ Phys.\ B {\bf 193}, 85 (1981).

\bibitem{oldstuff}
  J.~S.~Hagelin, 
  Nucl.\ Phys.\ B {\bf 193}, 123 (1981).

\bibitem{Buras_und_co}
  A.~J.~Buras, W.~Slominski, H.~Steger,
  Nucl.\ Phys.\ B {\bf 245}, 369 (1984);
  Nucl.\ Phys.\ B {\bf 238}, 529 (1984).

\bibitem{NiersteHerrlich}
  S.~Herrlich and U.~Nierste,
  Nucl.\ Phys.\ B {\bf 476}, 27 (1996).

\bibitem{BSS}
  M.~Bander, D.~Silverman, A.~Soni, 
  Phys. \ Rev. \ Lett. {\bf 43}, 242 (1979).

\bibitem{pdg_review}
  For a review of $CP$ violation phenomenology, see 
  D.~Kirkby and Y.~Nir in
  S.~Eidelman {\it et al.}, 
  Phys. Lett B {\bf 592}, 1 (2004).

\bibitem{sin2phi1}
  B.~Aubert {\it et al.} (BABAR Collaboration),
  Phys. Rev. Lett. {\bf 94}, 161803 (2005);
  K.~Abe {\it et al.} (Belle Collaboration),
  Phys. Rev. D {\bf 71}, 072003 (2005);
  Erratum-ibid. D {\bf 71}, 079903 (2005).
  
\bibitem{phi_ambig}
  Ya.~Azimov {\it et al.},
  JETP Lett. {\bf 50}, 447 (1989),
  Phys. Rev. D {\bf 42}, 3705 (1990),
  Z Phys. A {\bf 356}, 437 (1997);
  B.~Kayser, NSF-PT-97-2, hep-ph/9709382;
  H.R.~Quinn, T.~Schietinger, J.P.~Silva, A.E.~Snyder,
  Phys. Rev. Lett. {\bf 85}, 5284 (2000);
  Yu.~Grossman, H.~Quinn, 
  Phys. Rev. D {\bf 56}, 7259 (1997);
  J.~Charles, A.~Le~Yaouanc, L.~Oliver, O.~P\`ene, J.-C.~Raynal,
  Phys. Lett. B {\bf 425}, 375 (1998); Erratum-ibid. B {\bf 433}, 441 (1998);
  T.E.~Browder, A.~Datta, P.J.~O'Donnell, S.~Pakvasa,
  Phys. Rev. D {\bf 61}, 054009 (2000);
  J.~Charles, A.~Le~Yaouanc, L.~Oliver, O.~P\`ene, J.-C.~Raynal,
  Phys. Rev. D {\bf 58}, 114021 (1998).
  See also~\cite{babar_psikstar} and references therein.

\bibitem{babar_psikstar}
  B.~Aubert {\it et al.} (BABAR  Collaboration),
  Phys. Rev. D {\bf 71}, 032005 (2005).

\bibitem{bgk}
  A.~Bondar, T.~Gershon and P.~Krokovny,
  Phys. Lett. B {\bf 624}, 1 (2005).
  Similar methods were proposed in 
  D.~Atwood and A.~Soni,
  Phys.\ Rev.\ D {\bf 68}, 033009 (2003) 
  and 
  M.~Gronau, Y.~Grossman, N.~Shuhmaher, A.~Soffer and J.~Zupan,
  Phys.\ Rev.\ D {\bf 69}, 113003 (2004).

\bibitem{grossman_worah}
  Yu.~Grossman, M.P.~Worah,
  Phys. Lett. B {\bf 395} 241 (1997).

\bibitem{rd}
  D.A.~Suprun, C.-W.~Chiang and J.L.~Rosner,
  Phys. Rev. D {\bf 65}, 054025 (2002).

\bibitem{fleischer1}
  R.~Fleischer,
  Phys. Lett. B {\bf 562}, 234 (2003).

\bibitem{fleischer2}
  R.~Fleischer,
  Nucl. Phys. B {\bf 659}, 321 (2003).
     
\bibitem{anton}
  A.~Poluektov {\it et al.} (Belle Collaboration),
  Phys. Rev. D {\bf 70}, 072003 (2004).

\bibitem{ggsz}
  A.~Giri, Y.~Grossman, A.~Soffer and J.~Zupan,
  Phys. Rev. D {\bf 68}, 054018 (2003).

\bibitem{dqstl}
  I.~Dunietz {\it et al.}, Phys. Rev. D {\bf 43}, 2193 (1991).

\bibitem{belle_d0h0}
  K.~Abe {\it et al.} (Belle Collaboration),
  BELLE-CONF-0546, hep-ex/0507065.
  
\bibitem{LASS} 
  E-135 Collaboration (LASS Collaboration), 
  D. Aston {\it et al.}, Nucl. Phys. {\bf B296}, 493 (1988).

\bibitem{cosBELLE}   
  R.~Itoh {\it et al.}  (Belle Collaboration),
  Phys.\ Rev.\ Lett.\  {\bf 95}, 091601 (2005).

\bibitem{cosBABAR} 
  B. Aubert {\it et al.} (BABAR Collaboration),
  Phys. Rev. D {\bf 71}, 032005 (2005).

\bibitem{dstdexpect}
  Z.Z.~Xing, Phys. Rev. D{\bf 61}, 14010 (2000);
  X-Y.~Pham and Z.Z.~Xing, Phys. Lett. B {\bf 458}, 375 (1999).

\bibitem{DL}
  A.~Datta and D.~London, \plb {584}, 81 (2004).
\bibitem{ADL}
  J.~Albert, A.~Datta, and D.~London, \plb {605}, 335 (2005).

\bibitem{belle_psipi0_pub} 
  S.~U.~Kataoka {\it et al.} (Belle Collaboration),
  Phys. Rev. Lett. {\bf 93}, 261801 (2004).
\bibitem{babar_psipi0_pub} 
  B.~Aubert {\it et al.} (BABAR Collaboration),
  Phys. Rev. Lett. {\bf 91}, 061802 (2003).
\bibitem{belle_dspdsm_pub} 
  H.~Miyake {\it et al.} (Belle Collaboration),
  Phys. Lett. B. {\bf 91}, 34 (2005). 
\bibitem{belle_dsd_pub} 
  T.~Aushev {\it et al.} (Belle Collaboration),
  Phys. Rev. Lett. {\bf 93}, 201802 (2004).
\bibitem{belle_psipipi_pre} 
  K.~Abe {\it et al.} (Belle Collaboration),
  BELLE-CONF-0441, hep-ex/0408107.
\bibitem{babar_psipipi_pub} 
  B.~Aubert {\it et al.} (BABAR Collaboration),
  Phys. Rev. Lett. {\bf 90}, 091801 (2003). 
\bibitem{belle_dpdm_pub} 
  G.~Majumder {\it et al.} (Belle Collaboration),
  Phys. Rev. Lett. {\bf 95}, 041803 (2005).

\bibitem{Aubert:2001tu}
  B.~Aubert {\it et al.}  (BABAR Collaboration),
  Nucl.\ Instrum.\ Meth.\ A {\bf 479}, 1 (2002).

\bibitem{newresults}
  B.~Aubert {\it et al.}  (BABAR Collaboration),
  Phys.\ Rev.\ Lett.\  {\bf 95}, 131802 (2005)
  
\bibitem{YB}
  D.~Acosta {\it et al.},  Phys. Rev. D {\bf 71}, 032001 (2005).

\bibitem{RF_dsds}
  R.~Fleischer, Eur.\ Phys.\ J.\ C {\bf 10}, 299 (1999).

\bibitem{lattice}
  J. Simone, invited talk at {\it Lattice 2004}, Fermilab, June 2004:\\
  http://lqcd.fnal.gov/lattice04/presentations/paper252.pdf

\bibitem{ref:sin2bg_th}
  R.G.~Sachs, Enrico Fermi Institute Report,
  EFI-85-22 (1985) (unpublished);
  I.~Dunietz and R.G.~Sachs, Phys. Rev. D {\bf 37}, 3186 (1988)
  [E: Phys. Rev. D {\bf 39}, 3515 (1989)];
  I.~Dunietz, Phys. Lett. B {\bf 427}, 179 (1998);
  P.F.~Harrison and H.R.~Quinn, ed., SLAC-R-504 (1998), Chap. 7.6.
  
\bibitem{belle_dsp_full} 
  T.~Sarangi, K.~Abe {\it et al.} (Belle Collaboration),
  Phys. Rev. Lett. {\bf 93}, 031802 (2004).

\bibitem{belle_dsp_partial}
  T.~Gershon {\it et al.} (Belle Collaboration),
  Phys. Lett. B 624, {\bf 11} (2005).

\bibitem{ref:run1-2-breco}
  B.~Aubert {\it et al.} (BABAR Collaboration),
  \jprl{92}, 251801 (2004).

\bibitem{ref:run1-2-ihbd}
  B.~Aubert {\it et al.} (BABAR Collaboration), 
  \jprl{92}, 251802 (2004).

\bibitem{ref:abc} O. Long,
  M. Baak, R.N. Cahn, and D. Kirkby,
  Phys. Rev. D {\bf 68}, 034010 (2003).

\bibitem{ref:th-dsta0_a2}
  M.~Diehl, G.~Hiller, \plb{517}, 125 (2001).

\bibitem{ref:br-dsta0_a2}
  M.~Diehl, G.~Hiller, JHEP {\bf 0106}, 067 (2001).
  
\bibitem{GronauLondon}
  M.~Gronau and D.~London, \plb{253}, 483 (1991).

\bibitem{ref:th-dst0k0}
  B.~Kayser and D.~London, \jprd{61}, 116013 (2000); 
  A.I.~Sanda, DPNU-01-19, hep-ph/0108031.  See also~\cite{ads2}.

\bibitem{ref:dst0kst0}
  B.~Aubert {\it et al.}  (BABAR Collaboration),
  BABAR-CONF-04/006, hep-ex/0408052.

\bibitem{Aleksan:2002mh}
  R.~Aleksan, T.~C.~Petersen and A.~Soffer,
  Phys.\ Rev.\ D {\bf 67}, 096002 (2003).

\bibitem{Aleksan:2003fm}
  R.~Aleksan and T.~C.~Petersen,
  eConf {\bf C0304052}, WG414 (2003), hep-ph/0307371.

\bibitem{Aubert:2004at}
  B.~Aubert {\it et al.}  (BABAR Collaboration),
  Phys.\ Rev.\ Lett.\  {\bf 95}, 171802 (2005).

\bibitem{dspi0}
  I. Dunietz, Phys. Lett. B {\bf 427}, 179 (1998).

\bibitem{utfit} 
  http://www.utfit.org
\bibitem{freq}
  B.~Aubert {\it et al.} (BABAR Collaboration),
  Phys.Rev. D {\bf 71}, 112003 (2005).

\bibitem{rahatlou_bad}
  Sh.~Rahatlou, private communication.
  
\bibitem{path} 
  D.~Atwood and A.~Soni,
  Phys.\ Rev.\ D {\bf 71}, 013007 (2005).

\bibitem{ads2} 
  D. Atwood, I. Dunietz and A. Soni, 
  Phys. Rev. Lett. {\bf 78}, 3257 (1997); 
  Phys. Rev. D {\bf 63}, 036005(2001).   

\bibitem{GronauWyler} 
  M. Gronau and D. Wyler, Phys. Lett. B {\bf 265}, 172 (1991).

\bibitem{sbf12}
  S. Hashimoto {\it et al}, KEK-REPORT-2004-4, hep-ex/0406071;
  J. Hewett {\it et al}, SLAC-R-709, hep-ph/0503261. 

\bibitem{bg}
  A.~Bondar and T.~Gershon, Phys. Rev. D {\bf 70}, 091503(R) (2004).

\bibitem{babar_glw_btdk1} 
  B.~Aubert {\it et al.}  (BABAR Collaboration),
  Phys. Rev. Lett. {\bf 92}, 202002 (2004).
\bibitem{babar_glw_btdk2} 
  B.~Aubert {\it et al.}  (BABAR Collaboration),
  SLAC-PUB-10655, hep-ex/0408082.
\bibitem{babar_glw_btdstark} 
  B.~Aubert {\it et al.}  (BABAR Collaboration),
  Phys. Rev. D {\bf 71}, 031102 (2005).
\bibitem{babar_glw_btdkstar} 
  B.~Aubert {\it et al.}  (BABAR Collaboration),
  SLAC-PUB-10639, hep-ex/0408069.
\bibitem{babar_ads_btdk} 
  B.~Aubert {\it et al.}  (BABAR Collaboration),
  BABAR-CONF-04/13, hep-ex/0408028.

\bibitem{ref:pdg2004}
  Particle Data Group, S. Eidelman {\it et al.}, Phys. Lett. {\bf B592}, 1 (2004).

\bibitem{Grossman:2002aq}
  Y.~Grossman, Z.~Ligeti and A.~Soffer,
  Phys.\ Rev.\ D {\bf 67}, 071301 (2003).
  M.~Gronau,
  Phys.\ Lett.\ B {\bf 557}, 198 (2003).

\bibitem{Sinha:2004ct}
  N.~Sinha,
  Phys.\ Rev.\ D {\bf 70}, 097501 (2004)

\bibitem{Kayser:1999bu}
  B.~Kayser and D.~London,
  Phys.\ Rev.\ D {\bf 61}, 116013 (2000).
  D.~Atwood and A.~Soni,
  Phys.\ Rev.\ D {\bf 68}, 033009 (2003).

\bibitem{Aleksan:1991nh}
  R.~Aleksan, I.~Dunietz and B.~Kayser,
  Z.\ Phys.\ C {\bf 54}, 653 (1992).

\bibitem{Gronau:1998vg}
  M.~Gronau,
  Phys.\ Rev.\ D {\bf 58}, 037301 (1998).

\bibitem{Atwood:2003mj}
  D.~Atwood and A.~Soni,
  Phys.\ Rev.\ D {\bf 68}, 033003 (2003).
  
\bibitem{Aubert:2004kv}
  B.~Aubert {\it et al.}  (BABAR Collaboration),
  BABAR-CONF-04/043, hep-ex/0408088.
  
\bibitem{Abe:2004gu}
  K.~Abe {\it et al.} (Belle Collaboration),
  BELLE-CONF-0476, hep-ex/0411049.
  
\bibitem{Abe:2005ct}
  K.~Abe {\it et al.} (Belle Collaboration),
  BELLE-CONF-0502, hep-ex/0504013.
  
\bibitem{ref:cleomodel}
  S. Kopp {\it et al.} (CLEO Collaboration),
  Phys. Rev. D {\bf 63}, 092001 (2001);
  H. Muramatsu {\it et al.} (CLEO Collaboration),  
  Phys. Rev. Lett.  {\bf 89}, 251802 (2002);
  Erratum-ibid: {\bf 90} 059901 (2003).

\bibitem{Gronau:2004gt}
  M.~Gronau, Y.~Grossman, N.~Shuhmaher, A.~Soffer and J.~Zupan,
  Phys.\ Rev.\ D {\bf 69}, 113003 (2004).

\bibitem{work}
  Y. Grossman, A. Soffer, J. Zupan, work in progress.

\bibitem{Intro}
  D.~Atwood, these proceedings; 
  Y.~Grossmann, these proceedings.

\bibitem{Falk:2000ga}
  A.~F.~Falk and A.~A.~Petrov,
  Phys.\ Rev.\ Lett.\  {\bf 85}, 252 (2000);
  A.~F.~Falk,
  Phys.\ Rev.\ D {\bf 64}, 093011 (2001).

\bibitem{Silva:1999bd}
  J.~P.~Silva and A.~Soffer,
  Phys.\ Rev.\ D {\bf 61}, 112001 (2000).

\bibitem{Falk:2001hx}
  A.~F.~Falk, Y.~Grossman, Z.~Ligeti and A.~A.~Petrov,
  Phys.\ Rev.\ D {\bf 65}, 054034 (2002).

\bibitem{Bergmann:2000id}
  S.~Bergmann, Y.~Grossman, Z.~Ligeti, Y.~Nir, A.A.~Petrov,
  Phys.\ Lett.\ B {\bf 486}, 418 (2000).

\bibitem{Falk:1999ts}
  A.~F.~Falk, Y.~Nir and A.~A.~Petrov,
  JHEP {\bf 9912}, 019 (1999).

\bibitem{GGR}
  M.~Gronau, Y.~Grossman and J.~L.~Rosner,
  Phys.\ Lett.\ B {\bf 508}, 37 (2001).

\bibitem{GP}
  E.~Golowich and S.~Pakvasa,
  Phys.\ Lett.\ B {\bf 505}, 94 (2001).

\bibitem{AtwoodPetrov}
  D.~Atwood and A.~A.~Petrov,
  Phys.\ Rev.\ D {\bf 71}, 054032 (2005).

\bibitem{Nelson:1999fg}
  A.~A.~Petrov,
  eConf {\bf C030603}, MEC05 (2003), hep-ph/0311371;
  H.~N.~Nelson,
  in 
  {\it Proc. of the 19th Intl. Symp. on Photon and Lepton Interactions at High Energy LP99 } 
  ed. J.A. Jaros and M.E. Peskin, UCSB HEP 99-08, hep-ex/9908021; 
  see also
  A.~Datta, D.~Kumbhakar,
  Z.\ Phys.\ C {\bf 27}, 515 (1985);
  A.~A.~Petrov,
  Phys.\ Rev.\ D{\bf 56}, 1685 (1997);
  E.~Golowich and A.~A.~Petrov,
  Phys.\ Lett.\ B {\bf 427}, 172 (1998).

\bibitem{Inclusive}
  H.~Georgi, Phys. Lett. B297, 353 (1992);
  T.~Ohl, G.~Ricciardi and E.~Simmons, Nucl. Phys. B403, 605 (1993);
  I.~Bigi and N.~Uraltsev,
  Nucl.\ Phys.\ B {\bf 592}, 92 (2001).

\bibitem{Exclusive}
  J. Donoghue, E. Golowich, B. Holstein and J. Trampetic,
  Phys. Rev. D33, 179 (1986);
  L. Wolfenstein, Phys.\ Lett.\ B164, 170 (1985);
  P. Colangelo, G. Nardulli and N. Paver,  Phys.\ Lett.\ B242, 71 (1990);
  T.A. Kaeding,  Phys. Lett. B357, 151 (1995).

\bibitem{Rosner:2003yk}
  J.~L.~Rosner and D.~A.~Suprun,
  Phys.\ Rev.\ D {\bf 68}, 054010 (2003).

\bibitem{Georgi}
  H.~Georgi,
  Nucl.\ Phys.\ B {\bf 331}, 311 (1990);
  H.~Georgi and F.~Uchiyama,
  Phys.\ Lett.\ B {\bf 238}, 395 (1990).

\bibitem{bib:cleoc}
  R.A.~Briere {\it et al.}, CLNS-01-1742 (2001).

\bibitem{bib:asner}
  D.~Asner, in Review of Particle Physics, 
  Phys. Lett. B {\bf 592}, 1 (2004).

\bibitem{bib:Nelson}
  A.~A.~Petrov, \emph{Proceedings of Flavor Physics and $CP$ Violation} (2003).
  
\bibitem{asnersun} 
  D.~Asner and W.~Sun,
  Phys. Rev. D {\bf 73}, 034024 (2006).

\bibitem{bib:asner2}
  H.~Muramatsu {\it et al.} (CLEO Collaboration), 
  Phys. Rev. Lett. {\bf 89}, 251802 (2002).
  [Erratum-ibid.\  {\bf 90}, 059901 (2003).]

\bibitem{markiii-1}  
  R.M.~Baltrusaitis {\it et al.} (MARK III Collaboration),
  Phys.~Rev.~Lett.~{\bf 56}, 2140 (1986).

\bibitem{markiii-2} 
  J.~Adler {\it et al.} (MARK III Collaboration), 
  Phys.~Rev.~Lett.~{\bf 60}, 89 (1988).

\bibitem{cleo-c}
  Q.~He {\it et al.}  (CLEO Collaboration),
  Phys. Rev. Lett. {\bf 95}, 121801 (2005).

\bibitem{babargamma}
  B.~Aubert {\it et al.}  (BABAR Collaboration),
  Phys. Rev. Lett. {\bf 95}, 121802 (2005).

\bibitem{sarantsev}  
  V.V. Anisovich and A.V. Sarantsev, Eur. Phys. A {\bf 16}, 229 (2003).

\bibitem{focus3pi}
  J.~M.~Link {\it et al.}  (FOCUS Collaboration),
  Phys.\ Lett.\ B {\bf 585}, 200 (2004).

\bibitem{cleopipipi0}
  D.~Cronin-Hennessy {\it et al.}  (CLEO Collaboration),
  Phys. Rev. D {\bf 72}, 031102 (2005).

\bibitem{aumope} 
  K. L. Au, D. Morgan, and M. R. Pennington,
  Phys. Rev. D {\bf 35}, 1633 (1987).

\bibitem{bigi}
  I.~I.~Y.~Bigi and A.~I.~Sanda,
  Phys.\ Lett.\ B {\bf 211}, 213 (1988).

\bibitem{CDF-Bc}
  F. Abe {\it et al.}  (CDF Collaboration),
  Phys.\ Rev.\ Lett.~{\bf 81}, 2432 (1998).

\bibitem{D0-Bc}
  D0Note 4539-CONF (August 2004).

\bibitem{CDF-Bc-new}
  CDF Note 7438 (February 2005).

\bibitem{masetti}
  M. Masetti, Phys.\ Lett.\ B {\bf 286}, 160 (1992).

\bibitem{fw}
  R. Fleischer and D. Wyler,
  Phys.\ Rev.\ D~{\bf 62}, 057503 (2000).

\bibitem{RF-Phys-Rep}
  R. Fleischer,
  Phys.\ Rep.~{\bf 370}, 531 (2002).

\bibitem{D-NP}
  C.C. Meca and J.P. Silva,
  Phys.\ Rev.\ Lett.~{\bf 81}, 1377 (1998);
  J.P. Silva and A. Soffer,
  Phys.\ Rev.\ D~{\bf 61}, 112001 (2000).

\bibitem{IKP}
  M.A. Ivanov, J.G. K\"orner and O.N. Pakhomova,
  Phys.\ Lett.\ B~{\bf 555}, 189 (2003).

\bibitem{IKS}
  M.A. Ivanov, J.G. K\"orner and P. Santorelli,
  Phys. Rev. D {\bf 71}, 094006 (2005).
  
\bibitem{LHCWKSHP}
  For an overview of these channels' potential at the LHC see
  P. Ball {\it et al.}, CERN-TH-2000-101, hep-ph/0003238, in CERN report on
  {\it Standard Model Physics (and more) at the LHC}, CERN, Geneva, 2000.

\bibitem{RF}
  R. Fleischer, Nucl. Phys. B {\bf 671}, 459 (2003).

\bibitem{LHCBLITE}
  The LHCb Collaboration, 
  {\it LHCb Reoptimized Detector Design and Performance},
  CERN-LHCC-2003-030.

\bibitem{JONAS}
  J. Rademacker, 
  {\it Measuring the CKM Angle $\gamma$ with \dpistar},
  CERN-LHCb-2001-153.

\bibitem{RD}
  For an overview of strategies, see Cecilia Voena's talk at this conference.



\bibitem{hfag} 
  Heavy Flavor Averaging Group, \\
  hep-ex/0412073 and http://www.slac.stanford.edu/xorg/hfag.

\end{thebibliography}
\end{document}